 \documentclass[preprint2]{aastex}
\pdfoutput=1
\usepackage{graphicx}
\usepackage{epstopdf}

\usepackage{natbib}
\newcommand{\coone}{$\rm ^{12}CO(1-0)$}

\newcommand{\hh}{$\rm H_2$}
\newcommand{\nhh}{$\rm n(H_2)$}
\newcommand{\Nhh}{$\rm N(H_2)$}

\newcommand{\kkms}{\>{\rm K}\,{\rm km}\,{\rm s}^{-1}}
\newcommand{\kms}{\>{\rm km}\,{\rm s}^{-1}}

\newcommand{\mum}{\>{\mu {\rm m}}}

\newcommand{\msun}{\>{\rm M_{\odot}}}

\newcommand{\msunpc}{\>{\rm M_{\odot}\,pc^{-2}}}
\newcommand{\msunyr}{\>{\rm M_{\odot}\,yr^{-1}}}
\newcommand{\dg}{^{\circ}}

\newcommand{\as}{^{\prime\prime}}
\newcommand{\am}{^{\prime}}

\newcommand{\av}{\rm A_V}
\newcommand{\htwo}{\rm H_2}

\newcommand{\ha}{$\rm H\alpha$}

\newcommand{\shtwo}{\rm \Sigma_{M_{H_2}}}

\def\comm#1   {{\tt (COMMENT: #1) }}

\slugcomment{First draft}

\shorttitle{CO and other tracers in M\,51}
\shortauthors{Schinnerer et al.}

\begin{document}


\title{The PdBI Arcsecond Whirlpool Survey (PAWS). I. \\
A Cloud-Scale/Multi-Wavelength View of the Interstellar Medium in a Grand-Design Spiral Galaxy}


\author{
Eva Schinnerer\altaffilmark{1},
Sharon E. Meidt\altaffilmark{1},
J\'er\^ome Pety\altaffilmark{2,3},
Annie Hughes\altaffilmark{1},
Dario Colombo\altaffilmark{1},
Santiago Garc\'{i}a-Burillo\altaffilmark{4},
Karl F. Schuster\altaffilmark{2},
Ga\"{e}lle Dumas\altaffilmark{2},
Clare L. Dobbs\altaffilmark{5},
Adam K. Leroy\altaffilmark{6},
Carsten Kramer\altaffilmark{9},
Todd A. Thompson\altaffilmark{7,8},
and
Michael W. Regan\altaffilmark{10}
}

\altaffiltext{1}{MPI for Astronomy, K\"onigstuhl 17, 69117 Heidelberg, Germany}
\altaffiltext{2}{Institut de Radioastronomie Millim\'etrique, 300 Rue de la Piscine, F-38406 Saint Martin d'H\`eres, France}

\altaffiltext{3}{Observatoire de Paris, 61 Avenue de l'Observatoire, F-75014 Paris, France.}
\altaffiltext{4}{Observatorio Astron\'{o}mico Nacional - OAN, Observatorio de Madrid Alfonso XII, 3, 28014 - Madrid, Spain}
\altaffiltext{5}{School of Physics and Astronomy, University of Exeter, Stocker Road, Exeter EX4 4QL, UK}
\altaffiltext{6}{National Radio Astronomy Observatory, 520 Edgemont Road, Charlottesville, VA 22903, USA}
\altaffiltext{7}{Department of Astronomy, The Ohio State University, 140 W. 18th Ave., Columbus, OH 43210, USA} 
\altaffiltext{8}{Center for Cosmology and AstroParticle Physics, The Ohio State University, 191 W. Woodruff Ave., Columbus, OH 43210, USA}
\altaffiltext{9}{Instituto Radioastronom\'{i}a Milim\'{e}trica, Av. Divina Pastora 7, Nucleo Central, 18012 Granada, Spain}
\altaffiltext{10}{Space Telescope Science Institute, 3700 San Martin Drive, Baltimore, MD 21218, USA}



\begin{abstract}
The PdBI (Plateau de Bure Interferometer) Arcsecond Whirlpool Survey (PAWS) has mapped the molecular
gas in the central $\rm \sim 9\,kpc$ of M\,51 in its $^{12}$CO(1-0) line emission
at cloud-scale resolution of $\sim$40\,pc using both IRAM telescopes. We utilize this 
dataset to quantitatively characterize the relation of molecular gas (or CO emission) to other
tracers of the interstellar medium (ISM), star formation and stellar
populations of varying ages. Using 2-dimensional maps, a polar cross-correlation technique and pixel-by-pixel diagrams, we find: 
(a) that (as expected) the distribution of the molecular gas can be linked to different components of the gravitational potential,
(b) evidence for a physical link between CO line emission and radio continuum that seems not to be 
caused by massive stars, but rather depend on the gas density, 
(\,c) a close spatial relation between the PAH and molecular gas emission, but no predictive power of PAH emission for the molecular gas mass,
(d) that  the I-H color map is an excellent predictor of the distribution (and to a lesser degree the brightness) of CO emission, and 
(e) that the impact of massive (UV-intense) young star-forming regions on the bulk of the molecular gas in central $\sim$9\,kpc can not be significant due to a complex 
spatial relation between molecular gas and star-forming regions that ranges from co-spatial to spatially offset 
to absent. The last point, in particular, highlights the importance of galactic environment -- 
and thus the underlying gravitational potential -- for the distribution of molecular gas and star formation.

\end{abstract}



%%%%%%%%%%%%%%%%%%%%%%%%%%%%%%%%
%%%%%%%%  section 1 - introduction

\section{Introduction}

Understanding the distribution and properties of the cold molecular gas is fundamental for 
developing a coherent model for star formation as molecular gas is regarded the fuel out of 
which stars form.
The need for a physical model on the few parsec to kilo-parsec scale has become more urgent 
as theoretical simulations and models can now incorporate a more complete handling of the gas 
properties on these scales \citep[e.g.][]{robertson08,gnedin09,christensen12}.
In particular, Giant Molecular Clouds (GMCs) received tremendous attention as most massive star formation in the Milky Way is associated with them \citep[e.g.][]{mooney88,mead90}. This clearly suggests a very close link between (massive) star formation and molecular gas which has also been demonstrated in recent studies of external galaxies to hold at kpc-resolution \citep{leroy13,bigiel08,bigiel11}. However, at smaller scales it is not exactly clear what is causing or dominating the physical link, small scale physics, like local turbulence within clouds, or large-scale process such as density waves.

Most of the knowledge in this area 
is based on our understanding of molecular line emission observed in
the Milky Way where the role of galactic environment is harder to study than in
nearby galaxies due to the distance ambiguity. In the past few years there has been mounting evidence
that the properties of the molecular gas and GMCs in particular are not necessarily uniform in certain galactic environments 
\citep[e.g.][]{bolatto08,gratier12,rosolowsky03,donovan11,rebolledo12,hirota11,nieten06,espada12,wei12}. However, none of these studies covered a large enough sample of GMCs or environments within a single galaxy to obtain robust statistics and search for changes of the molecular gas or the environment that could quantify the deviations.

As the most abundant molecule \hh\ cannot be directly observed, therefore the line emission 
of the CO molecule serves as the most common substitute. Over the past decades several 
relations between CO line emission and line as well as continuum emission at other wavelengths 
have been observed. While the underlying cause for these relations is often not well understood, 
studies that aim to test on which spatial scales such relations might exist, i.e. globally versus locally, 
bear the potential to unravel the physical processes underlying the relations. Several relations between 
the CO emission and other tracers of the ISM as well as the stellar 
population have been established (mostly) on global scales, however, it is unclear if they persist on smaller 
spatial scales. Prominent examples are the relations between CO emission and (far-)infrared 
luminosity \citep[e.g.][]{young91}, CO emission and radio continuum \citep[e.g.][]{allen92}, and last 
but not least the relation between star formation surface density and neutral gas surface density, i.e. the
Kennicutt-Schmidt relation \citep[e.g.][]{kennicutt89,kennicutt98,bigiel08}.
The exact underlying causes for these relations are not fully understood, and several physical 
processes could lead to such relations. Therefore, observations at high spatial 
resolution are critical to shed light onto the underlying physical processes which might act at different
physical scales and potentially allow for the disentanglement of cause and relation. 

In this paper we combine new observations of the CO(1-0) line emission at 1'' resolution from 
the PdBI Arcsecond Whirlpool Survey (PAWS) with existing archival multi-wavelength data ranging 
from the radio to the UV to study the relation between molecular gas as seen in its CO emission and 
other galactic components (gas, dust and stars). The paper is organized as follows. After an introduction 
of the PAWS survey (\S \ref{sec:paws}), we describe the datasets used and the steps taken to obtain a 
common astrometry in \S \ref{sec:data}. In \S \ref{sec:comp} we present our results based on 2-dimensional 
maps, the polar cross correlation technique and a pixel-by-pixel analysis for the various ISM and stellar 
tracers used. The results are discussed in the context of processes acting and affecting the molecular gas 
on clouds scale (\S \ref{sec:cloud}) and the implications for changes in the galactic environment are presented in 
\S \ref{sec:env}. We summarize and conclude in \S \ref{sec:sum}. In appendix \S \ref{sec:co} we provide a 
short summary of the relevant molecular gas cooling and heating processes 
in relation to observations of the CO emission line. A description of the polar cross correlation technique is 
given in Appendix \S \ref{sec:polar_cross}.

This paper is the first of a series describing results from the PAWS survey. The PAWS data products will 
be publicly available from a dedicated web-page\footnote{{\tt http://www.mpia.de/PAWS}; see also 
{\tt http://www.iram-institute.org/EN/content-page-240-7-158-240-0-0.html}}.
\cite{pety13} find that about half of the CO emission arises from a diffuse component that is consistent 
with a thick molecular disk. In addition a detailed description of the observations as well as data reduction 
can be found there. Motivated by the results of this paper that shows environmental changes in the relations 
between the CO emission and other tracers, \cite{hughes13a} study the probability distribution functions 
of the molecular gas as a function of environment and find profound difference between center, arm and 
inter-arm regions. A follow-up of this result is the work by \cite{colombo13a} who identify about 1,500 
individual GMCs in the PAWS data and find that several GMC properties show a strong dependence on 
galactic environment, and thus the large-scale dynamics. A comparative study of GMC properties in M51, 
M33 and the LMC by \cite{hughes13b} shows that the GMC scaling relations found are very sensitive to 
methodological biases and that distinct difference in the GMC properties between the 3 galaxies studied 
exist. Finally, \cite{meidt13} provide an explanation for the varying GMC properties found using dynamical 
pressure arguments and propose environmental dependences of their star forming capabilities.

%%%%%%%%%%%%%%%%%%%%%%%%%%%%%%%%
%%%%%%%%  section 2 - PAWS motivation

\section{Motivation for the PdBI Arcsecond Whirlpool Survey (PAWS)}\label{sec:paws}

While most stars in the universe have been born in galaxies with
typical stellar masses of $\rm M_{\star}\,=\,10^{10.6\pm0.4}\,\msun$ \citep[e.g.][]{karim11},
detailed studies of the molecular ISM and GMCs -- the entities out of
which stars form -- have focussed on lower stellar mass systems in the
Local Group. Studies of more massive spiral galaxies often lack
resolution and/or areal coverage to identify statistically significant
numbers of GMCs in different galactic environments
\citep[e.g.][]{bolatto08,gratier12,rosolowsky03,donovan11,rebolledo12,hirota11,nieten06}. Thus
high spatial and spectral imaging of (a significant part of) the
molecular gas reservoir of a typical star forming galaxy has
not yet been undertaken.

We chose M\,51 (``The Whirlpool Galaxy'') as the target of our survey as
it is one of the closest \citep[$D \sim 7.6$~Mpc,][]{ciardullo02},
face-on \citep[$i \sim 22^\circ$,][Colombo et al. in prep.]{tully74},
grand-design spirals and one of the
most-observed galaxies in the sky resulting in an excellent coverage
across the full electromagnetic spectrum. A summary of the basic parameters assumed
for M\,51 is given in Tab. \ref{tab:m51}. Over 1,000 individual star
clusters have been observed and characterized based on HST imaging
\citep[e.g.][]{bastian05,scheepmaker07,scheepmaker09,haas08,hwang08,chandar11}
and deep, high-resolution imaging of the galaxy in the radio continuum
\citep{dumas11}, mid- to far-IR \citep{calzetti05,mentuch12},
P$\alpha$/H$\alpha$ \citep{scoville01}, and UV bands
\citep{calzetti05} trace young stellar populations with a wide range
of age sensitivities \citep[e.g.,][]{calzetti05,murphy11}. In
contrast to Local Group galaxies with resolved GMC populations, the
molecular gas clearly dominates the ISM over the 9\,kpc region PAWS
covers \citep[e.g.,][]{schuster07,leroy08} and the molecular gas is
concentrated into spectacular spiral arms
\citep[e.g.,][]{garcia-burillo93,helfer03,schuster07,koda09}. This offers the
unique opportunity to relate the physical properties of the molecular
gas and its clouds to different galactic environments, especially regarding
the galaxy dynamics and radiation field.

The fundamental goal of this large IRAM program is to resolve the CO
emission from M\,51 on the scales of individual GMCs. This enables a new approach 
to understanding star formation in
galaxies, one centered on studying the physical structures of interest
rather than averaging over large areas. The inferred total molecular
gas mass in NGC5194, the large spiral galaxy, is
$M_{gas}\,=\,6.2\times10^9\,\msun$. About 60\,\% of this gas mass
-- comparable to the total molecular mass of the Milky Way -- resides in the 
PAWS field-of-view (FoV) \citep{pety13}. Together with the existing
multi-wavelength data, PAWS was designed to provide the data to study the evolutionary
sequence for GMCs, place strong constraints on the
processes that drive cloud and star formation in disk galaxies, reveal
the physical underpinnings of one of the most commonly used scaling
relations in extragalactic astronomy 
\citep[the Schmidt-Kennicutt 'law';][]{kennicutt89,kennicutt98}, and add a crucial data point
regarding the lifetime of giant molecular clouds, a lingering unknown
that is critical to understanding the gas-star cycle in galaxies.

These goals resulted in the following considerations for the
observations with the IRAM Plateau de Bure Interferometer (PdBI) and
the 30m single dish telescope:

{\it Sensitivity:} The goal
of identifying individual GMCs in M\,51 determined the sensitivity required. In the Milky Way, most H$_2$ is
found in clouds with masses in the range $\rm 10^5 -- 10^6\,\msun$
\citep[e.g.][]{sanders85,blitz99}. In order to detect GMCs with $\rm
1\times10^5\,\msun$ at 5$\sigma$, a $1\sigma$ point mass sensitivity
$\approx 2\times10^4\,\msun$ is sufficient to recover clouds across
this entire range.

{\it Resolution:} The typical size of a Milky Way GMC is $\rm \sim
40\,pc$ \citep{solomon87} and thus a comparable spatial resolution is
required to identify and distinguish individual clouds. This
resolution also marginally resolves the dust lanes observed in the
optical in M\,51 (width $\sim 1''$). 

{\it Area surveyed:} The survey area was chosen to incorporate three
distinct environmental regions, the center encompassing the bulge plus
the star-bursting ring, the spiral arms and the
(relatively quiescent)  inter-arm region. A substantial area in each regime is
necessary to build up sufficient statistics on the molecular gas and
its GMC populations. 
%Other options (e.g., a radial strip or half the
%galaxy) would probe a wider range of environments at the expense of
%statistics which would severely hamper the major science goals for
%which it is critical to assemble the large enough number statistics,
%such as the largest possible cloud catalog.

These considerations resulted in imaging of the $^{12}$CO(1-0) line
emission from the central $\sim$9\,kpc of M\,51 between August 2009
and June 2010 using both IRAM facilities. The global properties
adopted for M\,51 throughout the PAWS survey are summarized in
Tab. \ref{tab:m51}.

%%%%%%%%%%%%%%%%%%%%
%%%%%% section 2 - data

\section{Data}\label{sec:data}

In addition to our own new CO(1-0) observations, we made use of
archival data that trace different phases of the interstellar medium
and the stellar populations as well as star formation present in the
disk of M\,51a. Below we describe in detail where the datasets used come
from and how they were aligned to a single absolute astrometric frame. A
summary of the multi-$\lambda$ data used here is provided in
Tab. \ref{tab:data}.

\subsection{IRAM PdBI+30m PAWS data}

We obtained a map of the molecular gas distribution in the central 
$\rm 11\times7$\,kpc of M\,51 as part of PAWS (PdBI Arcsecond Whirlpool Survey, PI Schinnerer). 
This IRAM
Large Program observed a 60-pointing mosaic with the Plateau de Bure
interferometer (PdBI) in all configurations and mapped the full galaxy
system with the 30m single dish telescope in the \coone ~line. The
resulting data cube of the combined PdBI and 30m data (i.e. short spacing corrected)
has a resolution
of $1.16\as \times 0.97\as$ (PA 73$\dg$) with a mean rms of 0.4\,K per
5$\, \kms$ wide channel. A detailed description of the data acquisition and reduction
is presented by \cite{pety13}. We also use the moment maps derived
from this PAWS datacube (for details see Colombo et al. in prep.). The conversion from integrated brightness 
temperature per beam into molecular hydrogen (\hh) gas surface density per 0.3$\as$ pixel assumed a 
Galactic conversion factor of $\rm X_{CO}\,=\,2\,\times\,10^{20}\,cm^{-2}K^{-1}km^{-1}s$. 

The PdBI observations consisted of two 30-pointing mosaics that were independently observed between 
August 2009 and April 2010. Each pointing within a mosaic was observed between two calibrator cycles 
ensuring good and similar $uv$ coverage for each pointing as well as the two mosaics. Standard 
observations of quasars were used for the passband and atmospheric calibration which was done using the 
GILDAS software package \citep{pety05}. The 3mm receivers were tuned to 115.090\,GHz corresponding to 
the redshifted \coone ~emission line with a LSR velocity of 471.3\,$\kms$. The 30m single dish was used in 
May 2010 to map a $\sim$60 square arcminute field that contains the entire M51 system. Seven separate 
regions were observed in position-switch on-the-fly mode; four of those cover the central $300\as \times 
300\as$, while the remaining three extend this map to the companion and the outer molecular spiral 
arms. The data was reduced in a standard manner and re-gridded onto a data cube with a pixel size of 4$
\as$. The final PAWS data cube is obtained from a joint deconvolution of the PdBI and 30m datasets. It 
has 120 channels of 5\,$\kms$ width and a pixel size of 0.3\,$\as$/px.

For the analysis presented here we use a 0th moment map that was derived using a combination of the 
dilated mask technique and HI priors \citep[e.g. Appendix B of ][]{pety13}. A detailed description of the moment calculation is presented by Colombo et al. (in prep.). Briefly, a dilated mask is created by identifying emission above a given threshold of $t\sigma$. Starting from these peaks the identified islands of emission are expanded to include fainter emission above $p\sigma$. If the central velocity of 
the islands was more than 30\,$\kms$ away from the central HI velocity (based on the 6'' resolution robust HI cube 
from THINGS, see \S \ref{subsec:hi}) the islands were excluded from the final mask. Due to the high angular resolution the PAWS CO(1-0) 1$\as$ data samples much better the rising part of the rotation curve than the HI data causing a lot of real emission to be dropped. Therefore two masks were necessary 
to capture significant emission in the disk and the center, i.e. we used $(t,p)$ of (4,1) and (10,1.5) for the disk and center, respectively, and had an HI prior for the disk and none for the center.

\subsection{HST archival data}

To compare our view of the molecular gas with M51's appearance in the
optical and near-infrared, we use HST
imaging that covers the PAWS FoV.  The HST ACS Heritage (GO-10452; PI Beckwith) final reduced
and drizzled mosaics \citep{mutchler05} in the F435W (B band), F555W
(V band), F658N (H$\alpha$ narrow band), and F814W (I band) filters
were obtained from the dedicated web-page\footnote{\tt
http://archive.stsci.edu/prepds/m51/}. No further processing was
applied to the combined mosaics. To obtain a stellar continuum
corrected map of the H$\alpha$ (plus NII) line emission we followed
the approach of \cite{gutierrez11} by subtracting a linear combination of F555W
and F814W scaled by 0.0878 from the F658N narrow line filter that contains the
H$\alpha$ line. Note that this leads to an over-subtraction in the
central 30$\as$ which is likely caused by the stellar bulge having a
different spectral energy distribution than the disk.

An HST $I-H$ color map was obtained from the ACS Heritage observations
and dedicated NICMOS H band observations in the F160W filter (GO-10501; PI
Chandar). The NICMOS observations were a set of observations at 18 positions. 
We dithered between two locations at each position. The NICMOS data was reduced 
in a standard manner. Because the NICMOS observations were made using three 
different guide stars, their relative astrometry is not consistent. We first corrected for the 
rotation angle offsets between visits using stars that were common between the ACS image 
and the NICMOS images. We then solved for the positional offset between the NICMOS images 
and the ACS mosaic by cross-correlating unsharp masked versions of the NICMOS images with 
the ACS images. We applied these determined offsets to each NICMOS image and used the 
standard DRIZZLE process to combine the images to form a mosaic. We formed an F814W-F160W 
($I-H$) color map by first smoothing the ACS F814W image to the resolution of the NICMOS F160W 
image and then dividing the F814W image by the F160W image. 

\subsection{Spitzer archival data}

M\,51 was imaged from the near- to far-infrared by all Spitzer
instruments as part of the SINGS Legacy project
\citep{kennicutt03}. The Spitzer IRAC 3.6\,$\mu$m and 4.5\,$\mu$m
images were reprocessed by the S$^4$G project
\citep{sheth10}. \cite{meidt12} applied and developed a separation
method of the stellar and non-stellar components present in these IRAC
images based on the Independent Component Analysis that requires two
adjacent IRAC bands as the only input. We utilized both the map of non-stellar
emission at 3.6\,$\mu$m and the contaminant-free old  stellar light from \cite{meidt12}. 
We used these maps of the old stellar light to subtract the stellar contribution to the 
8\,$\mu$m image from the SINGS DR5\footnote{\tt http://irsa.ipac.caltech.edu/data/SPITZER/SINGS/}.

To trace the recent star formation we use the Spitzer MIPS 24$\mu$m
image from SINGS DR5 processed with the HiRes algorithm
as described in detail by \cite{dumas11} resulting in an increased
angular resolution of $\sim 2.4\,\as$.

\subsection{Herschel archival data}

M\,51 has been imaged by PACS onboard of Herschel as part of a GTO
program \citep[PI Wilson, see also][]{mentuch12}. The archival photometric PACS data at 70 and 160\,$\mu$m 
were reduced to level one using HIPE v6.0, and then used SCANAMORPHOS v12.0 \citep{roussel12} to 
produce the final maps. As HIPE v6.0 was used, we converted the PACS images from flight model (FM), 5 to 
FM, 6 by dividing by the factors listed in the PACS Photometer Point Source Flux Calibration Report v1.0, which 
are 1.119 (70$\mu$m) and 1.174 (160$\mu$m). All images were converted to $\rm MJy\,sr^{-1}$. The mean full width 
at half-maximum (FWHM) of the PACS point response functions are $\rm \sim 5.6\as (70\,\mu m), and \sim 11.4\as 
(160\,\mu m)$. The actual images use pixels sizes of 1$\as$. 

The [CII] emission line has been imaged as well as part of this Herschel GTO program (PI Wilson; Parkin et al. subm.). 
The version used here was reduced using the KINGFISH pipeline for PACS spectroscopy \citep{kennicutt11} 
in HIPE v9.0.3063 using the calibration files in PACS calVersion 41. This is similar to the standard 
un-chopped pipeline, but includes some modifications for improved handling of transients in the data.
In addition to the calibration steps described in detail by \cite{croxall12} the latest pointing calibration 
and the newest drizzle algorithms were used to project these data using HIPE 10.0.2538. The final angular 
resolution is $\sim$11.5$\as$.

\subsection{GALEX archival data}

The FUV and NUV images are from GI program GI3-050 \citep[][and kindly provided to us by PI
Bigiel]{bigiel10}. M\,51 was observed for $\sim$10\,ks and details of the data reduction are provided in 
\cite{bigiel10}. {\it GALEX} simultaneously images in the far-UV (1350--1750$\AA$) and near-UV 
(1750--2800$\AA$) with an angular resolution (FWHM) of 4.0$\as$ and 5.6$\as$, respectively, and 
a field of view of 1.25$^\circ$ \citep{morrissey05}.

\subsection{THINGS HI data}
\label{subsec:hi}

For comparison to the atomic gas we use the robust weighted intensity map from THINGS \citep[The HI 
Nearby Galaxy Survey;][]{walter08}. The intensity map has an angular resolution of $\rm 5.78\as\,\times\,
5.56\as$ (PA -68.0) and fully covers M\,51a.

\subsection{Absolute Astrometry and Alignment}\label{subsec:astrom}

Unlike radio-interferometric images that naturally come with an
absolute astrometry, UV/optical/IR images need to be tied to an
external astrometric frame. Therefore we checked and, if necessary,
corrected their astrometry as detailed below. A summary of our
corrections applied is provided in Tab. \ref{tab:data}.

We used the VLA 20\,cm image of \cite{dumas11} as our absolute
astrometric reference since it encompasses the entire M\,51 galaxy system
with its angular extent of $\rm \sim 10\,\am$. Comparison to the 2MASS J band
image and a montage of the SDSS i band tiles (6 cover the full M\,51
system and are combined using the IRSA Montage
tool\footnote{http://montage.ipac.caltech.edu/}) obtained from the
respective archives, suggests that the agreement of (extragalactic)
point sources within the M\,51 system and around it is good. We find
small non-systematic offsets between the J band image and the i band
montage and therefore decided to use the J band image as our
astrometric reference for the optical/NIR images since it also has the
same geometric projection as the radio data. The final absolute
astrometry in 2MASS is $\le0.1\,\as$ depending on the source
brightness\footnote{See 'Explanatory Supplement to the 2MASS All Sky
Data Release and Extended Mission Products' at {\tt http://
www.ipac.caltech.edu/2mass/releases/allsky/doc/explsup.html}}. Before
deriving the absolute astrometry we changed the projection of all
images to the standard radio interferometric system, i.e. CD matrix
with rotation into CRPIX/CDELT/CRVAL description without rotation,
using the GIPSY task {\tt fitsreproj} \citep{hulst92,vogelaar01}.

As some of the stars present in the images are moving during the period
over which
the different datasets were obtained, we used a combination of
point sources (Galactic stars and stars/unresolved clusters within M\,51) using WCSTools
\citep{mink99}\footnote{\tt http://tdc-www.harvard.edu/wcstools/} and
the overall geometry of NGC\, 5194 and NGC\,5195 to obtain our best
alignment across the different wavelengths. We estimate that our final
astrometry is good to about 0.15\,$\as$ for the high resolution HST
images and slightly larger for the lower resolution IR images. In any
case, we do not expect any profound systematic effects in our analysis
(particular those done at 3.0\,$\as$ resolution) due to this small
error. A detailed discussion of the astrometric accuracy of the PAWS CO(1-0)
data is given by \cite{pety13}.

We verified that the relative astrometry of the HST images was
correct. Comparison to the 2MASS J band image showed that the HST
Heritage I band image (convolved to the same resolution as the J band
image) is offset from the J band image in both RA and Dec. direction
by 0.1$\as$ and -0.4$\as$, respectively. We verified this determined offset by
comparing the convolved HST ACS I band image to the SDSS i band
montage. The same offset was applied to all other HST ACS images as
well as the HST NICMOS H band image that had its relative astrometry
tied to the ACS I band image. Note that \cite{chandar11} find a
similar offset from a comparison to Chandra X-ray imaging.

The absolute astrometry of the IRAC images was found to agree very
well with the 2MASS J band image, thus no correction was applied. We
used a convolved version of the IRAC 8\,$\mu$m image to test the
absolute astrometry of the MIPS 24\,$\mu $m image, and find a small
offset of 0.25$\as$ and 0.5$\as$ in RA and Decl. that we corrected. The
same offset was applied to the HiRes version of the MIPS 24\,$\mu$m
image. Iteratively, we checked the absolute
astrometry of MIPS 70\,$\mu$m and MIPS 160\,$\mu$m images using the
corrected MIPS 24\,$\mu$m and MIPS 70\,$\mu$m images as reference,
respectively. Again, a small offset of 0.375$\as$ in Declination
was found and corrected.

For the Herschel PACS 70\,$\mu$m and 160\,$\mu$m we determined the
offset based on the corrected MIPS 24\,$\mu$m image (plus MIPS
70\,$\mu$m image for the PACS 160\,$\mu$m map) and found small offsets
of -2.5$\as$ and -2.0$\as$ that we corrected. The absolute astrometric uncertainty 
for the [CII] line map is larger ($\sim 1\as$) as the centering was achieved using both
the MIPS 24\,$\mu$m image as well as the SINGS H$\alpha$ image convolved to 3$\as$.
However, obvious common structures can not be easily identified between these maps leading to 
the higher uncertainty.

The astrometry of the GALEX NUV and FUV image was checked against a
corrected and convolved version of the IRAC 8\,$\mu$m as the overall
morphology of these tracers is very similar. In particular, coincident
point sources were utilized to determine an offset of 0.225$\as$ and 0.525$\as$ in 
RA and Declination that was corrected.

%%%%%%%%%%%%%%%%
%%%%% section 4 - results

\section{Comparison of CO(1-0) to ISM and Stellar
Tracers}\label{sec:comp}

The study of potential relationships between molecular gas line emission, namely CO, and 
other tracers of the ISM and stellar populations from various ages at cloud scales can provide new insights 
in the underlying physical processes linking these emissions. In particular, the wide range of galactic 
environments (i.e. bulge/disk, spiral arm/inter-arm) sampled by the PAWS FoV could help to discriminate
between different origins for existing correlations. After a description of the CO emission in the central $\sim$9\,kpc
as seen by PAWS (\S \ref{subsec:co}), we evaluate the correspondence between
CO emission and ISM (\S \ref{subsec:ism}) and stellar (\S \ref{subsec:sf}) tracers in this section. In the following 
section \S \ref{sec:cloud} we discuss potential physical causes for the relations and deviations seen. 
All the findings are placed in the wider context of galactic environment in section \S \ref{sec:env}.

In the following we compare the distribution and intensity of the
CO(1-0) molecular line emission to other prominent ISM tracers
such as the atomic gas emission lines, optical extinction, hot dust and
polycyclic aromatic hydrocarbons (PAHs) emission, and radio continuum.
In order to study the relation of the molecular gas to recent and
former star formation we compare the distribution of the CO(1-0) line
emission to line emission and thermal radio continuum of HII regions, the light from
(young) stellar clusters, the ultraviolet continuum from young to intermediate 
aged stars ($\sim 100$ \,Myr), as well as light emerging from the old 
stellar population. In order to capture the different aspects of the relationship between
these tracers we present the data in classical
2-dimensional maps using the images at their native resolutions, a polar 
projection including a cross-correlation analysis, and a pixel-by-pixel analysis after
convolving all data to a common resolution of 3.0$\as$.

The polar coordinate system is very powerful for
studies of spiral arms as it allows for an easy assessment of their
properties. We use a linear 4$\pi$ presentation of the azimuthal angle (with
'0' corresponding to west and a counter-clockwise direction) and a linear
projection for the radius. Before re-projection all images were
convolved to the resolution of the 3.0$\as$ PAWS data cube and re-gridded to its
pixel scale. We converted all flux scales to units of "per pixel area". The 
polar cross-correlation allows for a quantitative assessment of azimuthal offsets 
between two different tracers (see Appendix \ref{sec:polar_cross} for details). With this 
definition a $\Phi>0$ offset for, e.g., star formation tracers implies that gas is flowing 
through the spiral mode inside co-rotation for a given perturbation to the gravitational 
potential. Several patterns (or perturbations) have been identified in the disk of M51 with 
different pattern speeds \citep[e.g.][]{meidt08,vogel93,elmegreen89,tully74}. A detailed analysis
of the various patterns and their associated pattern speeds
\citep[see appendix of][]{meidt13} suggests that at any given radius (probed by the PAWS FoV) 
one is inside a corotation resonance and thus the expected reversal of gas flow beyond the
corotation resonance expected for a {\it single} pattern is not observed.

\begin{figure}
\begin{center}
\resizebox{1.0\hsize}{!}{\includegraphics[angle=-90]{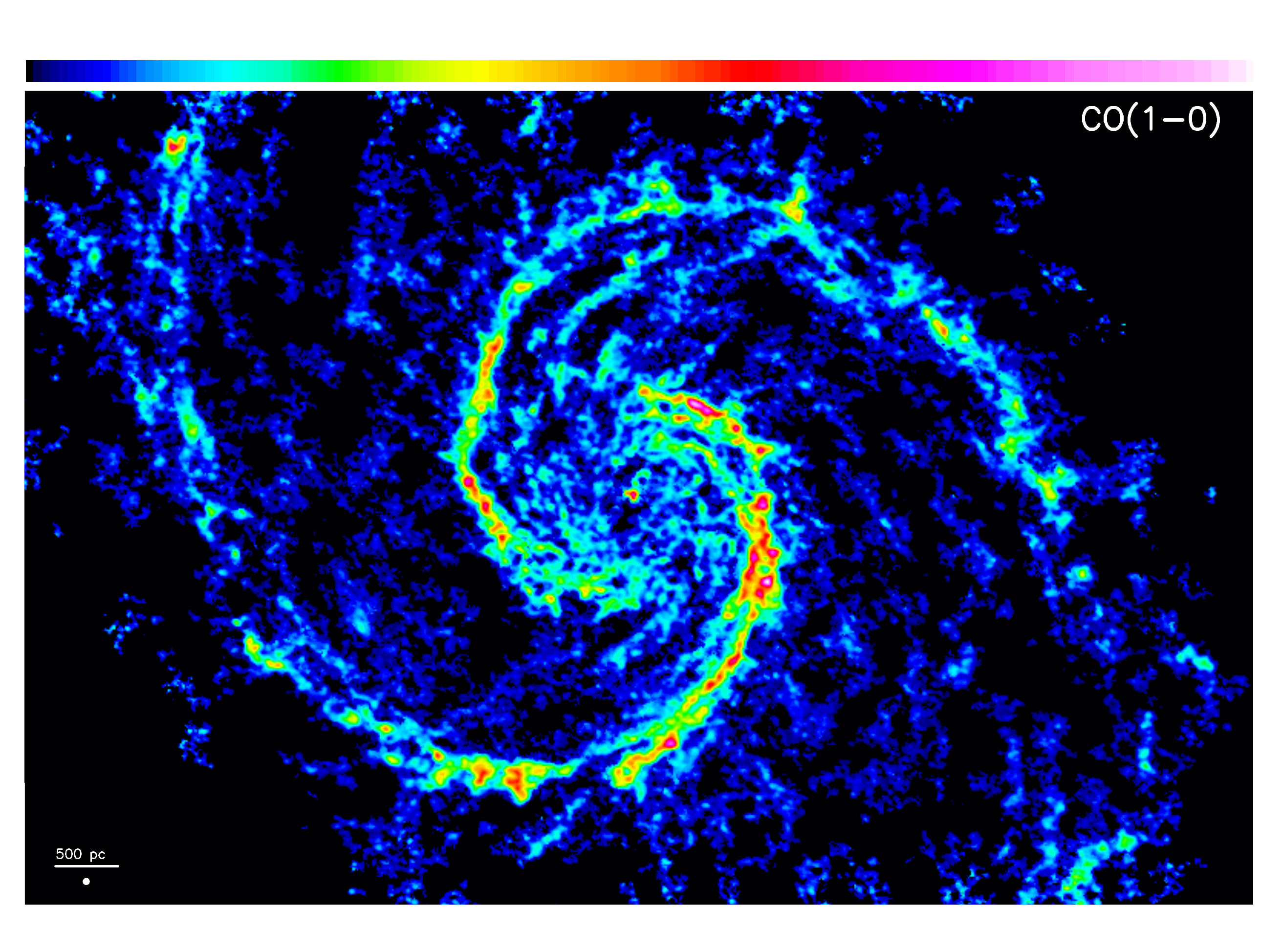}\includegraphics[angle=-90]{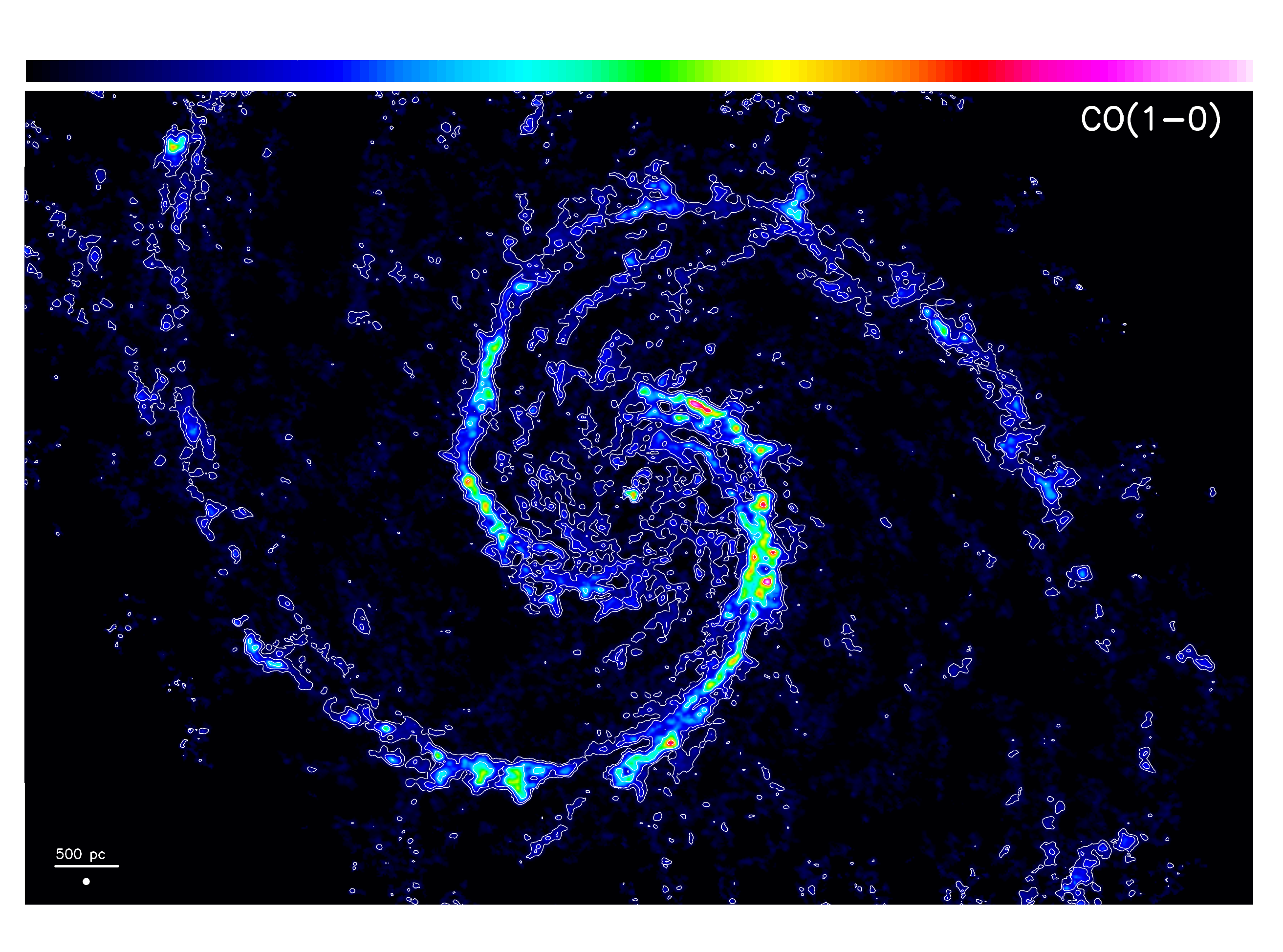}}\\
\resizebox{1.0\hsize}{!}{\includegraphics[angle=-90]{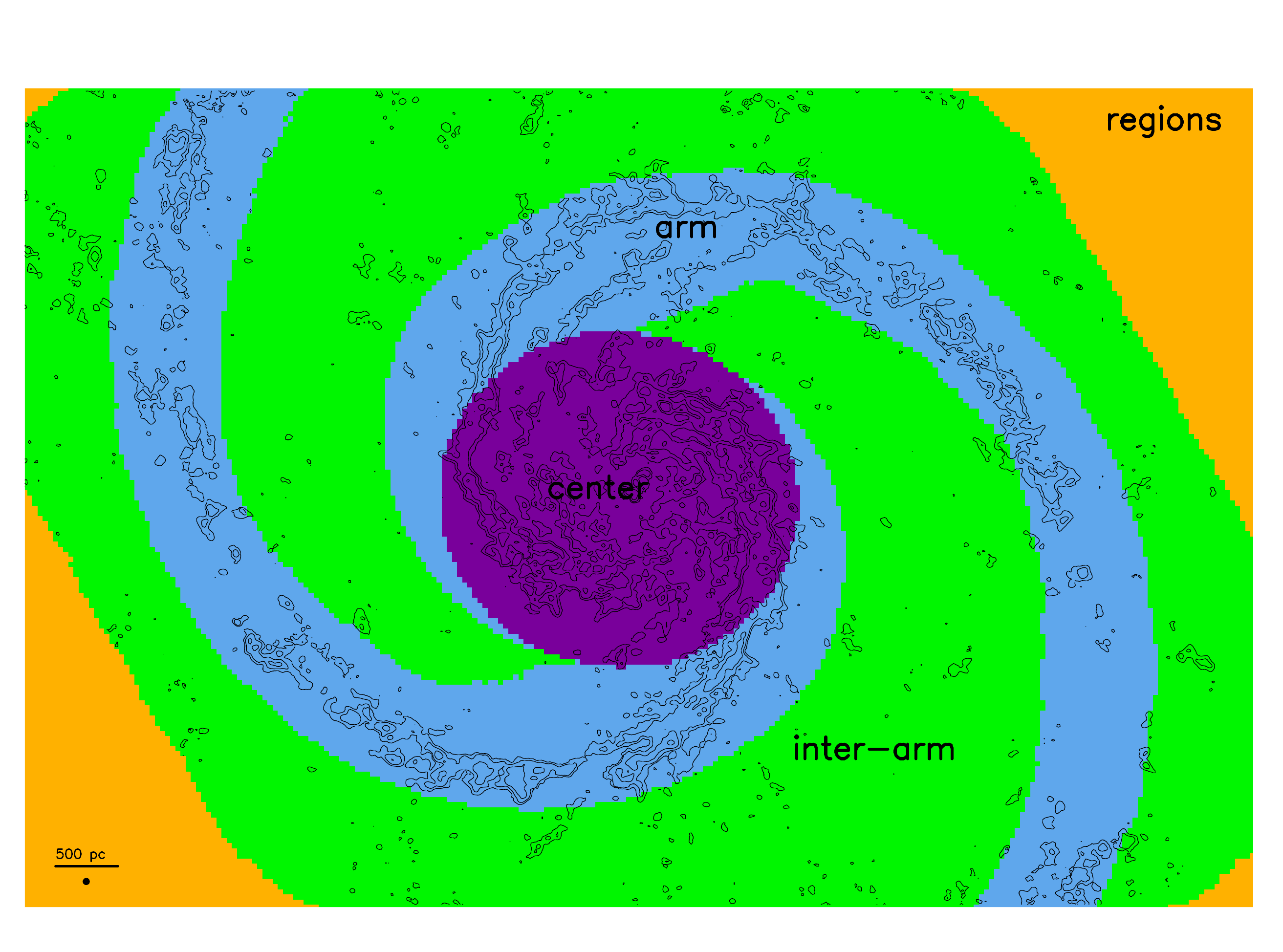}\includegraphics[angle=-90]{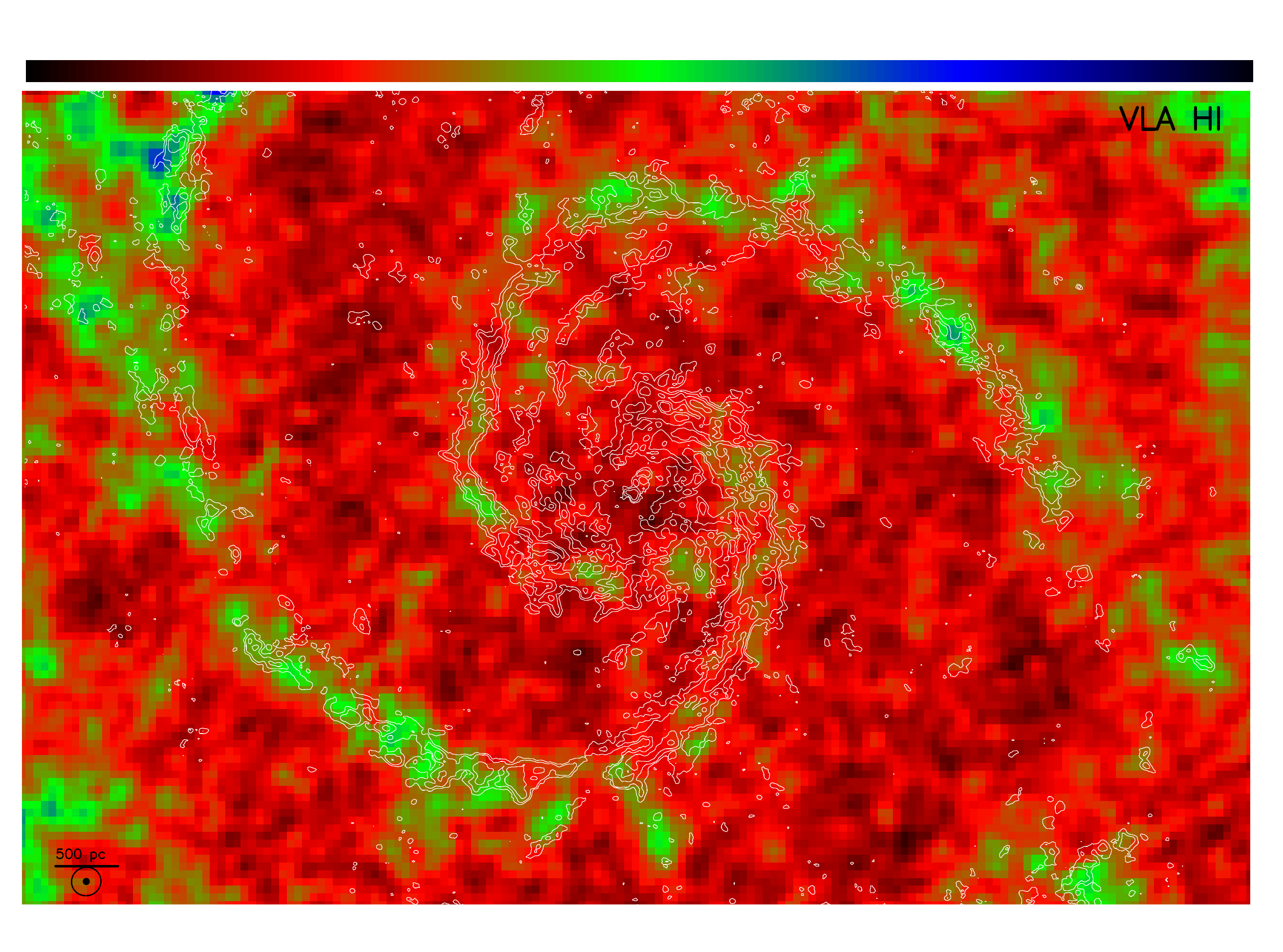}}
\end{center}
\caption{The CO(1-0) line emission in the central 9\,kpc of M\,51a as
observed by the PAWS project. The integrated intensity map is shown in a
square-root scaling to emphasize the distribution of the faint
emission ({\it top left}) and a linear scaling ({\it top right}; overlaid with contours at 40, 80, 160, and 320 $\kkms$ of the linear
representation that is used in the subsequent figures). The
location of the three region used for the pixel-by-pixel analysis is
shown in the {\it bottom left} panel. Comparison of the CO intensity map to the THINGS HI robust intensity map
tracing the atomic gas is presented in the {\it bottom right} panel. 
In the bottom left corner of each
panel a scale bar representing 500\,pc at the assumed distance of M51 and the CLEAN beam of the
CO(1-0) data are shown. 
\label{fig:co_2d}}
\end{figure}

To illustrate the differences between the different tracers and the CO emission
more quantitatively, we utilize pixel-by-pixel
diagrams plus an additional environmental mask
separating the central disk from the spiral arms and the inter-arm
region (Fig. \ref{fig:co_2d}, {\it bottom left}). The density of the data points
is presented on a logarithmic scale. Where appropriate we fit a linear relation using
the IDL procedure ROBUST\_LINEFIT, that
performs a best straight-line fit accounting the errors in both
coordinates as weights for the data-point. Lacking error maps for the
different tracers we use the same weight  of unity (rather than an error) 
for all data points/pixels.

\subsection{The Cloud-Scale Molecular Gas Distribution in M\,51 as seen
via CO(1-0)}\label{subsec:co}

The CO(1-0) line distribution exhibits two prominent spiral arms with
the northern arm being on average fainter than the southern arm
(Fig. \ref{fig:co_2d},{\it top}). After winding for about 180\,$\dg$
both gas spiral arms become significantly less prominent and even
indistinguishable from the inter-arm in the case of the northern arm.
In addition, the southern arm is showing a bifurcation exactly south
of the nucleus that persists for larger radii. On the other hand the
northern spiral arm shows a pronounced 'dent' north-east of the
nucleus. Taken together this strongly suggests that the spiral arms do not
represent a single pattern. The two spiral arms emerge from a
ring-like distribution close to the center. The disk inside this ring
shows fainter emission distributed in abundant smaller structures. CO
emission is present between the spiral arms, however, at an even fainter
level and in a more flocculent fashion.

In the polar representation, the CO(1-0) spiral arms are almost straight lines 
as expected for a log-normal spiral (Fig. \ref{fig:ism_pol}, {\it top left}). 
However, close inspections reveals three breaks with
differing prominence for both arms at radii of $\sim$27'',
$\sim$55'' and $\sim$90'' with observed pitch angles of about 40$\dg$, 30$\dg$, and 10$\dg$,
respectively. In addition, an azimuthal offset along the spiral arm at
 $\rm r \approx 35''$ is most evident
in the northern spiral arm. The overall slope of the two spiral arms is
not identical suggesting that the underlying gravitational potential can
not be described by a
pure m=2 mode. The molecular ring appears as an almost horizontal line
for radii of $\rm 15'' < r < 20''$. Below a radius of $\sim$10'' the
sampling in azimuth is becoming too sparse in our presentation of the
data for any firm conclusions. The difference in spiral arm pitch angle has 
already been noticed before by \cite{henry03} who analyzed
the BIMA SoNG data \citep{helfer03} and reported 'break radii' of 27''
and 60''. The latter break at 60'' is close to our inferred break at 55''. 
The discontinuity at 35'', on the other hand, appears 
strongly only at the high resolution of our data and is absent at the low resolution
of the BIMA SoNG map \citep[e.g. Fig. 1 and 2 of][]{henry03}. It is interesting to note 
that the radial profile of the azimuthally averaged torques \citep[see Fig. 1 of ][]{meidt13} 
changes direction at radii of
20'' (corotation resonance of the nuclear bar), 35'' (start of spiral pattern), 55'' (corotation
resonance of spiral pattern) and 90'' (start of second spiral pattern) consistent with
the breaks observed in the CO distribution at 35'', 55'' and 90''. Using a banana 
wavelet analysis \cite{patrikeev06} find values similar to ours for the pitch angles of the 
CO spiral arms and show that the pitch angles of the
two spiral arms differ by on average 10\,$\dg$ over the PAWS FoV. \cite{henry03} 
suggested the presence of an additional m=3 mode in the central 6\,kpc that could
explain both the angular offset between the two gaseous arms and their
asymmetry in intensity. 

\subsection{Comparison of CO(1-0) to other ISM
Tracers}\label{subsec:ism}

Two other tracers of the neutral gas can be compared to the distribution of the 
molecular gas at angular resolutions of $\sim 6\as$, the atomic gas in its HI line 
emission and from the strongest cooling line of the neutral gas, [CII] at 158\,$\mu$m 
as imaged by PACS.
Three different types of tracers of the dust in a galaxy have
sufficient angular resolution for a detailed comparison to the CO
maps: (a) an optical extinction map from HST, (b) maps of PAH emission
from Spitzer, and (c) the hot dust heated by young stars. We expect varying
degrees of correlation between the molecular gas distribution and these
dust maps, given that excitation conditions likely vary across our FoV. While the
cold dust is best traced at long infrared wavelengths \citep[e.g.][]{dale12,aniano12}, 
available maps have a fairly low angular resolution of,
e.g., $\ge$12'' for 210\,$\mu$m onward for the Herschel SPIRE
instrument. The non-thermal radio continuum is probing very
different components related to the ISM and it can have several origins (SN,
stellar remnants, radio jet, magnetic field etc.).

\subsubsection{Neutral Gas}\label{subsec:hicii_data}

The strongest far-IR cooling line of the cold ISM in nearby galaxies is [CII] at 158\,$\mu$m which 
can contain up to 1\% of the total far-IR luminosity \citep{stacey91,brauher08,gracia11}. Typically most of 
the [CII] emission in nearby galaxies arises from the surface layer of molecular clouds where the gas is warm,
dense and photo-dissociated by FUV emission from either nearby hot (OB) stars and/or the interstellar 
radiation field \citep[see, e.g., introduction of][]{stacey10}. Comparison to the CO(1-0) emission (Fig.  
\ref{fig:ism_2d}, {\it top left}) shows a close correspondence between the peaks of both emission lines in the 
molecular ring out to a radius of $\sim20\as$, except for a region in the north-east segment of the [CII] ring 
that has no counterpart in the CO(1-0) emission. Between the ring and the spiral arms the [CII] emission 
exhibits gaps, similar to the ones present in the host dust emission. Along the spiral arms the most intense 
[CII] emission is patchy and is offset to the convex side of the molecular gas spiral arms. 

The robustly weighted intensity map of the THINGS data that images the distribution of the atomic gas shows very 
little pronounced structure in the central region of M\,51; a few peaks coincide with the molecular gas 
ring (Fig. \ref{fig:co_2d}, {\it bottom right}). Along the molecular spiral arms the HI line emission becomes stronger 
reaching its maximum (within the PAWS FoV) roughly at the same locations where hot dust emission is 
observed. Taken together the distributions of [CII] and HI strongly suggest that along the spiral arms both 
emission lines trace the dissociation process of molecular clouds caused by the massive young star 
formation. Within the molecular ring the situation is more complex, as the [CII] emission is strongest, but no 
substantial HI emission is seen there.

\subsubsection{Optical extinction} \label{subsec:Av_data}

According to results for local Galactic clouds \citep[e.g.][]{dickman78,lada07}, maps of
(optical) extinction are a strong proxy for the cold dust distribution. Comparison to the HST $I-H$
image (Fig. \ref{fig:ism_2d}, {\it top right}) shows that the optical
extinction is a very good predictor of the location of strong CO(1-0)
emission and its underlying detailed fine-structured morphology. This
is most obvious in the spiral arms. In the inter-arm region the
correlation is less pronounced which in part is due to the presence of
stellar clusters along spurs \citep[see][] {lavigne06,chandar11} that
are especially prominent in the bluer bands observed with HST (see \S
\ref{subsec:ysc_data}). 

The polar cross-correlation of these 
two tracers shows no azimuthal offset nor any radial trend 
(Fig. \ref{fig:ism_cross}, {\it top right}), consistent with both tracers being 
spatially coincident. The prominent (dust) clump
in the inter-arm at an azimuthal angle of $\sim$240\,$\dg$ and a
radius of $\sim$39'' has a CO counterpart at a lower level. Inspection
of the $I$ and $H$ band images suggest that the color of this clump is
affected by a group of star clusters causing this apparent
inconsistency. 

Overall, the molecular gas distribution above a
$\htwo$ surface density $\shtwo$ level of $\sim 90-100\,\msunpc$ is
well predicted by the $I-H$ color map. Detailed comparison on a pixel-by-pixel basis at 3.0$\as$ resolution
shows that the relation between the HST $I-H$ color and
CO integrated intensity is linear before it appears to saturate (Fig.
\ref{fig:ism_pix}, {\it top row}). While this trend is present in all
environments, the slope in the central region is significantly shallower
suggesting that the parameters controlling the relation are changing. The similar
slope for the arm and inter-arm regions means that the gas-to-dust
ratio is not dramatically changing between arm and inter-arm.

\subsubsection{(Hot) dust emission} \label{subsec:hotdust_data}

\begin{figure}
\begin{center}
\resizebox{1.0\hsize}{!}{\includegraphics[angle=-90]{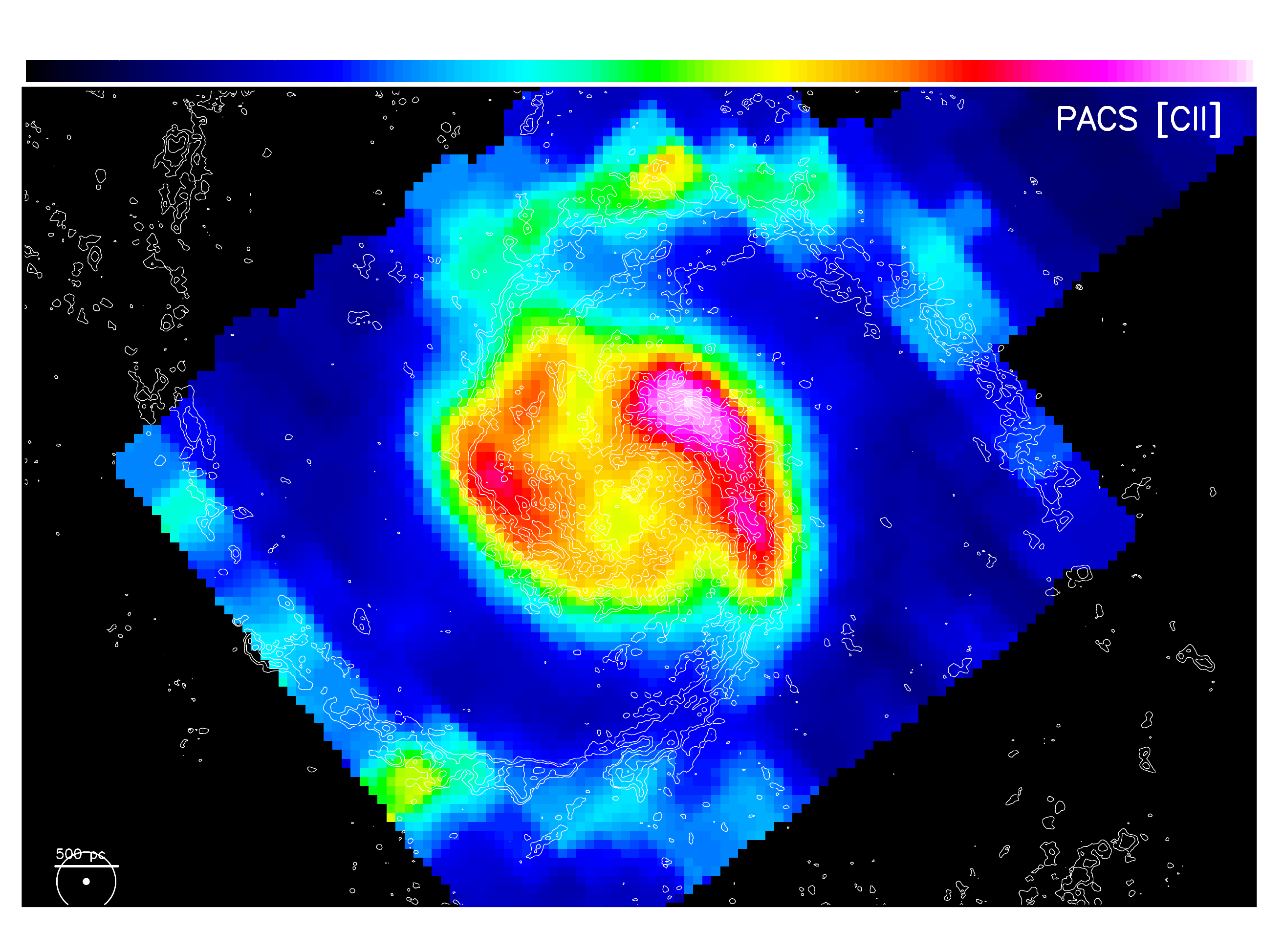}\includegraphics[angle=-90]{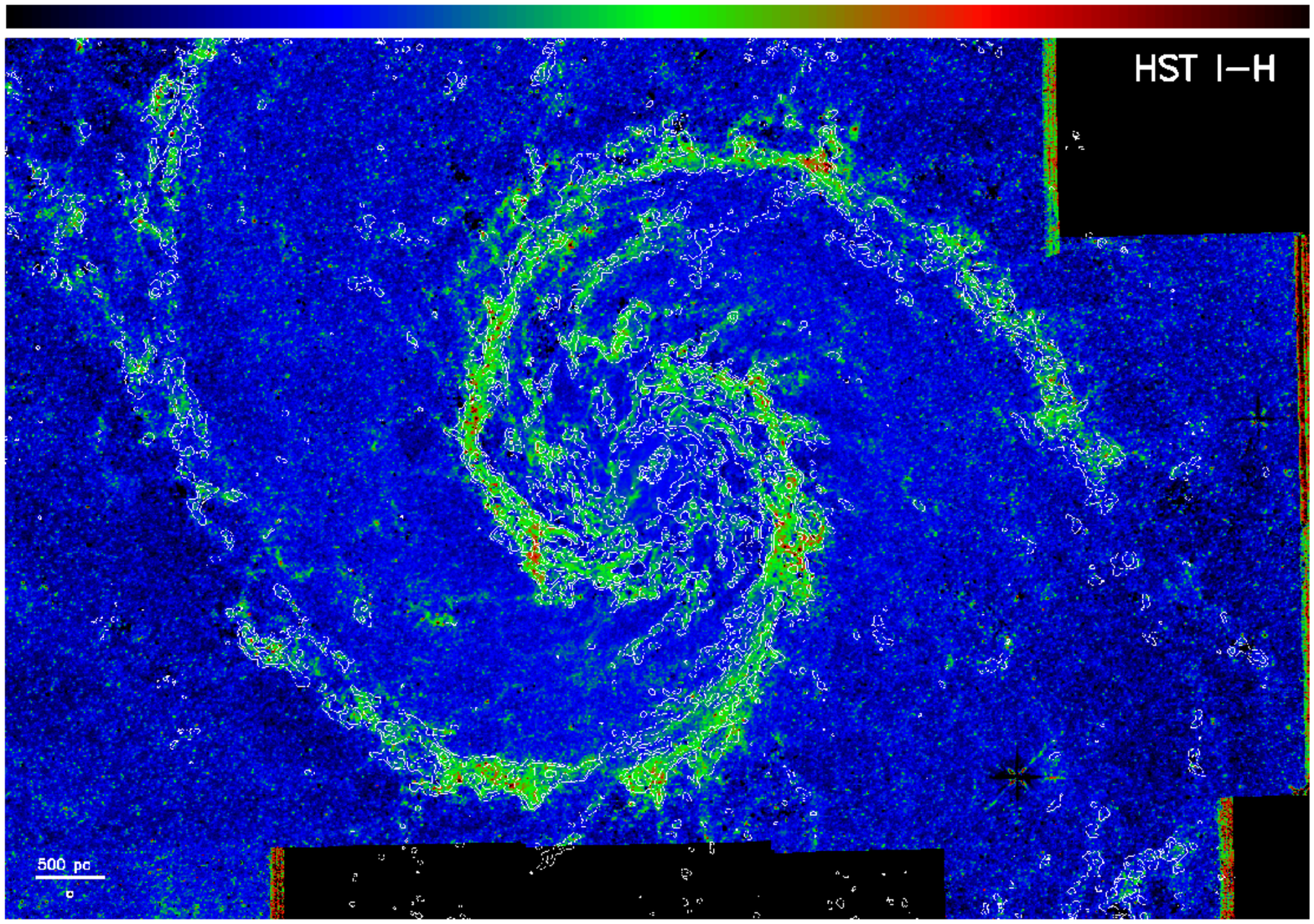}}\\
\resizebox{1.0\hsize}{!}{\includegraphics[angle=-90]{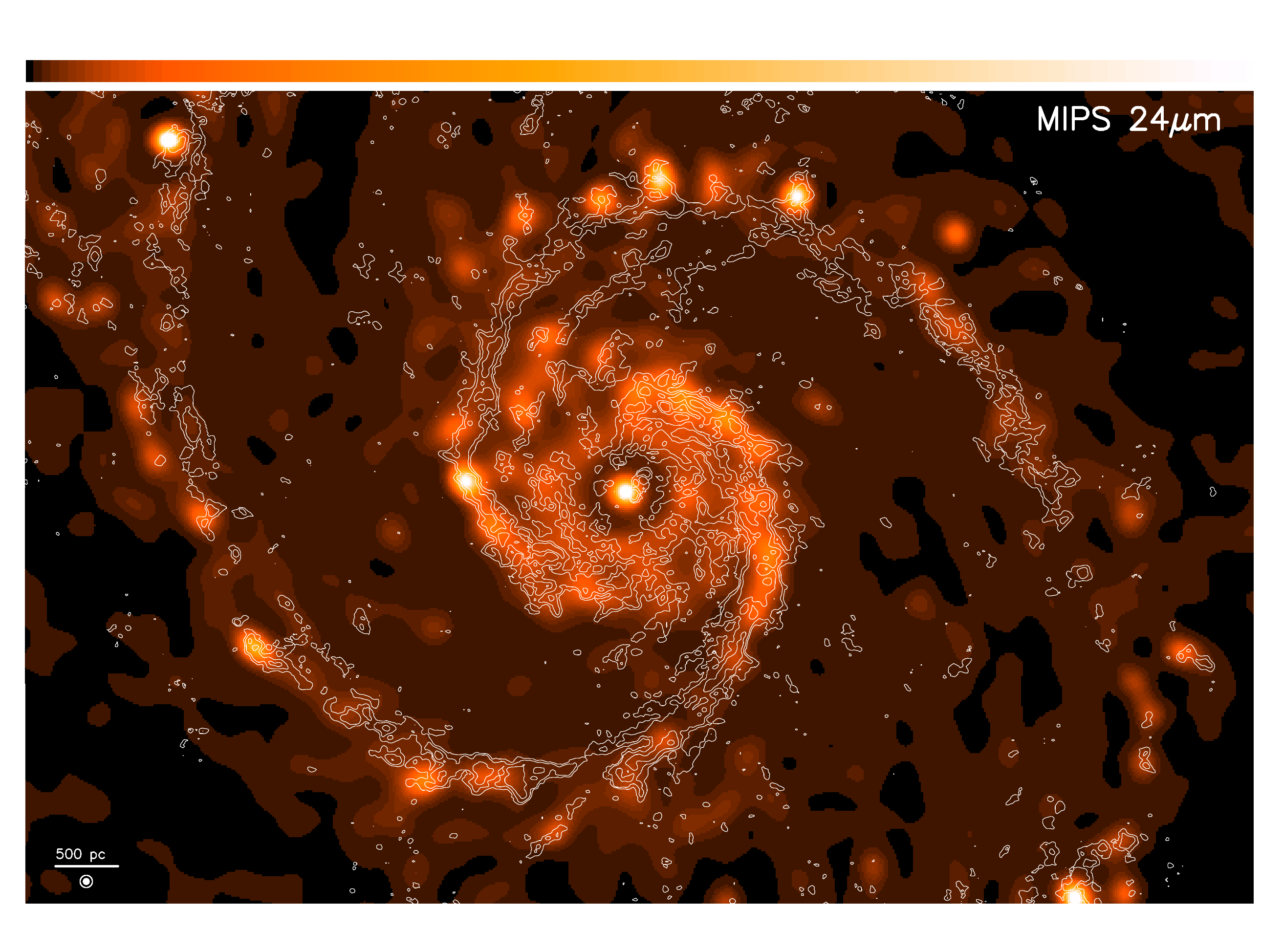}\includegraphics[angle=-90]{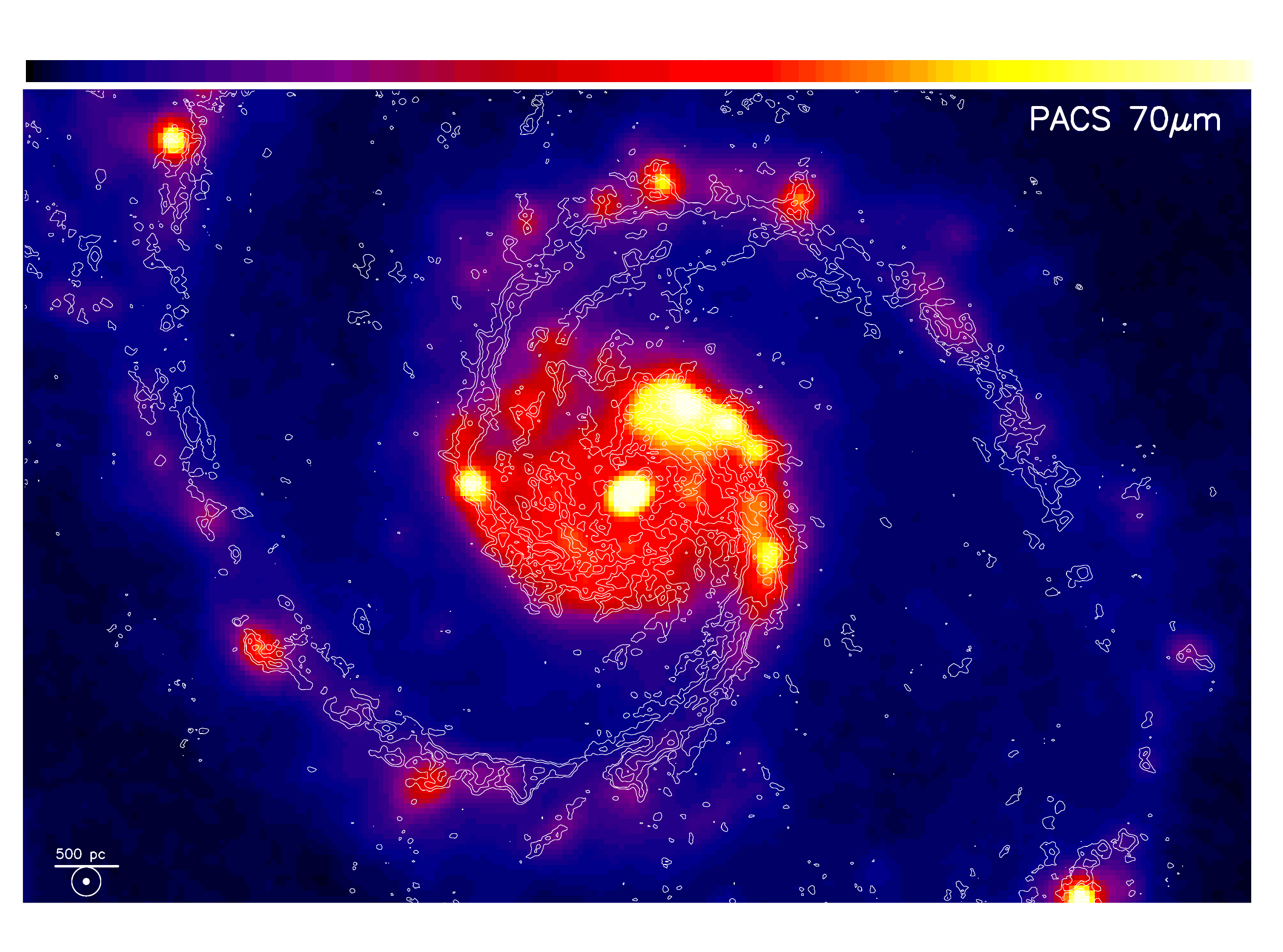}}\\
\resizebox{1.0\hsize}{!}{\includegraphics[angle=-90]{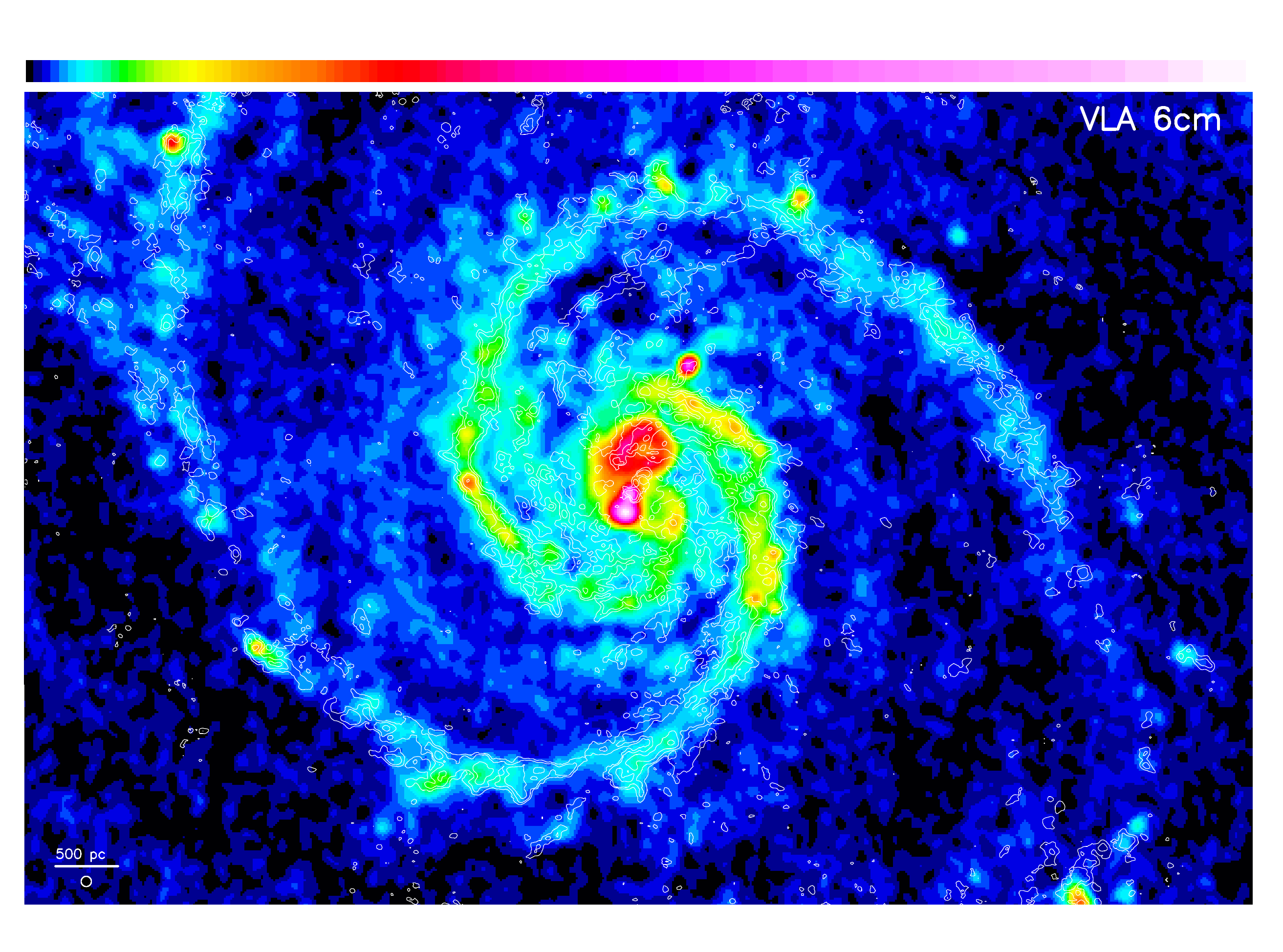}\includegraphics[angle=-90]{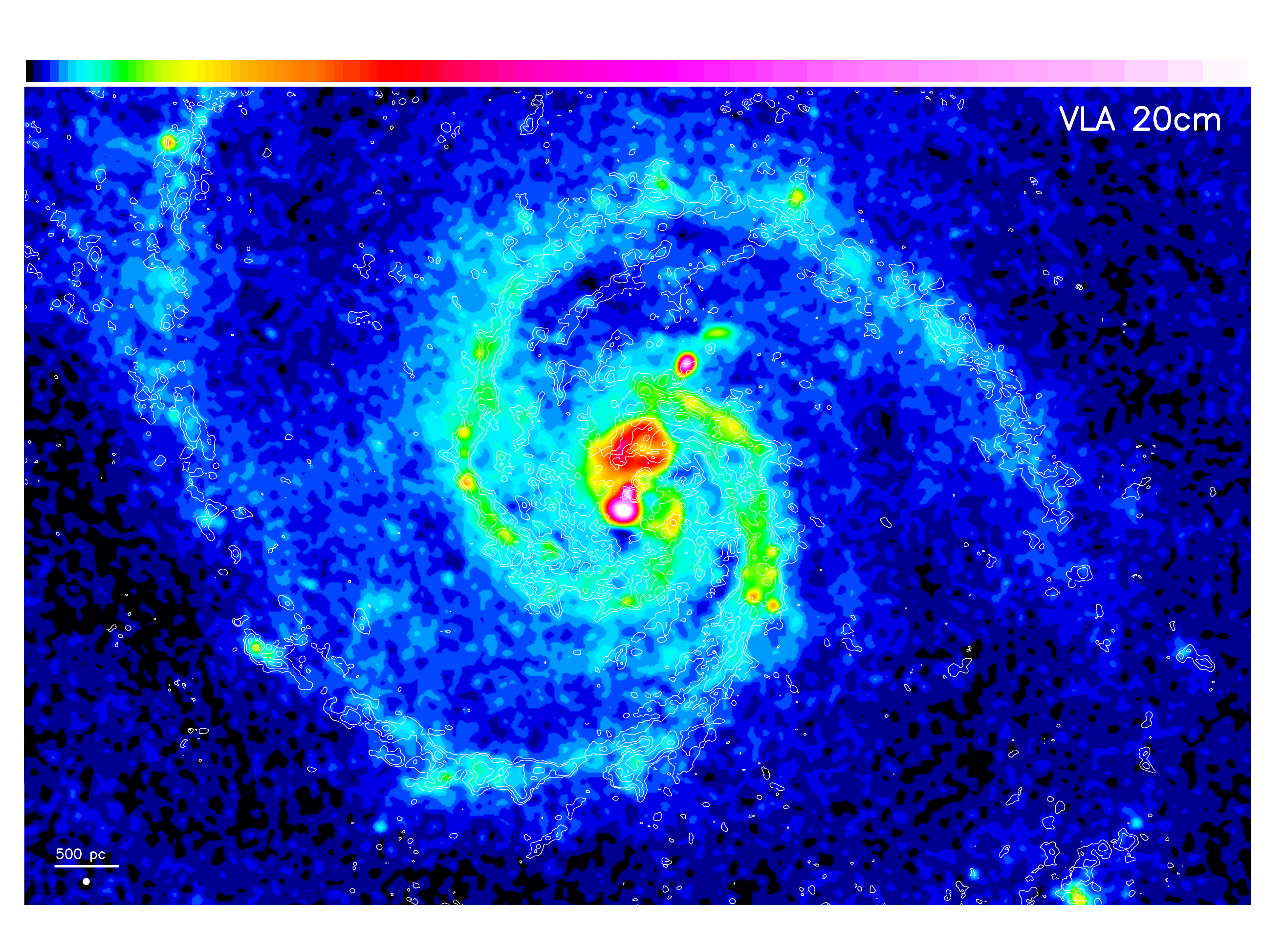}}
\end{center}
\caption{\footnotesize Comparison of the CO(1-0) line emission in the central 9\,kpc of M\,51a as observed by the 
PAWS project overlaid in contours (same as in Fig. \ref{fig:co_2d}, {\it top right}) 
on the Herschel [CII] line map tracing photo-dissociation ({\it top left}), 
the HST $I-H$ color image tracing extinction ({\it top right}),
the MIPS HiRes 24$\mu$m image tracing hot dust emission ({\it middle left}),
the PACS 70$\mu$m image tracing warm dust emission ({\it middle right}),
the VLA 6cm image tracing thermal and non-thermal radio continuum ({\it bottom left}),
and the VLA 20cm image tracing mainly non-thermal radio continuum ({\it bottom right}). In the bottom left corner of each panel a scale bar representing 500\,pc and the CLEAN beam of the CO(1-0) data as well as the resolution of the respective dataset are shown.
\label{fig:ism_2d}}
\end{figure}

The distribution of the emission from hot
dust probed at 24\,$\mu$m and 70\,$\mu$m by the Spitzer MIPS and
Herschel PACS instruments shows a bright central disk encompassing
the molecular gas ring (Fig. \ref{fig:ism_2d}, {\it middle row}). In addition,
several bright peaks are detected along the molecular spiral arms with
a slight azimuthal offset of $\Phi > 0$ (measured counter-clockwise) from the convex side 
of the spiral arm and an apparent gap close
to the inner disk. The active nucleus in M\,51a stands out as a bright
point source in the IR maps. Low level (mid-)IR emission is seen in
the inter-arm region (about one order of magnitude fainter than the
peaks). While the spatial coincidence between the CO(1-0) and hot dust emission 
is very good for the central disk, it starts to vanish for the spiral arms, especially when 
only the high surface brightness regions are considered. 

The polar representation (Fig. \ref{fig:ism_pol}, {\it
middle left}) clearly reveals gaps between the central disk and the spiral arms:  
in the northern spiral arm the gaps are slightly more
prominent ranging from 35$\as$ to 43$\as$ and in the
southern arm at slightly smaller radii of
$28\as\,<\,r\,<\,36\as$. After these gaps, the bright 24\,$\mu$m spots
along the spiral arms are typically azimuthally offset from the CO emission.
%by 15\,$\dg$ in the Northern arm and about half this value in the Southern arm (though
%with larger scatter). 
The polar cross-correlation (Fig. \ref{fig:ism_cross}, {\it middle left}) 
shows that the offset of the hot dust emission associated with the 
spiral arms starts at radii of $\rm r\,>\,50\as$ (i.e. after the gaps)
with an offset of $\sim 30\dg$ that is radially decreasing to $\sim 5\dg$ 
and rising again for $\rm r\,>\,80\as$ with a decline thereafter. A recent study by \cite{louie13} 
finds qualitative similar results while the lower (13'') resolution study of \cite{foyle11} sees 
no evidence for offsets. This finding can be easily explained by the spottiness of the bright 
24$\mu$m peaks that are clearly associated with young massive star formation and the 
enhancement of 24$\mu$m emission in the molecular spiral arms which become mixed 
at the 13" resolution that no longer resolves the spiral arms.
It is interesting to note that also the bright 24$\mu$m spots do not fall on a 
straight line suggesting a certain stochasticity when the initial star formation
event occurs with respect to the spiral arm passage.
% but appear to have a smaller 
%azimuthal offset at radii larger than $\rm r\,\simeq\,58\as$. 

A low level component of the 24\,$\mu$m
emission coincides with the CO emission in the spiral arms and extends
into the downstream inter-arm region (Fig. \ref{fig:ism_2d}, {\it middle row} 
and Fig. \ref{fig:ism_pol}, {\it middle left}). The factor 2$\times$
enhancement in the smooth component of the
24\,$\mu$m emission in the arm compared to the inter-arm region is similar
to the average arm/inter-arm difference in CO brightness.
This suggests that the enhancement simply reflects a
constant dust-to-gas ratio. Therefore the 24\,$\mu$m emission
detected in the (molecular) arms is likely not tracing any deeply embedded star
formation. Detailed modeling of the dust SED at an angular
resolution sufficient to exclude the influence of the bright star
forming regions associated with the arms would be required to firmly test this interpretation.

There appears to be only a very broad correlation between the 
hot dust emission traced by 24\,$\mu$m and the CO emission
when considering the entire PAWS FoV (Fig. \ref{fig:ism_pix}, {\it
upper middle row}). However, comparison of the three individual
environments provides more insights. The central disk shows a very
steep linear relation, though with significant scatter, while there might be
an additional loose relation with a significantly
shallower slope present in the spiral arms and the inter-arm region is almost
consistent with no relation at all. The second shallower relation is arising
from the low level 24\,$\mu$m emission seen in the 2D and polar
representation (Figs. \ref{fig:ism_2d}, {\it middle left}, and
\ref{fig:ism_pol}, {\it middle left}).

\subsubsection{(Non-thermal) radio continuum} \label{subsec:NT_data}

The radio continuum emission at 6
and 20\,cm in M\,51 is dominated by non-thermal synchrotron emission
\citep[]{dumas11}. Surprisingly the correspondence to the CO(1-0)
distribution is significantly better than for the hot dust emission
(Fig. \ref{fig:ism_2d}, {\it bottom row}). While bright radio continuum emission 
peaks exist along the spiral arms, the radio continuum
spiral arms coincide with the molecular gas spiral arms. However, the
radio continuum arms appear to be more contiguous and diffuse compared
to the CO(1-0) emission at comparable angular resolution. A wavelet decomposition of the radio
continuum suggests that the emission arising from the spiral arms is
more diffuse and not composed of many individual point sources
\citep{dumas11} confirming the visual impression of
Fig. \ref{fig:ism_2d} ({\it bottom row}). 

In the polar coordinate system the brightest knots in the 6 and 20\,cm radio continuum along the spiral arms 
(Fig. \ref{fig:ism_pol} {\it middle right}, only shows 20\,cm)
are often offset from the gas
arm and align with the brightest regions seen, e.g., in the hot dust
emission traced by the 24\,$\mu$m map (Fig. \ref{fig:ism_pol} {\it
middle left}). This suggests that recent star formation is dominating
the radio excitation here while another process must cause the
enhancement of the radio continuum in the spiral arms. This
behavior is only mildly reflected in the polar cross-correlation with the CO(1-0) 
emission (Fig. \ref{fig:ism_cross}, {\it middle right}) where the azimuthal 
offset is consistent with zero at most radii, 
except for the center where the radio jet is present and along the spiral 
arms for $\rm 57\as\,\le\,r\,\le\,80\as$. Interestingly, the azimuthal offset 
is only $\Phi \sim 8\dg$ (compared to $\Phi \sim 30\dg$ seen for the hot dust emission). 
The southern arm exhibits a bubble-like feature in the 
20\,cm continuum map in the polar coordinate system at a radius
of $\rm r\sim 65\as$ between an azimuthal angle of $\rm
220\,<\,\Phi\,<\,240\,\dg$.

The inner disk is dominated by the
well-known radio jet \citep[e.g.][]{crane92,maddox07} emanating from the central AGN,
while the molecular ring has a radio continuum
counterpart. Inter-arm radio continuum emission is detected,
but at a significantly lower level. At 6\,cm a second pair of
spiral arms appears mid-way between the prominent molecular arms. New
EVLA data taken at 20\,cm shows this feature at that wavelength as
well (priv. comm. J. Ott). The location of this second pair of arms is roughly
consistent with the location of 'magnetic arms' in the inter-arm region shown by \cite{fletcher11}. 
Already \cite{tilanus88} noted that
non-thermal radio emission associated with the spiral arms (after
removal of a 'base' disk in their 8\,$\as$ resolution 6 and 20\,cm
short spacing corrected VLA data) coincides with the dust lanes rather
than star forming giant HII complexes probed by H$\alpha$ imaging.

Two tentative correlations
between the radio continuum at 6 and 20\,cm and the CO emission
are apparent in the pixel-by-pixel comparison (Fig. \ref{fig:ism_pix}, {\it two bottom rows}): 
one steeper (present in the center region) and a more shallow one (in the arm region) which separates better in the
20cm continuum versus CO emission plot. The situation in the inter-arm regions 
seems to be more complex, judging from 
the significantly larger scatter in the respective diagrams. The steeper correlation is also
significantly stronger in the central disk while the shallower
relation dominates in the spiral arms and (tentatively) the inter-arm
region. It is interesting to note that in the spiral arm region of the 6\,cm vs. CO emission 
(Fig. \ref{fig:ism_pix} {\it middle right panel in second last row}) there is an indication for a down turn 
at large CO fluxes which might point toward synchrotron losses in very dense gas. 
Better data is required to confirm this tentative trend.
Slopes between the CO flux and 6\,cm, respectively 20\,cm, radio
continuum flux density are not identical. The few points scattered along an almost vertical 
branch (esp. evident in the center diagram of the 6\,cm emission) are
likely caused by the prominent radio jet. 

%We can use the estimated slopes $a$of the linear relation between the 20 and 6\,cm radio continuum and the CO emission to derive an average spectral index $\alpha_{20-6}$ (with $\rm S_{\nu} \sim \nu^{- \alpha}$) of the radio continuum associated with the CO emission. Using $\rm \alpha_{20-6} = - \frac{log(\frac{a_{20}}{a_6})}{log(\frac{1.4\,GHz}{4.8\,GHz})}$ we find $\alpha_{20-6}$ of 0.17, 0.34 and 0.11 for the central, spiral arm and inter-arm regions, respectively. 

\subsubsection{PAH emission} \label{subsec:PAH_data}

\begin{figure}
\begin{center}
\resizebox{1.0\hsize}{!}{\includegraphics[angle=-90]{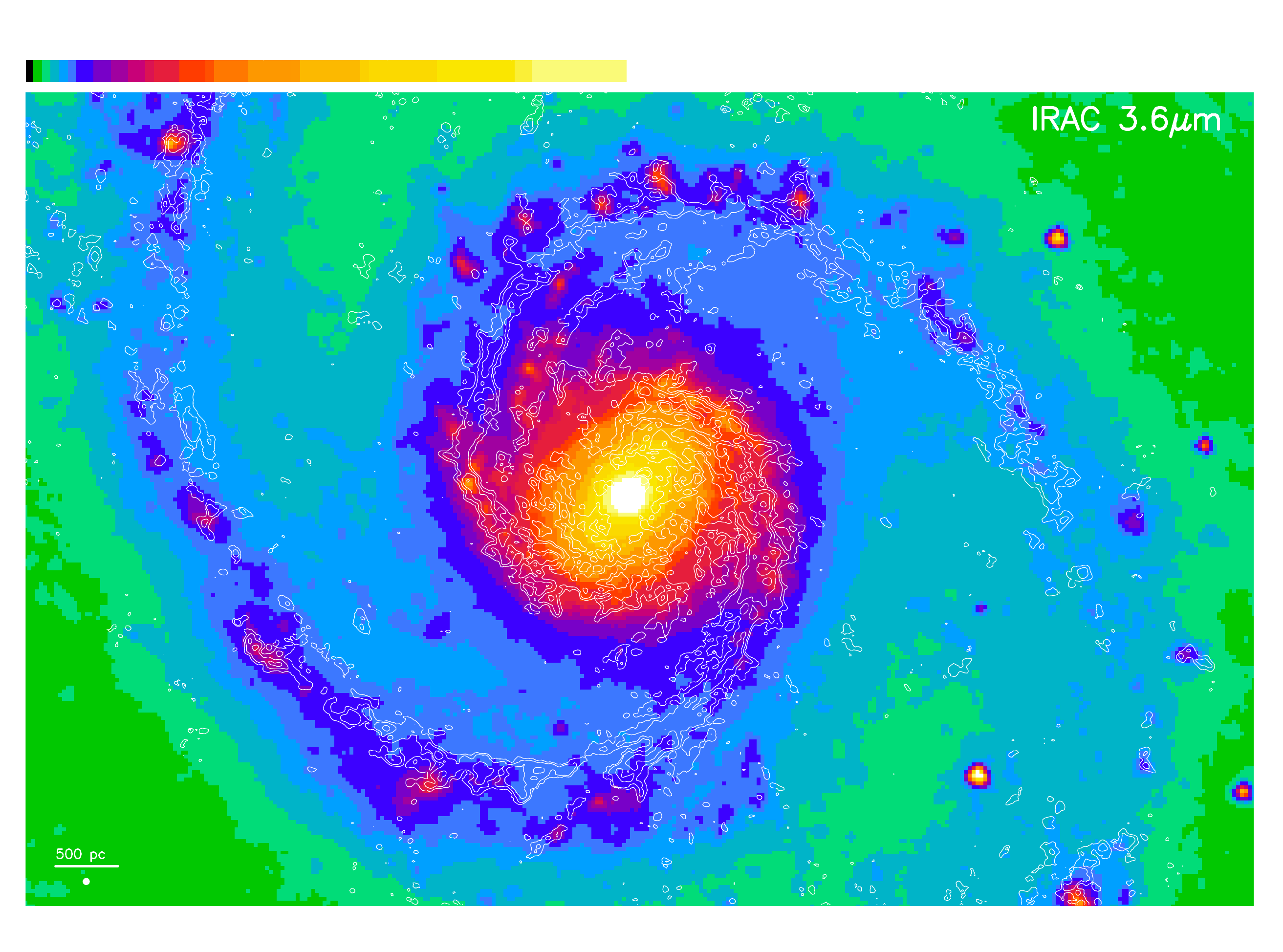}\includegraphics[angle=-90]{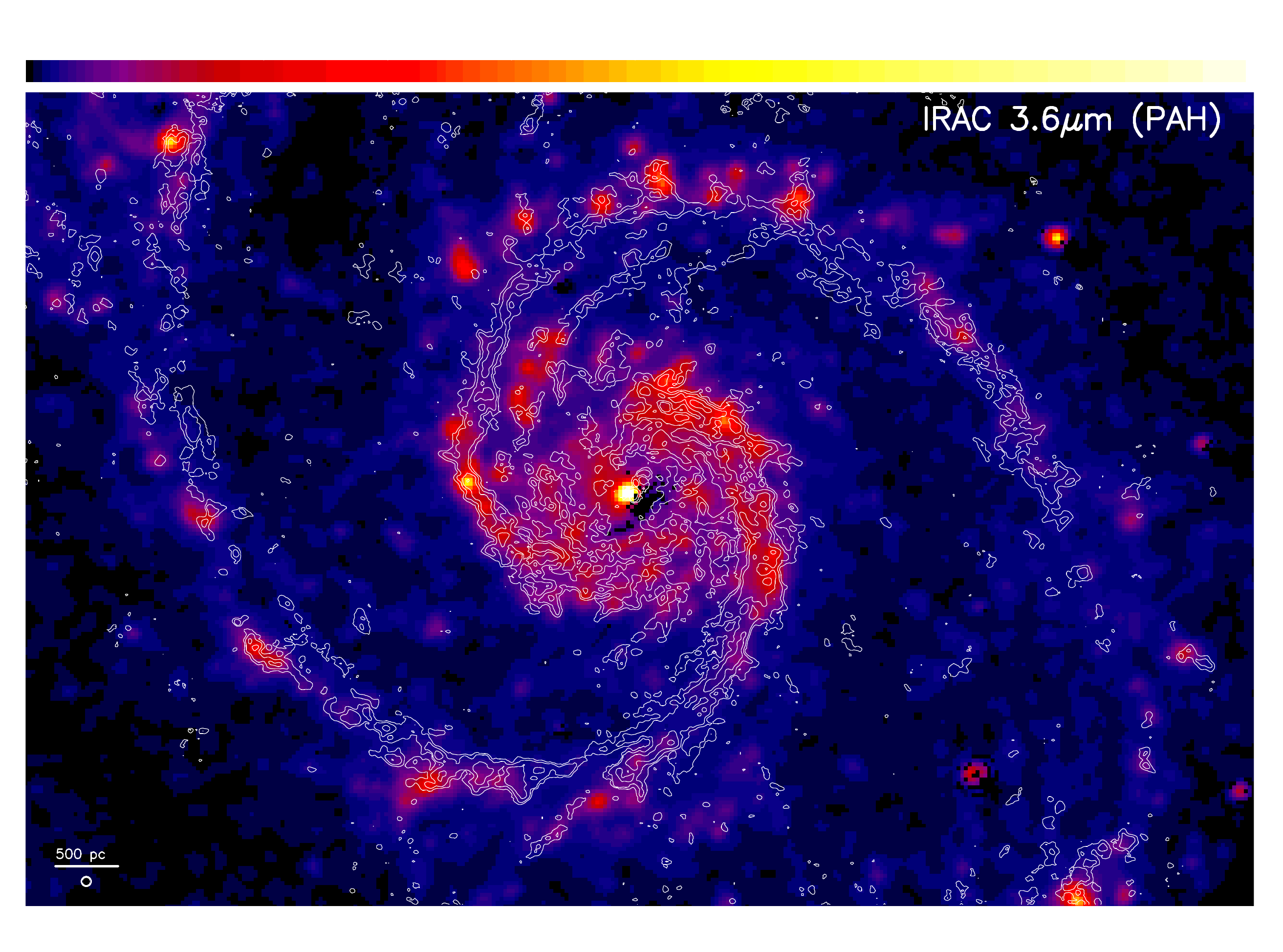}}\\
\resizebox{1.0\hsize}{!}{\includegraphics[angle=-90]{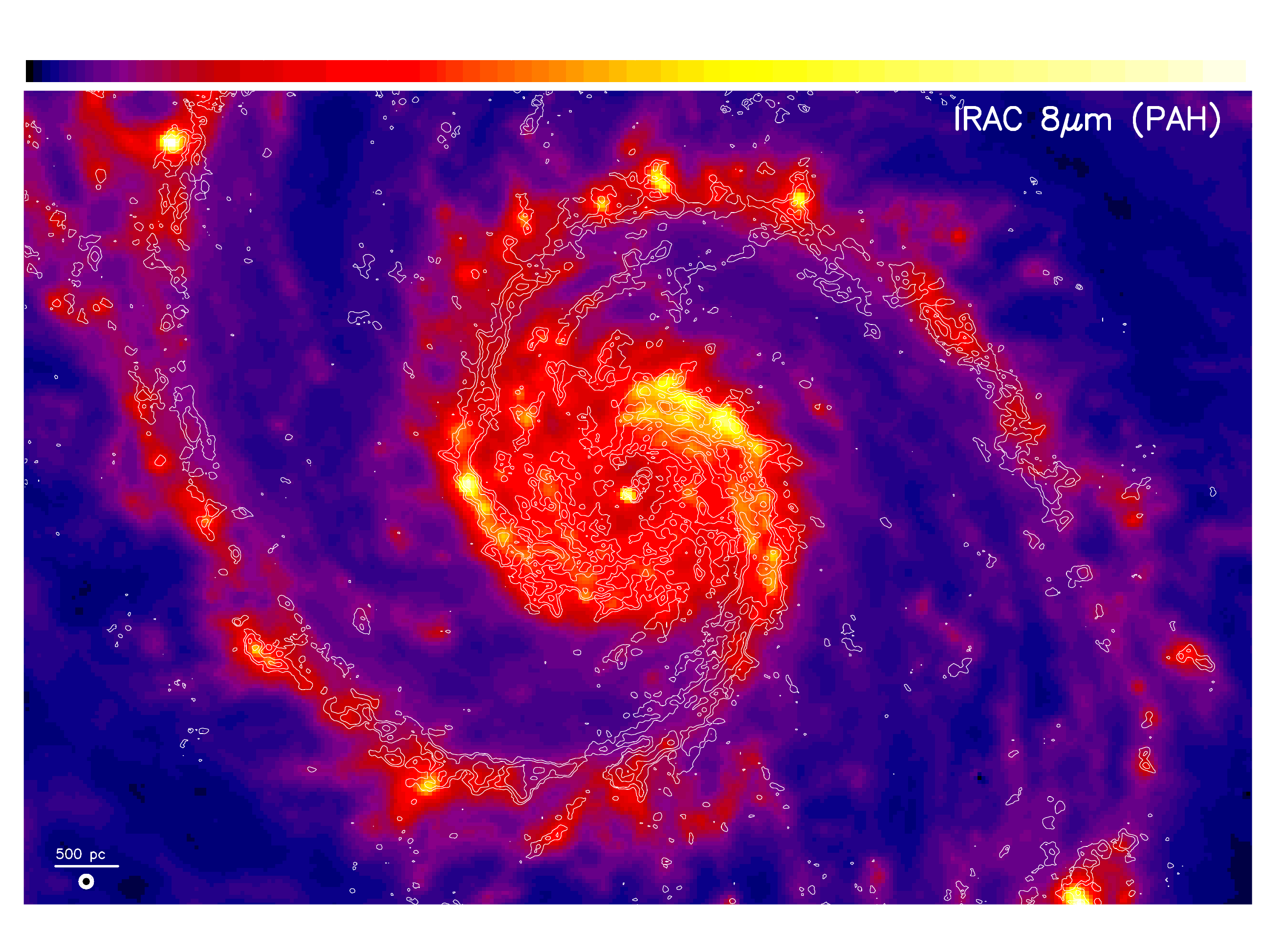}\includegraphics[angle=-90]{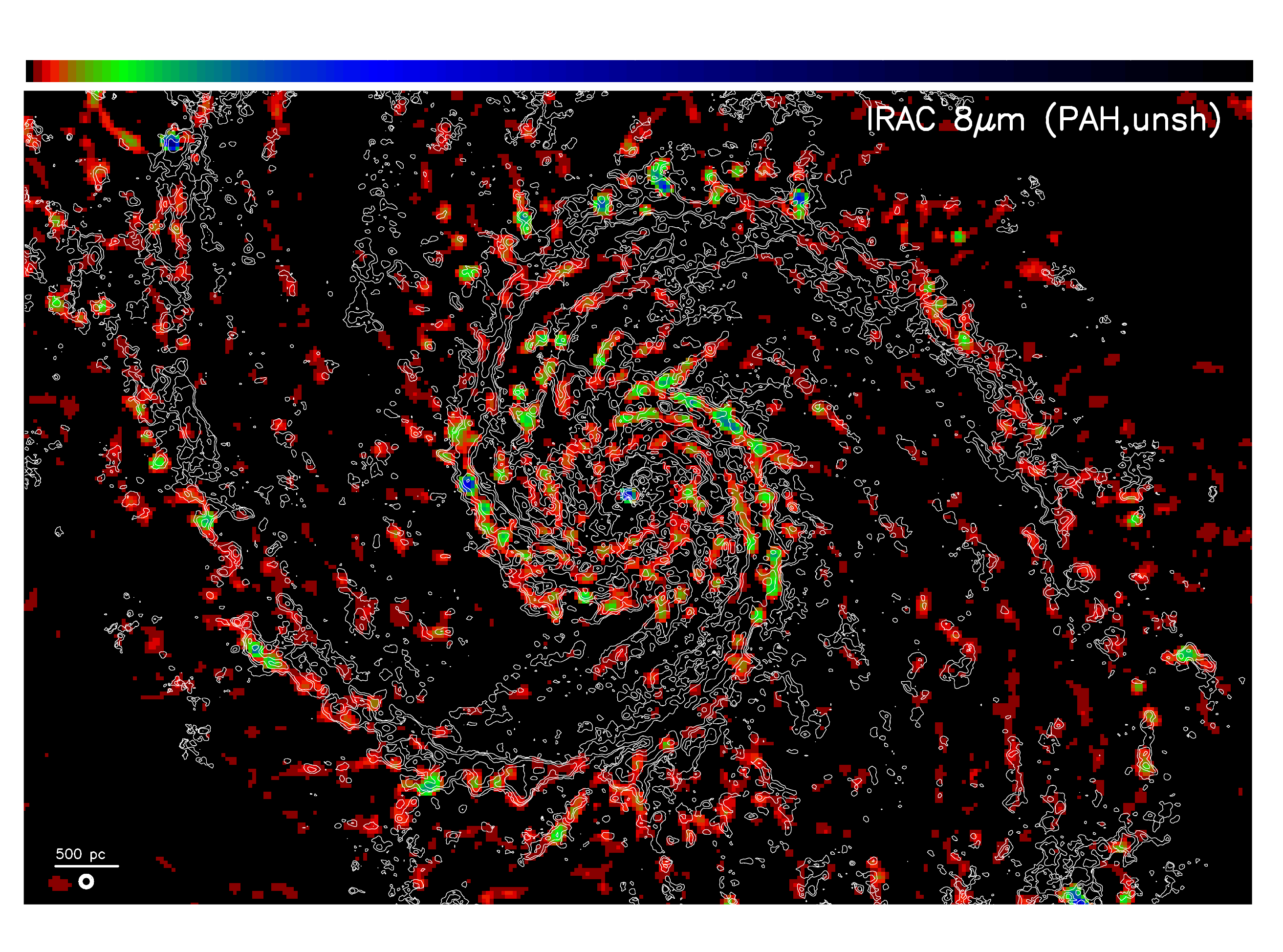}}
\end{center}
\caption{Comparison of the molecular gas distribution of M\,51a to tracers of PAH emission. 
\coone ~intensity distribution overlaid onto
the total 3.6$\mu$m emission ({\it top left}) and its non-stellar component only ({\it top right}),
the 8$\mu$m non-stellar component mainly representing PAH emission ({\it bottom left}) and its unsharped masked version 
({\it bottom right}) where an extra \coone ~contour at 20\,$\kkms$ is added.
In the bottom left corner of each panel a scale bar representing 500\,pc and the CLEAN beam of the CO(1-0) data as well as the resolution of the respective dataset are shown.
\label{fig:pah_2d}}
\end{figure}

Our stellar continuum corrected IRAC images at 3.6\,$\mu$m and 8\,$\mu$m contain two non-stellar emission 
components, namely PAH emission features and hot dust continuum. The 3.6\,$\mu$m image covers the 
3.3\,$\mu$m PAH feature while the 8\,$\mu$m image captures emission from the prominent PAH features at 
7.7 and 8.6\,$\mu$m. Comparison of the stellar continuum corrected 8\,$\mu$m (Fig. \ref{fig:pah_2d} {\it bottom left}) 
and 3.6\,$\mu$m images (Fig. \ref{fig:pah_2d} {\it top right}) to the distribution of the CO emission 
shows that the non-stellar 3.6 and 8\,$\mu$m emission is co-spatial with the CO emission in the ring and 
at smaller radii. Neglecting the nucleus (where
the AGN could contribute to or even dominate the emission seen), very
bright 3.6\,(8)\,$\mu$m non-stellar emission is found in the northwestern
part of the ring close to the brightest CO emission in the ring. As
this is along the direction of the radio jet, it is not clear if this
is just a coincidence. 

The spatial coincidence between non-stellar 3.6\,(8)\,$\mu$m and CO emission 
becomes less perfect along the spiral arms. The brightest emission 
regions at 3.6\,(8)\,$\mu$m are just situated outside the molecular spiral arms.  However, a roughly 
one order of magnitude fainter non-stellar 3.6\,$\mu$m emission ($\sim
0.15\,MJy\,sr^{-1}$) is associated with a significant
portion of the molecular spiral arms. The fainter 8\,$\mu$m PAH emission 
in the inter-arm region exhibits a filamentary structure connecting the two spiral
arms. We find that our CO emission aligns well with the brighter
regions, especially when looking at an unsharped version of the
8\,$\mu$m PAH image (Fig. \ref{fig:pah_2d} {\it bottom right}). The
non-detection of CO emission from the faintest filaments in the
inter-arm region might thus be due to our limited CO sensitivity. 
Overall, the spatial distribution of the PAH emission seems a
very good predictor of the CO emission while deviations are seen in
regions of massive star formation.

The high surface brightness peaks seen in both the non-stellar 3.6 and 8\,$\mu$m emission 
along the spiral arm (Fig. \ref{fig:ism_pol} {\it bottom left and right}) correlate very well with
the peaks present in the 24\,$\mu$m map (Fig. \ref{fig:ism_pol} {\it
middle left}) indicating that in addition to the PAH
features hot dust could contribute. The very similar radial profile of 
the polar cross-correlation of the non-stellar 3.6\,$\mu$m emission with the CO emission (Fig. \ref{fig:ism_cross}, 
{\it bottom left}) supports this notion. The cross correlation profile of the non-stellar 8\,$\mu$m emission (Fig. \ref{fig:ism_cross} {\it bottom right}) exhibits a different shape with an azimuthal offset typically 
smaller than the one seen in the hot dust emission
(see \S \ref{subsec:hotdust_data}). The average offset is only 
$\Phi \sim 15\dg$ and shows much less pronounced variations with 
radius in the spiral arm region of $r\,>\,55\as$.
The difference in the cross-correlation profiles of the non-stellar 3.6 and 8\,$\mu$m 
emission can be explained with different amounts of hot dust contribution. 
The distribution of the faint PAH
emission is more obvious in the 8\,$\mu$m PAH emission due to better
sensitivity (Fig. \ref{fig:ism_pol} {\it bottom right}). The faint PAH
emission follows very well the gas spiral arms and the CO distribution
in the ring. The gaps seen at 3.6\,$\mu$m in the spiral arms at roughly $\rm r \approx 35\as$ are not 
obvious at 8\,$\mu$m suggesting that the gaps are due to a lack of sensitivity and 
not necessarily a lack of PAH emission. It is interesting to note that the spiral arms are wider 
in the non-stellar 8\,$\mu$m emission than seen in CO, specially in the radial range where
prominent star formation is present.

Both non-stellar components at 3.6 and 8\,$\mu$m
have two broad relations with CO emission with slightly differing slopes
and a prominent vertical offset in PAH intensity (Fig. \ref{fig:pah_pix} {\it two top
rows}). However, there is no obvious dividing line between these two
relations. As shown by \cite{meidt12} the 3.6\,$\mu$m non-stellar
emission can be a combination of PAH features and hot dust,
especially, in regions of active massive star formation, thus a single
relation is not necessarily expected. As the upper relation is most
prominent in the central region (Fig. \ref{fig:pah_pix} {\it middle
left}), this strongly suggests that a contribution from hot dust is at least adding 
to this relation. The ratio of 24\,$\mu$m and non-stellar 8\,$\mu$m emission 
(Fig. \ref{fig:pah_pix} {\it  lower middle row}) exhibits also a steeper slope in the center,
consistent with enhanced hot dust contribution to the non-stellar 8\,$\mu$m
light. Basically no relation between the ratio of 
24\,$\mu$m and non-stellar 8\,$\mu$m and the CO emission is seen
for the spiral arms, suggesting that this might be indeed genuine PAH
emission that correlates with dense molecular gas. As the bulk of points in the
inter-arm region corresponds to higher PAH intensities than in the arm region, we speculate
that both regimes - hot dust mixed with PAH as well as PAH
only - are present, although, on a significantly lower level than observed for the center.  
The ratio of the non-stellar 8\,$ \mu$m/3.6\,$\mu$m emission
shows some dependence on CO emission as a function of galactic environment. 
The smaller scatter of this relation in the central disk could be interpreted as being 
dominated by a (constant) hot dust component, while in the spiral arms the scatter 
in this ratio becomes large reflecting the large range of physical conditions  
sampled in this region. The larger scatter implies
either a more varying hot dust contribution or an intrinsic variation of the
ratio of the PAH features sampled by the 3.6\,$\mu$m and 8\,$\mu$m
filters.

%The metallicity dependency in the 8/24\,$\mu$m ratio observed in a
%sample of nearby galaxies has been interpreted as a combination of PAH
%emission disappearance for low metallicities while a low level hot
%dust component becomes more prominent for the 8\,$\mu$m band
%\citep{engelbracht05}.

\subsection{Comparison of CO(1-0) to Star Formation and Stellar Tracers}\label{subsec:sf}

Comparison of the morphology and distribution of the molecular gas
which is viewed as the fuel for star formation \citep[e.g.][]
{bigiel08,bigiel11,leroy08,schruba11} to tracers sensitive to star formation
with varying age sensitivity is illustrative to study the relation
between the fuel and the end-product on cloud-scales. The H$\alpha$
emission from HII regions traces the youngest star forming sites
between a few and about 10\,Myr \citep[e.g.][]{whitmore11}. The
thermal radio continuum should trace the same population of stars
responsible for the ionization of the HII regions, however, it is basically
unaffected by the presence of dust
\citep{condon92,murphy11,murphy12}. The B and V band light from young
stellar clusters probes ages out to several 10\,Myr \citep[for results
on M\,51, see][]{chandar11,bastian05,scheepmaker07,scheepmaker09,haas08,hwang08},
similarly the far- and near-UV 
emission can arise from main sequence stars as late as A type, again
being sensitive to ages of up to $\sim$100\,Myr. Finally, the H band
and 3.6\,$\mu$m filters are most sensitive to a wavelength range where
the old stellar population consisting of red giants dominates the
spectral energy distribution (in the absence of red supergiants and/or
AGB stars which have typical lifetimes of a few 10 to 10$^3$\,Myrs and
should also exhibit a clustered distribution).

\subsubsection{HII regions} \label{subsec:hii_data}

Most of the H$\alpha$ emission is closely
associated with the molecular gas (Fig. \ref{fig:sf_2d}, {\it top left}). In
the center the H$\alpha$ emission coincides with the CO emission
out to a radius of $\rm r \sim 35\as$ (Fig. \ref{fig:sf_pol},
{\it top left}).
A partial gap is present adjacent to the inner gas
disk and continues out to a radius of $\rm r \sim 41\as$. After this
gap the H$\alpha$ emission tends to be offset toward the convex side of the CO spiral arm by
about (9 - 18)$\dg$ at a given radius. It is interesting to note that
the H$\alpha$ emission is not forming a straight line but is
oscillating randomly as a function of radius within this offset 
range.\footnote{The application of the polar cross-correlation was 
not successful due to the patchiness of the H$\alpha$ emission.}
Particularly, along the spiral arms the H$\alpha$
morphology exhibits large diffuse envelopes and shells signaling the
presence of giant HII regions \citep[see
also][]{lee11,gutierrez11}. These very bright and large HII regions
tend to overlap with the gas spurs emanating from the gas spiral
arms. Faint H$\alpha$ emission is also detected in the inter-arm
region with an apparent preference for being mid-way between the two
spiral arms.

There is no obvious relation between the H$\alpha$
emission related to HII regions and the CO emission when comparing it on a
pixel-by-pixel basis in the full PAWS FoV
(Fig. \ref{fig:sf_pix}, {\it top row}). A closer inspection of the individual
region shows that H$\alpha$ emission in the central disk exhibits some complex
relation, while we see basically no relation between H$\alpha$ and CO line emission
in the arm and inter-arm region.

\subsubsection{(Thermal) radio continuum} \label{subsec:T_data}

Unlike the radio
continuum at longer wavelengths (see \S \ref{subsec:ism}), the 3.6\,cm radio continuum
is distributed in a more patchy fashion and dominated by bright emission peaks
(Fig. \ref{fig:sf_2d}, {\it top right}). This suggests that the thermal
contribution in these regions is higher. The radio jet is still the
brightest feature seen in the central part. The continuum emission is
co-spatial with the CO emission in the central disk.
Diffuse, faint 3.6\,cm radio emission is associated
with the gas spiral arms (Fig. \ref{fig:sf_pol}, {\it top
right}), while the bright peaks along the spiral arms show
a similar offset from the CO emission as the H$\alpha$ emission. About
half of these peaks can be directly related to the brightest HII
regions. The polar cross-correlation with the CO emission (Fig. 
\ref{fig:sf_cross}, {\it top left}) gives a very similar radial profile to the one seen
for the 20\,cm continuum with a similar azimuthal offset of $\rm \sim 10\dg$.
The lack of strong 3.6\,cm emission within the CO spiral arms implies that no 
strong star formation is hidden in the molecular gas spiral arm. 
 
It is interesting to note that the
non-thermal/thermal separation of their 8\,$\as$ resolution 6 and
20\,cm radio imaging led \cite{tilanus88} to conclude that the thermal
emission solely arises from giant HII complexes without any evidence
for such giant HII complexes being hidden in the dust lanes. They did
not rule out the existence of significantly smaller HII regions based
on the sensitivity achieved in their radio maps. Given the fact that the radio imaging used
here has an rms 2.5$\times$ and 11$\times$ lower while achieving a
16$\times$ and 35$\times$ higher spatial resolution (in beam area) at
6 and 20\,cm, respectively, and that no prominent bright peaks have
appeared at our resolution inside the spiral arms suggests that not
much hidden massive star formation is occurring inside the gas spiral
arms.

Similar to the 6 and 20\,cm
radio continuum emission, the 3.6\,cm radio continuum shows three
broad relations with the CO emission in the pixel-by-pixel diagrams 
(Fig. \ref{fig:sf_pix}, {\it 2nd row from top}):
 a) a large range in radio flux density for a fixed low CO
intensity, b) a steep relation, and c) a possibly very
shallow or even no relation between (low) 3.6\,cm flux density and CO
intensity. The explanation is analogous to the 6 and 20\,cm radio
continuum: the radio jet is causing the large range of radio flux for
a given CO intensity in the center, while the 
steeper relation that is mainly present in the center is related to the larger amount
of star formation whereas the very shallow one is exclusively seen in the arm and
inter-arm regions and can be interpreted as non-thermal emission (see
discussion \S \ref{subsec:CR}).

\subsubsection{Young stellar clusters and stars} \label{subsec:ysc_data}

The presence and location of
young stellar clusters relative to the molecular gas is best seen in
the HST B and V bands (Fig. \ref{fig:sf_2d}, {\it middle panels}). The
stellar light is significantly stronger in the center reflecting the
presence of a stellar bulge. Prominent dust lanes are seen in both the
B and V band image tracing the spiral arms as well as an intricate web
in the center. As the CO emission very well follows these dust lanes, 
a clear anti-correlation with the distribution of young stellar clusters is seen 
(Fig. \ref{fig:sf_pol}, {\it middle left}). 
As the correspondence of CO emission to dust lanes is also very good in the center, dust 
lanes appear to be a superb predictor
of the location of CO emission. 

The B band emission and CO spiral arms appear to converge
and meet at a radius of $\rm r\sim 78\as$. There is a gap in the B
band emission at roughly $\rm 32\as < r < 42\as$ that is more
pronounced along the southern arm while along the northern arm a few
small stellar clusters are present. The distribution of stellar clusters
is significantly enhanced along the convex sides of both gas spiral arms. 
The distribution is wider and brighter along the southern spiral arm while
more clustering of stellar clusters is seen along the northern
spiral arm. Overall, the width of the arms as seen in the (young)
stellar clusters appears significantly wider than the CO
arms. Similarly to the H$\alpha$ emission, there are a few prominent
strings of stellar clusters mid-way between the spiral arms in the
inter-arm regions. 

The azimuthal offset between the bright
CO and B band emission is larger beyond a radius of $\rm r\approx
55\as$ and decreasing till a radius of $\rm r \approx 80\as$. (The negative offset
for radii $\rm r\le45\as$ is similar to that seen for the old stellar component and 
discussed there, see \S \ref{subsec:sfo_data}.) The V
band emission (Fig. \ref{fig:sf_pol}, {\it middle right}) shows a very
similar picture to the B band and has also gaps. It is interesting to
note that the V band emission is basically anti-correlated with the
24\,$\mu$m emission along the northern spiral arm. The overall
brightness in the V (and B) band falls rapidly out to a radius of $\rm
r \sim 32\as$ before leveling out to the disk value. The polar 
cross-correlation analysis of the V band continuum with the CO emission 
(Fig. \ref{fig:sf_cross}, {\it top right}) reveals a negative azimuthal offset 
for radii of $\rm r\,<\,42\as$ that becomes a positive azimuthal offset 
beyond $\rm r\,>\,50\as$ showing a similar radial dependence as the 
hot dust emission (Fig. \ref{fig:ism_cross}, {\it middle left}), however, 
with a much less pronounced amplitude, i.e. a smaller azimuthal offset.

Ultraviolet continuum arises from stars with ages
less than $\sim$100\,Myrs and is thus a good proxy for recent star
formation. The GALEX FUV and NUV images (Fig. \ref{fig:sf_2d}, {\it bottom
panels}) exhibit a bright, slightly patchy central disk with CO
emission often falling into regions of less bright UV
emission. The extent of the central UV disk is larger
than the CO disk.  In addition two prominent spiral arms are
present. Interestingly, these UV spiral arms do not connect to the
central disk but show a broad gap. The UV light along the spiral arms
is offset towards the convex side of the spiral arms seen in CO emission and 
shows several emission peaks that
are embedded in a fainter smooth component that is best visible in the
NUV continuum along the southern arm.

As is the case for the HII
regions there is no obvious relation between the emission from young
stellar populations (using the HST B band as a proxy) and the CO
emission (Fig. \ref{fig:sf_pix}, {\it middle row}). The large range in B
band fluxes in the pixel-by-pixel diagram for the center reflects the
presence of the stellar bulge. As the emission from the stellar disk
and individual stellar clusters is more homogeneous in intensity, the
range is much lower in the arm and inter-arm regions. It is interesting to
note that the anti-correlation seen in the images is reflected in the negative relation
between HST B band flux density and CO flux present in the arm region, i.e. for a 
higher CO flux less B band light is observed.

\subsubsection{Old stellar population}\label{subsec:sfo_data}

The old stellar population should be a
good proxy for the underlying stellar potential and thus the possible
driving force of the observed CO distribution. The HST H band image
exhibits a bright bulge that has roughly the extent of the molecular
disk (Fig. \ref{fig:sfo_2d}, {\it left}). Bright stellar clusters are
plentiful along the spiral arms, similar to their distribution in the
HST B and V band images (Fig. \ref{fig:sfo_2d}, {it middle panels}). 
Given the possibly large contribution from
younger stellar clusters, the contaminant-corrected 3.6\,$\mu$m map of
\cite{meidt12} provides a better view of the old stellar population
alone and its spatial relation to the molecular gas
(Fig. \ref{fig:sfo_2d}, {\it right}). The nuclear bar \citep[e.g][] {zaritsky93} 
is easily identified
and sits inside the molecular gas ring. The molecular gas ring, or
more precisely the tightly wound CO spiral arms, appear to start at its
tips. Both spiral arms are also visible in this image implying that
the gravitational potential should have at least one m=2 mode where
the molecular gas spiral arms are. The southern spiral arm is about
10-20\% brighter than the northern arm. Assuming a constant
mass-to-light ratio this would imply that the southern arm is about
1.1-1.2$\times$ more massive than the northern arm. The peaks
evident in the contaminant-corrected 3.6\,$\mu$m image are likely
artifacts due to imperfect correction in the regions of most intense
star formation \citep{meidt12}. 

In the polar representation a slight dip in H band surface intensity 
is evident at the location of the CO spiral arm (Fig. \ref{fig:sf_pol}, {\it bottom 
left})\footnote{The black points are artifacts caused by hot pixels.}. 
This suggests that even the H band light is affected by extinction there. 
The contaminant-corrected
3.6\,$\mu$m image (Fig. \ref{fig:sf_pol}, {\it bottom right}) exhibits at
small radii a sinusoidal profile which reflects the
geometry of the inner bar out to a radius of $\rm r \sim 17\as$ at an
angle of, e.g., $\rm \theta \sim 185\dg$, corresponding to a PA (measured 
north through east) of $\rm \sim 147\dg$ close to the value of $\rm PA = 139\dg$ 
for the nuclear bar derived by \cite{menendez07}.
A second modulation in the
radial range of $\rm 25\as \le r \le 50\as$ is prominent at, e.g., an
azimuthal angle of $\rm 40\dg \le \theta \le 160\dg$. This modulation
has a very wide peak ($\rm \theta \sim 100\dg$). The width of this
modulation decreases significantly at radii larger than $\rm r \approx 50\as$.
The most likely explanation is that in addition to the spiral arms (which start 
beyond  $\rm r \approx 50\as$) another component of the galactic potential, like
a wide oval, is present in addition to the spiral
arms. As the width of the modulation at the largest radii probed is
significantly smaller, the spiral arms likely affect
a small part of the disk at a given radius. 

As already seen for the polar cross-correlation of the HST B band, both the 
HST H band and non-stellar continuum corrected 3.6\,$\mu$m emission 
(Fig. \ref{fig:sf_cross}, {\it bottom panels})
exhibit a negative azimuthal offset of $\Phi \sim 5\dg$ compared to the 
CO emission for radii of $\rm r\,<\,42\as$. This negative offset can be easily understood,
as these radii are inside the corotations of the nuclear bar and the spiral arms 
(plus the potential oval) and therefore it is expected that the gas response is leading
the gravitational potential of these non-axisymmetric components that are most prominent
in the old stellar light. Beyond radii of $\rm r\,>\,50\as$ 
the profile is consistent with no azimuthal offset. We cannot exclude that 
there is a residual effect from light contributed by the younger stellar 
population that leads to the almost exact zero offset along the spiral 
arms as the azimuthal offset decreases from the B band to the stellar 
3.6\,$\mu$m to the H band emission.

As expected there is, in general, no correlation between CO emission and 
the light of the old stellar population as traced
by their H band and contaminant-corrected 3.6\,$\mu$m emission
(Fig. \ref{fig:sf_pix}, {\it two bottom rows}). Interestingly, the central region
shows evidence for a linear relation between CO flux and old stellar light, however, with 
a second plume of points that shows no correlation with CO flux. A possible explanation
could be excess emission caused by young(er) stars. This seems unlikely as the light distribution
is very smooth in the H band and the contaminant-corrected 3.6$\mu$m images (Fig. \ref{fig:sfo_2d}).
A further argument against contribution from young stars is the pixel-by-pixel diagram of the HST B 
band light vs. CO emission that shows no correlation at all in the center. Thus it seems
plausible that the stellar bulge contributes to an enhancement of CO emission, e.g. via heating.
On the other hand, no relation between the old stellar population and the CO
emission is detected in the two disk regions (arm,inter-arm).

%%%%%%%%%%%%%%%%%%%%%%%%%%%%%%%%%%%%%%%%%%%%%%%%%%
%%%%% section 5 - CO vs. other observables - physical implications/relations

\section{The molecular ISM on cloud-scales}\label{sec:cloud}

In the following we will place our findings in the context of the different galactic environments 
that are probed by the PAWS FoV. The high spatial resolution of 40\,pc of the PAWS data allows 
us to better probe the physical origins of the (cor-)relations seen among the different ISM and stellar 
tracers and the CO emission.

Broadly speaking the energy density fields that affect the ISM 
can be categorized in radiation fields, cosmic rays, kinetic energy fields (e.g. large-scale turbulence), 
and magnetic fields. We discuss only the relation between the distribution of the molecular gas and 
of the cosmic rays (\S \ref{subsec:CR}) as well as the radiation field (\S \ref{subsec:radiation}), i.e. 
from stars and the AGN. The role of the kinetic energy for organizing 
the molecular gas and the injection of (large-scale) turbulence is discussed in 
companion papers \citep{colombo13a,hughes13a,meidt13}. As no high angular 
resolution, high sensitivity studies of the magnetic field in M\,51 exist, we do not investigate the 
importance of the magnetic field for determining the molecular gas properties. As the gravitational 
potential basically organizes both the 
distribution of the interstellar medium and the energy density fields that affect it, we start with 
a brief description of the different galactic component of the gravitational potential present in the 
PAWS FoV (\S \ref{subsec:potential}). This will provide a guideline when we try to separate the different 
components of the energy density field(s) that affect the ISM.

\subsection{Gravitational potential and the ISM geometry}\label{subsec:potential}

The shape of the gravitational potential determines the distribution of both the ISM and the energy 
density fields, and thus the different galactic environments the molecular gas resides in.
The old stellar population can be considered the best tracer for the underlying
stellar potential \citep{rix93}. Based on the distribution of the old stellar
component (from \S \ref{subsec:sfo_data}) and literature results
\citep[e.g.][]{rix93,zaritsky93,beckman96,lamers02,henry03,menendez07} we identify the
following components in the PAWS field of view (from the center outwards):

\begin{enumerate}
\item[(a)] a central super-massive black hole that is actively accreting \citep[e.g.][and reference therein]{maddox07}
\item[(b)] a bulge with a radial extent of $\rm r\,\le\,16\as$ \citep{fisher10}, and
\item[(c)] a galactic disk that contains

\begin{enumerate}
\item[(i)] an inner nuclear bar with a major axis length of $\rm r\,\sim\,(15-17)\as$ and
an orientation of $\rm PA \sim 139\,\dg$ \citep{menendez07},
\item[(ii)] an indication of an oval with a radial extent of $\rm r\,\sim\,50\,\as$, basically
oriented in north-south direction, and
\item[(iii)] a $m=2$ spiral pattern that is consistent with a spiral density
wave in the inner part and a second spiral pattern that is often referred to as material wave in the outer part
\citep[for details see][]{meidt13}. 
\end{enumerate}

\end{enumerate}

At all galacto-centric radii the molecular gas distribution follows the stellar
spiral arms very well. Even for the section
$25\as\,\le\,r\,\le\,50\,\as$ where we see evidence for a stellar oval
the molecular gas is roughly resembling spiral arms. This radial range
is also where \cite{henry03} and \cite{patrikeev06} found deviations
from a single pitch angle and we see a lack of recent star formation
(see \S \ref{subsec:hotdust_data}, \ref{subsec:hii_data}, and
\ref{subsec:ysc_data}). Further inward the gas distribution is
reminiscent of a starburst ring with two tightly wound spiral arms
which start at the tip of the inner nuclear bar. The gas distribution
along the inner bar reflects the orientation of the nuclear bar. Taken
all together this strongly suggests that the observed molecular gas
distribution is heavily influenced by the galactic potential. In addition, 
kinetic energy, via non-circular motions or large-scale turbulence, will 
be injected in to the ISM altering its properties and relative composition. 
However, a detailed analysis of the dynamical effects is beyond the scope 
of this paper. Various aspects of how the
dynamical environment impacts the properties of the molecular gas 
are studied in a number of companion papers on the probability distribution 
functions (PDFs) of the CO emission \citep{hughes13a}, the GMC properties
themselves \citep{colombo13a} and possible dynamical implications have been 
derived \citep{meidt13}.

It is worth noting that the azimuthal offsets $\Phi$ measured by the polar cross-correlation 
(positive $\Phi$ for star formation tracers and negative $\Phi$ for tracers of the 
underlying stellar potential) are consistent with gas flow that is always inside 
the corotation resonance. This is independent evidence that multiple patterns with 
different pattern speeds must be present in the PAWS FoV.

\subsection{The relativistic ISM}\label{subsec:CR}

Since molecular gas is thought to be the fuel for star
formation, numerous studies have examined the relationship between
CO emission and empirical star formation tracers in galaxies. It is
well-established that the far-infrared emission is correlated with
the CO emission on large scales within galaxies 
\citep[e.g.,][and references therein]{devereux90,young91}, while a
correlation between the total CO and 20\,cm radio continuum (RC)
luminosities of galaxies has also been reported 
\citep{rickard77,israel84,adler91,murgia02}. The correlation between the CO and RC emission has
been shown to exist for surface brightness as well as intensity in
the Milky Way, and for galaxies of diverse Hubble types
\citep[e.g. dwarfs and ULIRGS,][]{adler91,allen92,leroy05,liu10}. Detailed studies on a
few 100\,pc resolution \citep{murgia05,paladino06} have shown that
the relation is linear within galaxies as well as between regions of
individual galaxies. The robust correlation down to small spatial scales
is intriguing since RC emission is only indirectly linked to star
formation. Ionized gas surrounding young high mass stars produces
thermal free-free radiation at radio frequencies, but approximately
90\% of the 1.4GHz emission in a normal star-forming galaxy is
synchrotron radiation emitted by cosmic ray electrons gyrating in
magnetic fields. While non-thermal radio emission is also linked to star
formation, since supernova remnants (SNRs) from short-lived high
mass stars are the primary site for cosmic ray production in
galaxies, less than 10\% of the non-thermal radio emission
is due to discrete SNRs; the remaining emission is from electrons
that escape their parent SNRs and diffuse into the disk and halo,
where they are accelerated by the galactic magnetic field
\citep[e.g.][]{lisenfeld00}.

%If the physical connection
%  between molecular gas and synchrotron radiation is limited to
%  molecular clouds being the birth place of high-mass stars, then a
%  tight relationship between the distributions of CO and 20cm RC
%  emission should not hold on scales shorter than the average cosmic
%  ray diffusion length.\\

In \S \ref{subsec:NT_data}, we showed that the morphology of the RC
emission closely resembles the CO distribution within the PAWS
FoV, even down to spatial scales of $\sim100$ to 200\,pc
(see Fig. \ref{fig:ism_2d}). Moreover, the pixel-by-pixel comparison 
(see Fig. \ref{fig:ism_pix} {\it two bottom rows}) between the
CO and RC emission in M51 reveals two relations with slightly
different slopes ($\sim0.7$ and $\sim0.5$ for the correlations at
20\,cm, and $\sim0.9$ and $\sim0.7$ for the correlations at 6\,cm), which
appear to originate from distinct environments within M51's inner
disk. Compared to the arm and inter-arm regions, the radio continuum
emission at both 20 and 6\,cm in the central region is brighter for a
given CO flux. This relationship is independent of the emission due to
the radio AGN, which is manifested in Fig. \ref{fig:ism_pix} as a cloud of points
that show no relation between radio continuum and CO emission. 

%The existence of distinct local correlations between the RC and CO
%emission within M51 might be due to two different physical mechanisms,
%which combine to produce the global radio-CO correlation in
%galaxies. As the origin of the RC-CO correlation -- both globally and
%for resolved measurements within galaxies -- is not well understood,
%here we review potential explanations for the origin of the
%correlation, and discuss whether they are supported by the trends that
%we observe in M51.\\

Potential explanations for the RC-CO correlation fall into
two broad categories: those in which the correlation is due to a
single stellar population that powers both types of emission, and
those where the correlation arises because the synchtrotron emissivity
is closely linked to the dense gas distribution. Among the former
class of models, \cite{adler91} argued for a direct link between CO
emission and non-thermal radio continuum via cosmic rays produced by
massive star formation. Supernovae inject new cosmic ray electrons
into the ISM leading to enhanced synchrotron emission. At the same
time the accompanying cosmic ray nuclei (i.e. protons) heat the
molecular gas \citep[for a detailed model see,
  e.g.,][]{suchkov93}. \cite{allen92} points out that this would imply
that the observed CO brightness is only sensitive to the amount of
excitation by cosmic rays and not the intrinsic gas density. As
discussed by \cite{murgia05}, moreover, the observed range of CO
intensities would imply gas temperatures above 1000\,K if this
scenario is applied. Our finding that the radio continuum in M\,51 shows
no spatial offset from the CO emission -- unlike the hot dust and
H$\alpha$ emission, which presumably trace sites of massive star
formation -- also argues against this model.

Another possibility is that the good correlation between RC
and CO emission in M51's spiral arms is due to enhanced synchrotron
emission that arises from the secondary cosmic ray electrons produced
in the interaction of cosmic rays with the dense molecular gas
\citep[as proposed by, e.g,][]{marscher78,murgia05,thompson07}. In this
scenario, interactions between cosmic ray protons and \hh\ molecules
produce (negatively) charged pions whose decay produces secondary
(cosmic ray) electrons. \cite{thompson07} calculate that for gas
surface densities of $\rm \Sigma_{gas} \sim 0.03\,-\,0.3\,g\,cm^{-2}$
(i.e. $\rm \approx 150\,-\,1500\,M_{\odot}pc^{-2}$) the timescale for
cosmic ray protons to lose all their energy is significantly shorter
than the diffusion timescale to leave the medium. As the observed
\hh\ gas densities in M51 are in this range, this might be a plausible
explanation for the very good spatial correlation between CO emission
and radio continuum in the spiral arms, although the smoother
appearance of the radio continuum would point toward some diffusion of
the cosmic ray electrons from their production sites.  
While the global spectral index for the entire disk of NGC\,5194 is $\alpha_{20-6} = 0.9$ 
\citep[for $\rm S \sim \nu^{-\alpha}$; ][]{dumas11},  there are pronounced differences at 
15$\as$ resolution with an average $\langle \alpha_{20-6} \rangle$ of about 0.6, 0.8 
and 1.0 for center, spiral arm and inter-arm region, based on an spectral index map 
presented by \cite[][; their Fig. 7b]{fletcher11}. 
%This value of
%$\alpha$ is shallow compared to the index expected for emission from a
%CR population that is entirely cooled via synchrotron emission, and
%the trend between the arm and inter-arm region is opposite to what
%would be expected (i.e. stronger cooling in environments with higher
%gas densities). 

Since the contribution of secondary electrons to the
total RC emission (and the ratio between the total RC and CO emission)
depends sensitively on parameters such as the efficiency of energy
transfer from supernovae to CRs, the ratio of primary electrons to
nuclei in CRs, and the fraction of CO emission that arises from dense
rather than diffuse molecular gas, it is also uncertain whether such a
model can explain why such a tight linear RC-CO correlation is
observed for galaxies with a diverse range of Hubble types. Further
analysis, e.g. estimates for the dominant CR cooling processes and
timescales for the typical ISM conditions (gas densities, radiation
field, magnetic field strength) in different M51 environments, and a
detailed comparison between the RC-CO correlation and the non-thermal
radio spectral index, would be required to test whether secondary
electron production in molecular clouds helps to explain the close
relationship between CO and RC emission in some regions of the PAWS
field. It is interesting to note that attempts to detect non-thermal radio continuum 
caused by secondary CR electrons from
Galactic clouds has not been successful so far \citep[e.g.][and references therein]{protheroe08}.

%This would imply that the molecular gas in M\,51 is acting
%like a 'cosmic ray proton calorimeter' \citep{thompson07}, i.e. the
%cosmic ray cooling time is much shorter than their escape time
%\citep[calorimeter theory, see][]{voelk89}. 
%\cite{paglione12} show that the $\gamma$-ray emission from M\,82
%and NGC\,253 are consistent with such a model and the observed 
%molecular gas density in these two galaxies. They suggest that the 
%product of gas volume density and
%cosmic ray diffusion length is constant, thus for the derived \hh\
%~densities of \nhh of $\rm 100-300\,cm^{-3}$ in the spiral arms of M\,51 
%\citep{schinnerer10} this would imply diffusion lengths of $\rm 30-10\,pc$ 
%well below the median derived effective size of GMCs in M\,51 of $\rm 40\,pc$
%\citep{colombo13a}.

An alternative explanation for the RC-CO correlation is that
the enhanced radio continuum emission is due to a coupling between the
gas density $\rho$ and magnetic field strength $B$, via $B \propto
\rho^{\beta}$. That is, since the flux density of the non-thermal radio
emission $S_{\rm 1.4,nth}$ at 1.4\,GHz may be written (e.g. Hoernes et al 1998) as

\begin{equation}
S_{\rm 1.4,nth} \propto n_{\rm CR}l_{\rm CR}B^{1-\alpha_{\rm nth}},
\label{eqn:nthemission}
\end{equation}

an increase in the RC emission may be linked to an increased magnetic
field strength ($B$), rather than an increase in the number of
CR electrons. In this equation, $l_{\rm CR}$ is
the path length through the synchrotron-emitting region along the
line-of-sight, and $n_{\rm CR}$ is the volume density of cosmic ray electrons
with energy between $E$ and $E + \delta E$, for an injection cosmic
ray energy spectrum $N(E) = N_{0}E^{-p}$, where $\alpha_{\rm nth} =
-p/2$. The expression in Equation~\ref{eqn:nthemission} is fairly
general, as it allows for variations in the scale-height of the
synchrotron disk, and does not assume a relationship between $B$ and
$n_{\rm CR}$. 
%On large scales, equipartition between the magnetic and
%cosmic ray energy densities and a constant $l_{\rm CR}$ are often assumed
%\citep[e.g.][]{niklasbeck97}. In this case, $n_{\rm CR} \propto B^{2}$ and
%$S_{\rm 1.4, nth} \propto B^{3-\alpha_{\rm nth}}$
%\citep{pacholczyk70}.\\

As an explanation for the local RC-CO correlation, the
advantage of $\rho$-$\beta$ coupling models is that the physical
mechanism responsible for the coupling between the magnetic field and
gas volume density is truly local in nature, involving a relationship
between the magnetic field and charge density within (nearly) neutral
gas. Observations indicate $\beta\sim0.5\pm0.1$ globally and within
galactic disks, even on the scale of individual clouds
\citep[$\sim$100\,pc,
  e.g.][]{fiebig89,berkhuijsen93,niklas97}. This range
of $\beta$ values is in good agreement with results from turbulent MHD
simulations \citep[$\beta=0.4$ to 0.6,
  e.g.][]{ostriker01,cho00,groves03} and with
equipartition between magnetic and turbulent energy densities in the
ISM \citep[$\beta=0.5$, e.g.][]{ko89}. 
%Dynamo theory and static
%pressure equilibrium between the magnetic field and the interstellar
%gas also predict $\beta=0.5$ \citep{becketal96,kulkarniheiles88}. In
%principle, other scalings are possible, and may provide evidence for
%less common ISM processes: $\beta\sim1$ is expected for
%shearing/compression of magnetic fields in diffuse gas, for example,
%while $\beta\lesssim0.4$ could occur for a cosmic ray population
%dominated by strong synchrotron/inverse Compton losses
%\citep{beckkrause05}. 
Regardless of $\beta$'s exact value, $B-\rho$ coupling yields

\begin{equation}
S_{\rm 1.4,nth} \propto n_{\rm CR}l_{\rm CR}\rho^{\beta(1-\alpha_{\rm nth})}
\end{equation}

for the correlation between the non-thermal 1.4\,GHz emission
and the gas volume density. The gas surface density $\Sigma_{\rm gas}$
is simply the projection of the volume density through the disk, so for
a gas disk with constant scale height $l_{\rm gas}$ we would expect to
observe a power-law scaling with a similar exponent, i.e.

\begin{equation}
S_{\rm 1.4, nth} \propto n_{\rm CR}l_{\rm CR}\Sigma_{\rm gas}^{\beta(1-\alpha_{\rm nth})}.
\label{eqn:nthobserved}
\end{equation}

Since the interstellar gas in M51's inner disk is almost
entirely molecular, we can replace $\Sigma_{\rm gas}$ in
Equation~\ref{eqn:nthobserved} with $I_{\rm CO}$. The relationship
between the surface density of CRs, $N_{\rm CR}=n_{\rm CR}l_{\rm CR}$, 
and $\Sigma_{gas}$ is somewhat uncertain \citep[see
e.g. the discussion in][]{murgia05}. In general terms, regions
with high gas surface density are linked to regions with high levels
of star formation activity. On one hand, high star formation rates
would tend to increase the supply of cosmic ray electrons, but on the
other hand may lead to higher rates of convective escape. Assuming
that the surface density of CRs $N_{\rm CR}$
is independent of $\Sigma_{\rm gas}$, then for $\beta = 0.5$ and
$\alpha_{\rm nth} = -0.7$, we obtain a slightly sub-linear RC-CO
correlation, i.e. $S_{\rm 1.4, nth} \propto I_{\rm CO}^{0.85}$, in
general agreement with our pixel-by-pixel correlations in
Fig. \ref{fig:ism_pix}. The different slope and vertical offset between the RC-CO
correlations in the center and disk of M51 could be due to the higher
level of ongoing massive star formation in the central region. That
is, fresh cosmic ray electrons from recent supernovae increase $N_{\rm
  CR}$ \citep[e.g.][]{tabatabaei07}, while the shallower slope is
consistent with the shallower radio spectral index in the central
region (which itself may be due to a higher rate of convective escape
via a galactic wind, or the fact that the CR electrons are still
relatively 'warm'). The presence of a few points scattered along the
shallower relation in the spiral arm plot is consistent with this
interpretation, since the brightest radio continuum peaks coincide
with the brightest 24\,$\mu$m emission (see \S \ref{subsec:NT_data}).

The detailed analysis of the magnetic fields and radio continuum emission 
in M51 by \cite{fletcher11} found that compression of the magnetic field
by a spiral arm shock would significantly over-predict the observed moderate
contrast between arm and inter-arm region. Therefore they concluded that 
no strong spiral arm shock is compressing the magnetic field, consistent with
the analysis of \cite{meidt13} who, based on a kinematic analysis of the CO emission,
find no evidence for a strong dynamical shock. However, higher angular resolution
and more sensitive radio continuum observations as now possible with the up-graded
Jansky Very Large Array would be required to further test if compression of the magnetic
field is indeed not important even on the small scales probed by our CO data.

\subsection{Radiation field}\label{subsec:radiation}

The radiation field within the PAWS FoV of M\,51 contains - broadly speaking - three main 
sources: the interstellar radiation field (ISRF) dominated by the old stellar population, young 
active star forming regions with a significant energy increase at short wavelengths (i.e. UV 
continuum) and the AGN itself that displays a jet feature \citep[e.g.][]{maddox07}. The spatial 
distribution of these three components varies not only radially, but also with respect to the 
geometry of the non-axisymmetric components (e.g. bar, spiral) of the gravitational potential.

The impact of radiation onto molecular gas is typically described in the form of a 
photo-dominated (or -dissociation)
region (PDR) that is nowadays defined as a region where far-UV photons dominate the energy 
balance or chemistry of the gas \citep[e.g.][]{tielens08}. Thus PDRs are not only associated 
with HII regions, but also the diffuse ISM, i.e. anywhere there FUV photons penetrate. PDRs in 
a broad sense are the transition region from the ionized to the molecular phase of the gaseous 
ISM. Energetic UV photons from nearby hot stars (e.g. within a HII region) cause 
photo-ionization and -dissociation in the outer layer of a molecular cloud. Due to attenuation of these 
photons by atoms, molecules and dust grains the following (stratified) structure exists from 
outside to inside, i.e. lower to higher gas densities: (a) a H$^+$/H transition zone followed by 
an atomic HI layer, and (b) the HI/\hh ~transition zone after which Hydrogen is molecular, 
within the molecular \hh ~gas the transition from ionized Carbon to CO (CII/CI/CO) occurs at 
slightly higher gas densities. However, the structure of GMCs is far from being simple spheres, recent 
Galactic observations show that GMCs rather contain filaments of varying density 
structures \citep[e.g][]{andre10,pety00} consistent with expectation from the ISM's 
turbulent nature \citep[e.g.][]{hennebelle07,hennebelle08}. Therefore it is observationally very challenging
on the size scales considered here to discriminate between GMCs that are full PDRs \citep[like in the center of Maffei\,2;][]{meier12} or only have PDR surface layers \citep[as in the center of IC\,342;][]{meier05}.
(A more detailed overview on the heating and cooling processes occurring in the molecular gas 
is provided in Appendix \S \ref{sec:co}.)

\subsubsection{The old stellar population (ISRF)}

The ISRF is mainly caused by the old underlying stellar population(s). As the stellar density of the 
old population is changing from the bulge to the disk as evident from, e.g., the optical-to-near-IR 
radial profiles \citep{munoz11}, so must the incident radiation field. In the following we discuss the
role of the ISRF component that is dominated by the old stellar population in the relationship between 
the molecular gas and dust emission (i.e. 24$\mu$m, PAH).

{\it Hot dust (Very small grains):} As the radiation of the old stellar population is significantly less energetic than that emitted by hot OB stars
it is not clear how much it can affect the properties of the CO-identified 
molecular clouds that  typically have extinction values of A$_V > 1$ (see \S \ref{subsec:Av_data} for details).
The radial variation in the old stellar population 
can explain the different slopes seen between 24\,$\mu$m emission 
and CO flux. The shallow slope seen in the spiral arms
(Fig. \ref{fig:ism_pix} {\it middle right panel in second row}) can be attributed to diffuse 24\,$\mu$m
'cirrus' emission \citep[for a detailed discussion see Section 5
of][]{leroy12} that is caused by heating from the ISRF. That is we see
the short wavelength tail of the blackbody emission of the cold dust
component that is well-mixed with the molecular gas and not hot dust
heated by young massive stars. This immediately implies that in the
center and in certain areas of the spiral arms the 24\,$\mu$m emission
is 'over'-luminous compared to the CO emission.  \cite{munoz09}
studied the radial profile of the dust emission in the SINGS sample,
including M\,51. They find that M\,51's surface density of the dust
luminosity and of the derived dust mass vary independently with
radius, with the dust in the central 2\,kpc being significantly more
luminous for a given mass surface density (their Fig. 7.44). Young star 
forming regions are very good candidates for causing this 'over'-luminosity.

However, the situation in the central disk is more complex. While it is very likely that the ongoing
massive star formation (providing significant UV photons) is the main heating source for the dust,
and thus the enhanced 24$\mu$m emission, in the molecular ring, it is not obvious what is 
causing the excess emission inside the molecular ring (best seen at 70$\mu$m in 
Fig. \ref{fig:ism_2d} {\it middle right panel}). This enhanced 24(70)\,$\mu$m emission is
indicative of dust hotter than that seen in the spiral arms and indeed
\cite{mentuch12} find that the dust temperature in the center is about
25\,K, only slightly lower in the (dust) spiral arms and about 3-4\,K colder
in the disk unaffected by recent star formation. The increase in 
dust temperature inside the molecular ring could either come from an increase in
stellar UV emission or the AGN. Unfortunately, the (F)IR imaging from
Herschel used by \cite{mentuch12} has not sufficient spatial
resolution to spatially resolve the central region. An obvious
candidate for the stellar radiation field would be the older stars that
make up the bulge of M\,51 that dominates over the central disk for
radii $\rm r\,\le\,15\as$ \citep{beckman96}.  \cite{lamers02} derive an
age of more than 5\,Gyrs for the prominent 'smooth, yellowish'
component which they attribute to be the bulge. Recently \cite{groves12} 
showed that the old stellar population in the bulge of M\,31 alone can
explain the observed increase in dust temperature without the need for
any young ($\le 100$Myr) stars. Similarly, \cite{bernard08} observe an increase
in dust temperature at the center of the LMC where the density of old stellar
population is enhanced due to the stellar bar. The relation of
CO emission and emission from the old stellar population (as traced by its H band and 
contaminant-free 3.6$\mu$m emission; see \S \ref{subsec:sfo_data}) shows a
remarkable difference for the center and the disk (arm and inter-arm region): The first region exhibits 
a clear linear relation, while the latter are basically consistent with no correlation. This could potentially
imply that the molecular gas might experience additional heating by the enhanced ISRF caused by the
higher density of old stars.

{\it PAH:} Given the excellent correlation between CO and PAH emission we suggest in the following that the ISRF
that is dominated by the old stellar population is the underlying cause for this correlation. 
Modeling of the infrared spectral energy distribution with 
the \cite{draine07} model using the PACS 70 and 160\,$\mu$m data in conjunction with the 
SPIRE 250\,$\mu$m imaging following the methodology of \cite{aniano12} shows that the ISRF 
in M\,51 is (within a factor of two) about 11$\times$, 7$\times$ and 6.5$\times$ the Galactic value 
in the central, spiral arm and inter-arm region of the PAWS area (B. Groves; priv. comm.), 
so likely sufficient to excite PAH emission.

While the exact excitation channels and conditions for PAH emission
observed in the MIR spectra of nearby galaxies are not well-understood
\citep[e.g.][]{smith07}, it is well-established that significant PAH
emission (similarly to mid-IR continuum arising from heated small
grains) is related to star forming regions ranging from PDRs to HII
regions \citep[see review by][] {tielens08}. Photo-electric heating by
UV photons from these star forming regions or UV photons present in
the ISRF is very likely the main excitation mechanism
\citep{tielens08}. 
\cite{heitsch07} investigated the predictive power of PAH emission for cloud structure and its 
underlying density distribution using a single cloud illuminated by a single O star. They find that 
for diffuse gas where UV photons can penetrate the entire cloud, a good correlation between PAH 
emission and gas density distribution is expected. However, at higher gas densities, UV photons no 
longer illuminate the full cloud but only its 'rims' and thus a mis-interpretation 
of the underlying gas distribution will become possible and most likely. Based on these theoretical 
findings the correlation between PAH and CO emission will depend on both the exact density distribution 
of the (GMC) structure and the distribution of the incident radiation field. 
This implies that a spatial correlation between CO
emission and PAH emission should exist, however, the observed
intensities might not agree, particularly in regions of intense star
formation. 

Our observational results showing a very good spatial relation between
CO and PAH emission corroborate this interpretation. Two of our
findings are interesting in this context. PAH emission is strong even
in regions where no excess UV flux from nearby massive star formation
is present, like the 'gaps' in the spiral arms seen in, e.g., UV
emission (see \S \ref{subsec:ysc_data}). In addition, there is tentative
evidence that the geometry of the CO distribution, especially in the inter-arm region, is
very similar to the geometry of the PAH emission (se \S \ref{subsec:PAH_data}). However,
the pixel-by-pixel comparison of both PAH emission features shows no predictive
power in the disk, i.e. the PAH strength is varying a lot for a given CO flux. Only in the
center, a tighter correlation is present. This points towards a strong link between PAH emission 
and molecular clouds. Although excess UV radiation from
slightly (up to $\sim$ 100\,Myr) older stars could increase the PAH
emission by providing a stronger (local) radiation field, we find no difference between the arm and inter-arm 
region when considering the range of PAH emission probed.
Thus, the absence of a correlation in the disk (arm,inter-arm) suggests
that the PAH emission is not tracing the true gas density, but rather arises from cloud 
surfaces in the disk of M51 while the incident radiation field is not very important or PAHs are readily
destroyed. 
The apparent correlation in the center between PAH and CO emission
is at least partially due to the contribution of hot
dust emission to the non-stellar emission at 3.6\,$\mu$m and 8\,$\mu$m as the ratio of 24$\mu$m
and non-stellar 8$\mu$m versus CO flux exhibits a shallower slope.

\cite{sandstrom10} investigated the spatial relation between PAH
emission and other dust and gas tracers in the Small Magellanic Cloud,
they noted that PAH emission shows the best correspondence with CO
emission tracing molecular gas. They offered two explanations for this
correlation either PAHs are formed in molecular clouds or the
shielding offered by the dense gas environments prevents destruction
of the PAH features. Our good spatial correlation between PAH and CO emission
in all environments underlines the close relation between PAHs and
molecular gas. A similar conclusion has been reached for 3 clouds in
IC10 by \cite{wiebe11}. Taking all this evidence together this strongly
suggests that PAHs might indeed form in molecular gas or at
least the dense phase of the neutral interstellar medium and that the ISRF is illuminating the 
PAHs in the outer shell of GMCs. Reversely,
it directly implies that maps of, e.g., the 8\,$\mu$m PAH emission
have a very good predictive power of the expected spatial distribution
of the molecular line emission traced by CO(1-0), however, not the expected line strength. 
As a corollary this implies that 8\,$\mu$m PAH emission should have a poor predictive
power of the recent star formation rate (at high spatial resolution). It should work fine at  $\sim$kpc
resolution where CO vs. star formation is robust.

\subsubsection{The OB stars (massive star formation regions)}

The distributions of the molecular gas and the sites of ongoing and/or recent 
star formation, as traced by HII regions, 
young stars and stellar clusters show a correlation on the scale of galactic
structures, e.g. the spiral arms and the central gas ring. However,
they do on average not spatially coincide when compared at $\sim 1\as$, i.e. 37\,pc, resolution
(see \S \ref{subsec:hii_data} and \ref{subsec:ysc_data}). 
This is expected as  the strong UV radiation from the young stars and their stellar winds are
believed to quickly disrupt the parent cloud \citep[e.g.][and references
wherein]{hopkins10}. In this context it is interesting to distinguish between 
the star forming sites offset from the dense molecular gas along the gas spiral arms and the star forming ring 
where regions of young stars and dense molecular gas are well-mixed on the 100\,pc 
scale. In the former case the expectation is that one observes the properties of a 
PDR not much diluted by contributions from the cold ISM component whereas the ring 
provides an ideal situation to obtain global properties of PDRs mixed together with cold molecular gas.
	
{\it Spiral arms:}	
The azimuthal offset of all star formation tracers from the CO spiral arms is on average at least 
$\Phi \sim 10\deg$ for radii of $\rm r > 50\as$ (i.e. past the 'gap'; see Fig. \ref{fig:ism_cross}).
This translates into a spatial separation of over 300\,pc for radii beyond 1.84\,kpc between the sites of massive
star formation, and thus OB stars, and the reservoir of molecular gas. Although the impact radius of giant HII regions onto
surrounding GMCs is not very well known, e.g. \cite{wilson97} quote an upper limit of $\rm < 120\,pc$ based on a detailed
study of GMCs with associated HII regions in M\,33, this immediately 
implies that the radiation of young stars is not affecting the bulk of the molecular gas in 
the spiral arms. It is interesting that emission from the expected dissociation products such 
as [CII] and HI is found off the dense gas arms, especially the northern arm 
(see \S \ref{subsec:hicii_data}). Similarly to the [CII] (and HI) distribution, the hot dust 
emission is present in distinct locations also providing further evidence that the UV radiation 
of the young stars is impacting the ISM, as evidenced by the higher dust temperature, 
offset from the CO arms \citep[see Fig. 10 of][]{mentuch12}. The spatial resolution of the 
infrared data is, unfortunately, not high enough to derive exact source sizes. However, 
comparison to the hot dust distribution and the location of the HII regions suggests that the 
impact is fairly localized - consistent with the previous findings by \cite{wilson97}. 
If we take the \ha\ emission as our best predictor of the location where radiative and mechanical 
energy released by young OB stars is impacting the molecular gas, stellar feed-back
should not strongly alter the properties of the bulk of the molecular gas located in the spiral 
arms. However, this does not rule out the existence of individual regions in the disk where stellar feed-back 
has likely significantly impacted the molecular gas, e.g. north-east of the molecular ring inside the northern spiral arm.

{\it Star forming ring:}
In the molecular ring hot young stars and cold molecular gas are very close to each 
other, if not spatially coincident. Therefore it is interesting to test if this massive ongoing 
star formation leaves a significant imprint on the molecular gas. The northwestern ring 
segment is brightest in CO, [CII], and 24\,$\mu$m emission suggesting that star formation 
might have the strongest impact onto the properties of the molecular gas. The spatial 
coincidence of these three tracers means that photo-dissociation and dust heating by 
young stars is occurring. How much this excess in UV radiation is affecting the cold molecular 
gas is not straight-forward to judge. Therefore, it is illustrative to compare the surface density of 
the cold molecular gas to that of the warm and hot \hh\ as derived from Spitzer/IRS imaging  by 
\cite{brunner08}. They find that the \hh\ surface density of both the warm (100-300\,K) and hot 
(400-1000\,K) phase is highest at the northwestern part of the molecular ring. Comparison to the 
cold \hh\ surface density (probed by our CO data and derived assuming a Galactic conversion factor) of 
$\rm \sim 1000\,M_{\odot}pc^{-2}$ shows that the warm phase makes up about 1\% while 
the hot phase is less than 0.02\% of the cold molecular gas surface density. We find for the 
outer northern spiral arm in the PAWS field that is covered by the IRS \hh\ footprint 
\citep[see Fig. 1 of][]{brunner08} similar values within a factor of two. This implies that the OB stars
are not sufficient to heat large fractions of the molecular gas above 100\,K. For instance, using
observations from multiple CO transitions \cite{weiss01}
derive an average kinetic temperature of $\rm T_{kin} \ge 100\,K$ for the molecular gas associated
with the central starburst in the nearby galaxy M\,82. Thus we speculate that most 
of the molecular gas in the star forming ring of M\,51 is still in the cold ($\rm < 100\,K$) phase despite 
the presence of strong UV heating by 
young stars. This would imply that the massive star formation is not significantly altering the 
properties of the cold molecular 
gas. However, a more detailed and proper analysis using multiple CO transitions is required
to firmly test this.

In summary, we find no strong evidence that recently formed massive stars are significantly impacting the bulk 
of the molecular gas reservoir in the PAWS FoV.

\subsubsection{AGN}

There is no strong evidence that the AGN is altering the properties of the cold molecular gas on 
large scales. \cite{scoville98} already noted that the nuclear molecular gas reservoir as seen 
in its CO(1-0) emission line is concentrated in a ring-like distribution with a size of $\sim$\,
100pc. This nuclear gas concentration brightens in higher J transitions of the CO molecule 
indicating higher excitation temperatures \citep{matsushita04}. Strong HCN(1-0) line emission 
is detected from this reservoir \citep{kohno96,helfer97} further indicating that the AGN is 
altering the molecular gas properties here. Based on a detailed kinematic analysis of high angular 
resolution CO(2-1) data, \cite{matsushita07} suggested that the molecular gas is entrained 
by the jet. 

A jet-like structure has been observed emanating from the nucleus at X-ray and radio 
wavelengths with an extent of about 6$\as$ to the south and 13$\as$ to the north \citep[e.g.][]
{crane92,terashima01,maddox07}. \cite{ford85} suggested based on optical line ratios and 
their unusual width that the line emission is caused by shocks where the bi-directed jet interacts 
with disk material. This interpretation is consistent with the properties of the X-ray gas and the 
radio continuum that are both suggesting a non-stellar source for their excitation 
\citep{maddox07}. Careful comparison of the distribution of the CO line emission, the optical 
extinction and the 20\,cm radio continuum (Fig. \ref{fig:ism_2d}) shows that the CO emission is 
closely following the optical dust lanes that are reminiscent of dust lanes regularly found along 
the leading side of large-scale bars. However, no obvious spatial correlation to the jet 
geometry (as seen in its 20cm continuum) is found. We see a lack of CO emission at the 
location of the southern bubble which could be interpreted as the jet causing dissociation of 
molecular disk material. Targeted sensitive molecular line observations
would be necessary to judge if the jet has indeed an impact on the molecular gas in the disk. 
No evidence for an impact of the jet is found outside a radius of $\sim$6'' and off the 
jet-features. Thus we conclude that the jet is basically not affecting the molecular gas in 
the central region despite its extra X-ray and UV emission contribution to the radiation field.

It is worth noting that \cite{brunner08} find that the hot (400-1000\,K) \hh\ gas is peaking at 
the nucleus. Interestingly, the nuclear distribution of the hot gas phase is elongated along the 
radio jet axis in the northern direction. The authors speculate that heating by the AGN is causing 
this excess emission.

\subsection{The relation between molecular gas and (cold) dust}

The excellent correspondence between the distribution of the CO
emission and the optical extinction as traced by the $I-H$ color
implies that the molecular gas and cold dust are at least co-spatial and very likely 
well-mixed. This is in
agreement with the large body of work on Galactic molecular
clouds. Previous work \citep{bohlin78,lombardi06} found that for extinction values
below $\sim(1-2)\,mag$ no CO emission is detectable (on GMC scales), in agreement with
the expectations from models for photodissociation of the CO molecule
\citep[e.g.][]{dishoeck88} where the shielding of CO requires certain
column densites. Studies of resolved Galactic clouds using extinction
measures of $\av$ from individual stars with NIR colors typically find that the
$^{12}$CO(1-0) emission has an almost linear relation to $\av$ for
$2\,mag\,\le\,\av\,\le\,(4-6)\,mag$ while for large extinction values
the $^{12}$CO emission appears to saturate
\citep{lombardi06,pineda08,heiderman10}.

While within the PAWS FoV the
morphology is very similar, the correlation between the CO intensity
and $I-H$ color is not constant (see Fig. \ref{fig:ism_pix} {\it top row}). The
slope is changing between the environments, most notably from center
to disk. A change in the color of the underlying stellar population can be
excluded to explain the different slopes as no radial change in color
has been observed \citep[e.g. I-H color by][]{rix93} nor a trend in
age for the (old) stellar population \citep{mentuch12}. Thus the
observed change in slope can be either caused by 
a change of the
properties of the obscuring material, i.e. the dust, or a change in
the excitation conditions of the CO emission line. Indeed the dust
temperature as well as the interstellar radiation field (ISRF) are
higher in the center than in the disk \citep[with the exception of
the spiral arms,][]{mentuch12}. \cite{pineda08} also observed
different slopes as a function of environment within one single GMC
(Perseus cloud), while \cite{heiderman10} find slightly different
dependencies between $\av$ and CO intensity for the two GMCs they
analyzed. This might indicate that the relation between $\av$ and CO
intensity depends on several parameters of the local environment. 

In order to better
understand and quantify these differences detailed modeling of the
stellar spectral energy distribution is important. In any case the
observed linearity between $I-H$ color and CO intensity shows that the
gas and dust are well-mixed and that there is a preferred gas-to-dust
ratio within M\,51 as found by \cite{mentuch12}.

Finally, we note that the correlation between the tracers of hot dust and the CO
emission is not very good in every environment. Since the dust-to-gas
ratio is basically constant around a value of 105 over the PAWS FoV
\citep{mentuch12}, the simplest explanation is that the heating source
of the dust as observed at 24\,$\mu$m (and 70\,$\mu$m) is varying and
therefore the luminosity of the dust. Obvious examples are the sites
of young star formation that can explain the azimuthal offset between the
molecular gas arms and the 24\,$\mu$m (70\,$\mu$m) emission along these arms.

%%%%%%%%%%%%%%%%%%%%%%%%%%%%%%%%
%%%%%%%%  section 6 - discussion: environment

\section{The importance of galactic environment for molecular gas properties}
\label{sec:env}

The detailed comparison of CO emission to other tracers of the ISM and stellar population(s) 
at cloud-scale resolution reveals that the relations are changing as a function of galactic 
environment, i.e. between disk and bulge, but also arm and inter-arm regime. 
It is interesting that we find no close or simple correlation between the presence of star formation 
and the distribution of the cold molecular gas. Two interesting aspects are the gaps observed 
between the central disk
and the spiral arms, especially in the UV emission, and the 'clumping'
of stellar clusters (clearly) seen along the northern spiral arm with
roughly equidistant separations. Close
inspection of the locations of the clusters of stellar clusters show
that they are associated with spurs in the molecular gas
distribution. This will be discussed in detail in a forthcoming paper.

We find that massive star formation is not always found at or even close to the brightest CO 
flux peaks (presumably also peaks in gas density), i.e. the gaps in star formation tracers between the center and the spiral arms. As the gaps are also apparent in
tracers of younger stellar populations (HII regions, embedded star
formation) this suggests that no massive star formation has been
occurring at this particular radial range for several 10\,Myrs.
This situation is reminiscent of the prominent dust and gas lanes found along bars, that also 
show no strong star formation \citep{sheth02}. However, in the case of M\,51 strong shocks cannot 
be evoked to explain this difference. The paper by \cite{meidt13} explains this behavior by pressure 
support of GMCs that prevents the collapse and thus subsequent star formation.

Clearly the underlying gravitational potential is significantly affecting the molecular gas distribution.
The spatial offset from the convex side of the spiral arms can be explained by the presence of a spiral density wave
\citep[e.g.][]{vogel88}. To some level an offset between HII regions and their birthplace, molecular gas clouds, is expected. Both \cite{schruba10} and \cite{onodera10} report that the relation between
(extinction-corrected) \ha\ emission and CO emission in M\,33 breaks on scales
below $\approx$ 300\,pc - similar to our observed angular separation between molecular gas and HII regions. They argue that the time evolution of
the GMCs themselves is adding significant scatter, as spatial offsets caused
by a spiral density wave can be ignored in the case of M\,33. Another point raised by
\cite{onodera10} is that the \ha\ emission does not necessarily trace
the ionizing stars but rather the shell of the HII regions and
therefore offsets are to be expected. As Galactic GMCs also show a
wide spread of more than 2 orders of magnitude in their star formation
activity \citep{mooney88,mead90} the
non-existence of a relation between the \ha\ and CO emission at
110\,pc resolution is not that surprising. However, the current data does not allow us to discriminate between an evolutionary delay or the fact that only a fraction of the molecular gas will form stars.

Despite the massive star formation present in the PAWS FoV, we find no strong evidence for a general impact of 
this star formation onto the cold molecular gas as probed by the CO emission. In particular, the relations between
CO flux and ISM/stellar tracers are always very similar for the arm and inter-arm region arguing against 
an strong effect from stellar feed-back, both radiative as well as mechanical, onto the molecular gas over the scales probed here (of about 40-100\,pc). Alternatively, the high molecular gas surface density in M\,51 could prevent a large propagation/penetration into the 
molecular gas phase. In order to thoroughly test such feed-back scenarios more detailed studies, 
also of other nearby galaxies with high 
quality, high angular resolution data, are required.

%%%%%%%%%%%%%%%%%%%%%%%%%%%%%%%%
%%%%%%%%  section 7 - summary

\section{Summary and Conclusion}\label{sec:sum}

We present a multi-wavelength comparison between the molecular gas distribution 
and other tracers of the ISM and stellar components at cloud-scale resolution in the central 
9\,kpc disk of the grand-design spiral galaxy M\,51. For our study we utilize new $\sim 1\as$ 
resolution CO(1-0) imaging from the PAWS project in conjunction with existing archival data 
from the UV to radio regime. In particular, we find that

\begin{itemize}

\item Based on the good spatial correlation of low frequency radio continuum and CO line emission, 
we argue for a physical relation between radio continuum and CO emission that does not directly depend 
on the formation/presence of massive stars. This immediately implies that (at least in certain galactic 
environments) the relation between CO line emission and radio continuum is more fundamental than 
between CO emission and IR emission. We show
that the relation between magnetic field strength and gas density expected from equipartition can explain our 
observations.

\item The good spatial correlation between CO and PAH emission seen in the entire PAWS FoV 
suggests that PAH molecules are closely associated with molecular gas and that even the old stellar population
is providing sufficient excitation in the inter-arm region. However, the poor correlation of flux between
CO and PAH emission implies that PAH emission is a poor predictor of the gas mass and most of 
the PAH emission is only arising from the surfaces of GMCs.

\item The spatial correlation between molecular gas and ongoing massive star formation 
is complex and strongly depends on galactic environment (i.e. the underlying gravitational 
potential), showing coincidence, offsets and even absence.
For the spiral arm, we conclude that due to the large average minimum spatial separation 
of $\rm \sim 300\,pc$ between the molecular gas arms and the sites of massive star formation 
no profound impact onto the properties of the molecular gas is expected, consistent with 
other observations presented here. The situation in the star forming ring is less clear. 
Comparison to Spitzer mid-IR \hh\ observations hints that the conditions are not as extreme as 
in the centers of nearby starburst galaxies.

\item The impact of the AGN onto the molecular gas in the disk of M\,51 is not strong and 
appears to be restricted to a small region around the nucleus.

\item The optical-near-IR color $I-K$ provides a very good mapping of the distribution 
of the molecular gas and shows a good correlation to the CO emission, however, with a 
changing slope as a function of galactic environment.

\end{itemize}

This study is the first in a series of papers investigating the molecular gas at GMC resolution in the disk of M\,51 
from PAWS. It clearly demonstrates that future studies of galactic disks with ALMA will provide new insights into 
the physics of the ISM as well as the star formation process.

%%%%%%%%%%%%%%%%%%%%%%%%%%%%%%%%
%%%%%%%%  acknowledgements & facilities

\acknowledgments

We thank the IRAM staff for their support during the observations with
the Plateau de Bure interferometer and the 30m telescope. 
ES thanks Kevin Croxall for reducing and providing the Herschel PACS [CII] map, Hendrik Linz for the reduction of the Herschel PACS imaging at 70 and 160\,$\mu$m and Frank Bigiel for providing the GALEX images used. ES is grateful for many useful discussions with Brent Groves.
DC and AH acknowledge funding from the Deutsche Forschungsgemeinschaft (DFG) via grant SCHI 536/5-1 and SCHI 536/7-1 as part of the priority program SPP 1573 'ISM-SPP: Physics of the Interstellar Medium'.
CLD acknowledges funding from the European Research Council for the FP7 ERC starting grant project LOCALSTAR.
TAT acknowledges support from NASA grant \#NNX10AD01G.
During this work, JP was partially funded by the grant ANR-09-BLAN-0231-01 from the French {\it Agence Nationale de la Recherche} as part of the SCHISM project (\url{http://schism.ens.fr/}).
ES, AH and DC thank NRAO for their support and hospitality during their visits in Charlottesville and Socorro.
ES thanks the Aspen Center for Physics and the NSF Grant \#1066293 for hospitality during the development and writing of this paper. This research has made use of the NASA/ IPAC Infrared Science Archive, which is operated by the Jet Propulsion Laboratory, California Institute of Technology, under contract with the National Aeronautics and Space Administration.

{\it Facilities:} \facility{IRAM (PdBI)}, \facility{IRAM (30m)}, \facility{HST (ACS)}, \facility{HST (NICMOS)}, \facility{GALEX}, \facility{NRAO (VLA)}, \facility{Herschel (PACS)}, \facility{Spitzer (IRAC)}, \facility{Spitzer (MIPS)}.

\clearpage
%%%%%%%%%%%%%%%%%%%%%%%%%%%%%%%%
%%%%%%%%  appendix A - summary of H2 and CO excitation

\bibliography{multi_pap}

\appendix
%%%%%%%%%%%%%%%%%%%%%%%%%%%%%%%%
%%%%%%%%  appendix A - summary of H2 and CO excitation

\section{Heating and Cooling of Molecular Gas}
\label{sec:co}

Here we provide a brief summary of the heating and cooling mechanisms
of molecular gas in (star forming) galaxies, discuss implications for
the observed CO line emission and the relation between CO intensity
and \hh ~mass. For the purpose of this paper we focus on the cold
dense molecular gas (with \nhh $> 10^2\,cm^{-3}$, $\rm T < 100\,K$)
that is normally assumed to be present in GMCs.
We will make use of these findings in the discussion of our
cloud-scale relations (see \S \ref{sec:cloud}) and environmental
effects (see \S \ref{sec:env}).

\subsection{Molecular Hydrogen}

There are several excitation avenues for \hh ~that contribute
differently depending on the exact physical properties of the gas
(e.g. density, temperature, metallicity, dust-to-gas ratio) and its
surrounding (radiation field, cosmic ray density, shocks) which are
both expected to vary across a galactic disk.

\subsubsection{Cosmic ray heating}

The interaction between cosmic rays -- mainly protons and electrons
accelerated to GeV energies and beyond -- and \hh ~molecules is the
prime heating source for the molecular ISM (in the absence of close-by ongoing star
formation) and proportional to the \hh ~density \nhh ~$\times$ the
cosmic ray ionization rate $\zeta_{CR}$ \citep{goldsmith78}. Cosmic
ray electrons and protons (above 10\,MeV) photo-ionize \hh ~molecules
(as these are the most abundant particles). The free electrons collide
with other \hh ~molecules and transfer some of its kinetic energy. In
addition, $\rm H_3^+$ is formed from the reaction of $\rm H_2^+ + H_2
\rightarrow H_3^+ + H$. The dissociative recombination into \hh ~is a
major source of heating for the case of dense molecular gas
\citep[see, e.g., Fig. 2 of][]{maloney96}. As most cosmic ray protons and
electrons can completely penetrate molecular clouds as evidenced by
$\gamma$-ray emission from molecular clouds
\citep[e.g.][]{bloemen86}\footnote{The interaction between cosmic
rays and (atomic and) molecular gas can be observed via the
$\gamma$-ray emission that results from $\pi^o$-decay and
Bremsstrahlung of the comic ray nuclei and electrons.} this heating
mechanism is expected to dominate (or even be the only one) in the
most dense and cold GMCs.

Galactic cosmic rays are thought to be accelerated by shocks from SN
remnants or even Wolf-Rayet stars \citep[e.g.][]{kotera11} and have a
fairly homogenous distribution throughout the MW with a
galacto-centric decrease \citep[e.g.][]{bloemen89}. It is conceivable
that the production of cosmic rays is significantly increased above
the solar value in the presence of massive star formation. It has been
shown for the nearby starburst galaxy NGC\,253 that enhanced cosmic
rays emission originating from the massive nuclear star formation with
$\rm SFR \sim 0.1\,\msunyr$ can explain the observed substantial
heating of the molecular gas \citep{bradford03,hailey08}; a similar
scenario has been proposed by \cite{papadopoulos10} to explain the CO
emission in star forming ULIRGs. These regions are called cosmic ray
dominated regions (CRDR) and might have a distinct chemical
signature \citep{bayet11}.

\subsubsection{\hh ~formation heating}

The formation of \hh ~on grains \citep[e.g.][]{hollenbach71} provides
also heating of the gas, as some fraction of the released binding
energy goes into kinetic energy and thus subsequently heats the gas
\citep{goldsmith78}. The heating rate is proportional to the total
Hydrogen density (in molecular and atomic form) in quadrature times
the fractional abundance of atomic Hydrogen. \cite{goldsmith78} note
that if \hh ~destruction by cosmic rays (see above) is taken into
account the maximum heating rate from \hh ~formation could be 1/3 of
the cosmic ray heating rate, however, in reality the steady-state
assumption might not be fully correct. Additionally the exact
properties of dust (grain sizes, PAHs, temperature etc.) might
affect the \hh ~formation rates, especially in diffuse, i.e. low
density, gas \citep{wolfire08}.

\subsubsection{Dust heating}

The predominant heating mechanism for dust (in molecular clouds) is
via the photo-effect \citep{draine78}. As dust couples better to the
interstellar radiation field (ISFR) than the molecular gas the dust kinetic temperature $\rm
T_{dust}$ is typically larger than the gas kinetic temperature $\rm
T_{gas}$ within a molecular cloud \citep{maloney88}. Thus collisions
of \hh ~with dust grains could increase the kinetic energy of the
gas. As the dust heating is governed by the available radiation field,
it is expected that increasing the local 
ISFR by adding an intense radiation field from, e.g. young (OB)
stars or an AGN, will lead to higher dust and thus potentially higher gas
temperature if both are well mixed and coupled.

The energy transfer between dust and gas via collisions can
be described by the energy difference of the \hh ~molecule before and
after a collision with a dust grain \citep{burke83}. As
it is possible that the dust temperature $\rm T_{dust}$ is higher than
the gas temperature $\rm T_{gas}$ as well as the other way round, both
heating or cooling of the gas is possible via this mechanism. The
heating/cooling rate is proportional to $\rm
n(H_2)^2\,\times\,\Delta\,T\,\times\,T_{gas}^{0.5}$ with $\rm
\Delta\,T\,=\,T_{gas}-\,T_{dust}$ \citep{goldsmith01}.

The radiation from the diffuse UV-visible-IR ISRF will heat dust
grains to temperatures higher than the gas temperatures observed for
the dense cores of dark clouds, however, for dust grains inside the
clouds this ISRF will be attenuated. \cite{goldsmith01} calculates
that in order to achieve dust temperatures of 6-10\,K a visual
extinction to the surface of the cloud of 10-12\,mag is required, consistent with
observations \citep{kramer99}. He notes that a further drop in dust
temperature is not expected due to heating provided by the re-emission
from dust grains themselves. Thus in molecular clouds (without star
formation) the heating by dust is proportional to the attenuated ISFR
times the \hh ~density \nhh ~divided by the gas-to-dust ratio. At the
same time the dust cools with $\rm T_{dust}^6$.

Thus significant heating from dust is only expected to occur at high
\hh ~densities \nhh ~above $\rm 10^4\,cm^{-3}$ where the coupling of
gas and dust via gas-dust collisions is good and can significantly
reduce the gas temperature while only slightly increasing the dust
temperature \citep{goldsmith01}. Examples might be dense clouds close
to or with embedded star forming regions (with OB stars) as such
clouds have elevated temperatures of 40 - 80\,K
\citep[e.g.][]{goldsmith78}. Recent simulations confirm this picture
\citep[e.g.][]{juvela11}, but the authors caution that a varying grain
size distribution within clouds can affect the gas-dust coupling and
thus the resulting gas temperature by 1-2\,K in dense core.

\subsubsection{UV heating}

It is expected that the intense (UV) radiation from young stars in
star forming regions heats the gas. Two mechanisms are generally
considered: heating via the grain photo-electric effect
\citep{tielens79} and direct UV pumping of \hh ~molecules followed by
collisional de-excitation
\citep{hollenbach79,black76,sternberg89}. Detailed models of these
so-called Photon Dominated Regions (PDRs) \citep{tielens85} have been
derived where UV photons penetrate through the surface into the cloud
while being attenuated and produce layers of ionized, atomic and warm
molecular gas around the cold molecular gas. How far a UV photon can
penetrate into a cloud depends on the dust opacity.

Gas heating via the photo-electric effect has been shown to occur by
PAH molecules and small graphite grains \citep[grains with size $a <
100 \AA$; ][]{bakes94}. About 50\,\% of the UV photon energy is absorbed
by these small dust particles. Absorption of a far-UV photon by such a
dust grain may lead to ionization and thus ejection of an
electron. The excited electron loses energy through inelastic
collisions with Carbon atoms inside the grain on its way to the
surface and will leave the grain with some kinetic energy. Via
inelastic collisions with gas molecules the thermal energy of the gas
will be raised \citep[e.g.][]{tielens85,bakes94}. The photo-electric
effect on interstellar dust grains should dominate the heating of the
neutral atomic ISM and play an important role in the warm inter-cloud
medium and in PDR regions of cold molecular clouds (i.e. UV
illuminated regions such as HII regions, planetary nebulae etc.). The
heating efficiency could be up to 3\,\% (for neutral grains; lower in
grains that are charged).

Dust plays an important role in attenuating the UV radiation (at
optical depths $\rm A_V \sim 1$) that controls gas heating in order to prevent
higher gas temperatures \citep{maloney88}. For a simple slab geometry
of the molecular gas, the grain-photo-electric heating is only important in
the outermost layer of a molecular cloud in the presence of radiation from the
ISRF (dominated by old stellar populations), its contribution could
become significant in the case of intense radiation fields from young
stars \citep[e.g. for the dense clouds in the center of
M\,82,][]{maloney88}. At low column densities as they are present at
the outer surface of molecular clouds photo-ionization heating can
become important. A comparison of contemporary PDR codes summarizing
the main characteristics deemed important is presented by \cite{roellig07}.

\subsubsection{X-ray heating}

X-ray dominated regions (XDRs) are expected to be present close to
nearby AGN where the strong X-ray radiation from the accreting black
hole is impinging surrounding molecular clouds. A recent model of XDRs
is presented by \cite{meijerink06}. While XDR modeling is analogous
to PDR models, it is interesting to highlight some profound
differences: the assumed input spectrum is much harder for an
AGN than for hot, e.g. OB, stars, and the main heating source is direct
photo-ionization of the gas which produces fast electrons. These
electrons collisionally excite the H and \hh ~gas which emits
Lyman\,$\alpha$ and Lyman-Werner photons that ionize atoms and ionize
and dissociate molecules (e.g \hh ~and CO) \citep{maloney96}. As hard
X-rays can penetrate deeply into dense clouds the heating efficiency
in XDRs is close to unity, however, the X-ray heating and ionization
rates depend on the actual slope of the X-ray spectrum
\citep{meijerink06}.

In this context it is interesting to note that \cite{lepp83} were
among the first to consider the impact of a X-ray source inside a
molecular cloud, as most stars are sources of soft X-ray emission and
a number of accreting white dwarfs should be present in each
GMC. While the contribution from X-ray heating via the old stellar
population is hard to assess separately from cosmic ray heating,
\cite{glassgold12} revisited the impact of cosmic ray and X-ray
heating on cold dense clouds based on new findings of the $\rm H_3^+$
ion abundance in Galactic clouds. They find that the combined cosmic
ray and X-ray heating (also referred to as chemical heating)
significantly changes with the properties of the medium
(molecular/electron fraction, total density of H nuclei) but has only a weak
temperature dependence.

\subsection{Carbon-Monoxid (CO) Line Emission}

It is generally assumed that the \hh ~and CO molecular gas is in
thermal equilibrium so that the kinetic temperatures are the
same. Detailed calculations for dense, fully shielded molecular regions
(i.e. $\rm A_V \ge 10)$ assuming a steady-state chemical composition
predict that the CO molecules dominate the cooling in the low density
($\rm n(H_2)\,<\,10^5\,cm^{-3}$) and low temperature ($\rm
T_{kin}\,\le\,40\,K$) regime while other molecules become the dominant
coolants at high densities and high temperatures \citep{neufeld95}. 

The radial distribution of \hh ~and CO molecules across the surface
into the center of molecular clouds is varying as the dissociation
conditions of CO and \hh ~are not the same. \cite{dishoeck88} provide
a detailed description and modeling of CO photo-dissociation: While
the unattenuated photo-dissociation from stellar UV light for CO is
high, substantial reductions happen inside the clouds including: a)
self-shielding, i.e. growth toward saturation of the absorption in the
line itself with increasing depth and column density, b) mutual
shielding, i.e. the blending of lines of the isotopic species with
those in the same band of the more abundant $^{12}$CO molecule, c)
shielding by coincident lines of H and \hh, and d) attenuation of
radiation by dust particles and Carbon atoms which significantly
decreases the photo-dissociation of CO even inside diffuse clouds.
Recently, \cite{visser09} up-dated the calculations from \cite{dishoeck88}
using the latest laboratory data and fully treating the heavier CO isotopologues
$\rm ^{13}CO, C^{18}O, and ^{13}C^{17}O$. In addition, they extended the
calculations to higher excitation temperatures as well as densities covering PDR
and circumstellar disk environments.

However, the exact relation between CO line emission (particular the
low-J transition) and \hh ~gas properties is still not very well
understood. Recently, \cite{glover10} modeled the CO formation. They
found that both molecular species are expected to form quickly,
however the relationship between CO abundance and gas density is more
complex. The CO abundance primarily depends on the photodissociation
rate rather than the gas density \citep{glover11}. \cite{clark12}
investigate the timescales required for a molecular cloud to become CO
detectable starting with a pure atomic phase in a turbulent medium. They
find that the timescales depend critical on the conditions of the
medium out of which the cloud forms with CO typically being only
detectable about $\sim$ 2\,Myr before the onset of star formation. 
As the neutral gas disk of M\,51 covered by the PAWS observations is molecular dominated
\citep[e.g.][]{schuster07} it is unclear how to adapt these modeling
results to the specific case of M\,51.

\subsection{The relation between CO line emission and Molecular Hydrogen Mass}

In order to be able to relate the observed CO line emission to the
amount of \hh ~mass present one needs to understand how the molecular
gas is cooling. As the type of main coolants is changing with
temperature, gas density, and relative abundance of the coolants
(which themselves depend on the ionization state of the gas, the
intrinsic metallicity of the gas as well as other chemical processes),
the observed CO emission per \hh ~is expected to vary as well
\citep[e.g.][]{goldsmith78}. Typically one expects the
following type of coolants (ignoring metallicity effects): The cold
(with temperatures of 10 - 40\,K or below 100\,K) molecular gas cools
via atomic and molecular line emission, however, the main cooling
lines shift from atomic lines such as CI, OI etc. in the low density
regime (\nhh $\rm <\,10^3\,cm^{-3}$) to CO as the main cooling channel
(for \nhh $\rm < 3\times10^4\,cm^{-3}$) and a wide range of species
including water, hydrides, molecular ions etc. for the highest
densities where $\rm ^{12}CO$ can freeze out \citep{goldsmith78}.

In addition to the physical heating and cooling conditions of the
molecular gas described above, the abundance of CO to \hh ~molecules
becomes crucial when determining the relation between CO line emission
and \hh ~mass present. Metallicity has the most profound effect on the
abundance ratio, while other effects such as CO freeze-out for very
high density gas or shocks could also alter the chemical composition
of the molecular gas \citep{goldsmith01,meijerink10,bayet11}. 
However, it is anticipated that metallicity has a fairly
uniform effect while the others mentioned are expected to be much more
localized across a galaxy.

For the case of average molecular clouds,
\cite{maloney88} showed that the conversion factor $\rm X_{CO}$
between CO intensity and \Nhh ~column density is depending on the
kinetic temperature $\rm T_{kin}$ and velocity dispersion of the
molecular cloud $\sigma_c$ which in turn is related to the \hh
~density such that: $\rm X_{CO} \propto T_{kin} ^{-1} \sigma_c \sim
T_{kin}^{-1}\sqrt{n(H_2)}$ \citep[for a slightly different dependence,
see][]{shetty11}. This dependency can be easily understood as the line
intensity is directly proportional to the excitation temperature
(i.e. $\rm I_{CO} \sim T_{ex}$) in optically thick clouds (where $\rm
T_{ex} \sim T_{kin}$ is valid, especially at the lower transitions),
thus the ratio $\rm \frac{I_{CO}}{T_{kin}}$ is basically constant. For
a gravitationally bound cloud the velocity dispersion is proportional
to $\rm \sqrt{M}$, with the cloud mass $\rm M$ being proportional to the \nhh
~density for a homogenous cloud \citep[see ][who showed that this
assumption is also correct for ensembles of clouds]{dickman86}.

The abundance of CO relative to \hh ~is another critical parameter in
determining the $\rm X_{CO}$ factor. If the metallicity is smaller the dust
abundance is smaller thus the attenuation of the UV radiation with
depth into the cloud is smaller. While \hh ~self-shielding will
prevent a drastic reduction of the \hh ~zone, the destruction of CO
will lead to an increase in cloud temperature \citep{maloney88}. The
importance of this effect will depend on the column density. The
implication is that a lowered CO abundance will cause the CO line to
be no longer optically thick except at the cloud cores for clouds with
similar column density but different CO abundance (e.g. only CO emission from the
core of LMC clouds vs. the full cloud for MW clouds). For more details on the
latest observational findings on the conversion factor in external galaxies 
we refer the reader to \cite{sandstrom12} and
\cite{leroy13}.

Based on the continuous conversion factor of $\rm X_{CO} =
\frac{min[4,675\times\,I_{CO}^{-0.32}]\,\times\,10^{20}}{Z^{\prime\,-.65}}$
derived by \cite{narayanan12} we expect a range of $\rm (1 -
3)\times\,10^{20}\,cm^{-2}\,(K\,km\,s^{-1})^{-1}$ for our measured
$I_{CO}$ per pixel and assuming solar metallicity as observed for the
PAWS FoV \citep{bresolin04,moustakas10}, i.e. $Z^{\prime}=1$. This is
similar to the results of \cite{schinnerer10} using LVG analysis and the
virial mass measurements of individual GMCs by \cite{colombo13a}.

%%%%%%%%%%%%%%%%%%%%%%%%%%%%%%%%%
%%%%%  appendix B - polar cross-correlation

\section{Calculating angular offsets via polar cross-correlation}
\label{sec:polar_cross}

In order to quantify the radial variations between two different
tracers along the spiral arms, we employ a polar cross-correlation. 
This method allows for easy
identification of variation in the position of the maximum correlation
as a function of radius.

\begin{figure}
\begin{center}
%includegraphics[]{f1.eps}
\resizebox{1.0\hsize}{!}{\includegraphics[angle=-90]{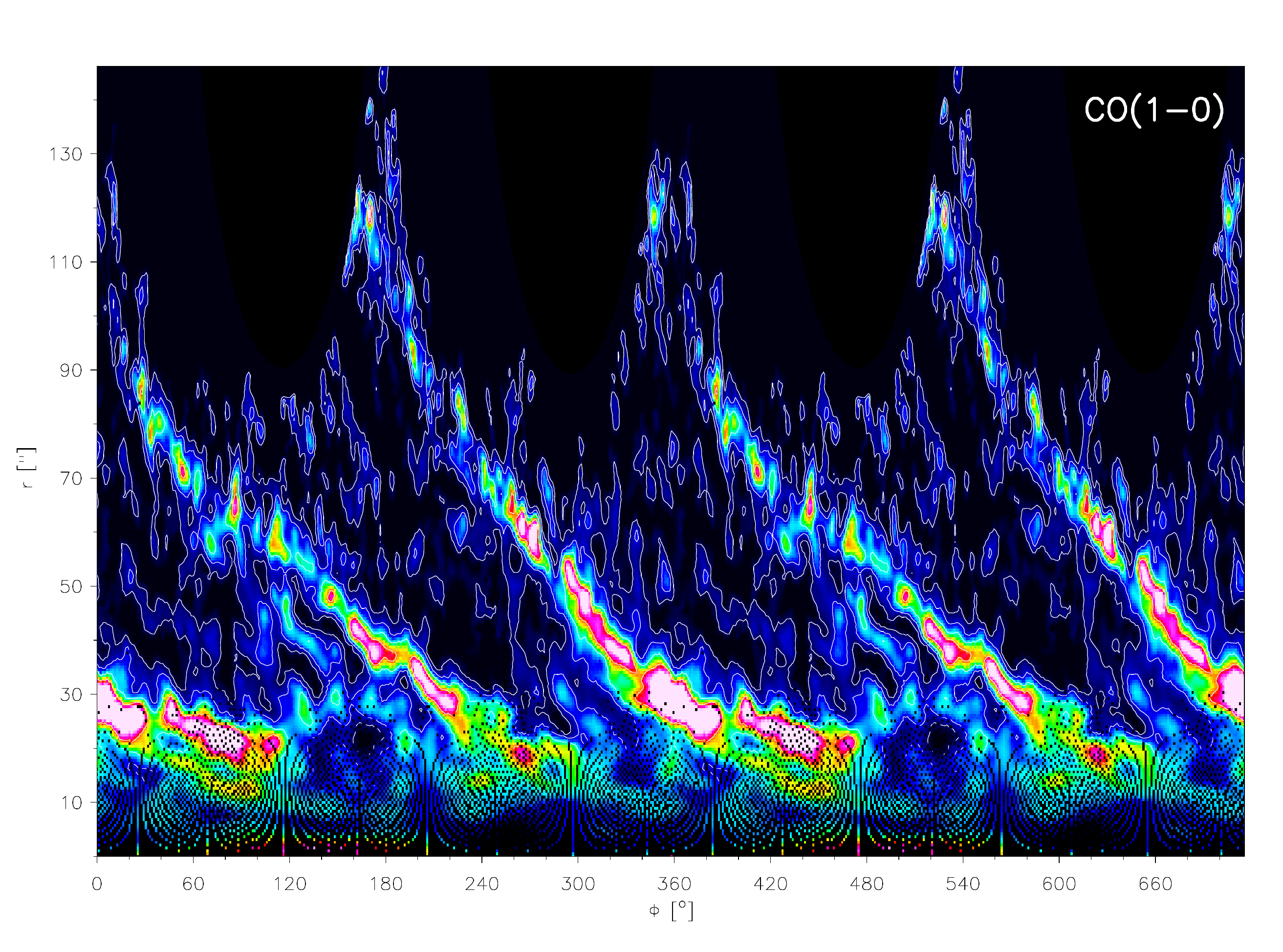}\includegraphics[angle=-90]{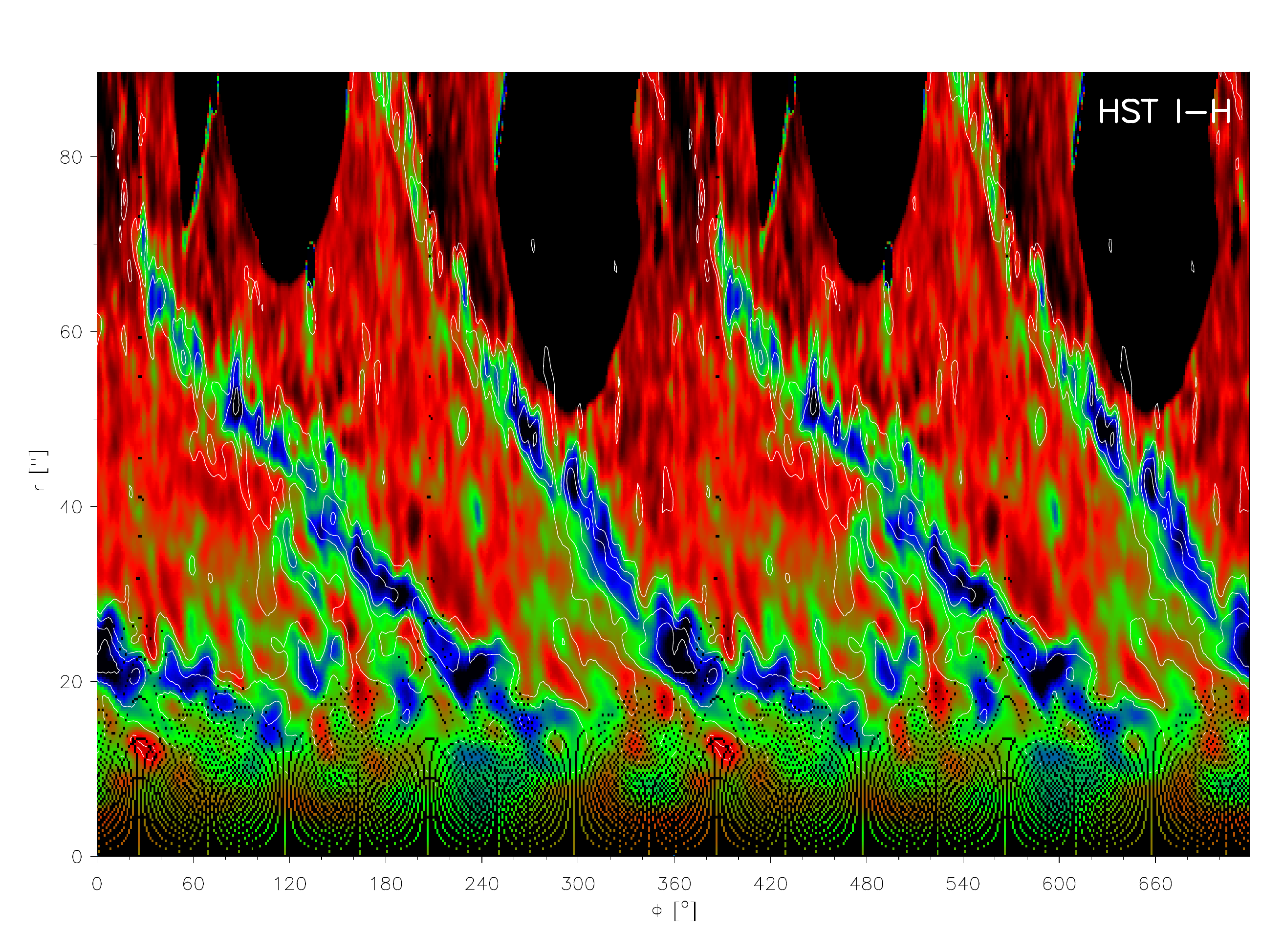}}\\
\resizebox{1.0\hsize}{!}{\includegraphics[angle=-90]{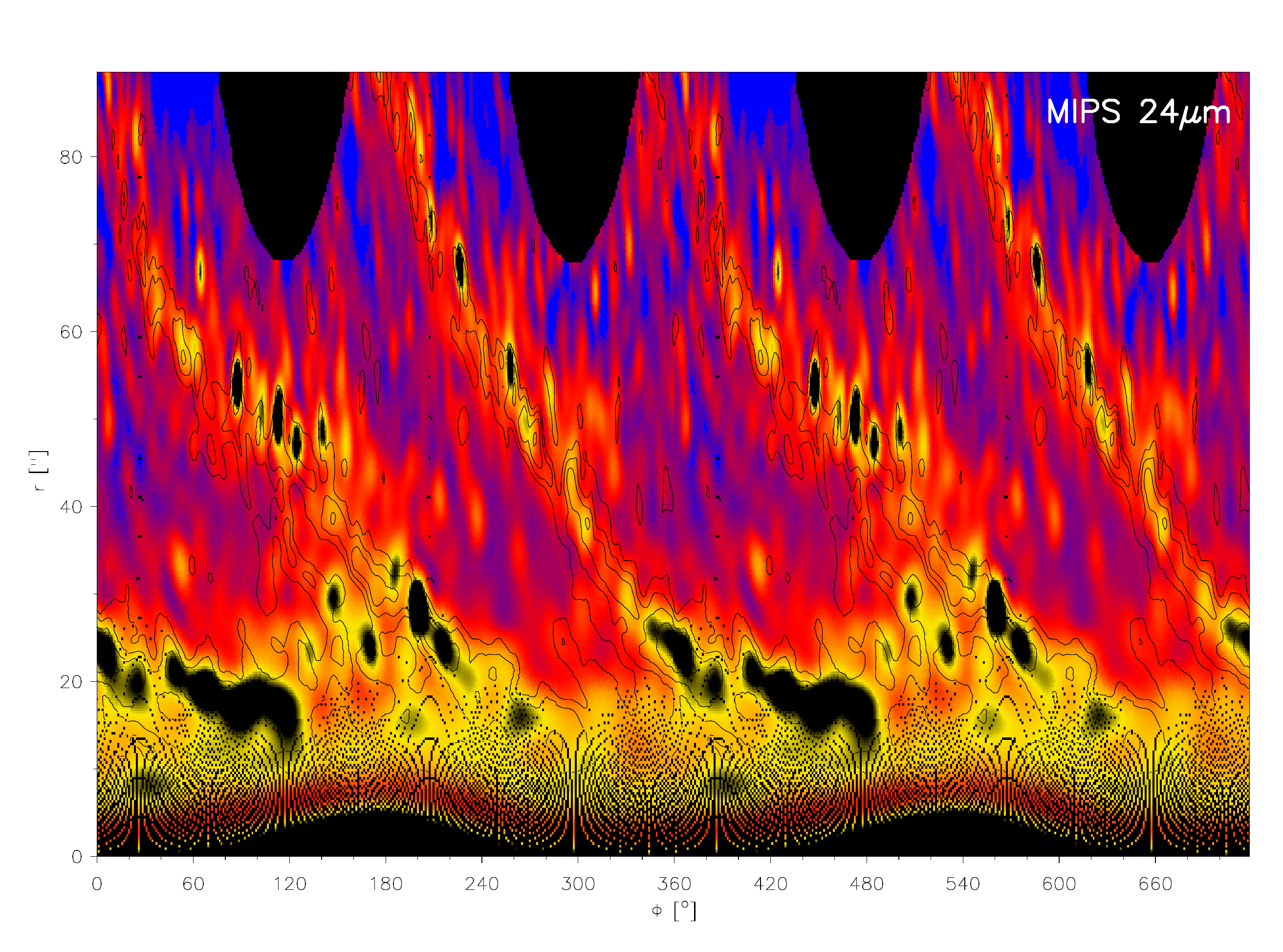}\includegraphics[angle=-90]{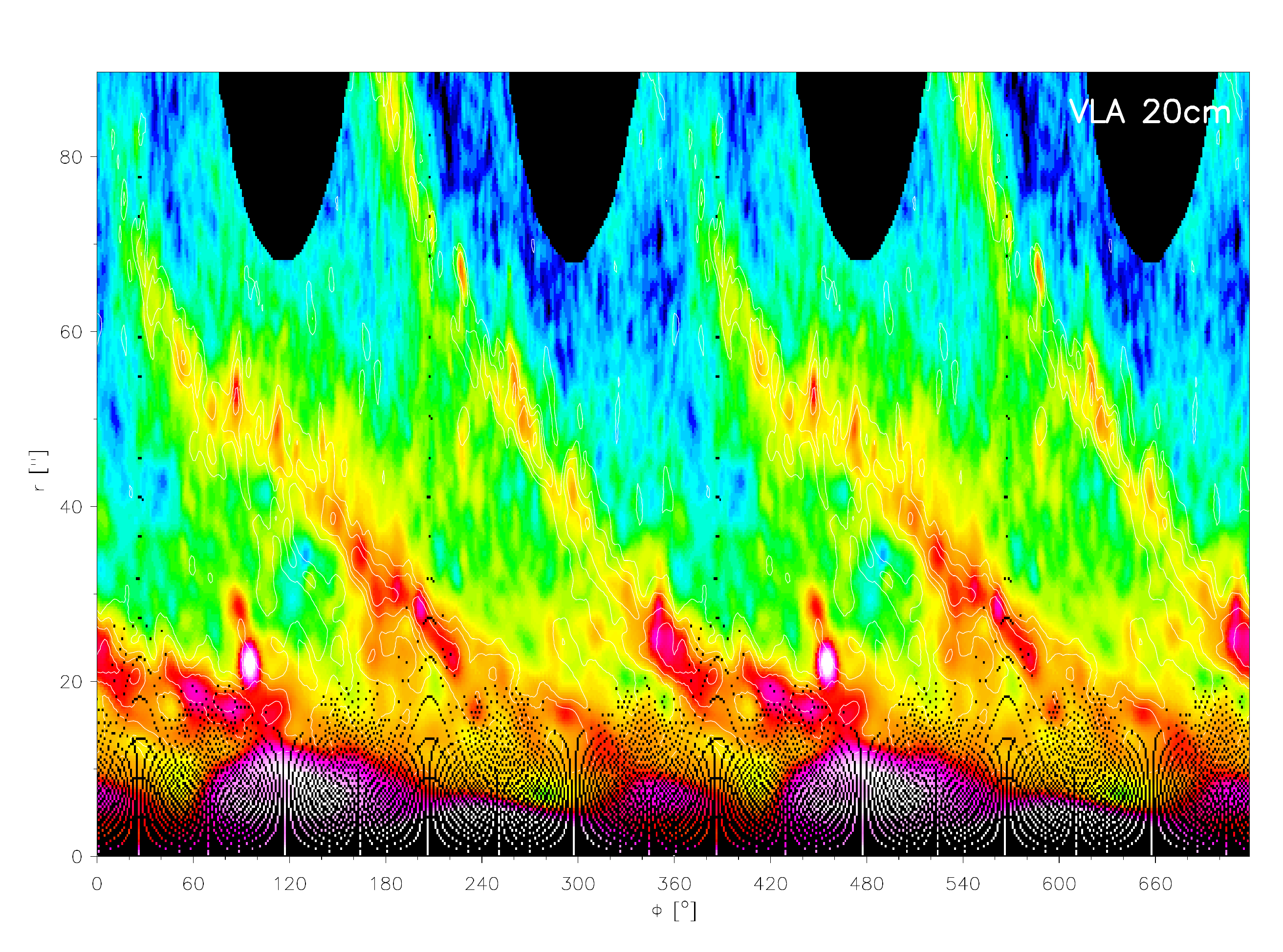}}\\
\resizebox{1.0\hsize}{!}{\includegraphics[angle=-90]{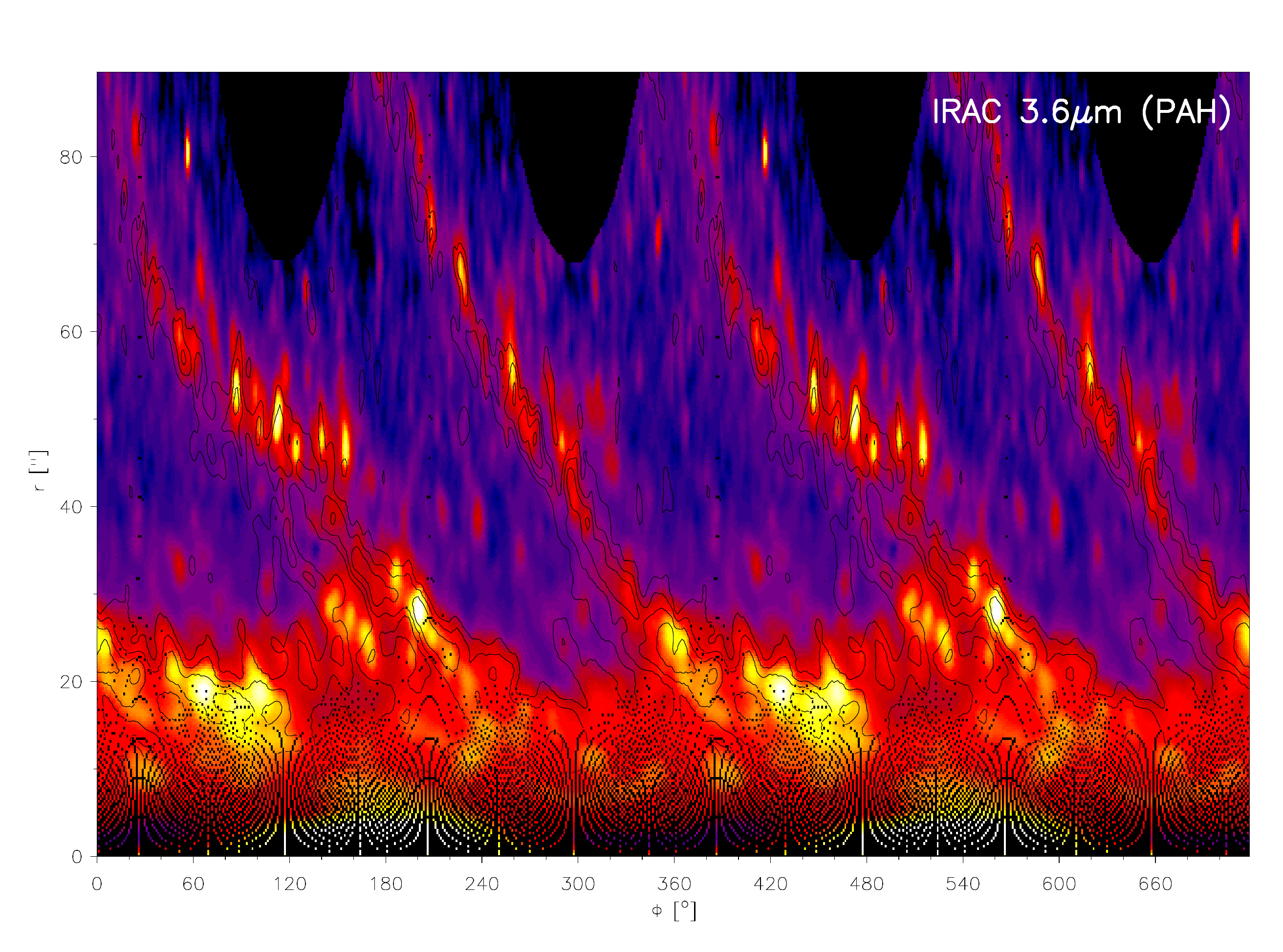}\includegraphics[angle=-90]{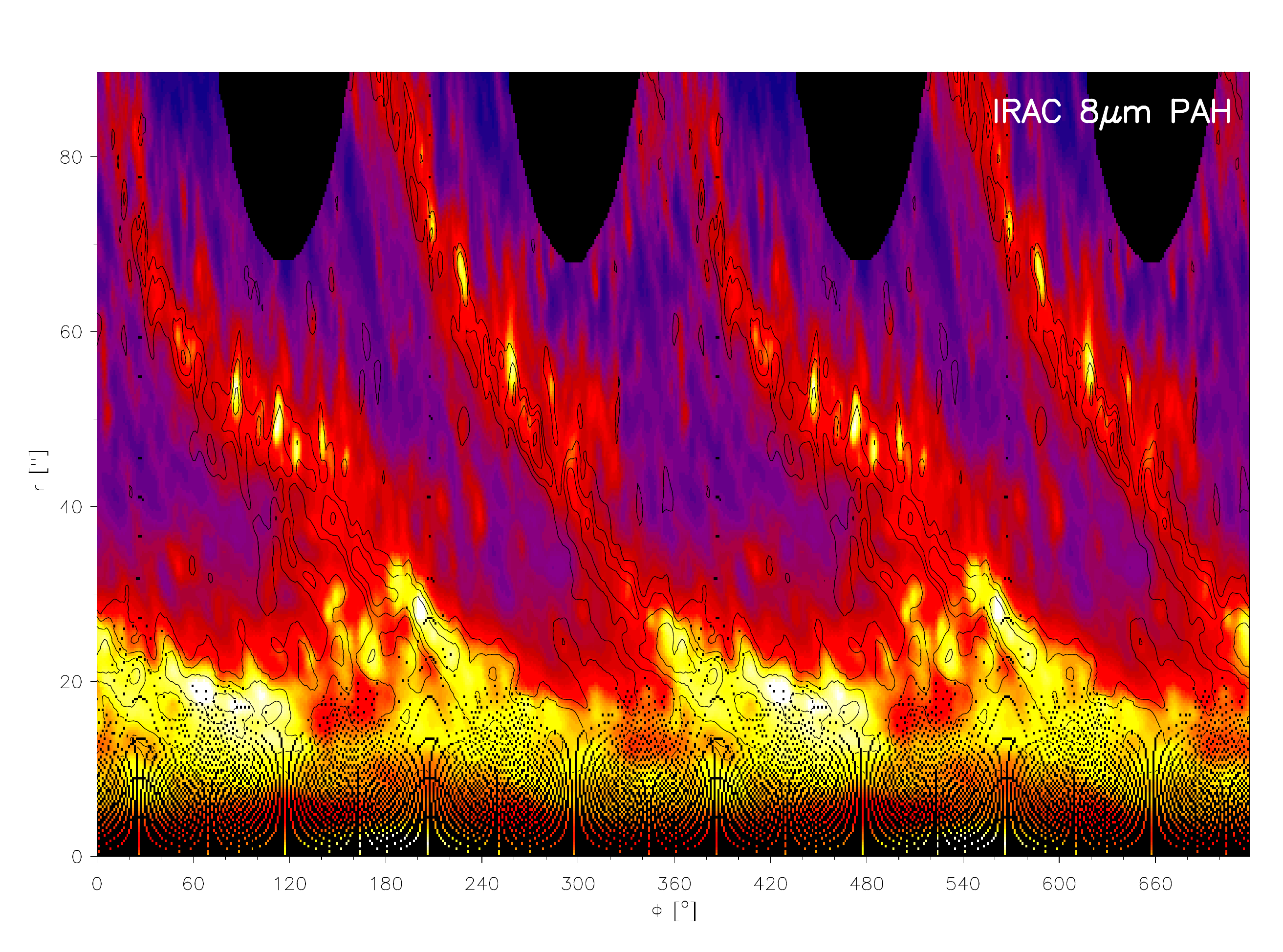}}
\end{center}
\caption{Polar representation of the \coone ~line emission ({\it top left}, and contours) and other ISM tracers: 
HST $I-H$ color image ({\it top right}), MIPS 24$\mu$m ({\it middle left}), VLA 20\,cm ({\it middle right}),
the non-stellar continuum at IRAC 3.6\,$\mu$m and IRAC 8\,$\mu$m ({\it bottom}).
\label{fig:ism_pol}}
\end{figure}

We calculate the angular offset $\phi$ between the molecular gas
spiral arms traced in CO and ISM, stellar as well as star formation tracers at a
second wavelength $\lambda$ via convolution of the two corresponding
maps. All images have the same astrometry, angular resolution and
pixel sampling (as described earlier in \S \ref{sec:data}).  Prior to
the convolution, the images were deprojected from the sky plane into
the galaxy plane using an inclination of $\rm i\,=\,21\dg$ and a
position angle of $\rm PA\,=\,173\dg$ (as determined from analysis of
the PAWS velocity field; Colombo et al. in prep.). Next the images were
projected onto a polar coordinate system (see \S \ref{subsec:ism},
e.g. Fig. \ref{fig:ism_pol}) and placed into the spiral arm frame
assuming a logarithmic spiral arm pitch angle of $\rm i_p\,=\,21\dg$. The
convolution is performed in log-polar coordinates. The spiral arm
frame has been chosen for ease in presentation as the maximum of the
auto correlation should occur very near the same azimuthal position in
each radial bin.

As our measure of the angular offset $\phi$ between the ridge of the
spiral arm traced by CO and the second tracer at wavelength $\lambda$,
we adopt the intensity-weighted first moment of the cross-correlation
(cc) profile tracing the location of maximum correlation. This is in
contrast to using polynomial fits to the profile or to using the
maximum itself. We prefer this approach for several reasons. Moment
measurements made directly from the cross-correlation profile are less
susceptible to systematic uncertainties introduced by, e.g., the
choice of the order of or limits of polynomial fits to the profile, which
typically limit the reliability of the measured offset
\citep{foyle10}. The first moment, in
particular, is preferred for its reduced sensitivity (compared to the
maximum) to local variations in the flux tracing structure at spatial
scales smaller than the arm width (e.g. brightness peaks in CO
emission tracing individual clouds)\footnote{We note that compared
to the maximum, the first moment also conveys additional information
regarding the symmetry of a given arm, or how the emission is
distributed relative to the spiral ridge; asymmetric arms at one or
both wavelengths results in a skewed profiles. This skewness can also
be measured directly from the profile as the third moment.}.

\begin{figure}
\begin{center}
\includegraphics[width=180mm]{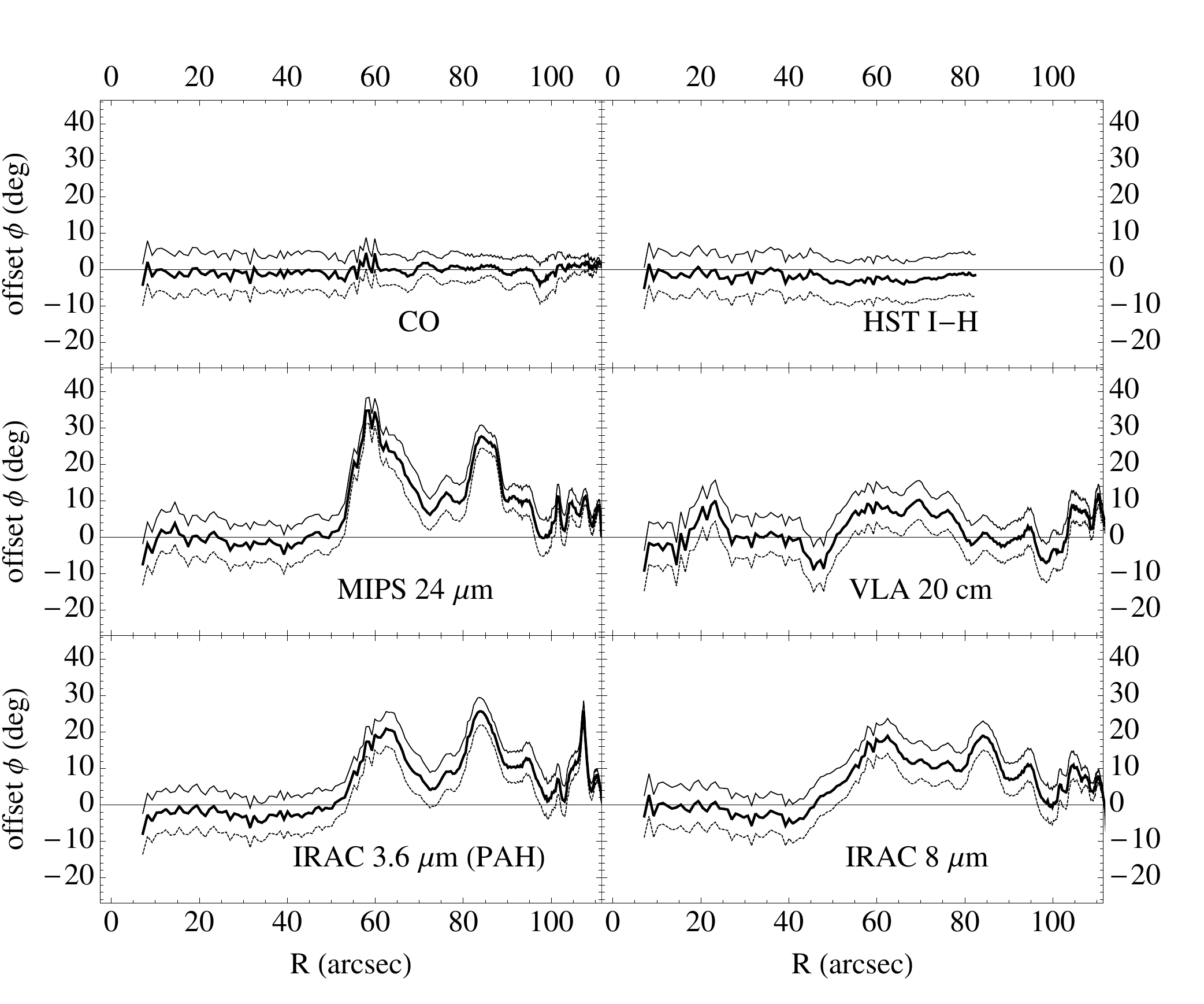}
\end{center}
\caption{Results of the polar cross-correlation between the \coone ~line emission ({\it top left}) and other ISM tracers. 
The panels show the radial profiles of the location of maximum correlation between the \coone ~intensity and
the emission from other tracers, tracing the azimuthal offset $\Phi$ between the two:
\coone ~intensity ({\it top left}), HST $I-H$ image ({\it top right}), 
MIPS 24\,$\mu$m image ({\it middle left}), VLA 20\,cm continuum ({\it middle right}),
non-stellar continuum at IRAC 3.6\,$\mu$m and 8\,$\mu$m ({\it bottom}). 
The thin lines represent  the uncertainty defined as the width of the cross-correlation profile at 95\% maximum correlation. 
\label{fig:ism_cross}}
\end{figure}

\begin{figure}
\begin{center}
%includegraphics[]{f1.eps}
\resizebox{0.9\hsize}{!}{\includegraphics[angle=0]{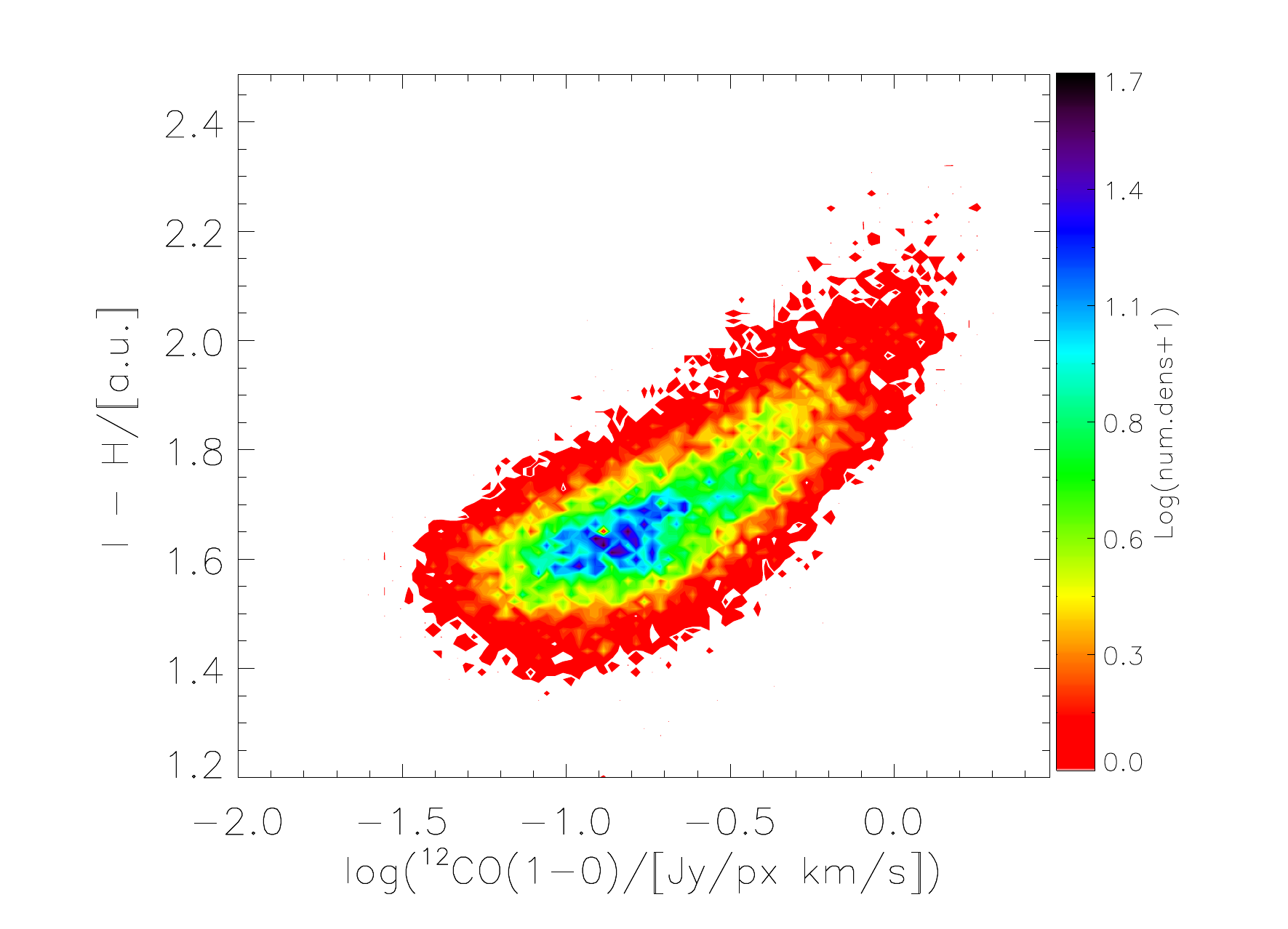}\includegraphics[angle=0]{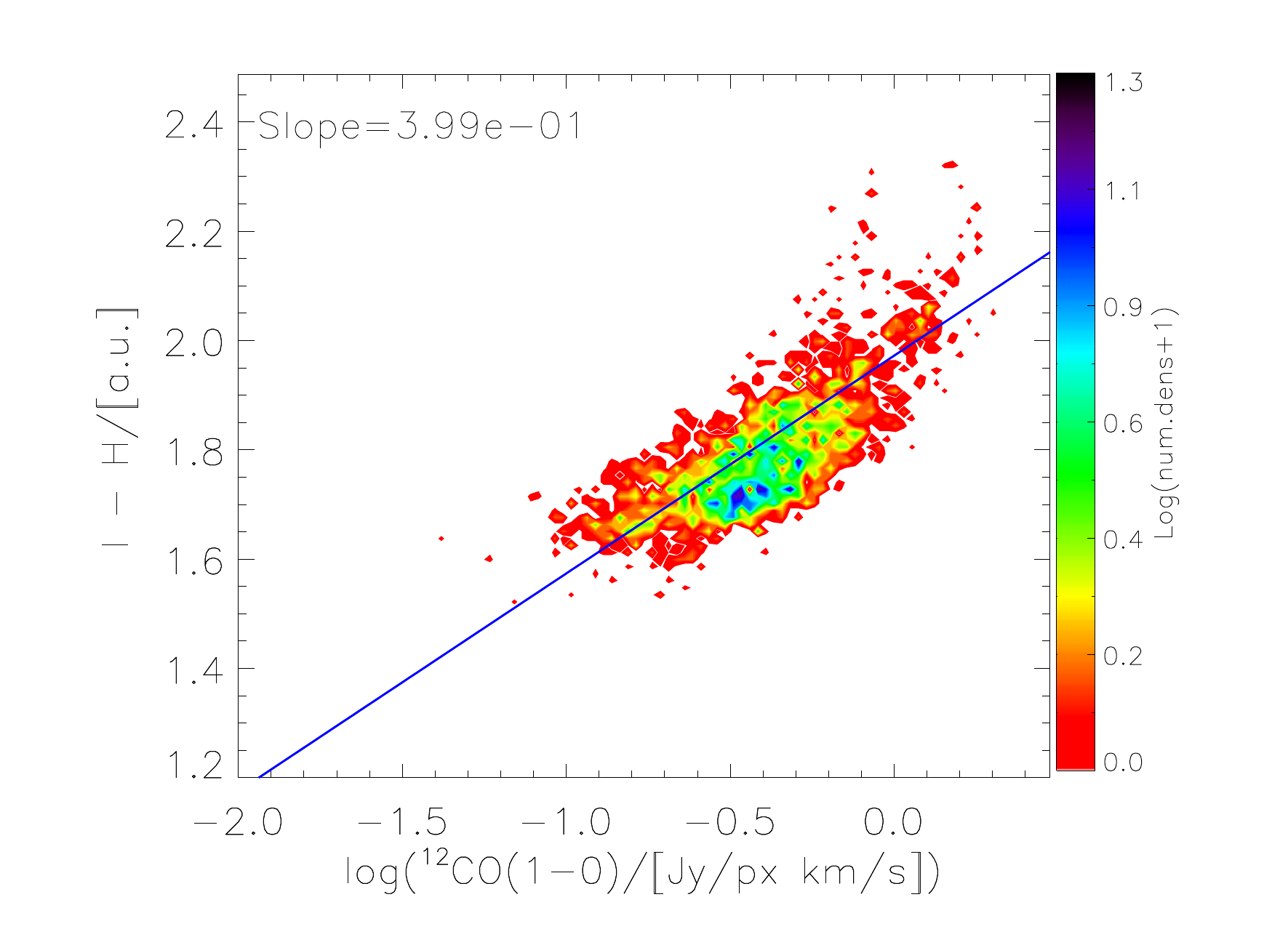}\includegraphics[angle=0]{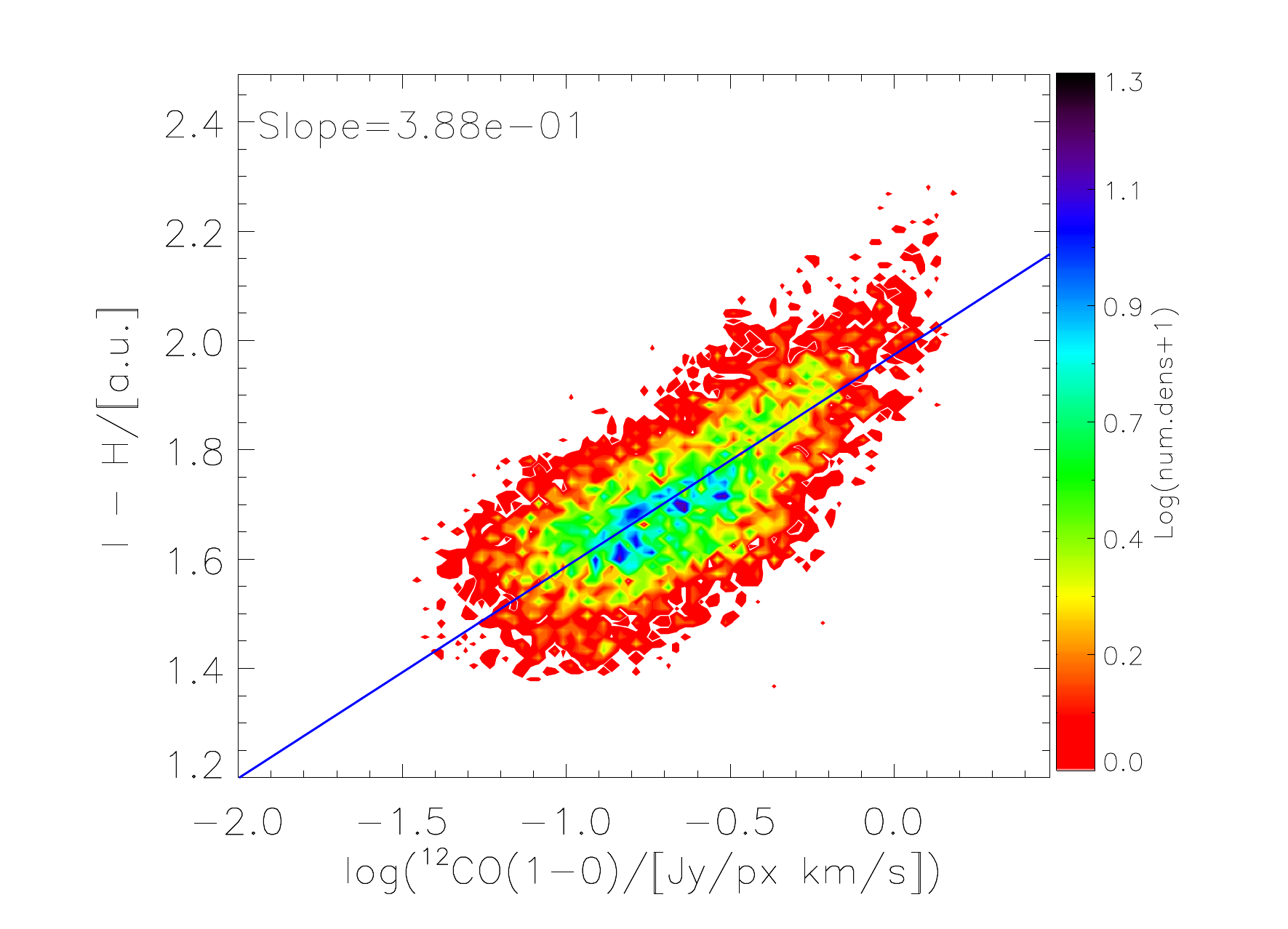}\includegraphics[angle=0]{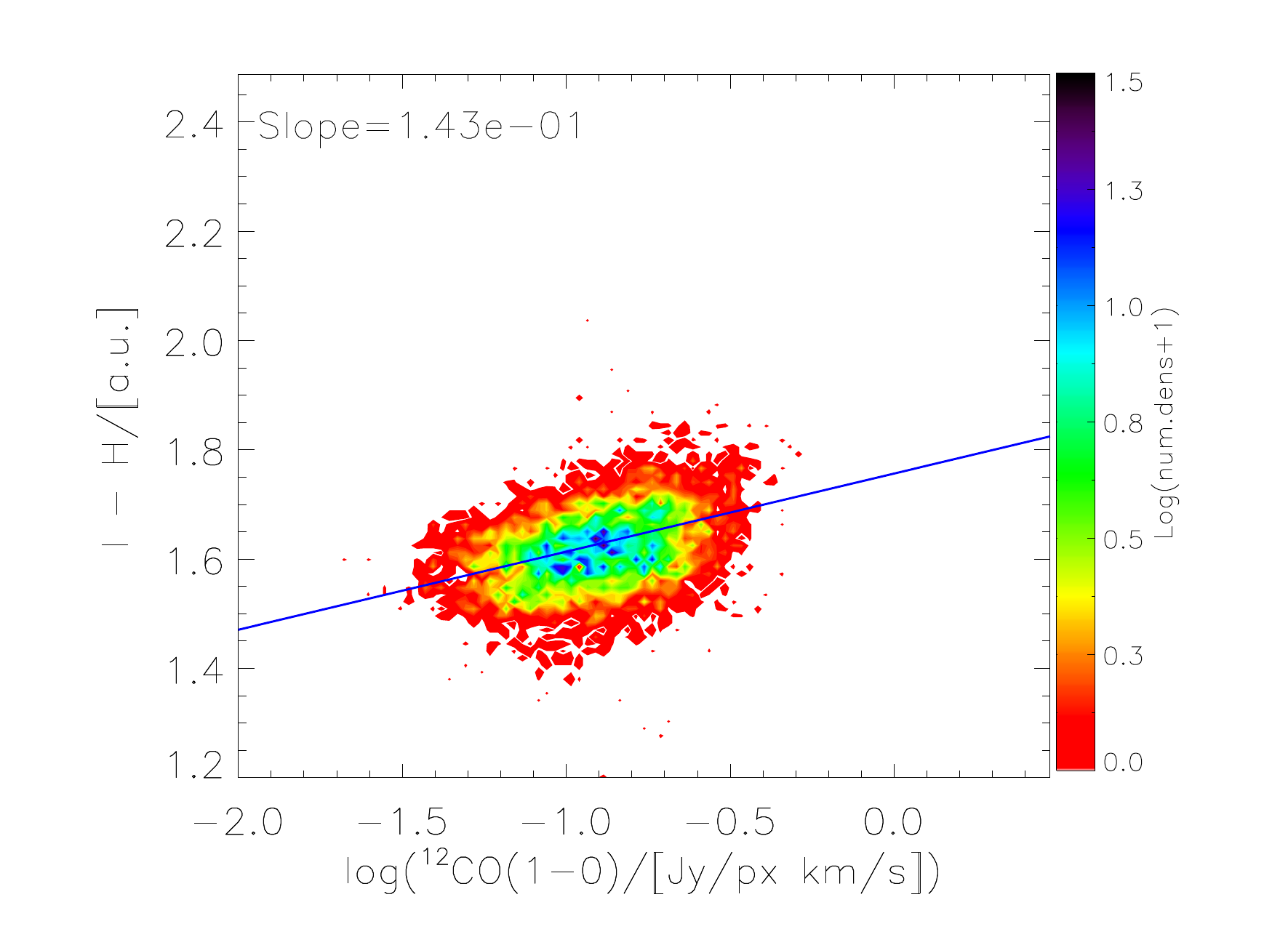}}\\
\resizebox{0.9\hsize}{!}{\includegraphics[angle=0]{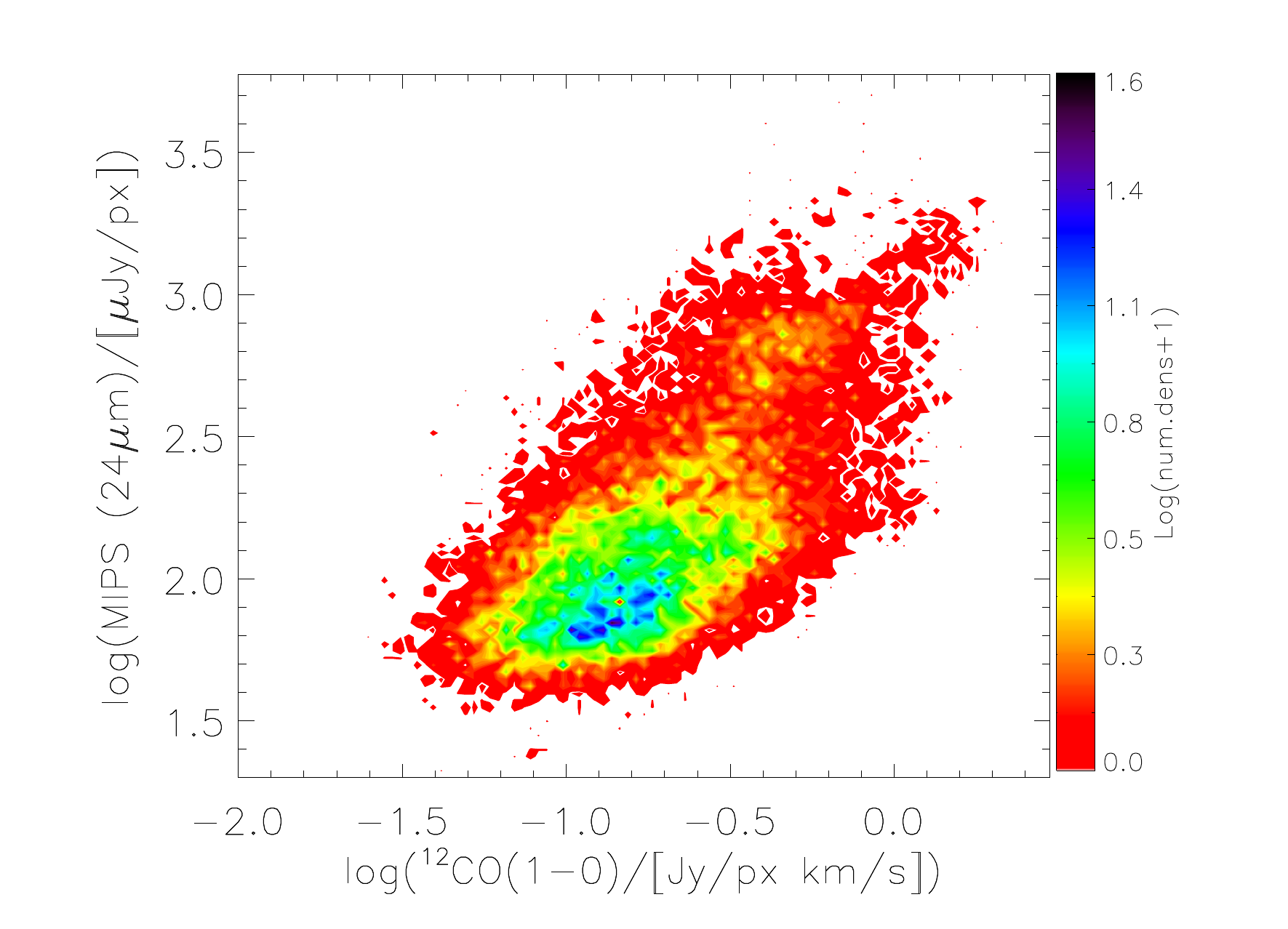}\includegraphics[angle=0]{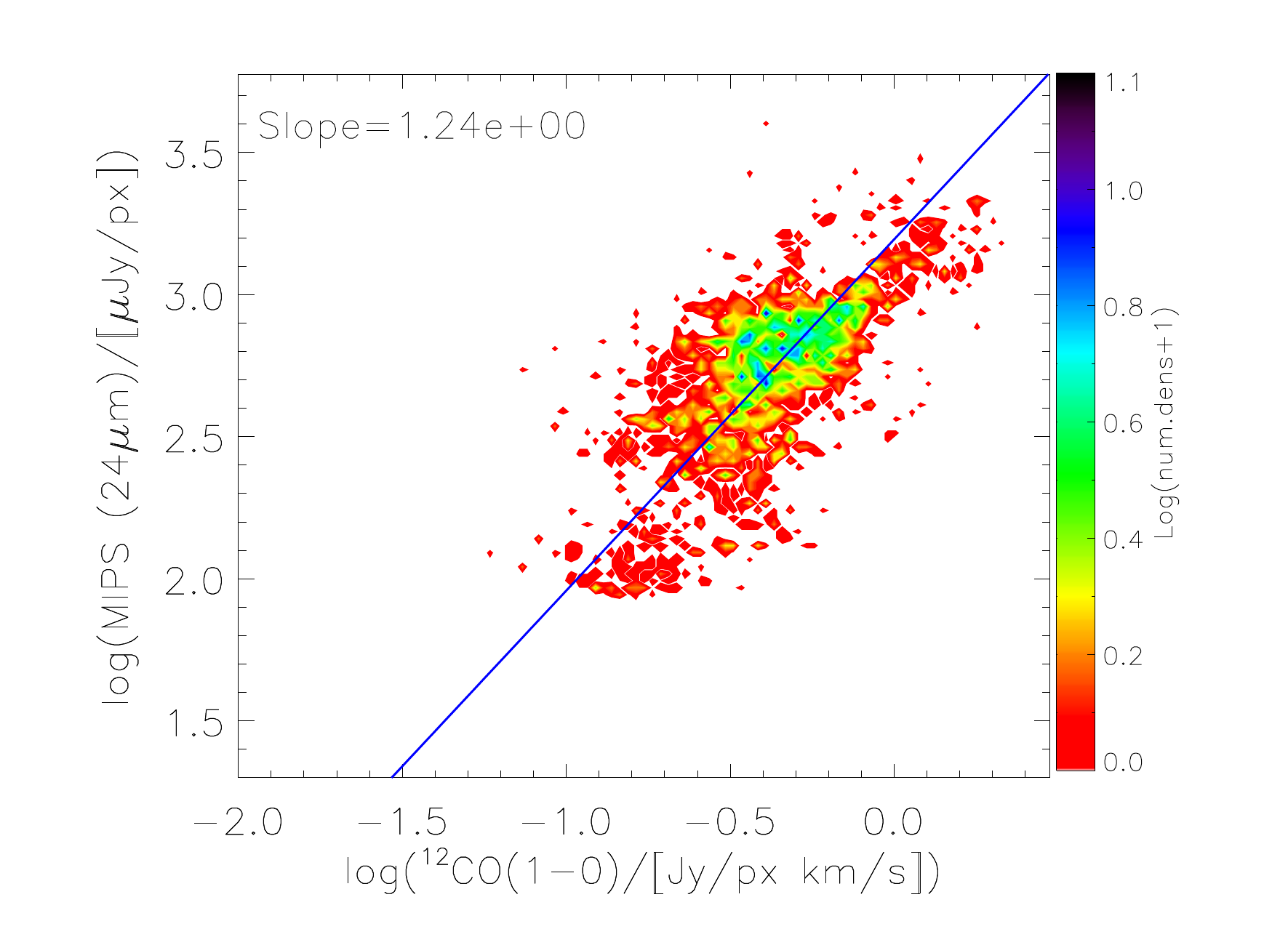}\includegraphics[angle=0]{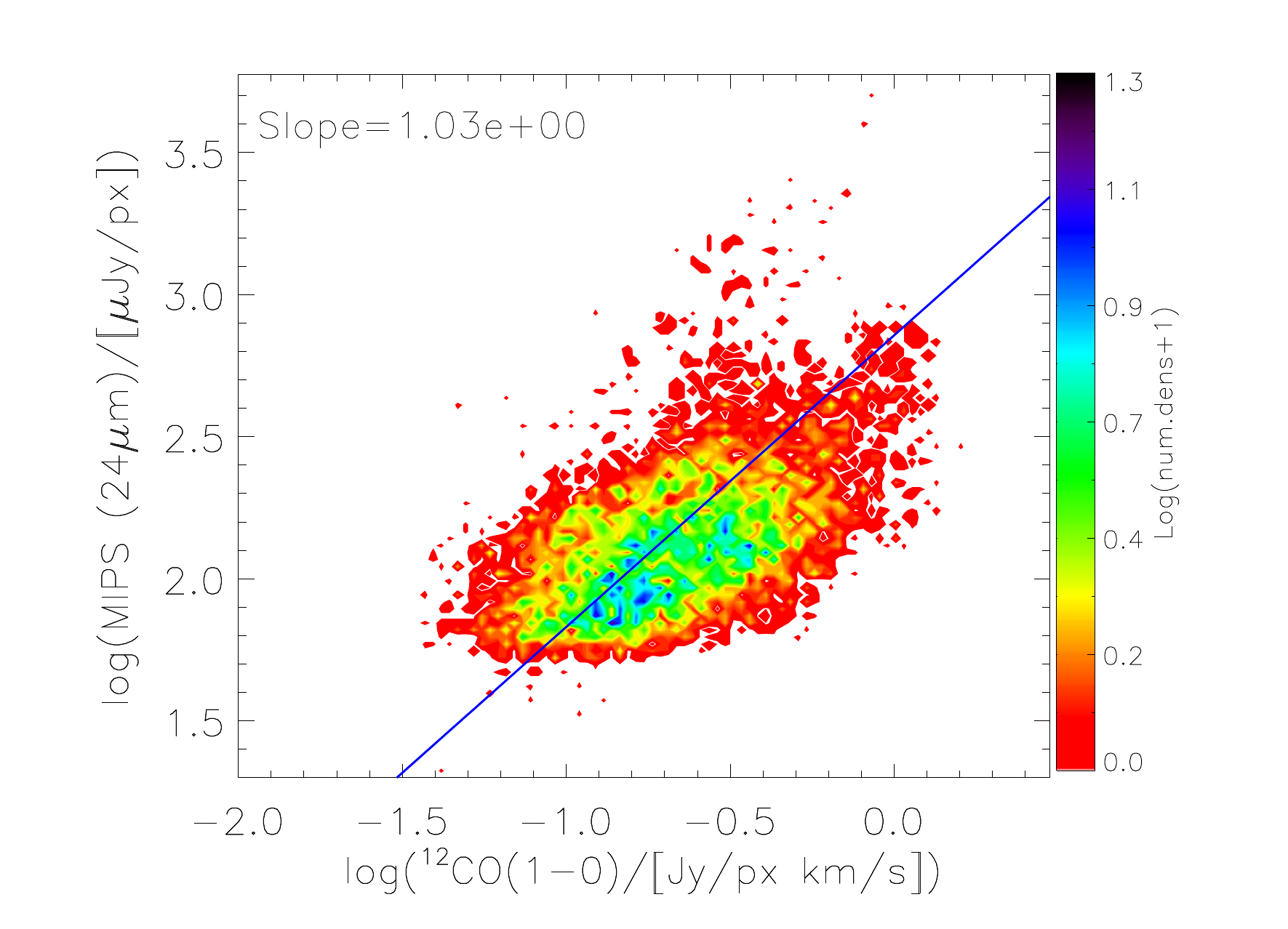}\includegraphics[angle=0]{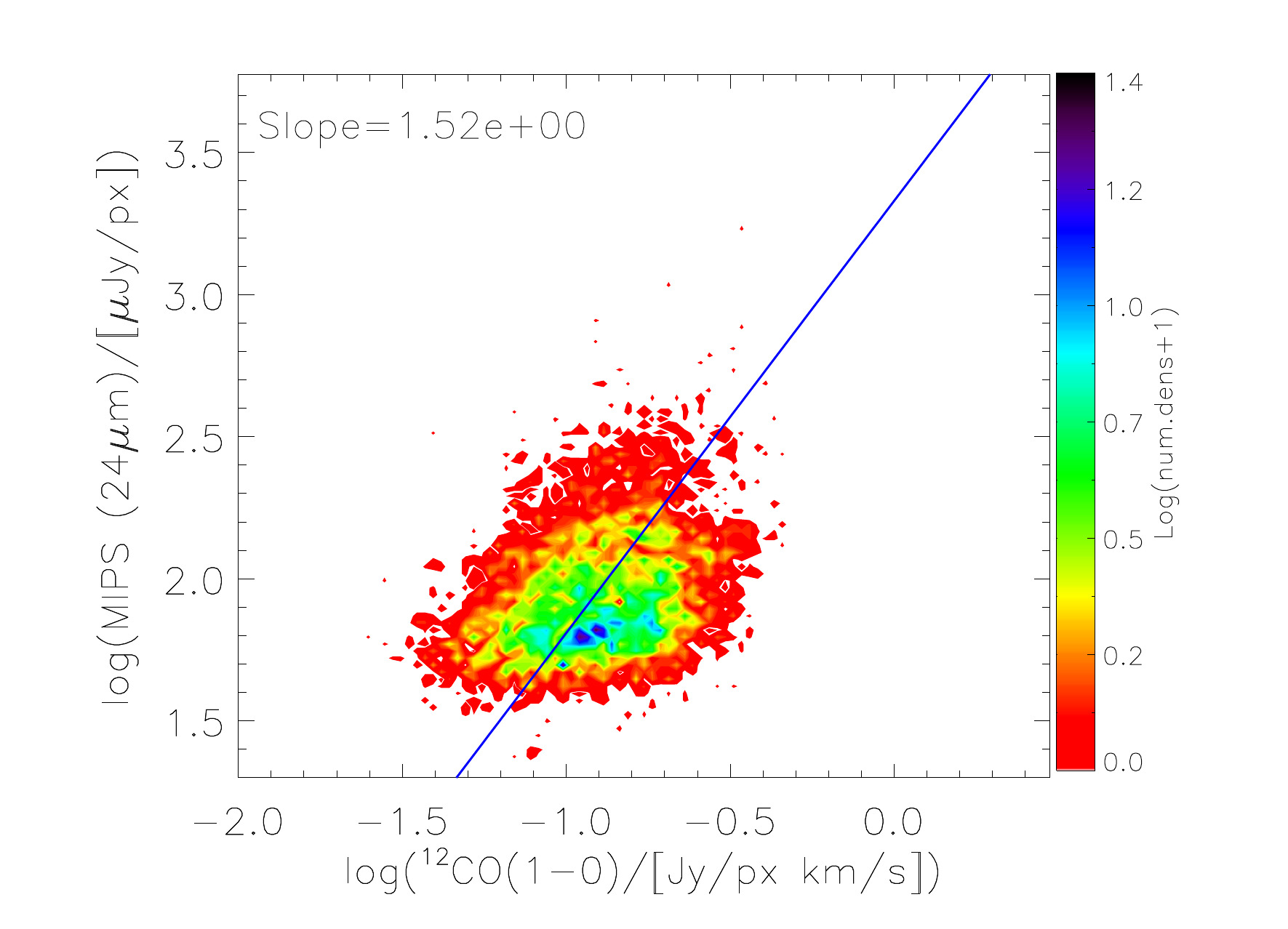}}\\
\resizebox{0.9\hsize}{!}{\includegraphics[angle=0]{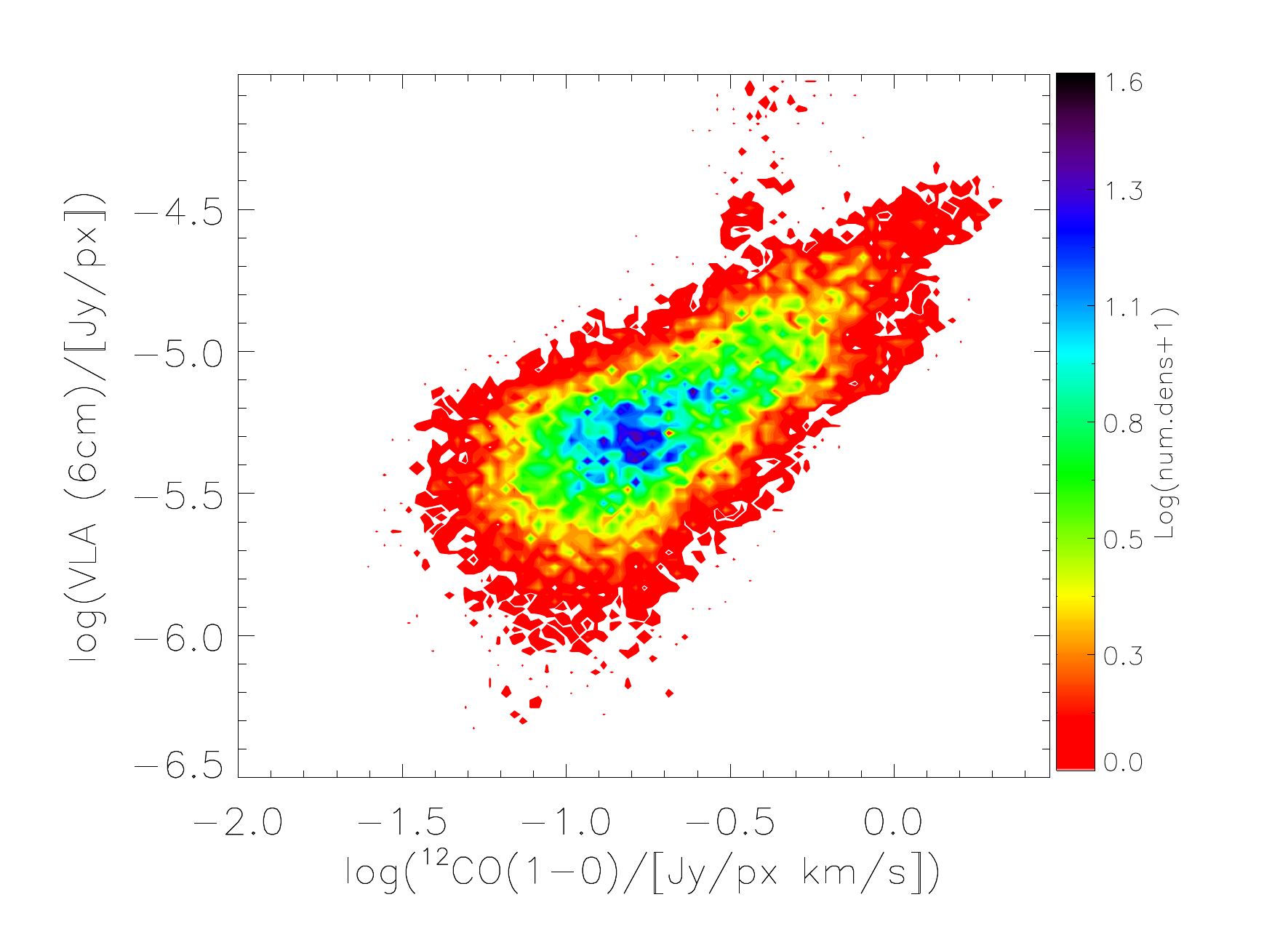}\includegraphics[angle=0]{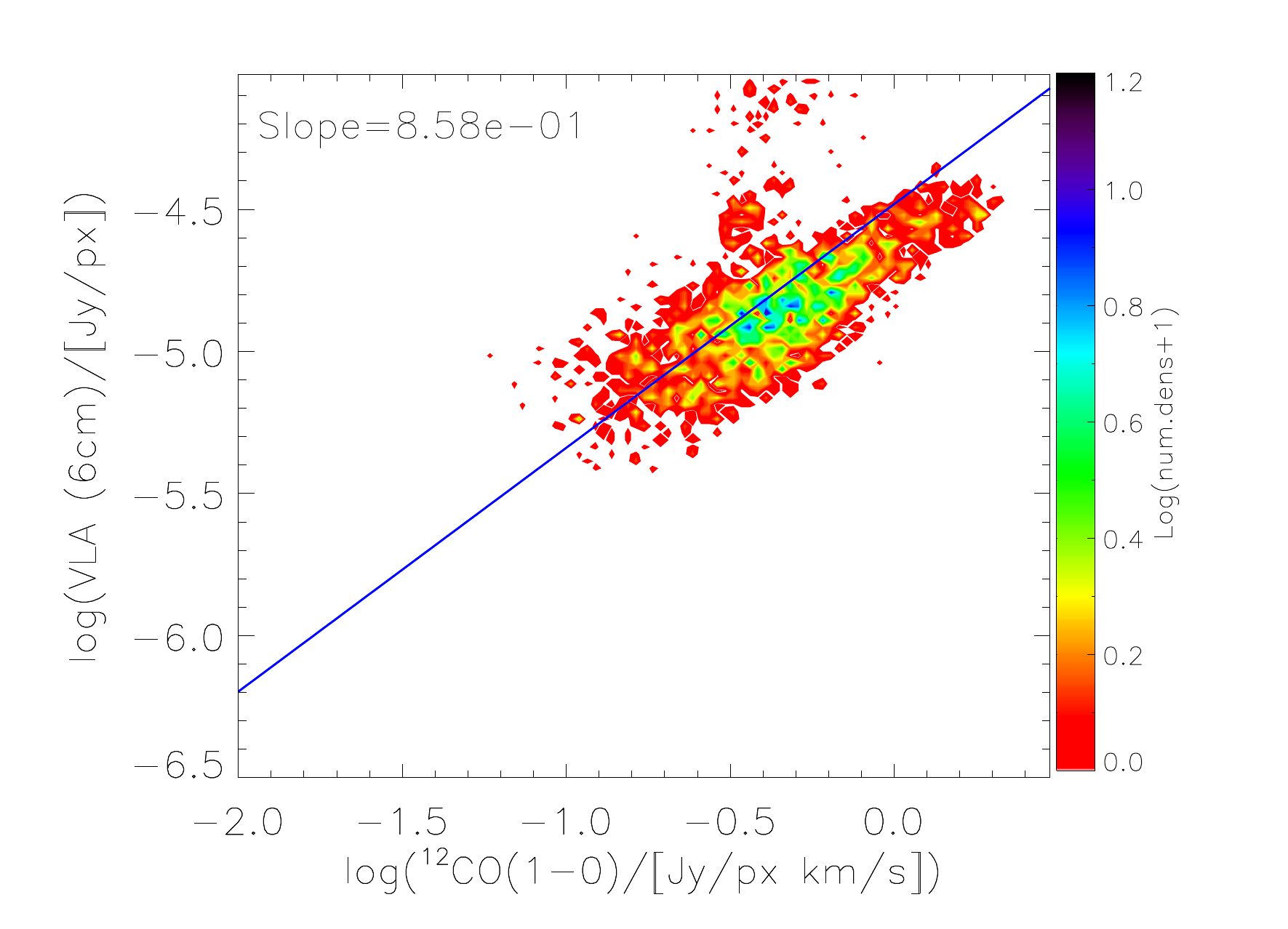}\includegraphics[angle=0]{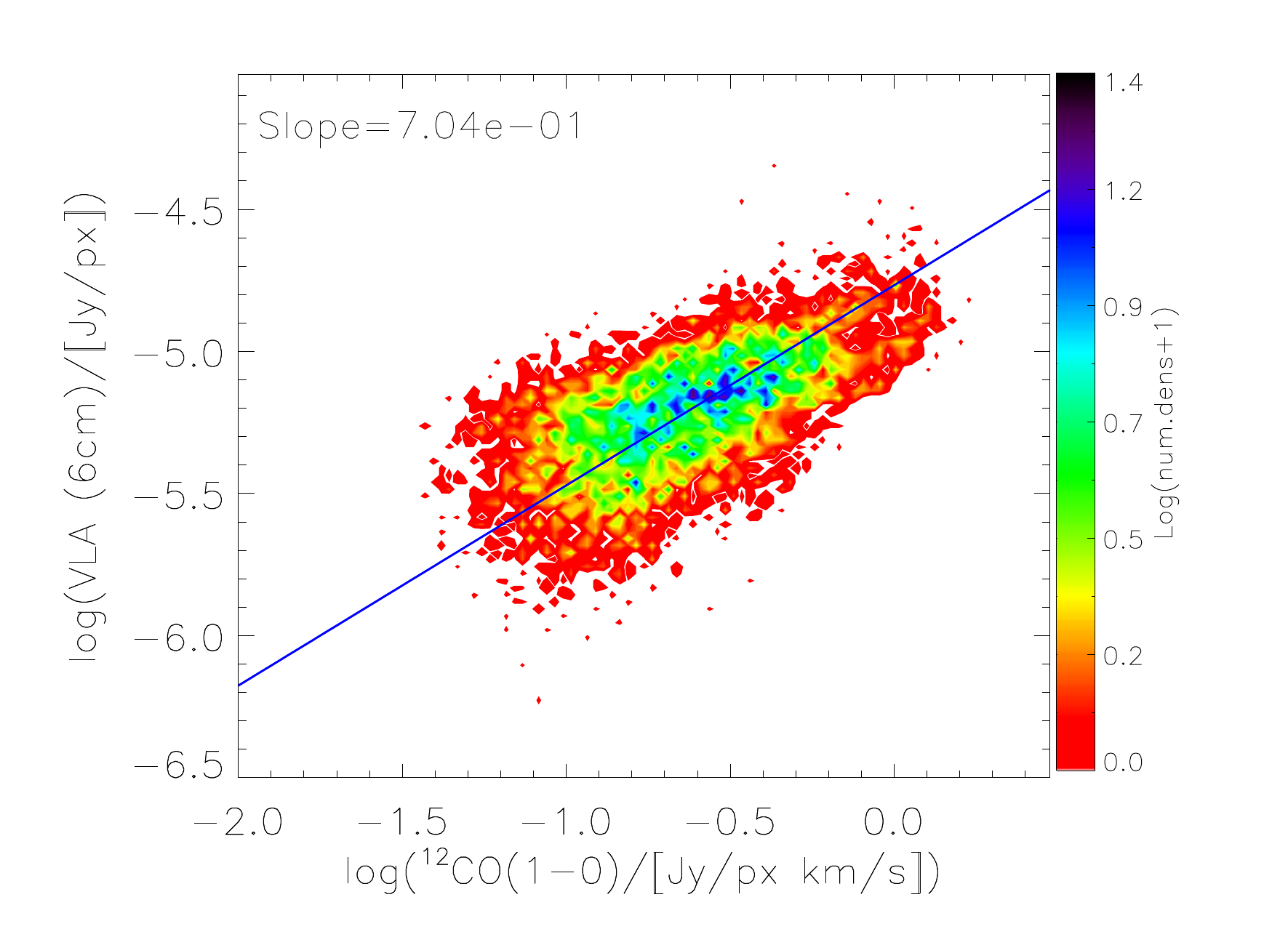}\includegraphics[angle=0]{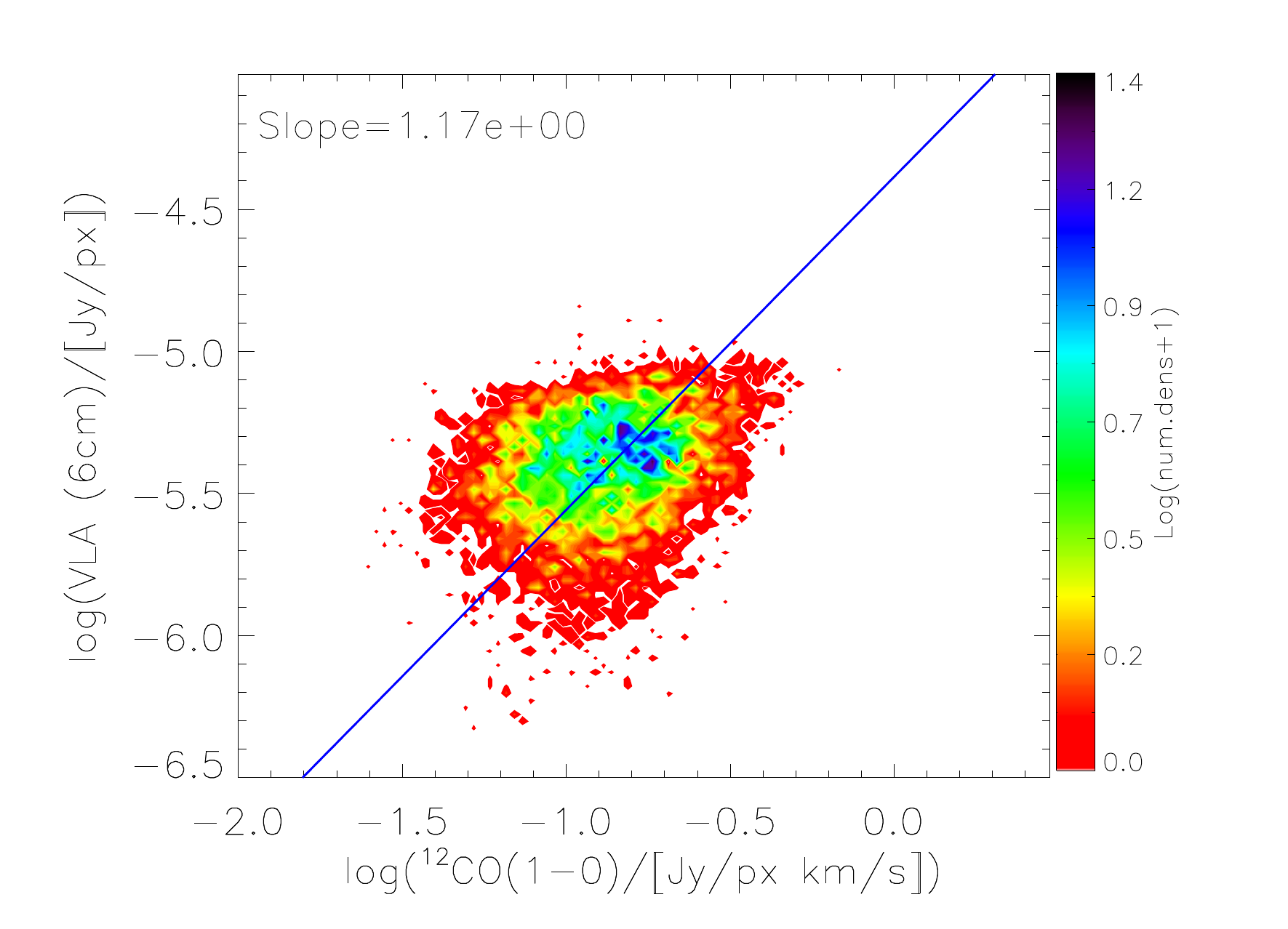}}\\
\resizebox{0.9\hsize}{!}{\includegraphics[angle=0]{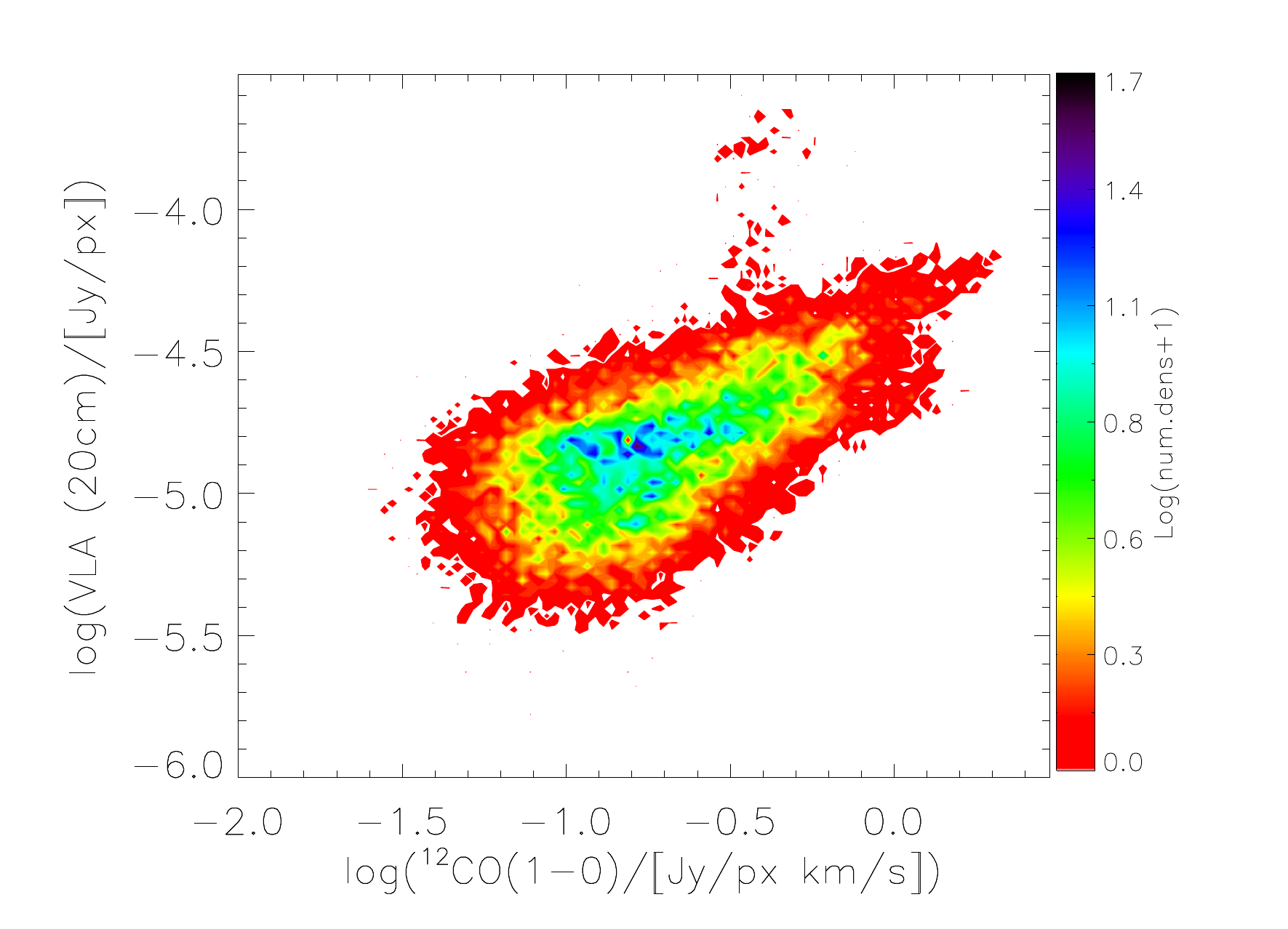}\includegraphics[angle=0]{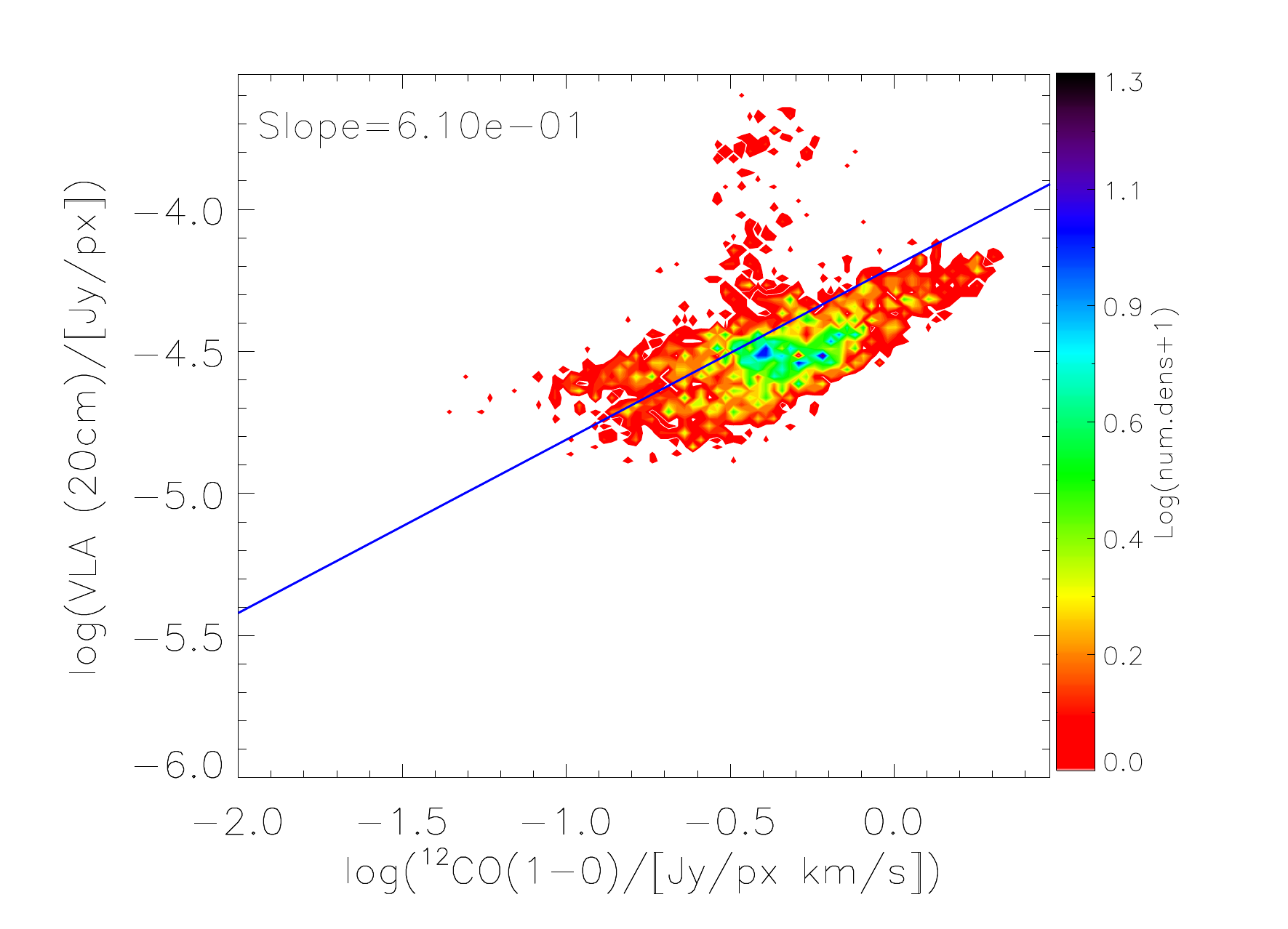}\includegraphics[angle=0]{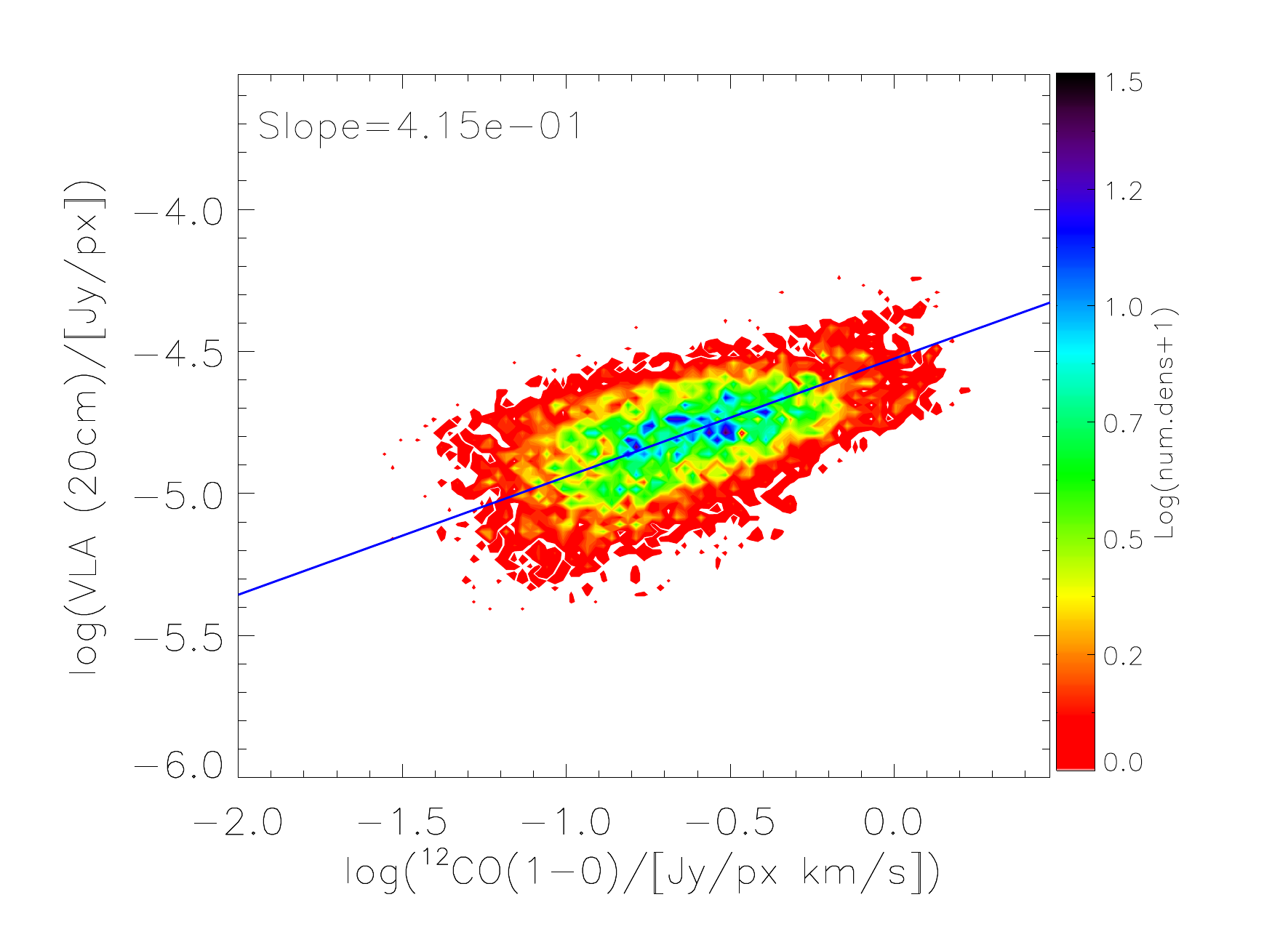}\includegraphics[angle=0]{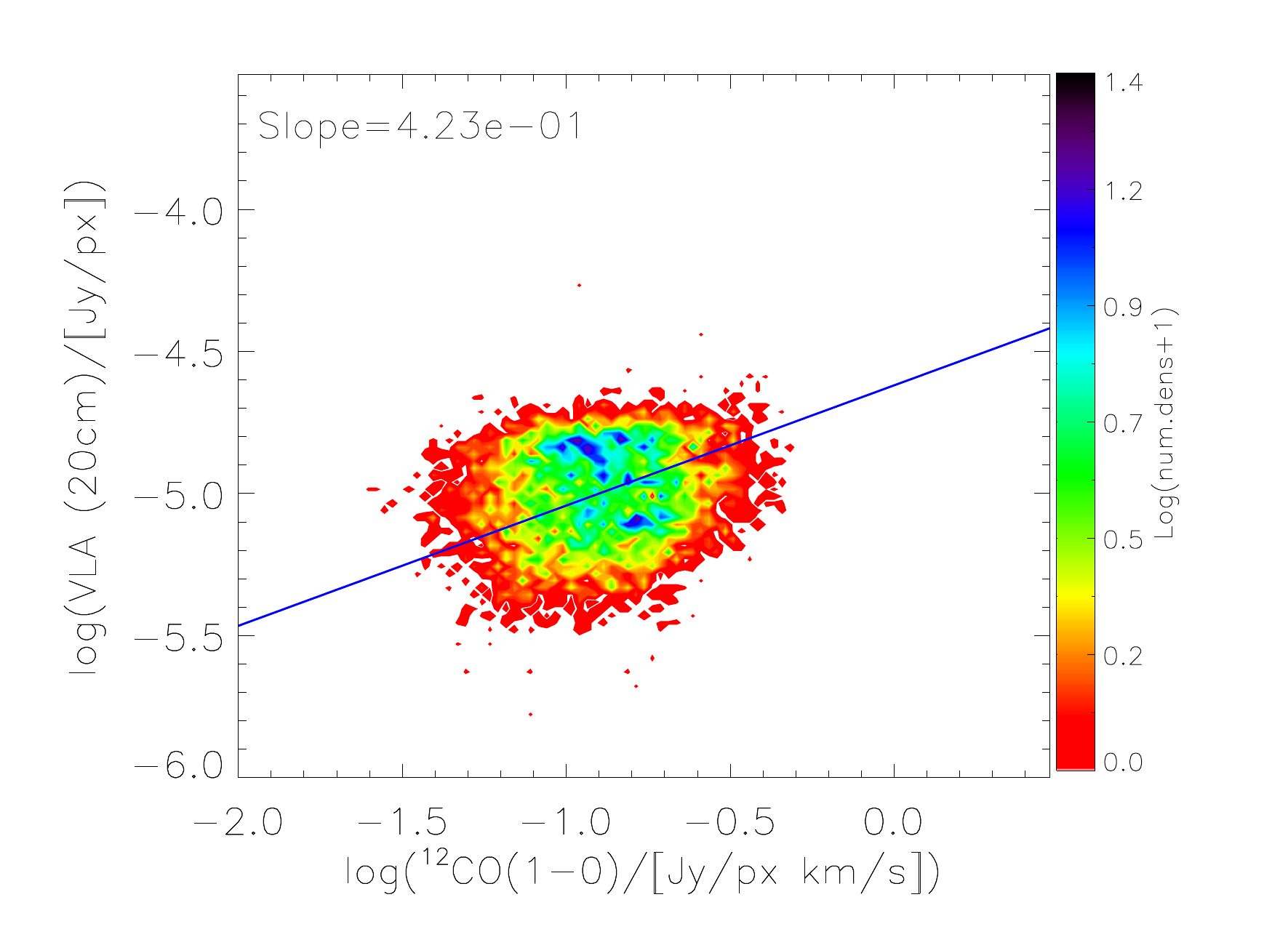}}
\end{center}
\caption{Pixel-by-pixel comparison at 3.0'' resolution of the \coone ~emission versus tracers of the ISM
({\it from top to bottom}): 
HST $I-H$ color, 
MIPS 24$\mu$m emission,
VLA 6\,cm radio continuum, and
VLA 20cm radio continuum. 
The distribution is shown ({\it from left to right}) 
for the full PAWS FoV, and separately for the central 40'',
the spiral arms and the inter-arm region (as defined in Fig. \ref{fig:co_2d}). The density is given on a logarithmic color scale.
\label{fig:ism_pix}}
\end{figure}

\begin{figure}
\begin{center}
%includegraphics[]{f1.eps}
\resizebox{0.9\hsize}{!}{\includegraphics[angle=0]{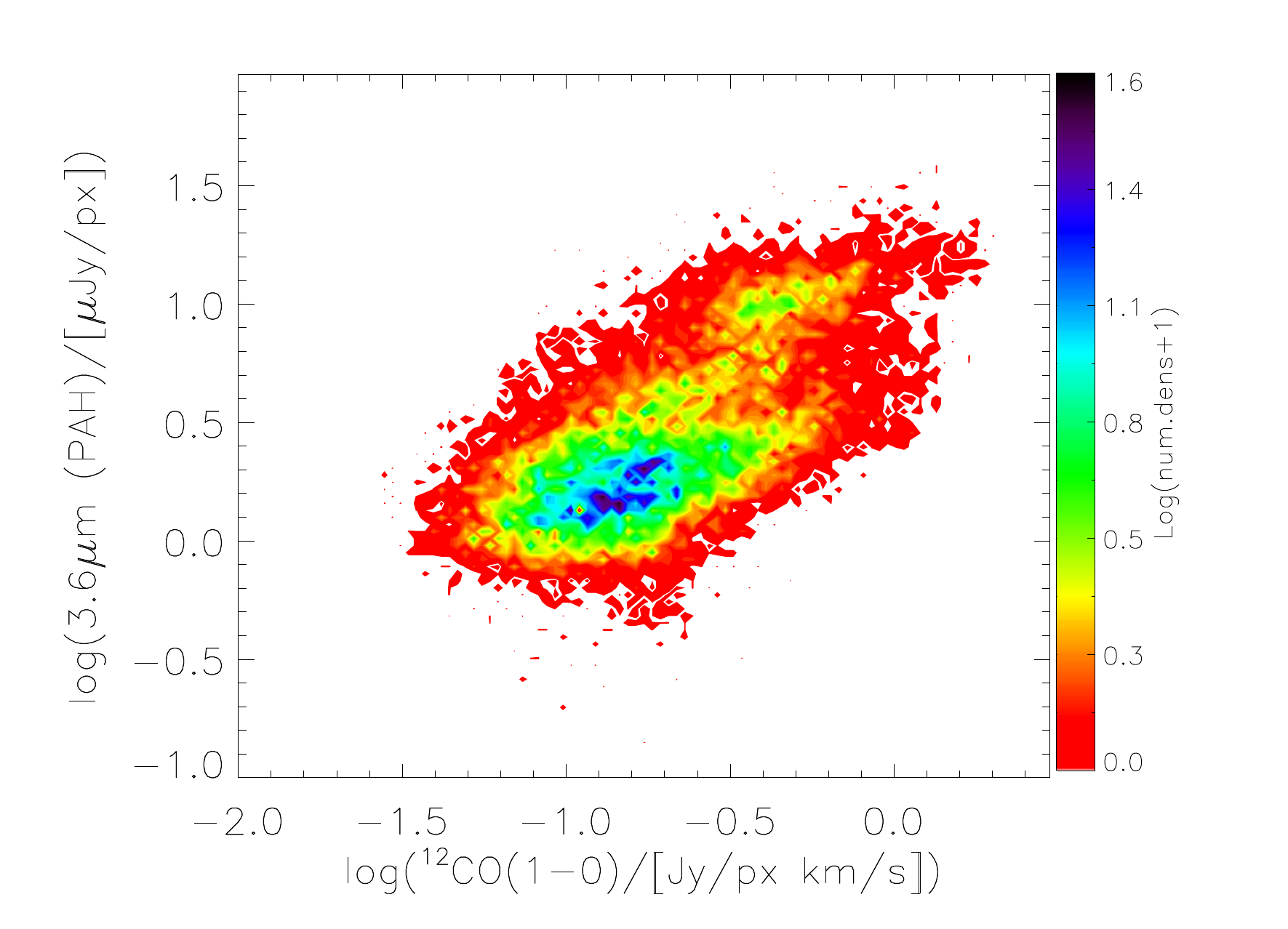}\includegraphics[angle=0]{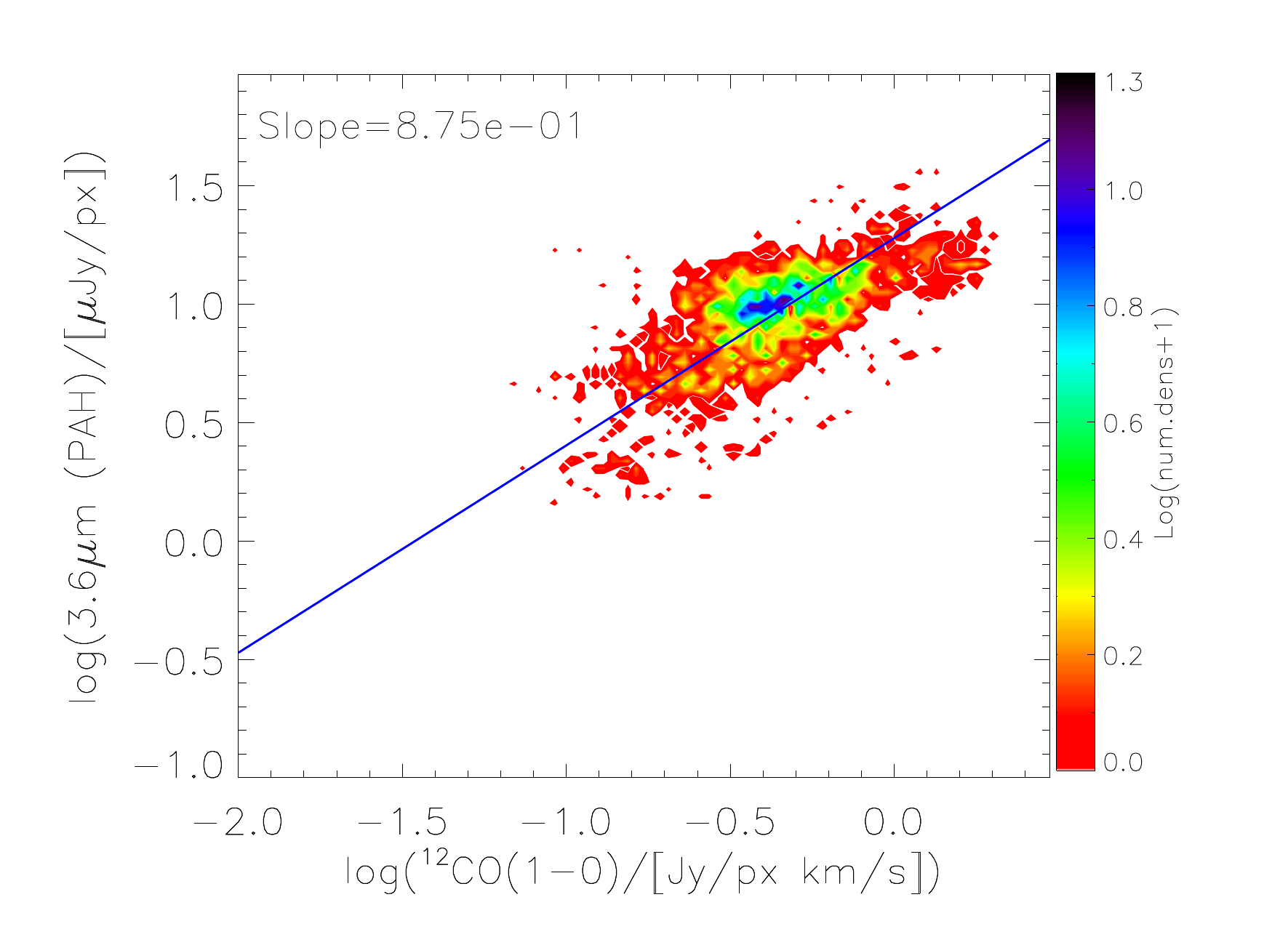}\includegraphics[angle=0]{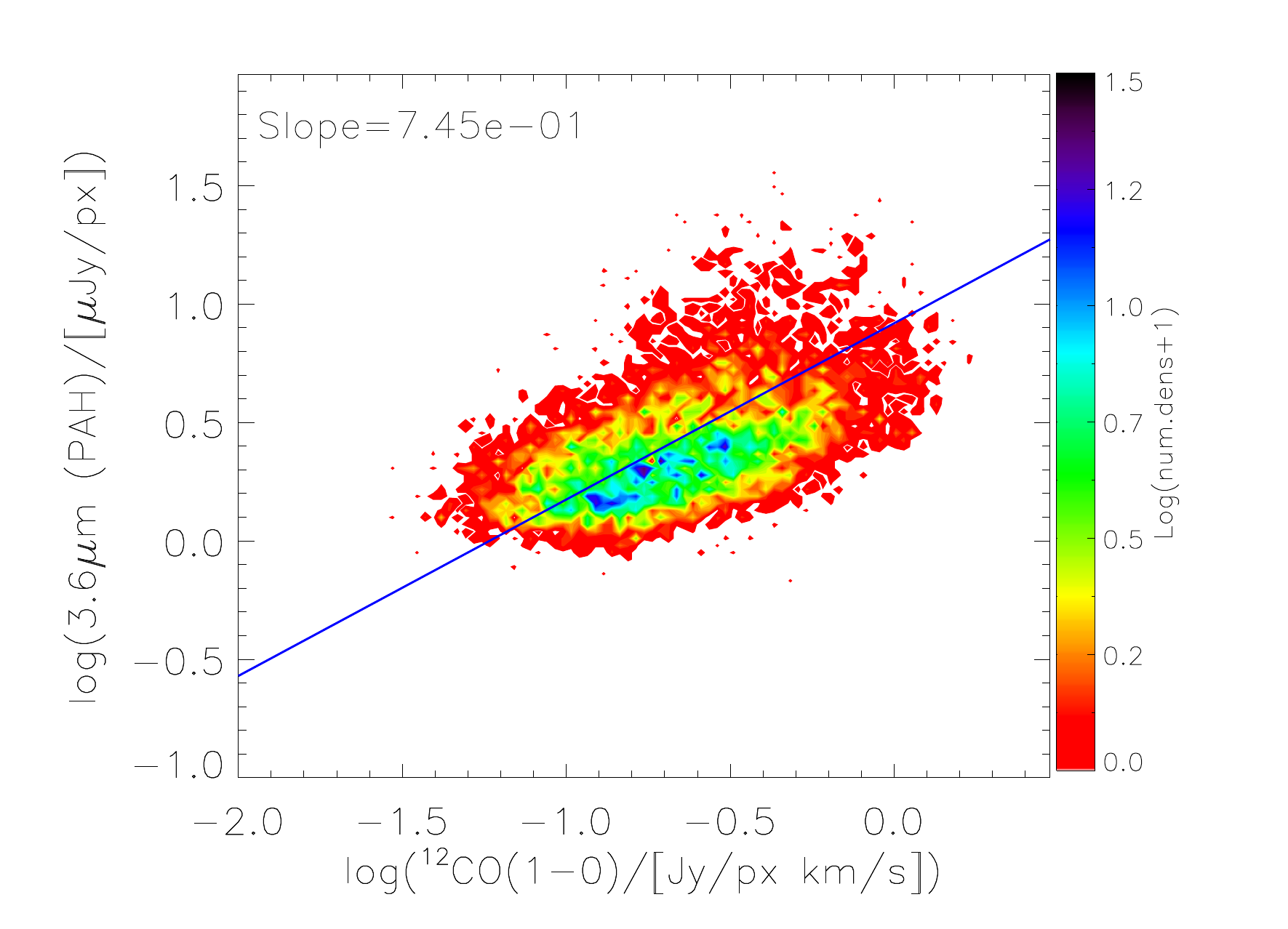}\includegraphics[angle=0]{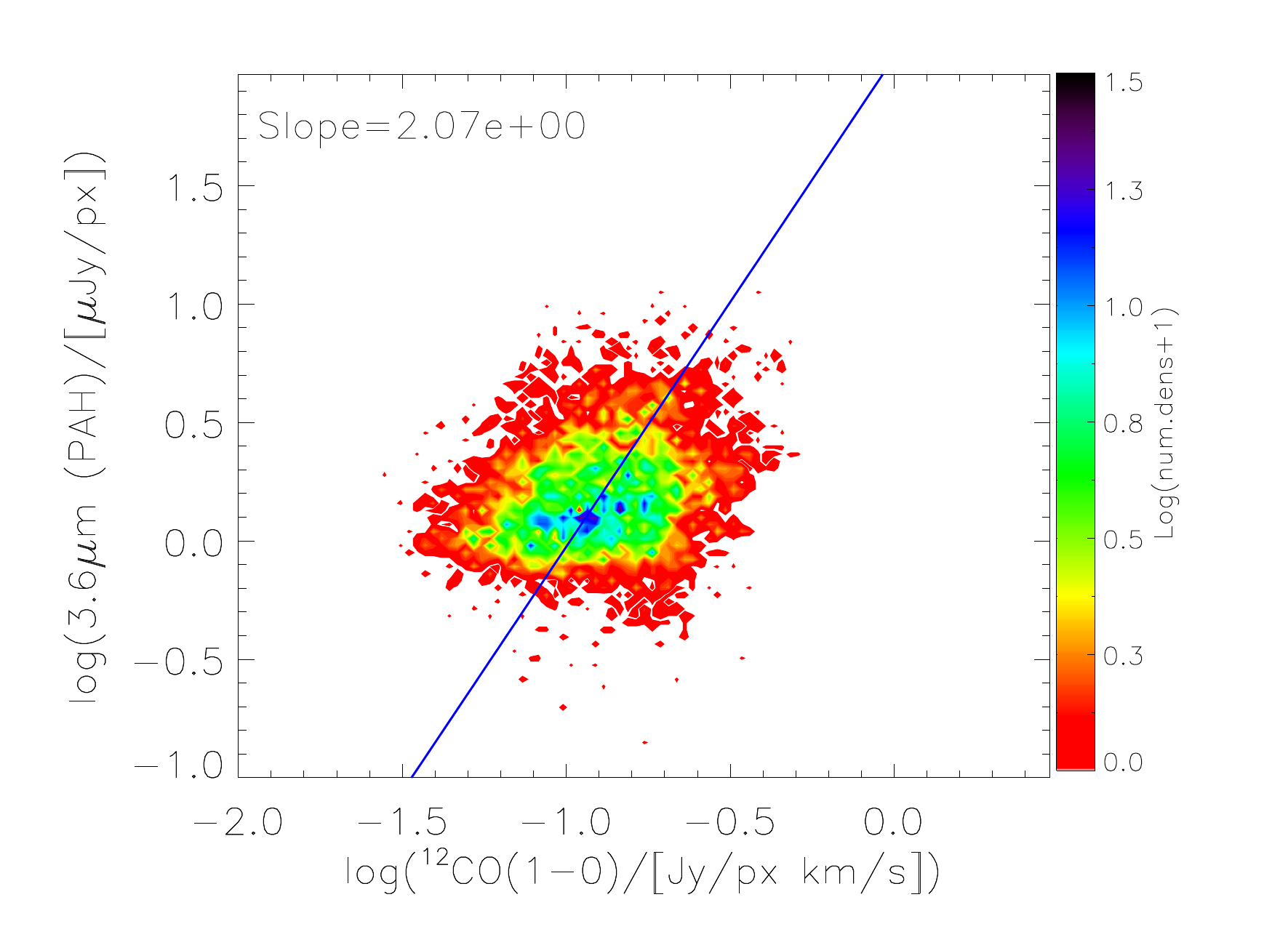}}\\
\resizebox{0.9\hsize}{!}{\includegraphics[angle=0]{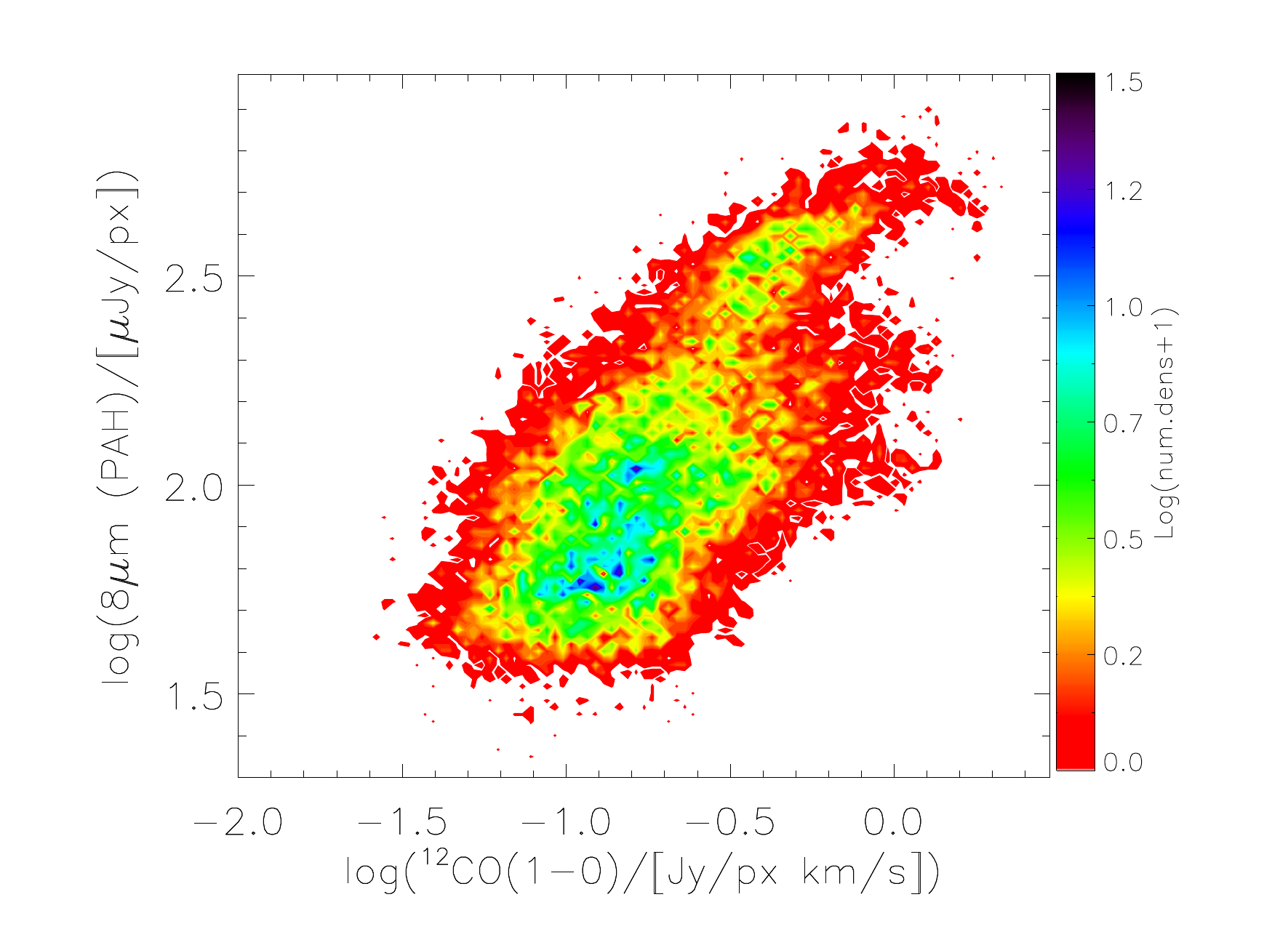}\includegraphics[angle=0]{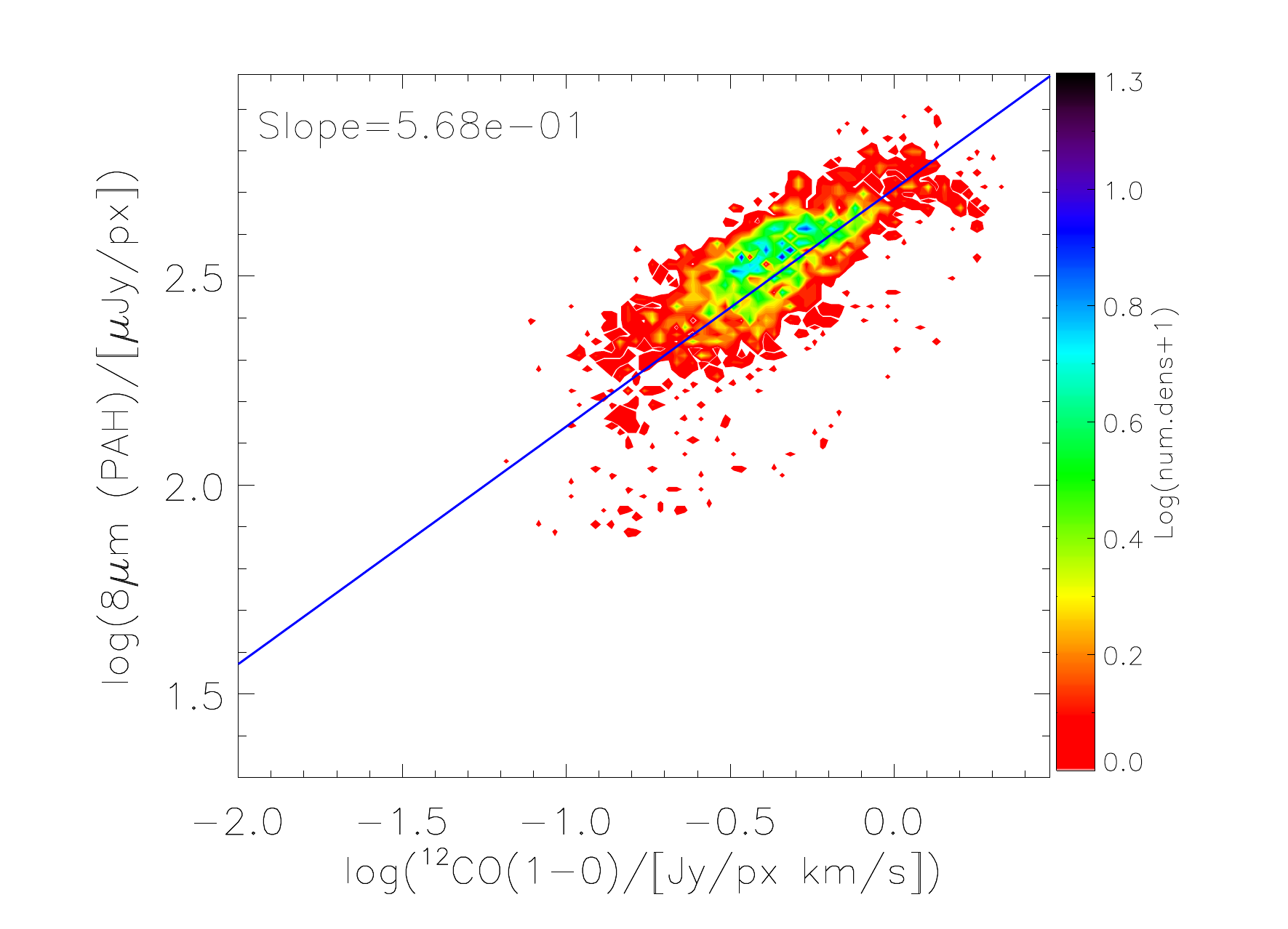}\includegraphics[angle=0]{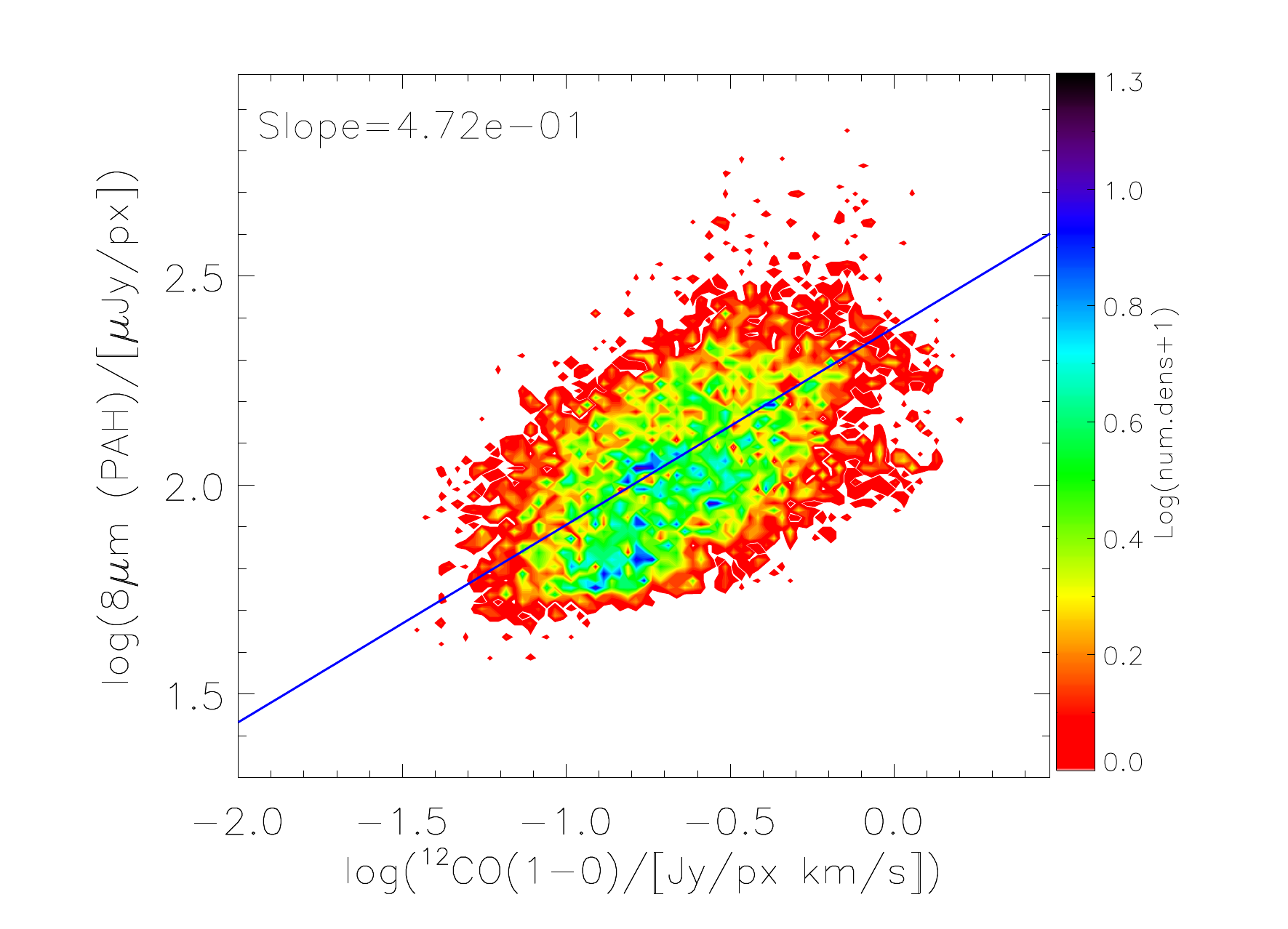}\includegraphics[angle=0]{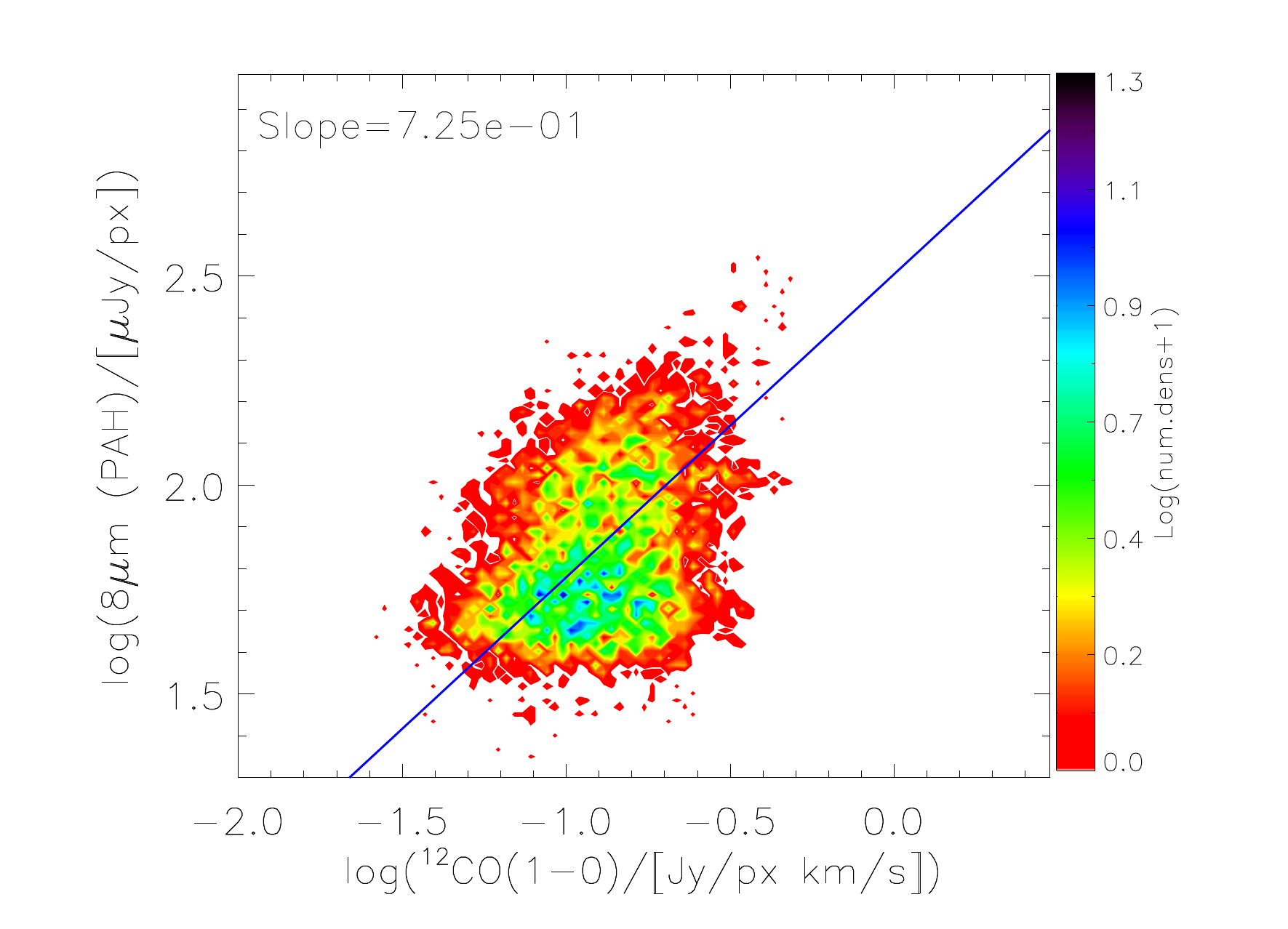}}\\
\resizebox{0.9\hsize}{!}{\includegraphics[angle=0]{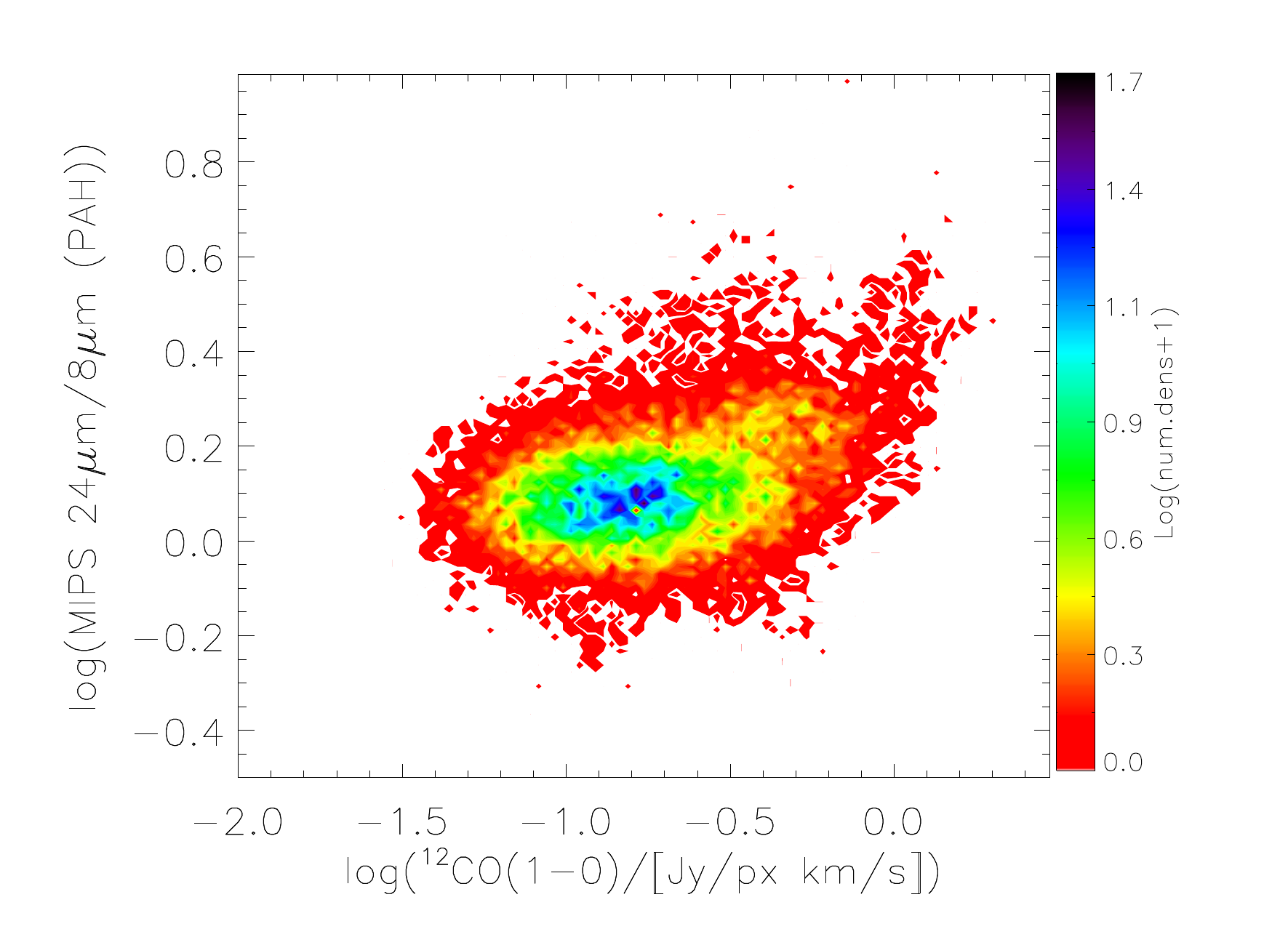}\includegraphics[angle=0]{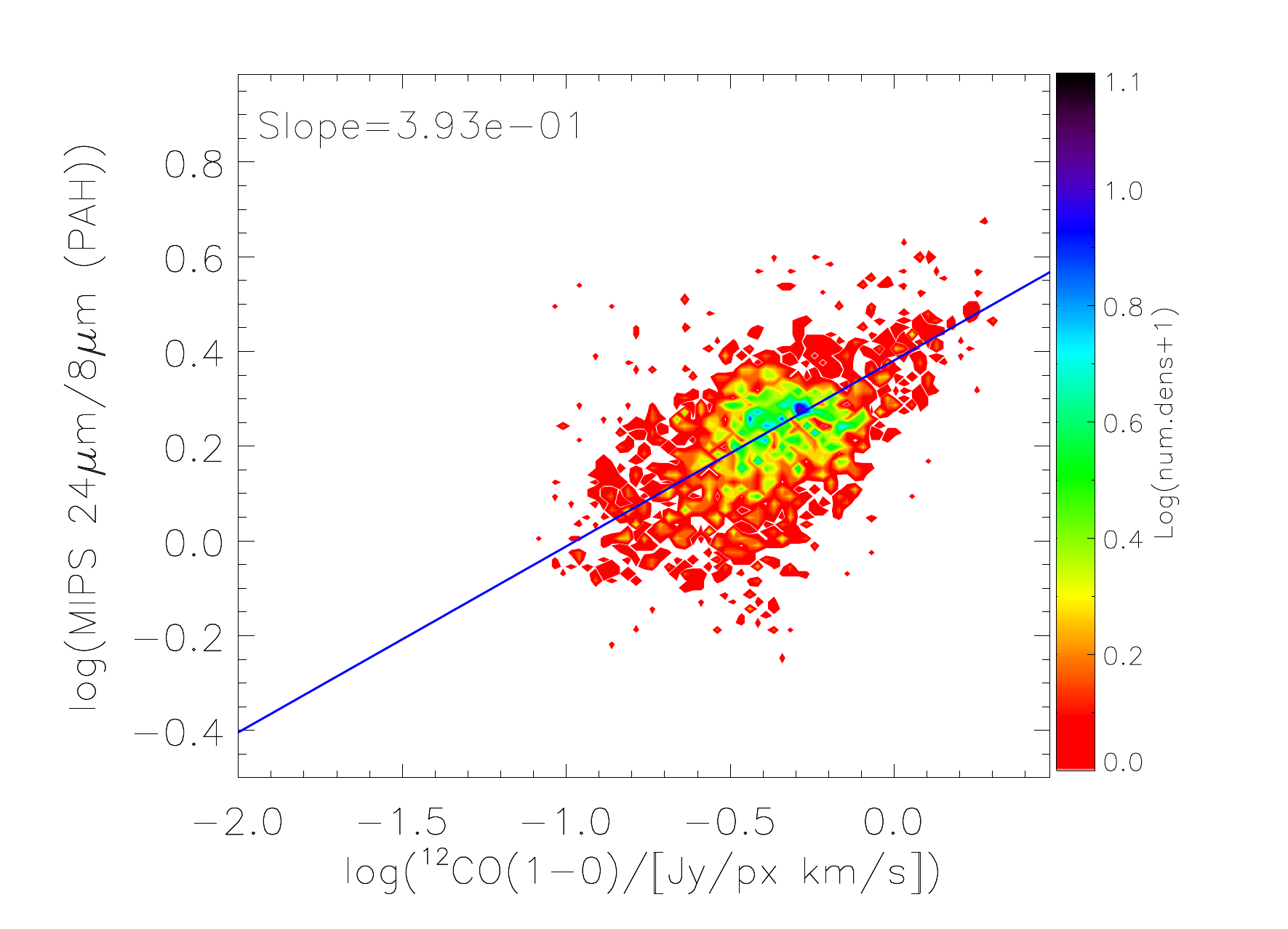}\includegraphics[angle=0]{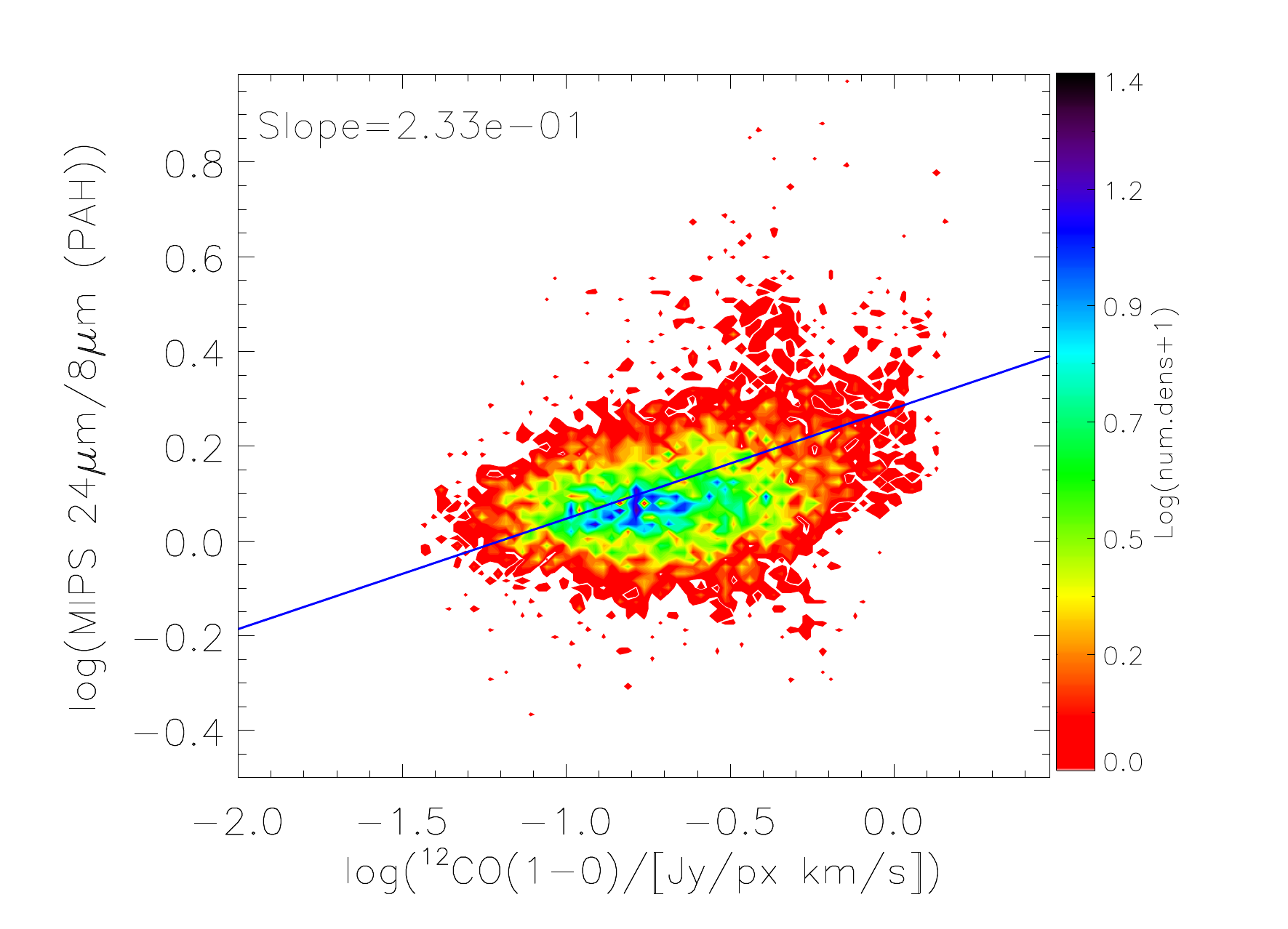}\includegraphics[angle=0]{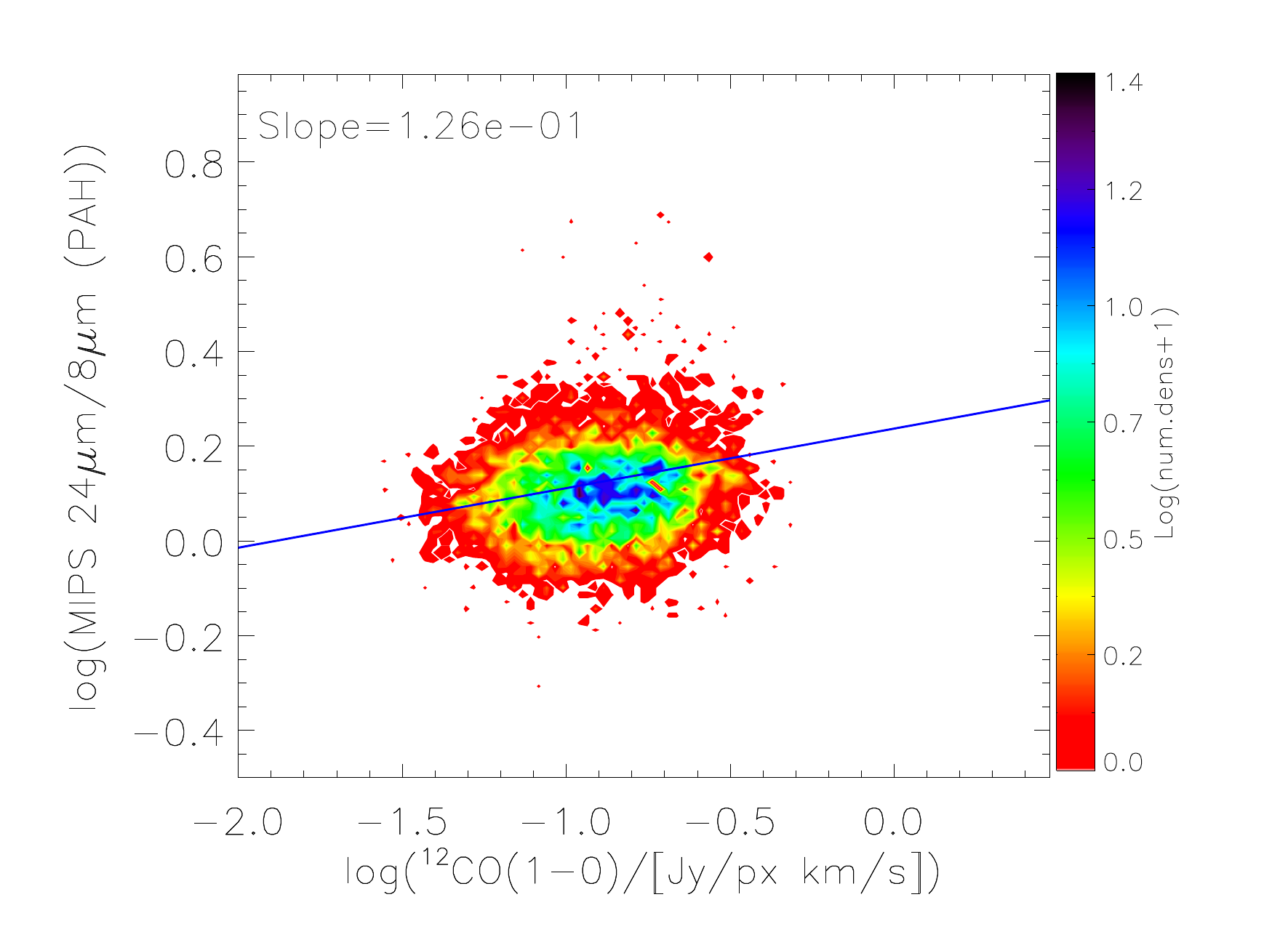}}
\resizebox{0.9\hsize}{!}{\includegraphics[angle=0]{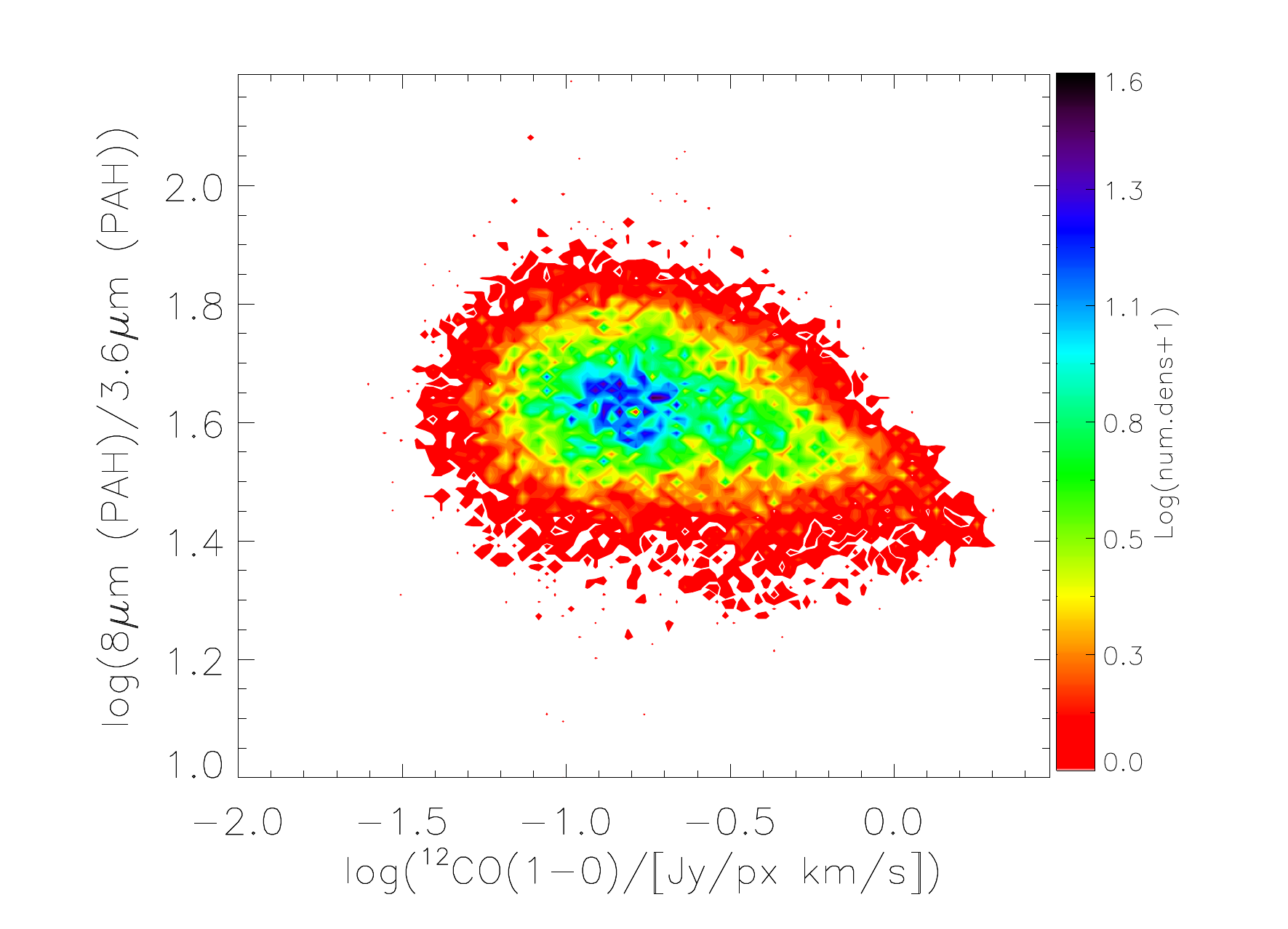}\includegraphics[angle=0]{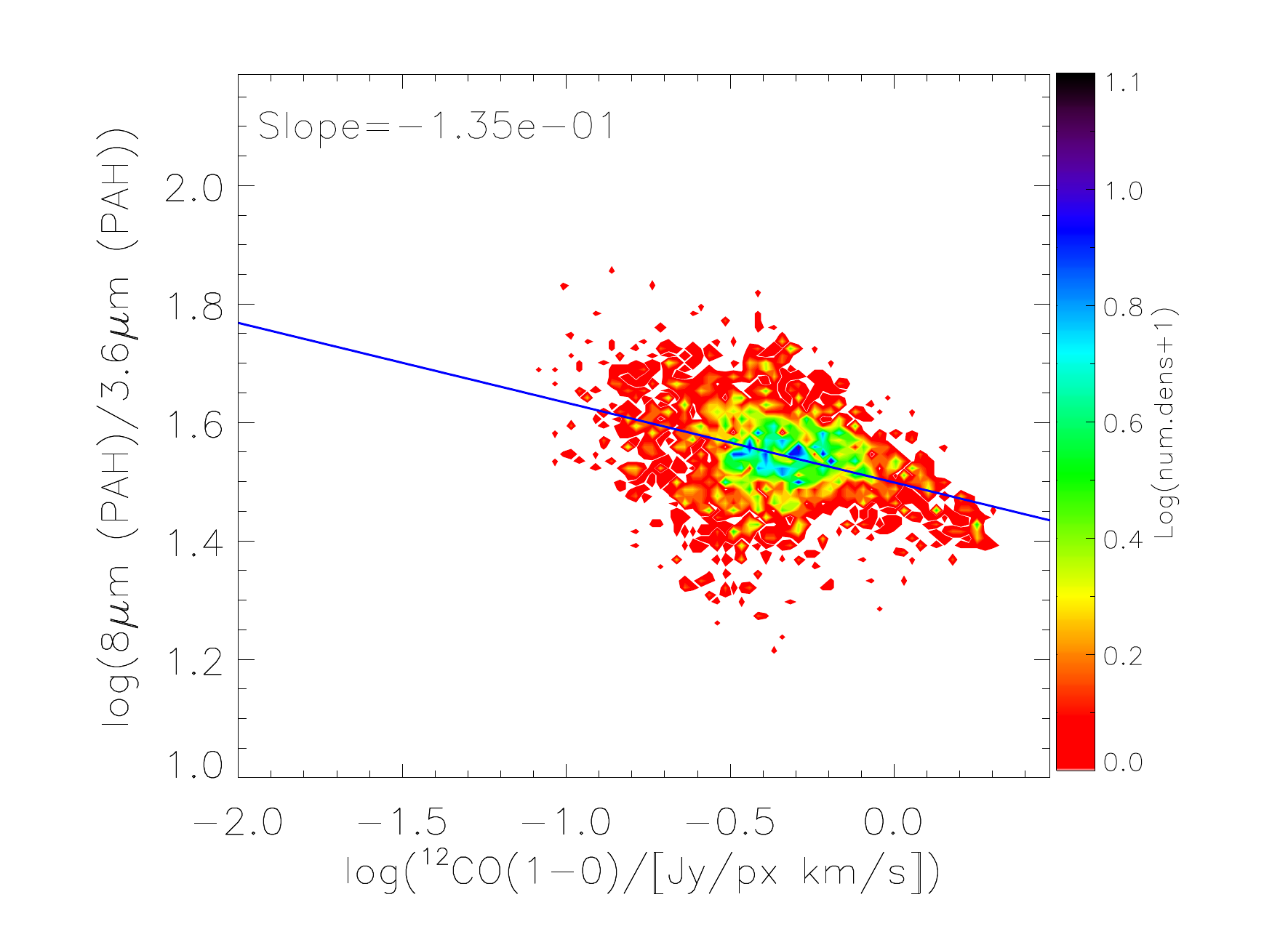}\includegraphics[angle=0]{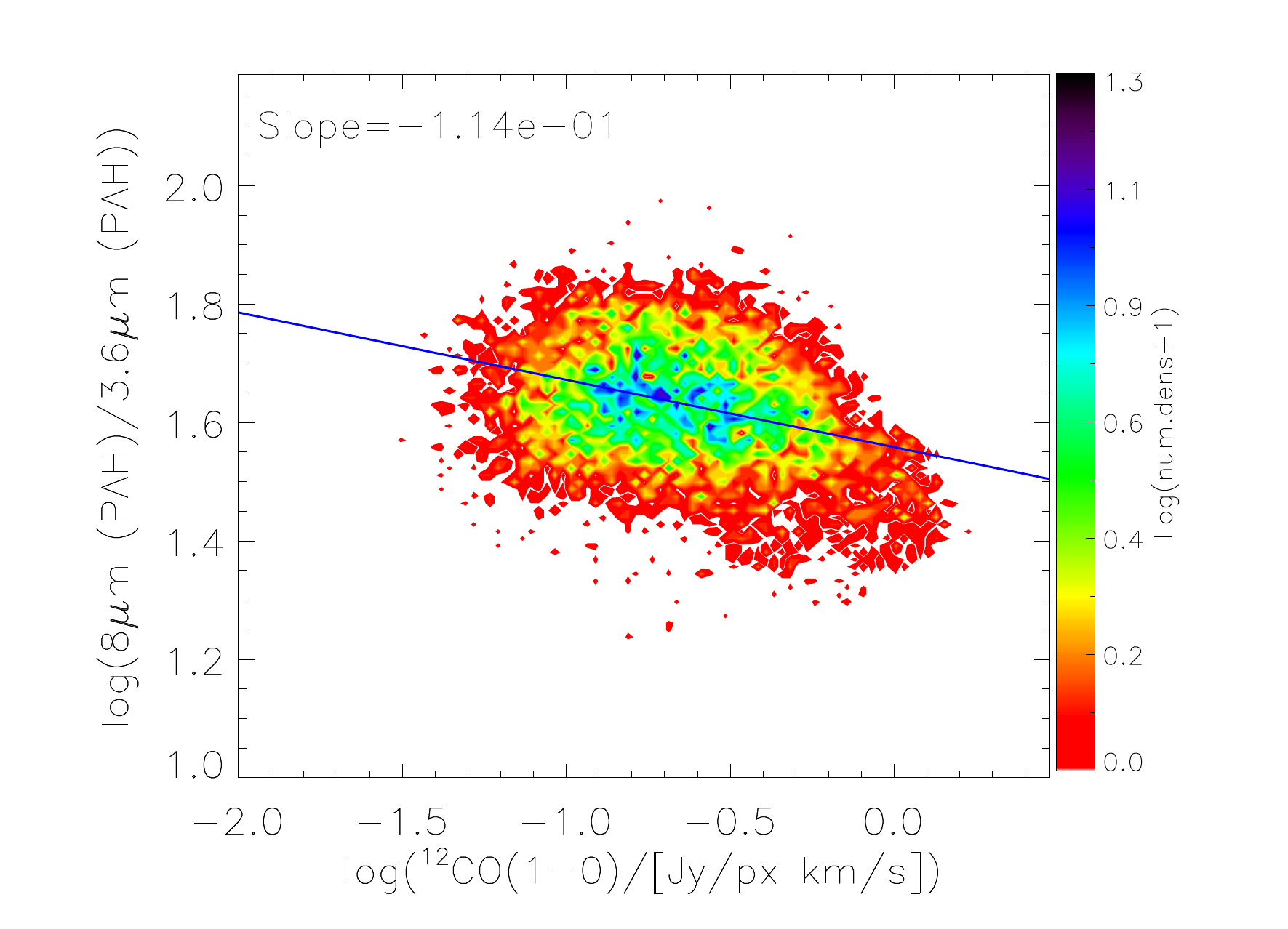}\includegraphics[angle=0]{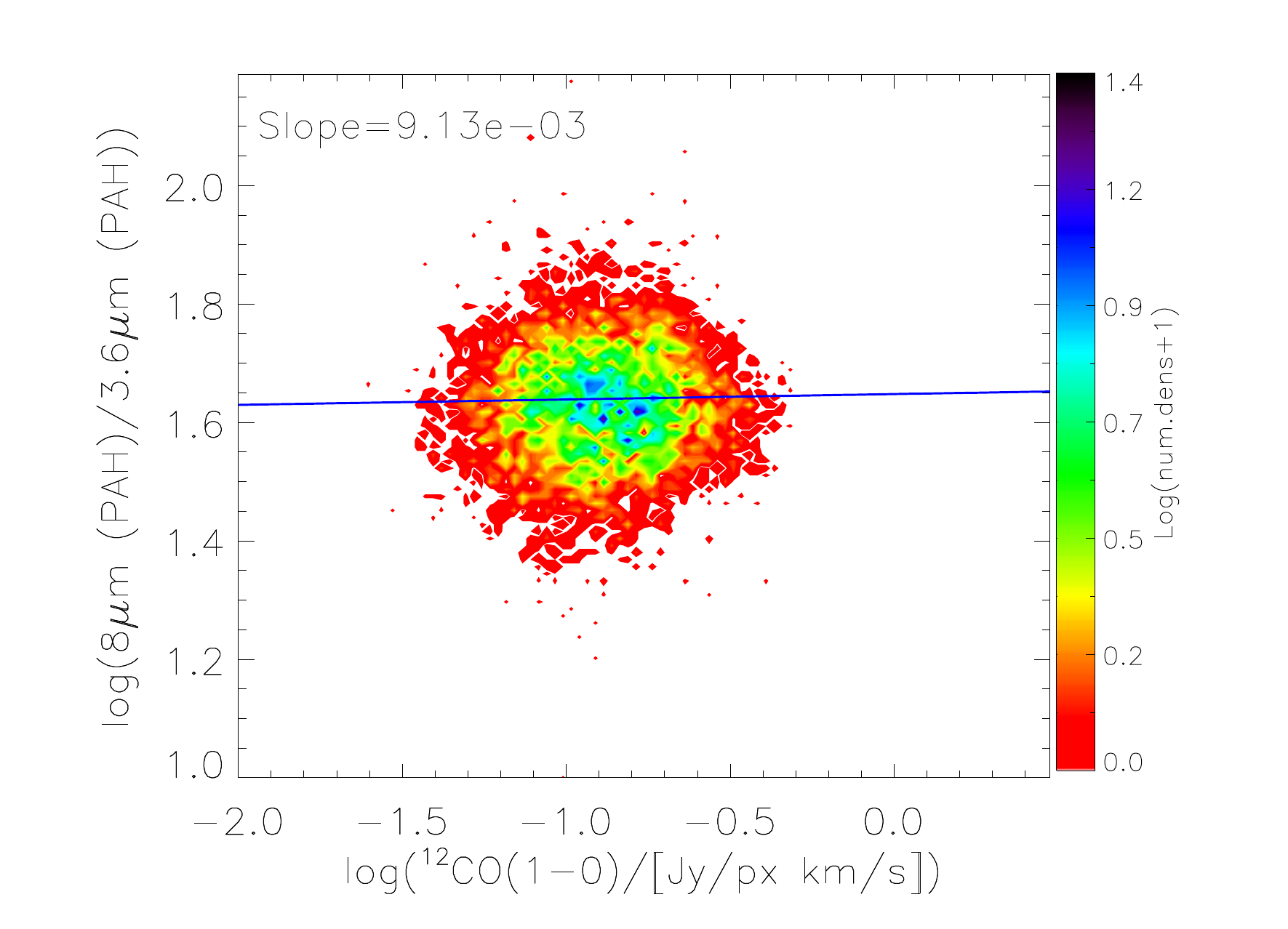}}
\end{center}
\caption{Pixel-by-pixel comparison at 3.0'' resolution of the \coone ~emission versus tracers of the PAH emission 
({\it from top to bottom}):
non-stellar emission at 3.6$\mu$m, non-stellar emission at 8$\mu$m,
the ratio of 24$\mu$m/8$\rm \mum_{non-stellar}$, and
the ratio of non-stellar emission at 8$\mu$m over 3.6$\mu$m. 
The distribution is shown ({\it from left to right}) 
for the full PAWS FoV, and separately for the central 40'',
the spiral arms and the inter-arm region (as defined in Fig. \ref{fig:co_2d}). The density is given on a logarithmic color scale.
\label{fig:pah_pix}}
\end{figure}

\begin{figure}
\begin{center}
%includegraphics[]{f1.eps}
\resizebox{1.0\hsize}{!}{\includegraphics[angle=-90]{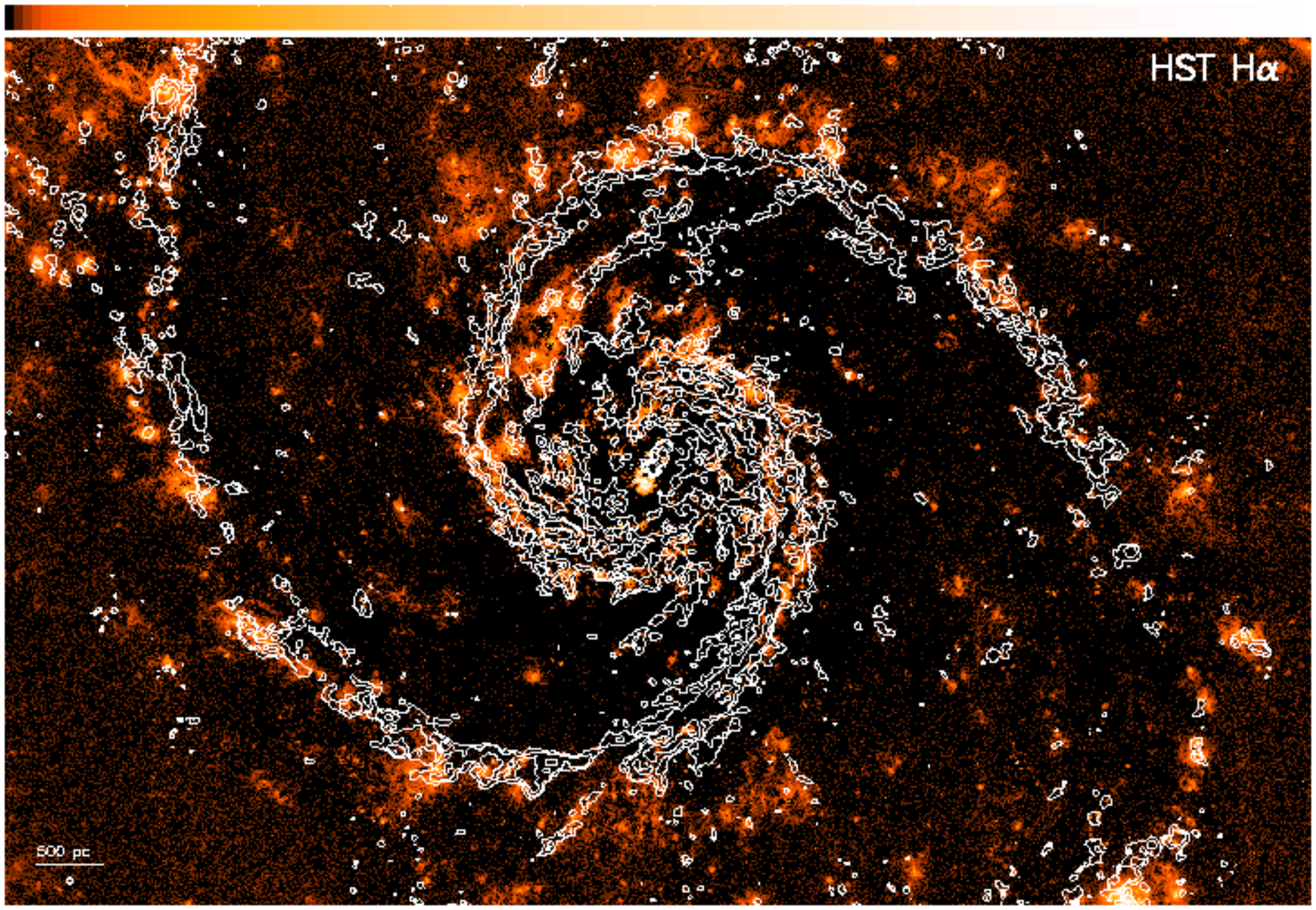}\includegraphics[angle=-90]{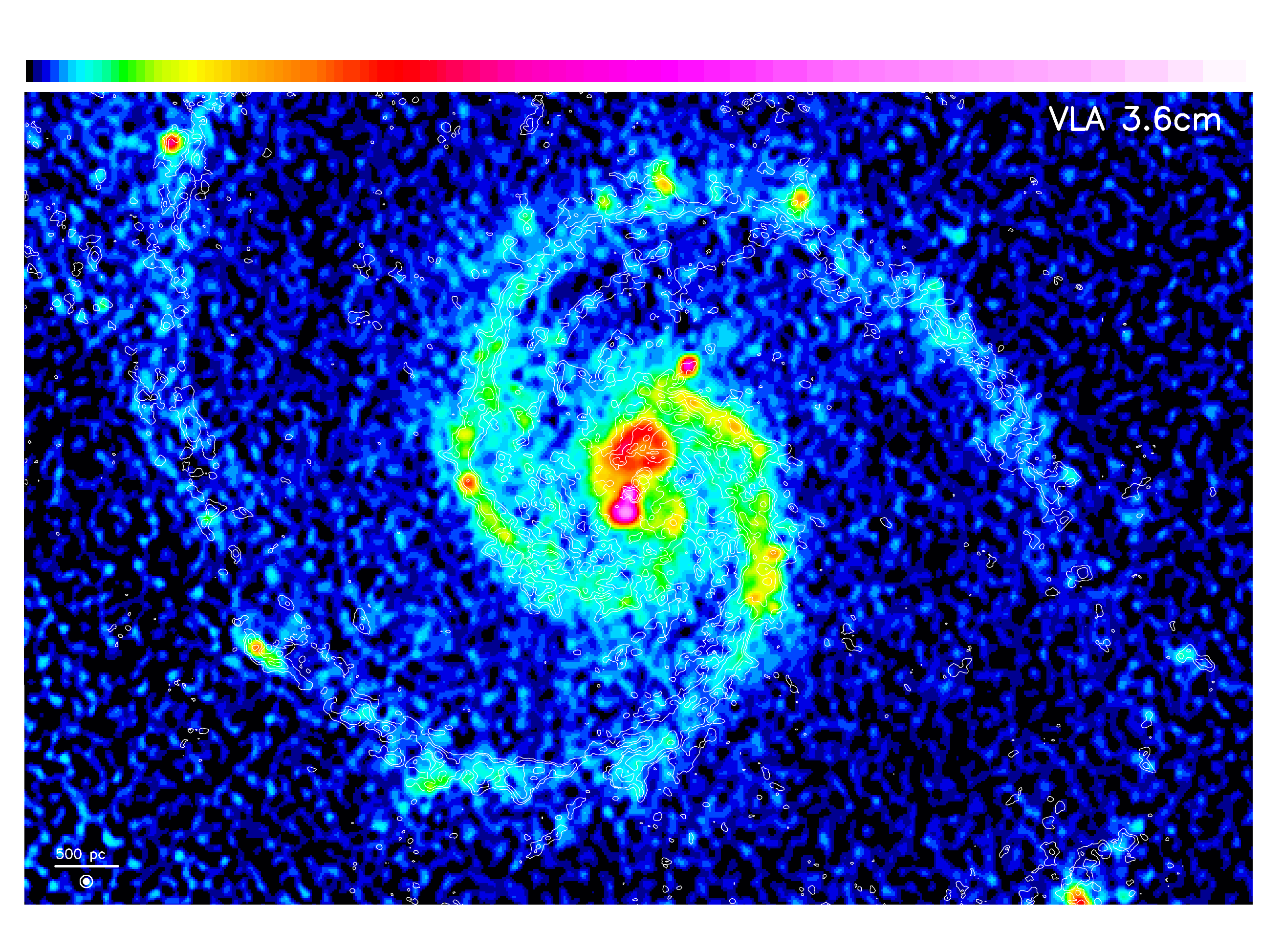}}\\
\resizebox{1.0\hsize}{!}{\includegraphics[angle=-90]{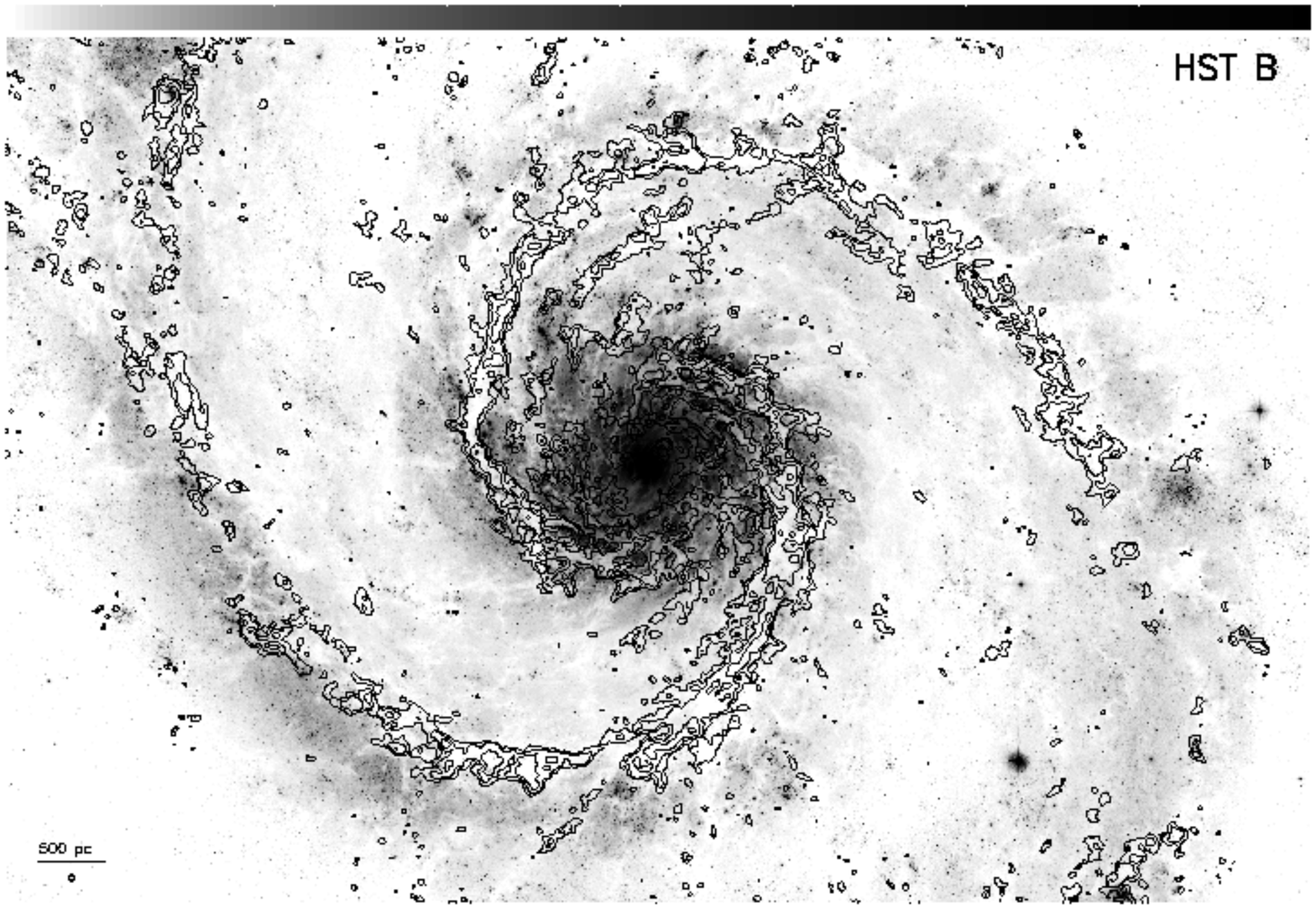}\includegraphics[angle=-90]{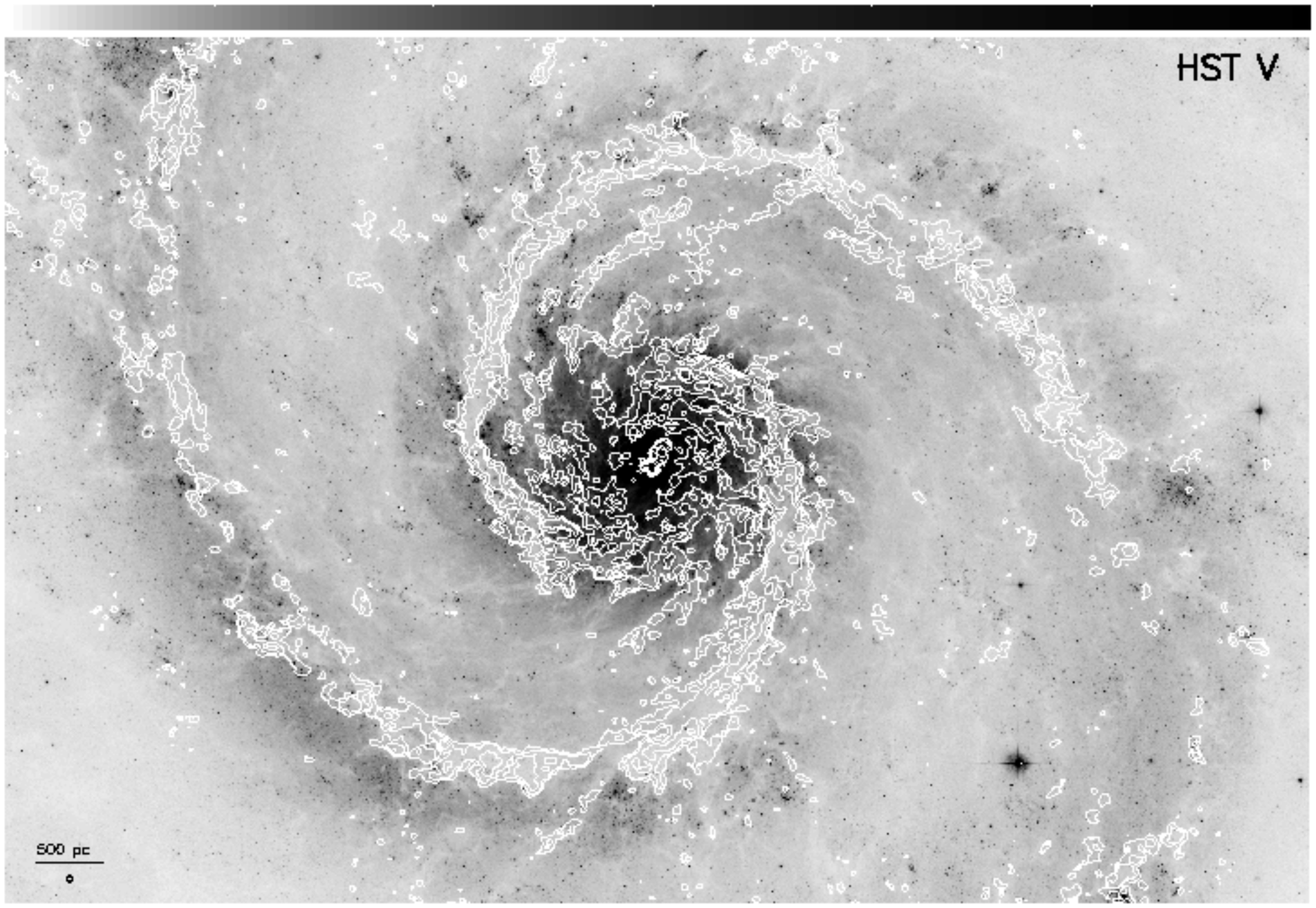}}\\
\resizebox{1.0\hsize}{!}{\includegraphics[angle=-90]{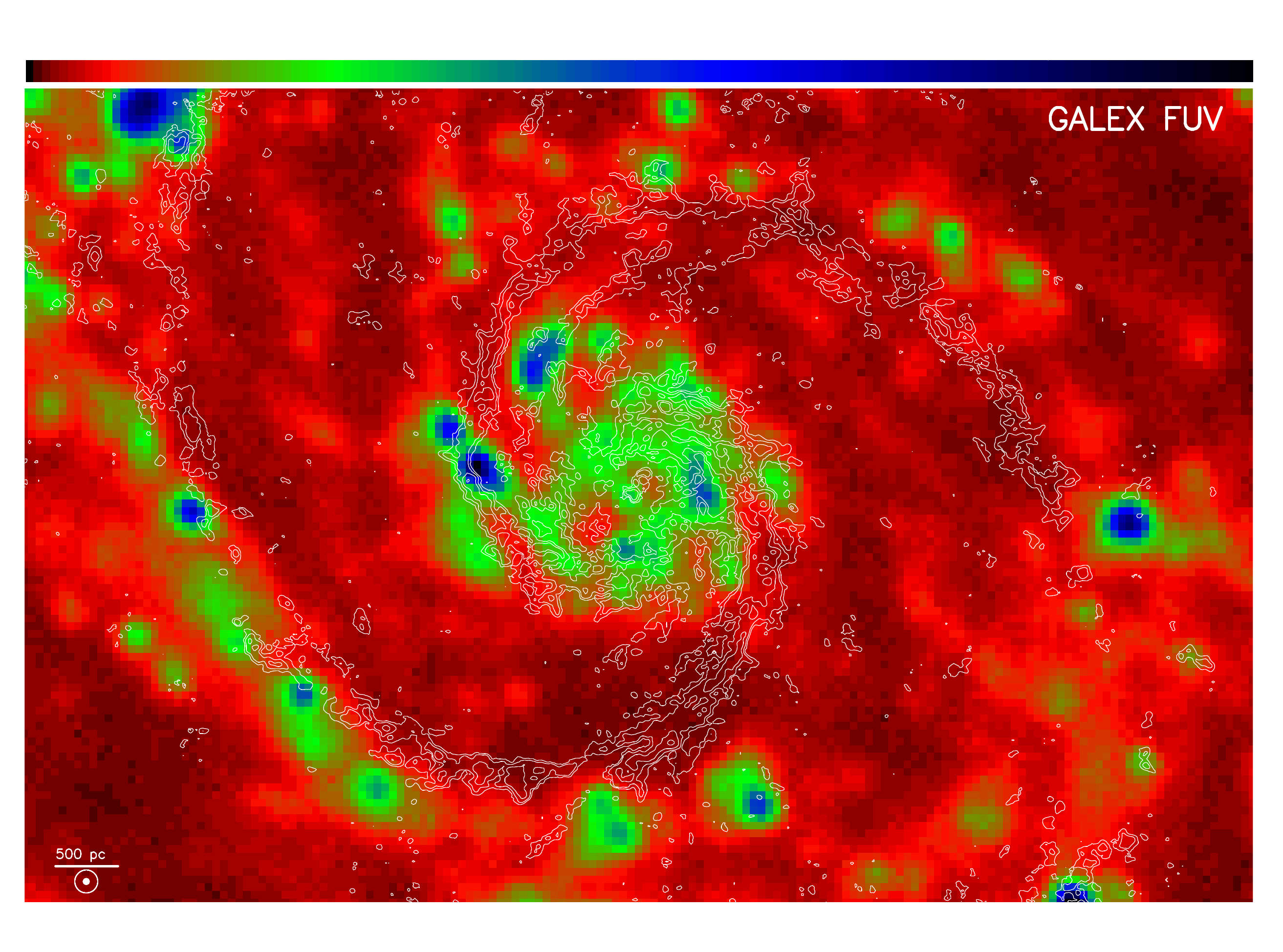}\includegraphics[angle=-90]{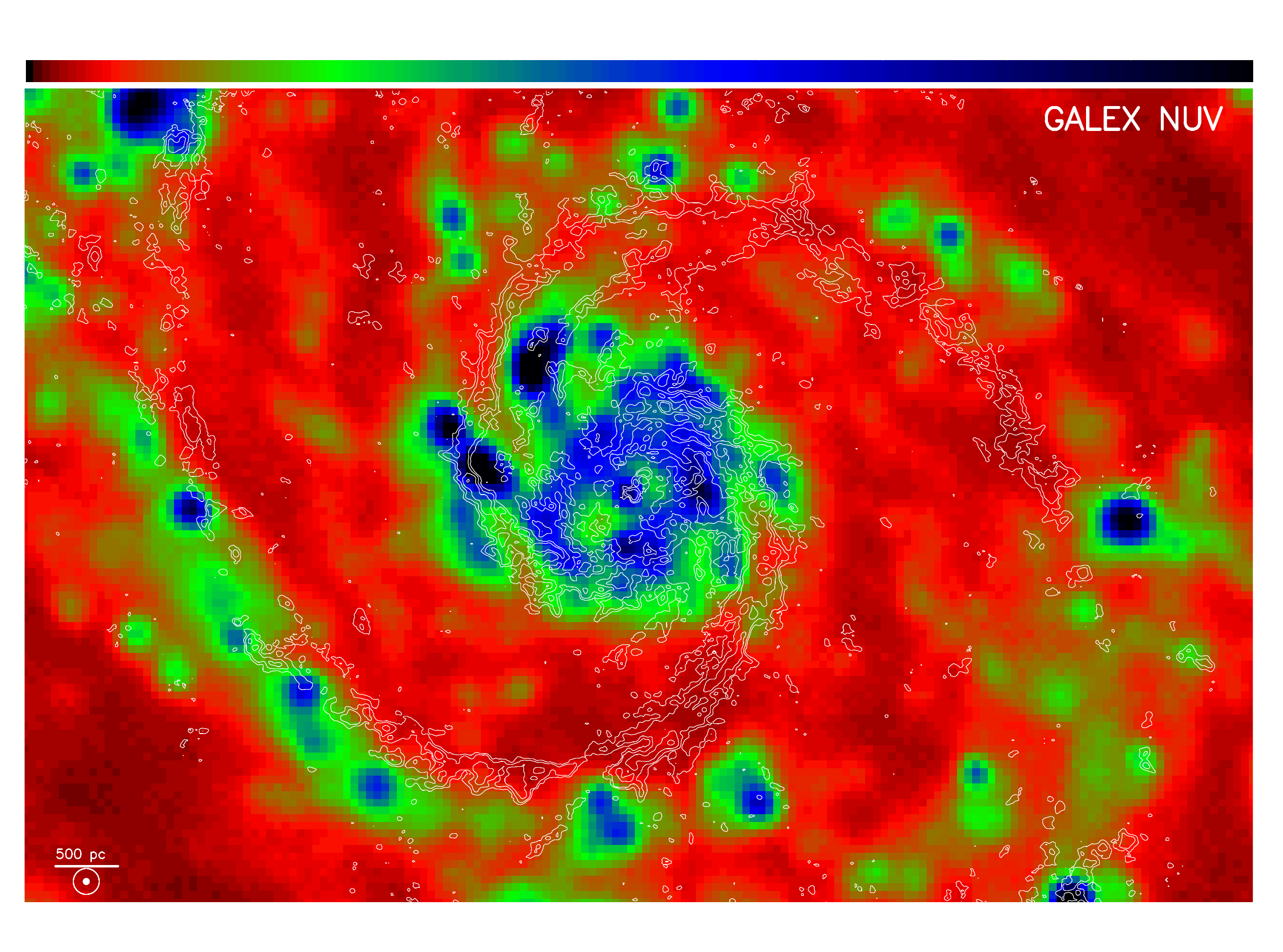}}
\end{center}
\caption{\footnotesize Comparison of the CO(1-0) line emission of M\,51a to SFR tracers.
\coone ~intensity distribution with contours overlaid onto
the HST H$\alpha$ line emission showing prominent HII regions ({\it top left}),
the VLA 3.6cm image tracing thermal and non-thermal radio continuum ({\it top right}),
the HST B band {\it middle left}) and V band ({\it middle right}) image tracing young clusters,
the GALEX FUV ({\it bottom left}) and NUV ({\it bottom right}) image tracing UV emission 
from young ($\rm \le 100\,Myrs$) stars. 
In the bottom left corner of each panel a scale bar representing 500\,pc and the CLEAN beam of the CO(1-0) data as well as the resolution of the respective dataset are shown.
\label{fig:sf_2d}}
\end{figure}

\begin{figure}
\begin{center}
\resizebox{1.0\hsize}{!}{\includegraphics[angle=-90]{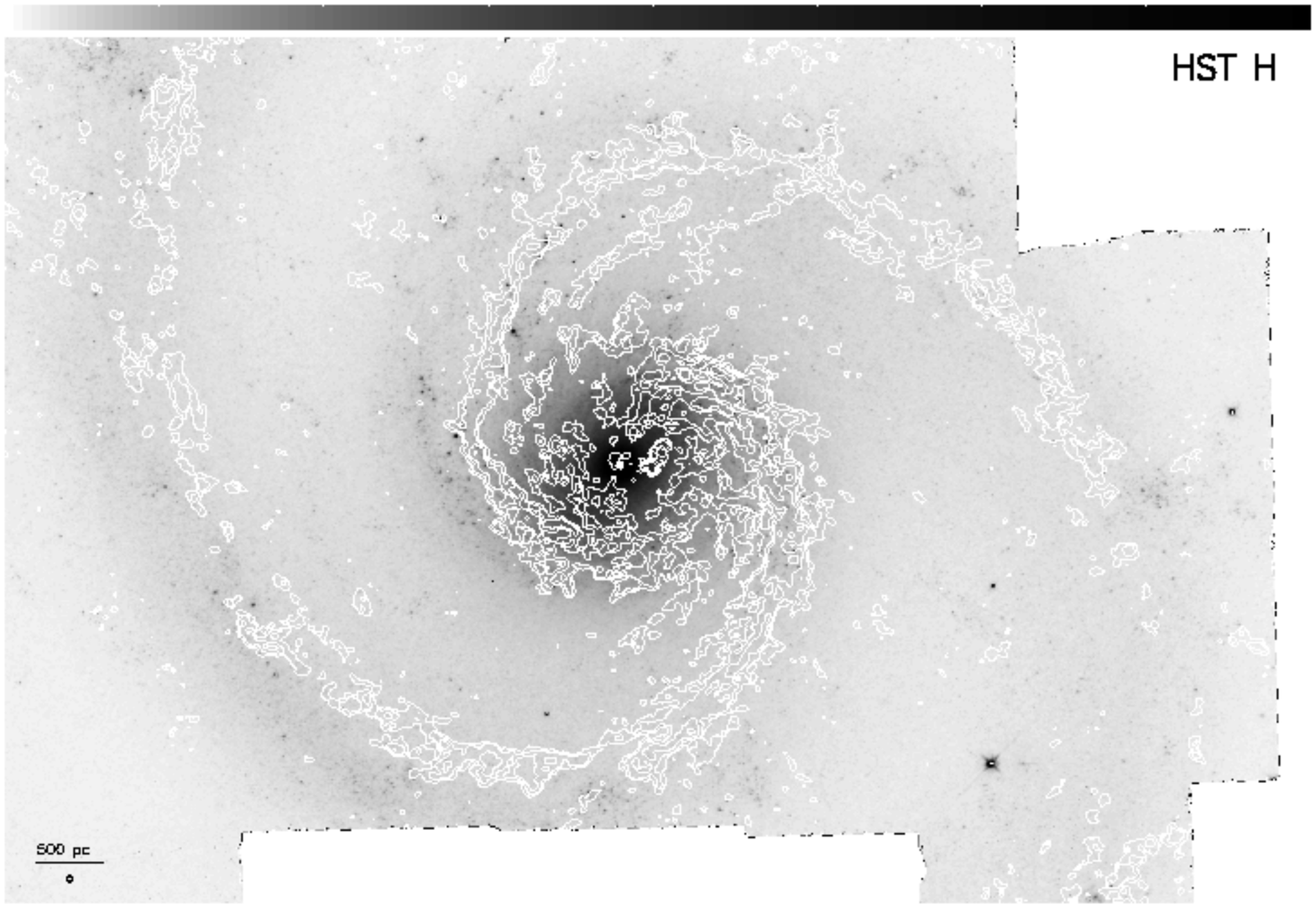}\includegraphics[angle=-90]{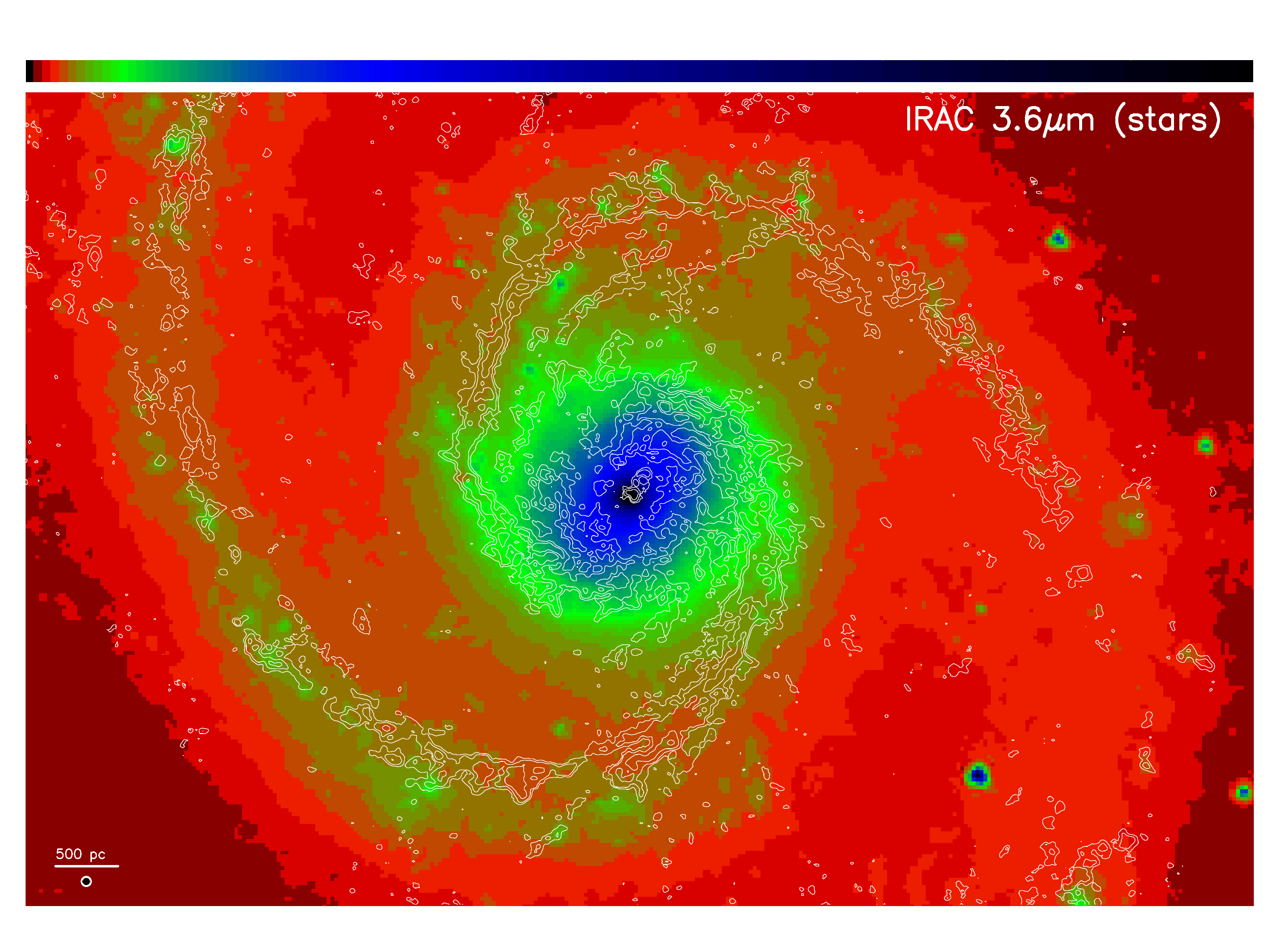}}\\
\end{center}
\caption{Comparison of the CO(1-0) distribution to the old stellar populations as seen in the HST NICMOS H band 
image ({\it left}) and the contamination corrected stellar light at 3.6\,$\mu$m \citep[{\it right}, from ][]{meidt12} that 
represents the underlying gravitational potential.
In the bottom left corner of each panel a scale bar representing 500\,pc and the CLEAN beam of the CO(1-0) data as well as the resolution of the respective dataset are shown.
\label{fig:sfo_2d}}
\end{figure}

\begin{figure}
\begin{center}
%includegraphics[]{f1.eps}
\resizebox{1.0\hsize}{!}{\includegraphics[angle=-90]{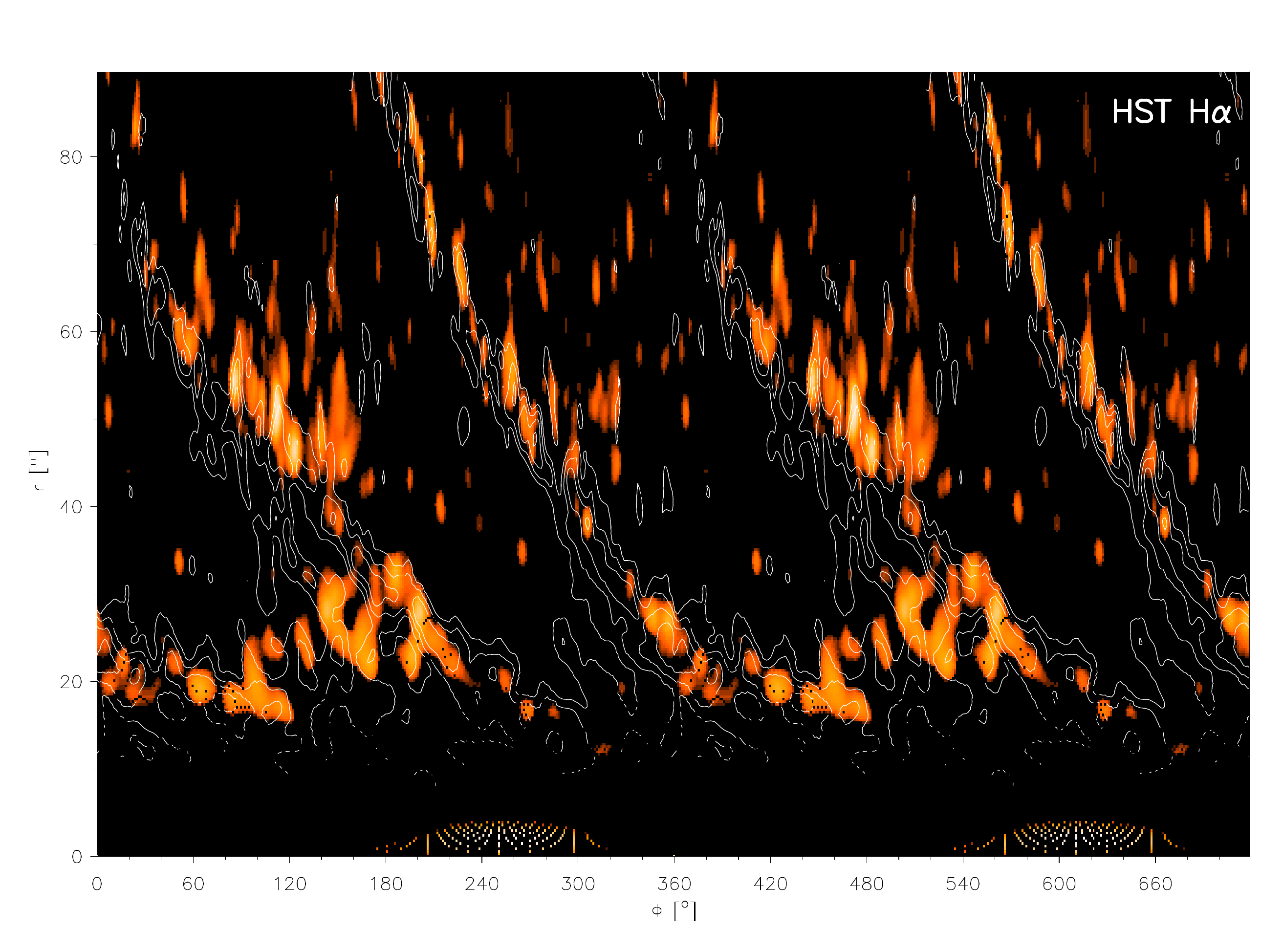}\includegraphics[angle=-90]{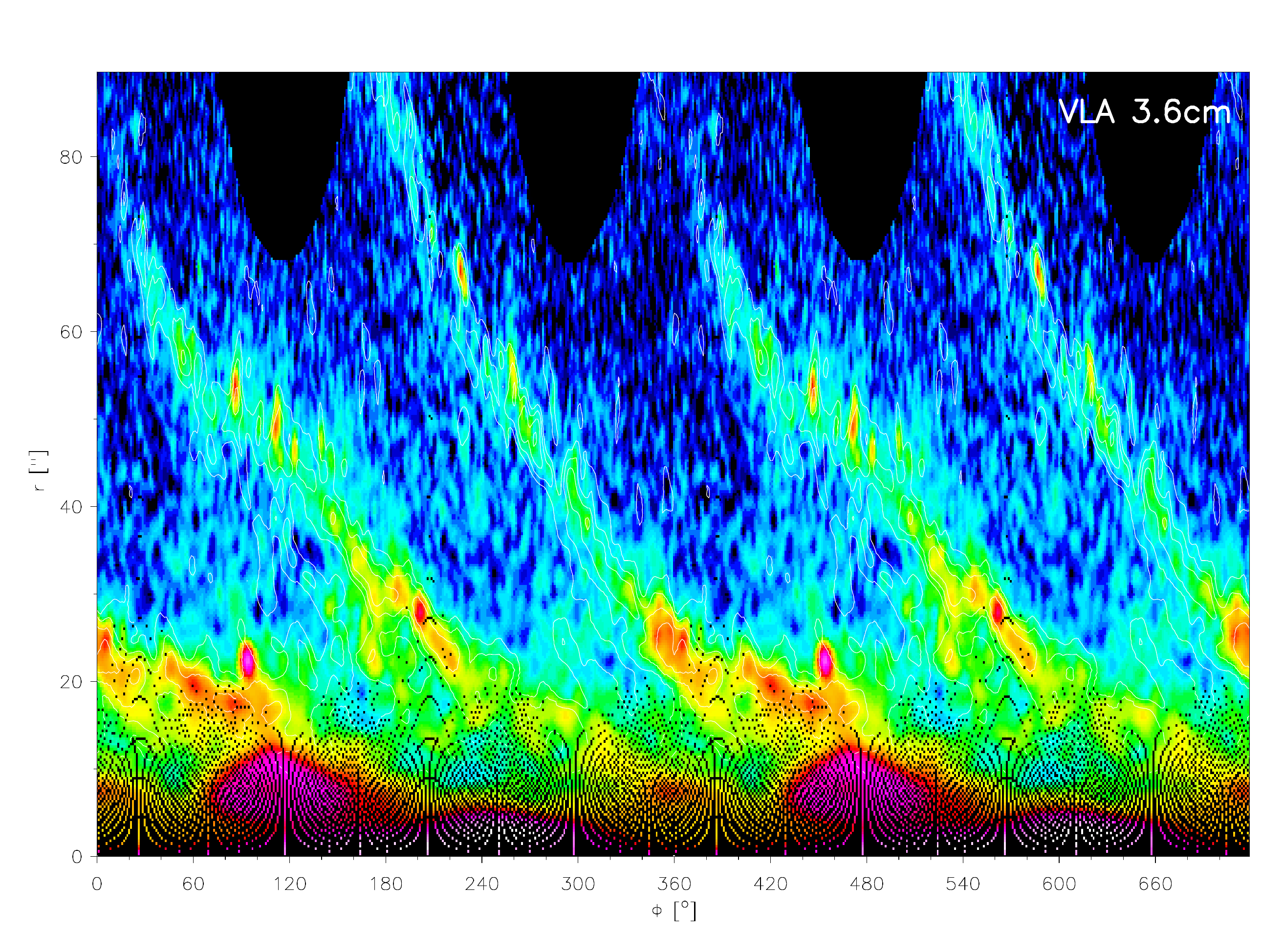}}\\
\resizebox{1.0\hsize}{!}{\includegraphics[angle=-90]{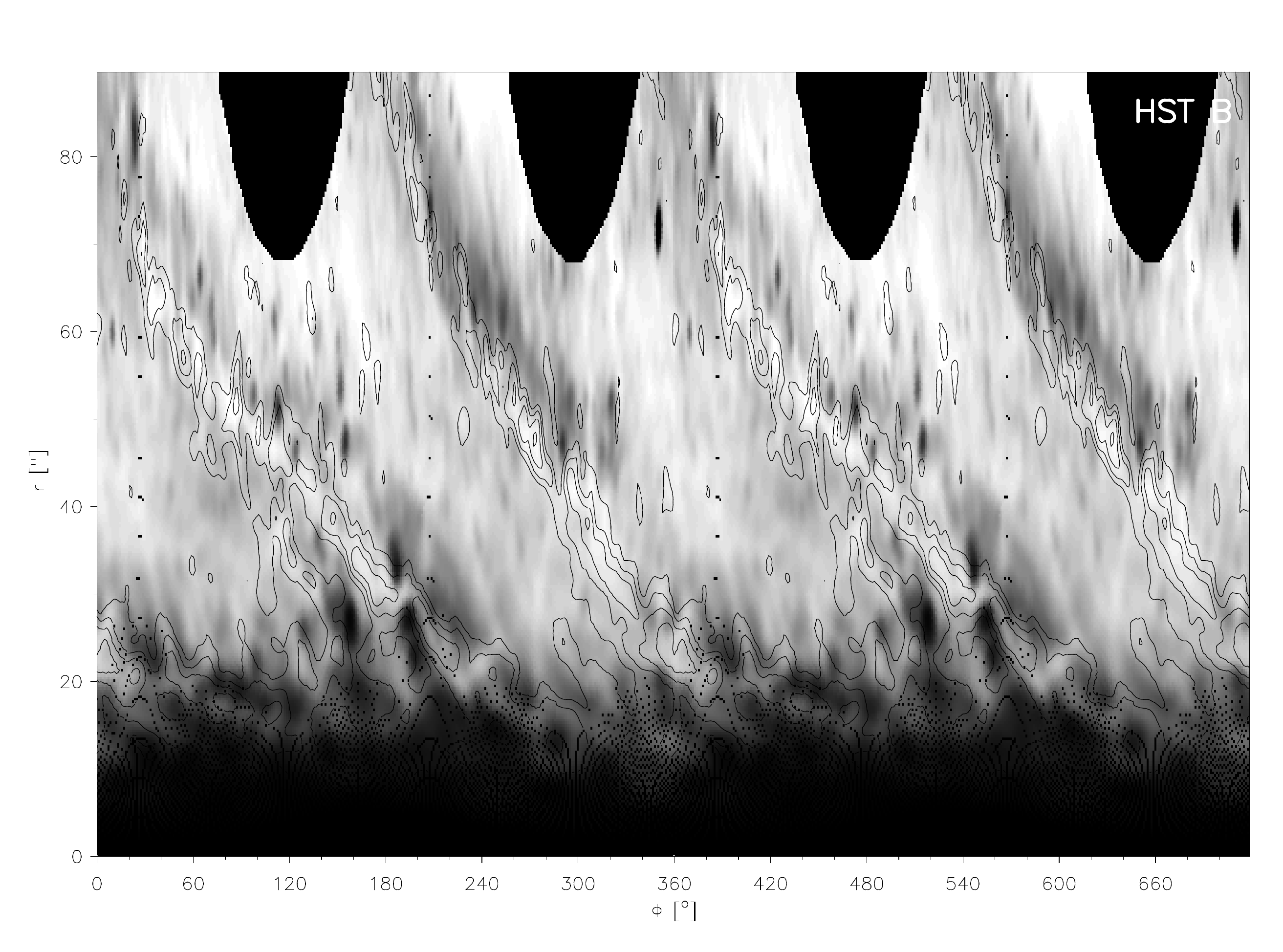}\includegraphics[angle=-90]{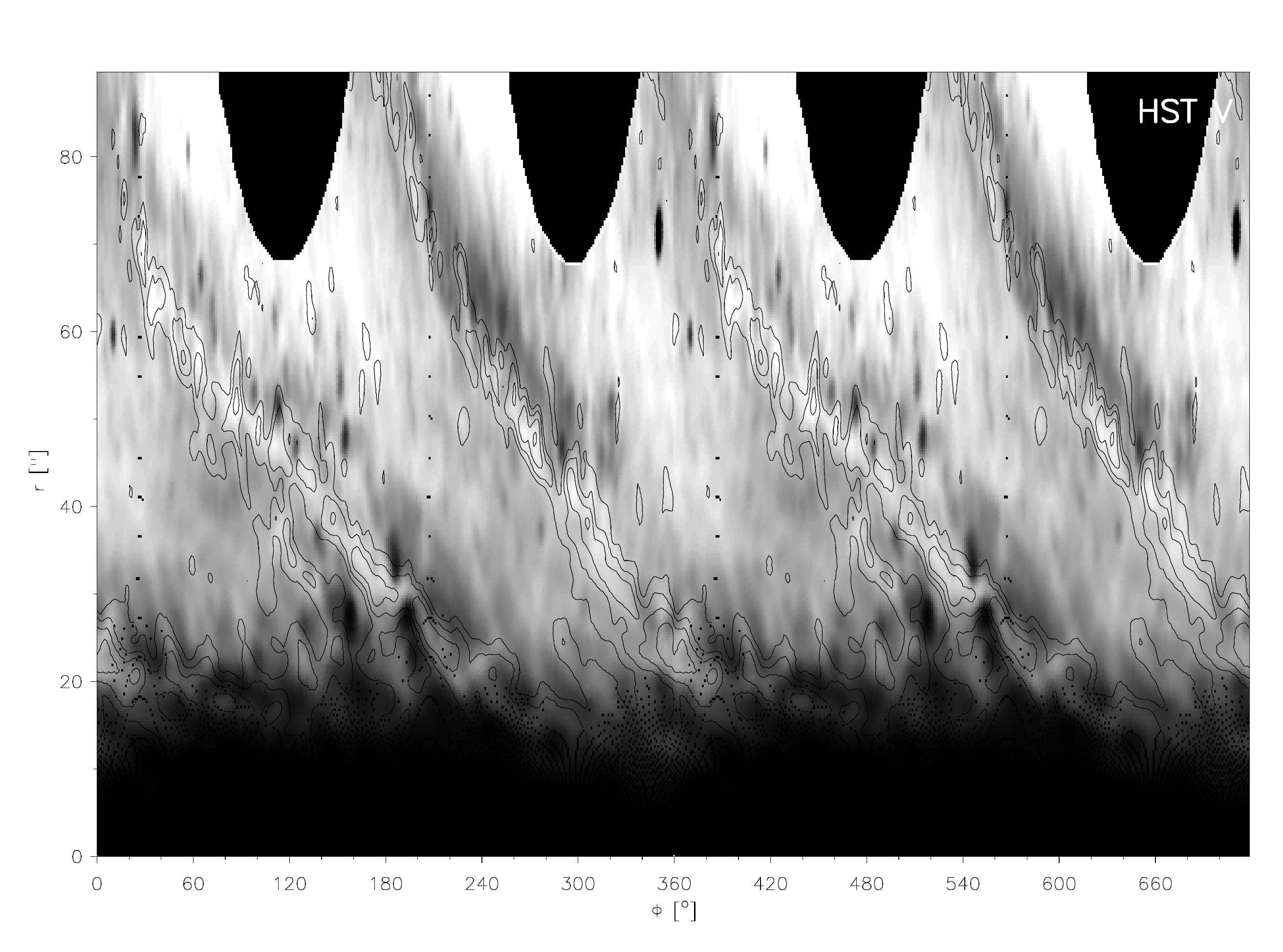}}\\
\resizebox{1.0\hsize}{!}{\includegraphics[angle=-90]{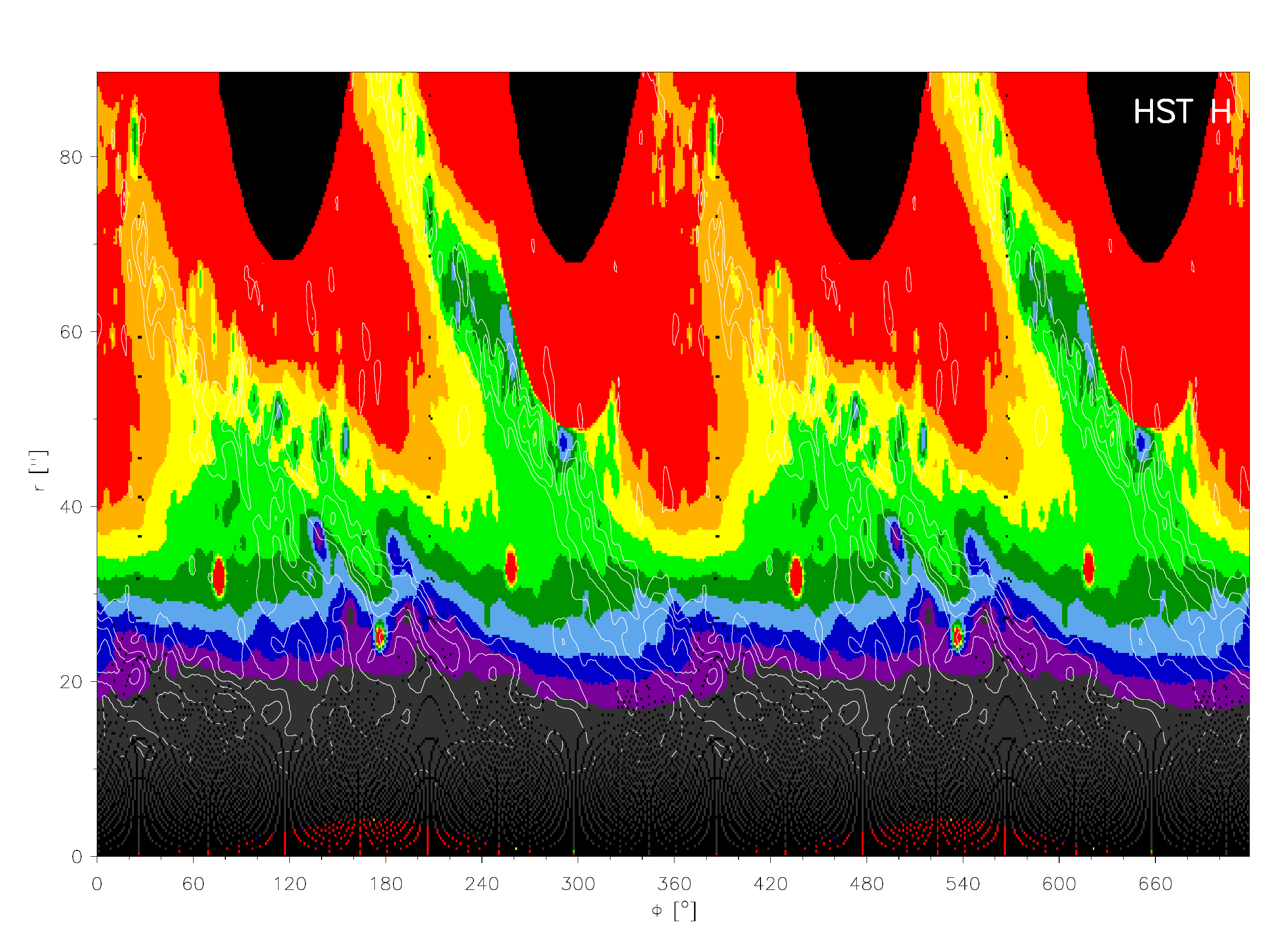}\includegraphics[angle=-90]{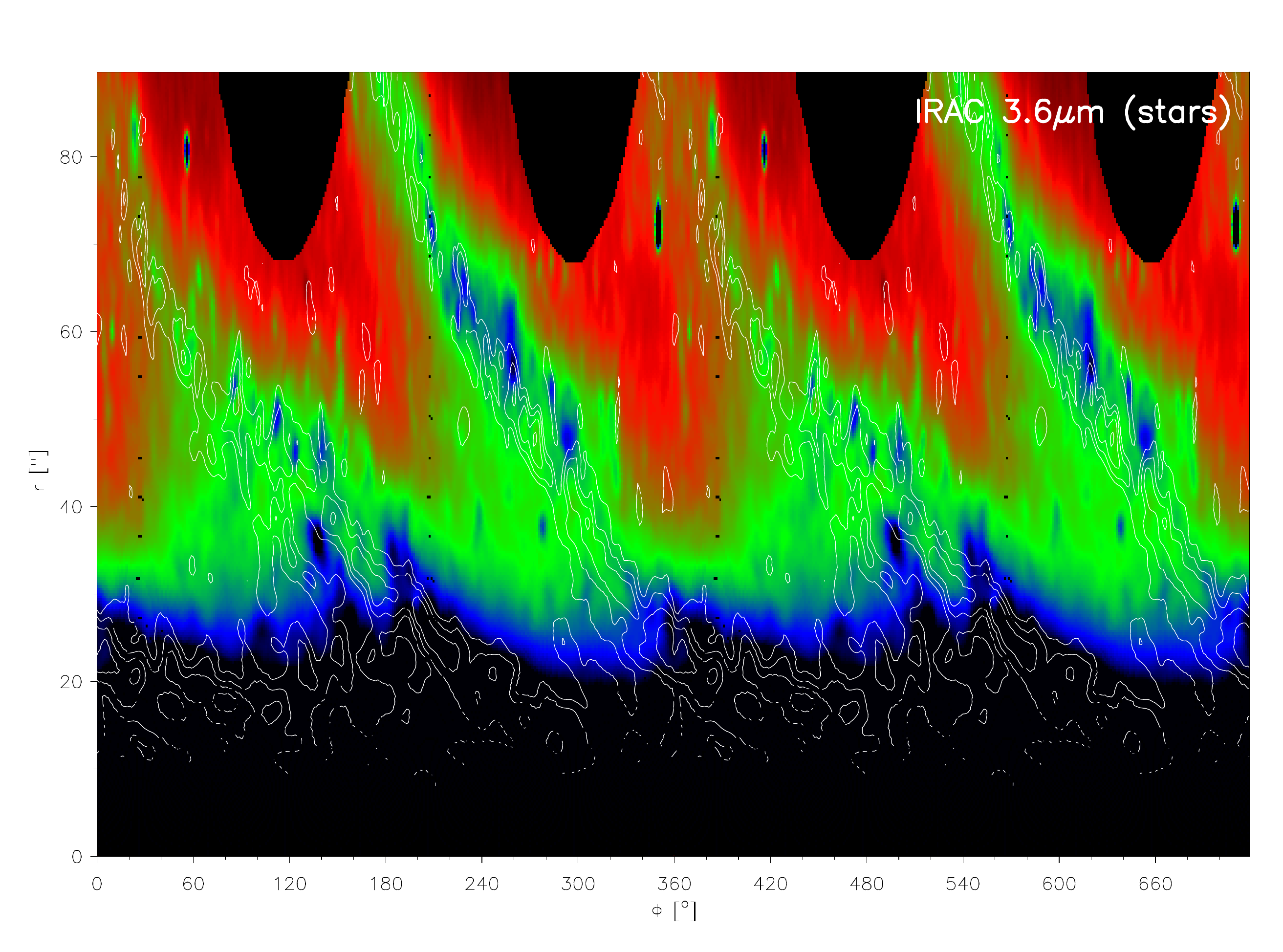}}
\end{center}
\caption{Polar representation of the \coone ~line emission ({\it contours}) overlaid onto SFR and stellar population tracers: 
HST \ha ~image({\it top left}), VLA 3.6\,cm ({\it top right}),
HST B and V band image ({\it middle}), 
HST H band image ({\it bottom left}), and the contaminant-corrected IRAC 3.6\,$\mu$m image ({\it bottom right}).
\label{fig:sf_pol}}
\end{figure}

\begin{figure}
\begin{center}
\includegraphics[width=180mm]{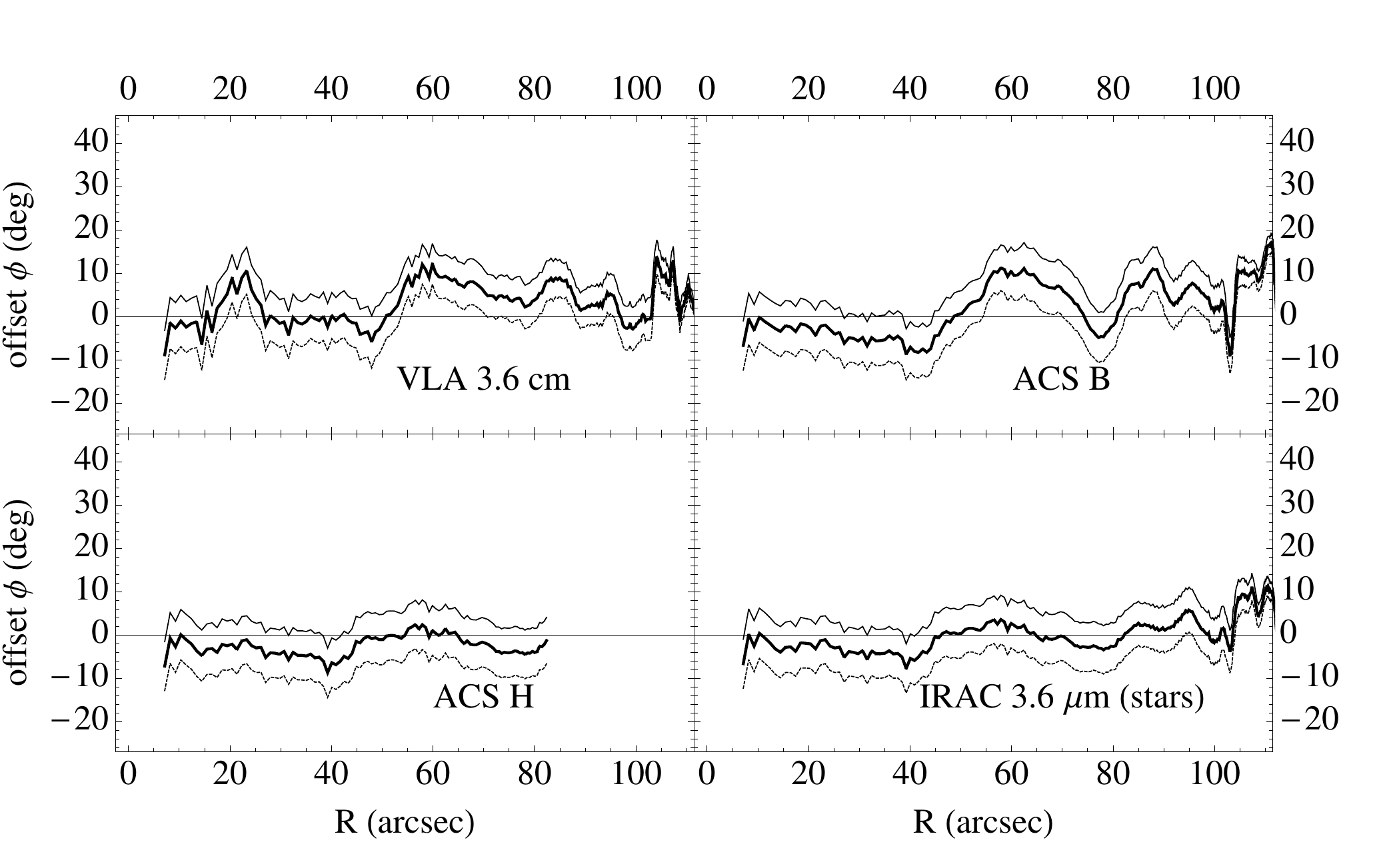}
\end{center}
\caption{Results of the polar cross-correlation between the \coone ~line emission and SFR and stellar population tracers. 
The panels show the radial profiles of the location of maximum correlation between the \coone ~intensity and
the emission from the other tracers, tracing the azimuthal offset $\Phi$ between the two:
VLA 3.6\,cm continuum ({\it top left}), HST B band image ({\it top right}), HST H band image ({\it bottom left}),
 and the contaminant-corrected IRAC 3.6\,$\mu$m image ({\it bottom right}). The thin lines represent 
the uncertainty defined as the width of the cross-correlation profile at 95\% maximum correlation. 
\label{fig:sf_cross}}
\end{figure}

\begin{figure}
\begin{center}
%includegraphics[]{f1.eps}
\resizebox{0.9\hsize}{!}{\includegraphics[angle=0]{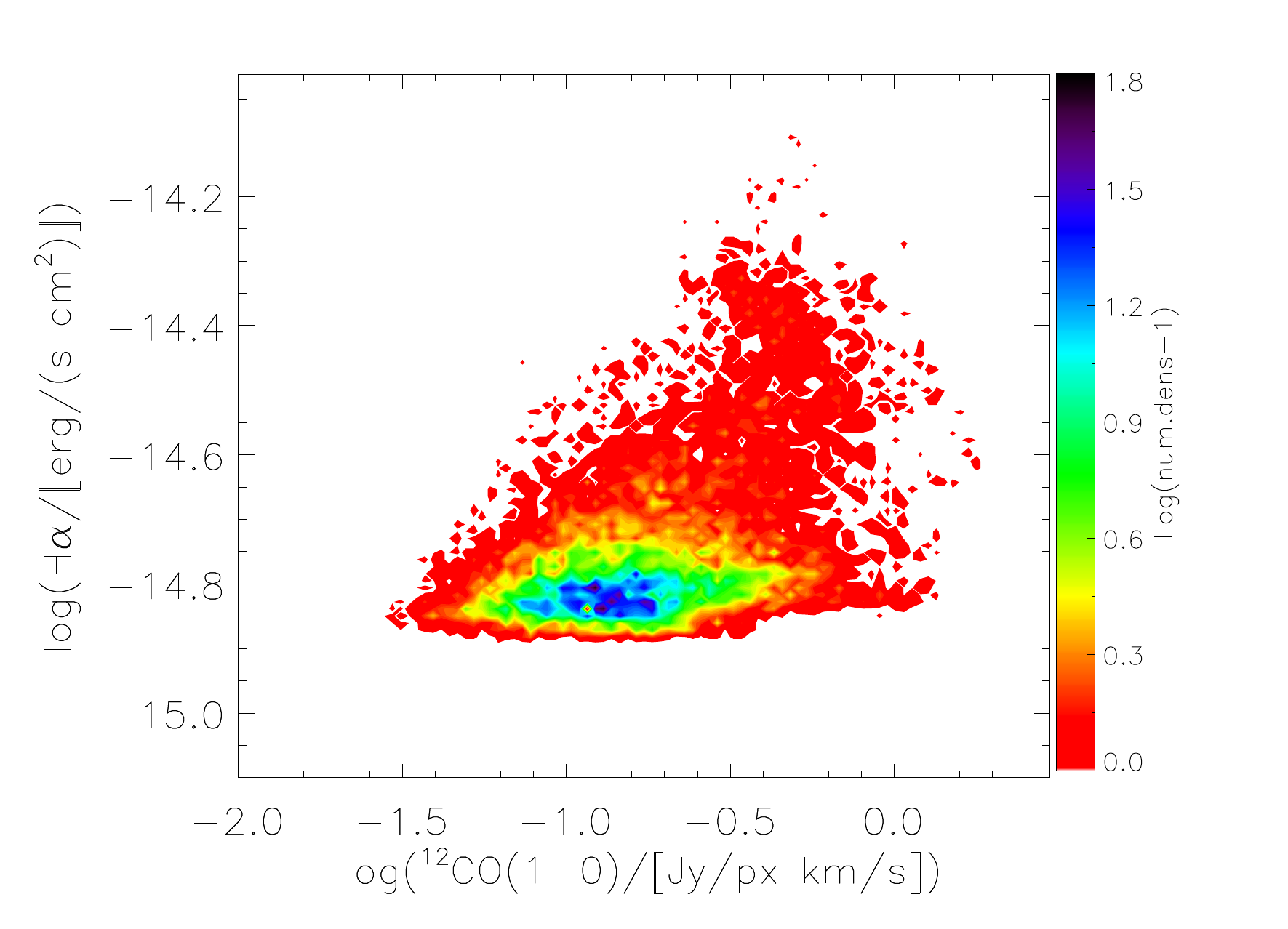}\includegraphics[angle=0]{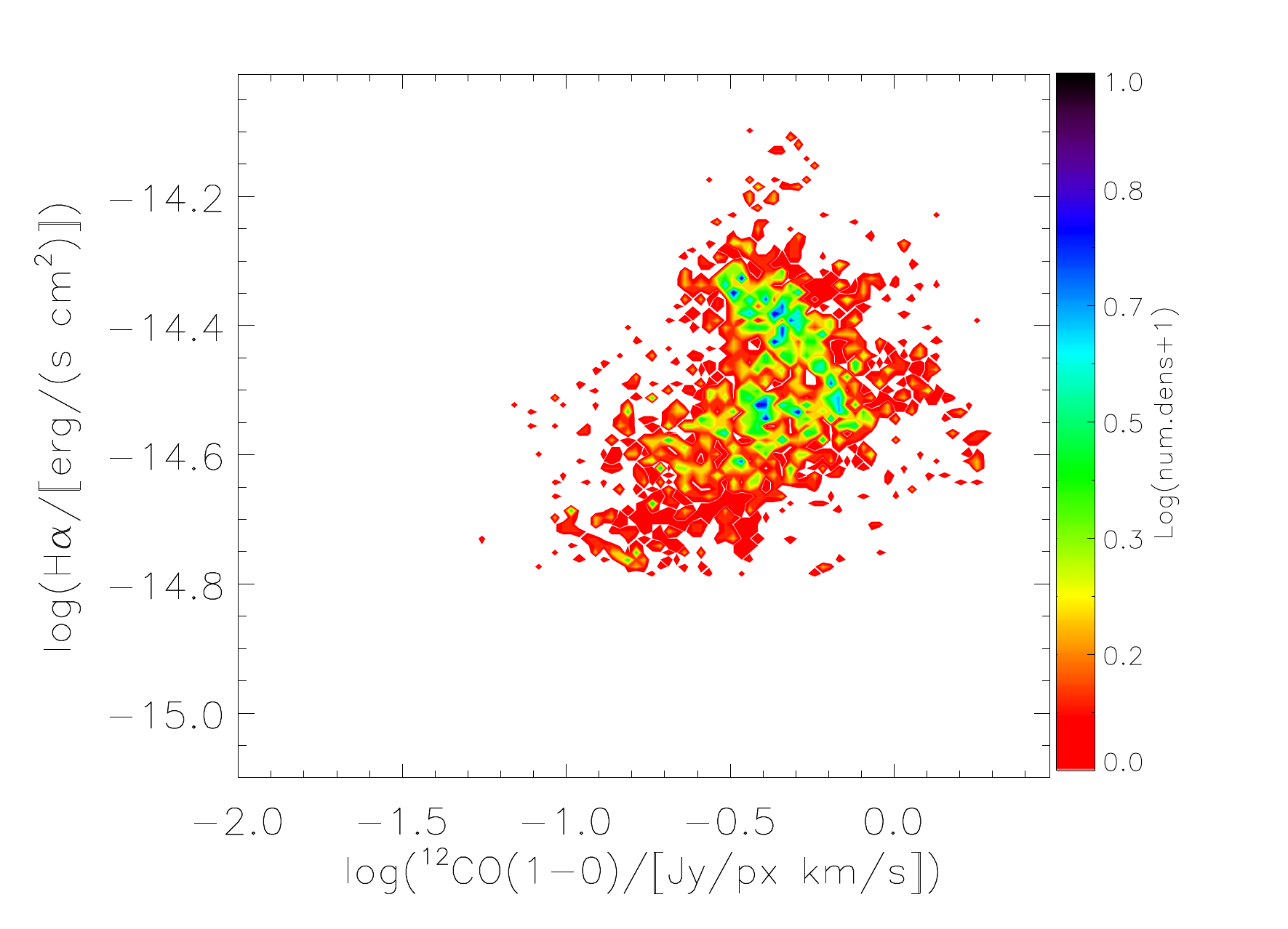}\includegraphics[angle=0]{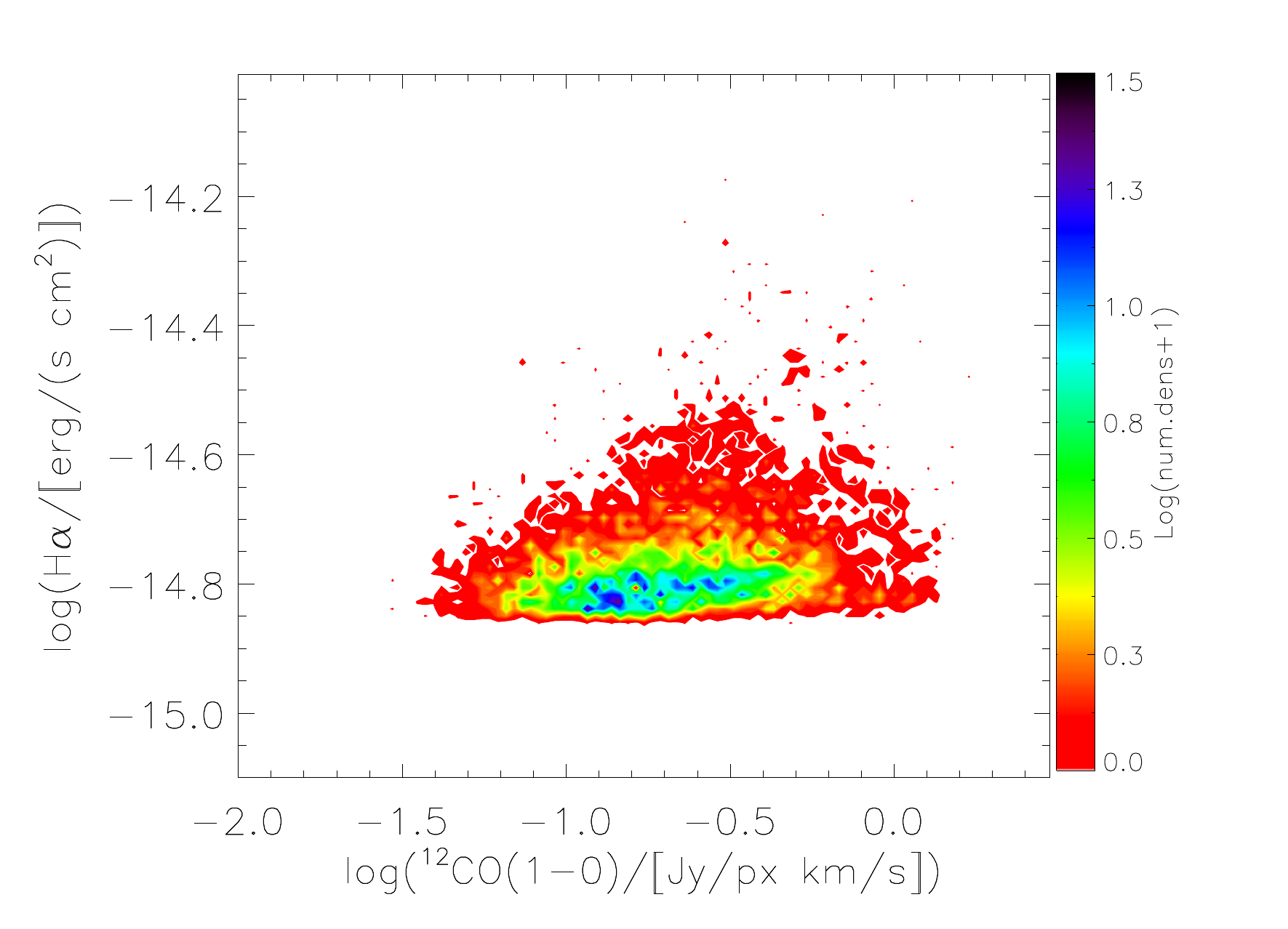}\includegraphics[angle=0]{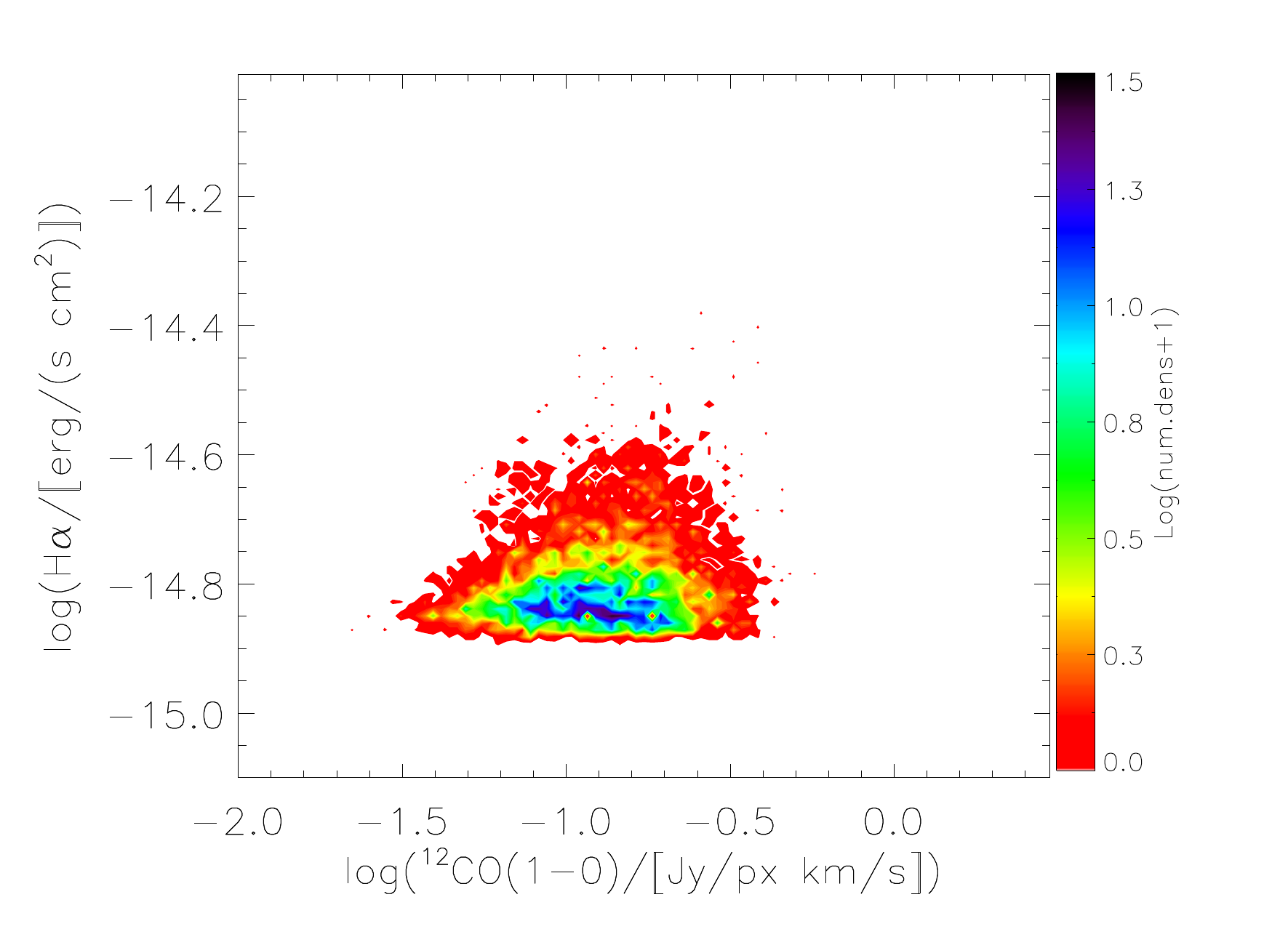}}\\
\resizebox{0.9\hsize}{!}{\includegraphics[angle=0]{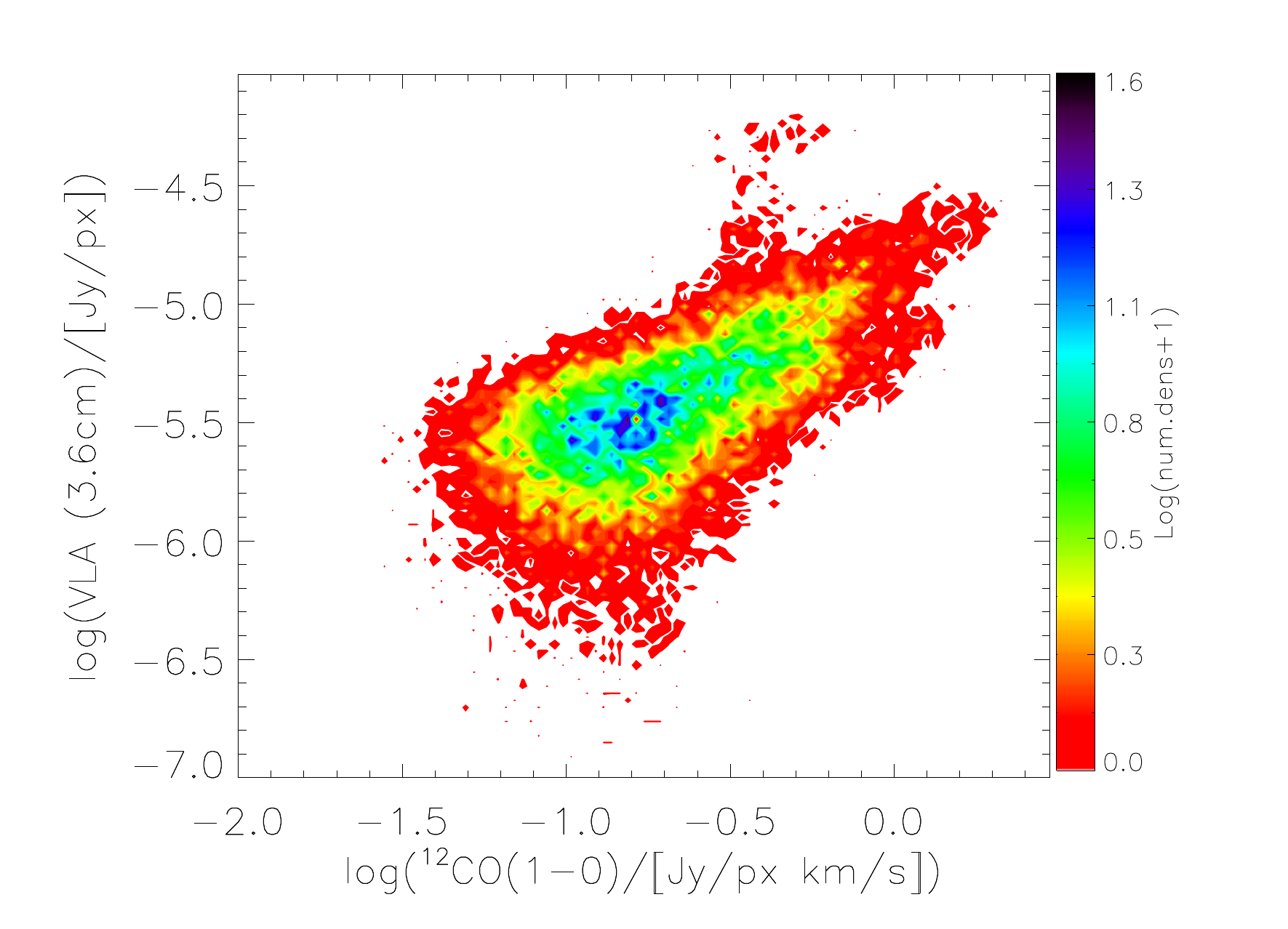}\includegraphics[angle=0]{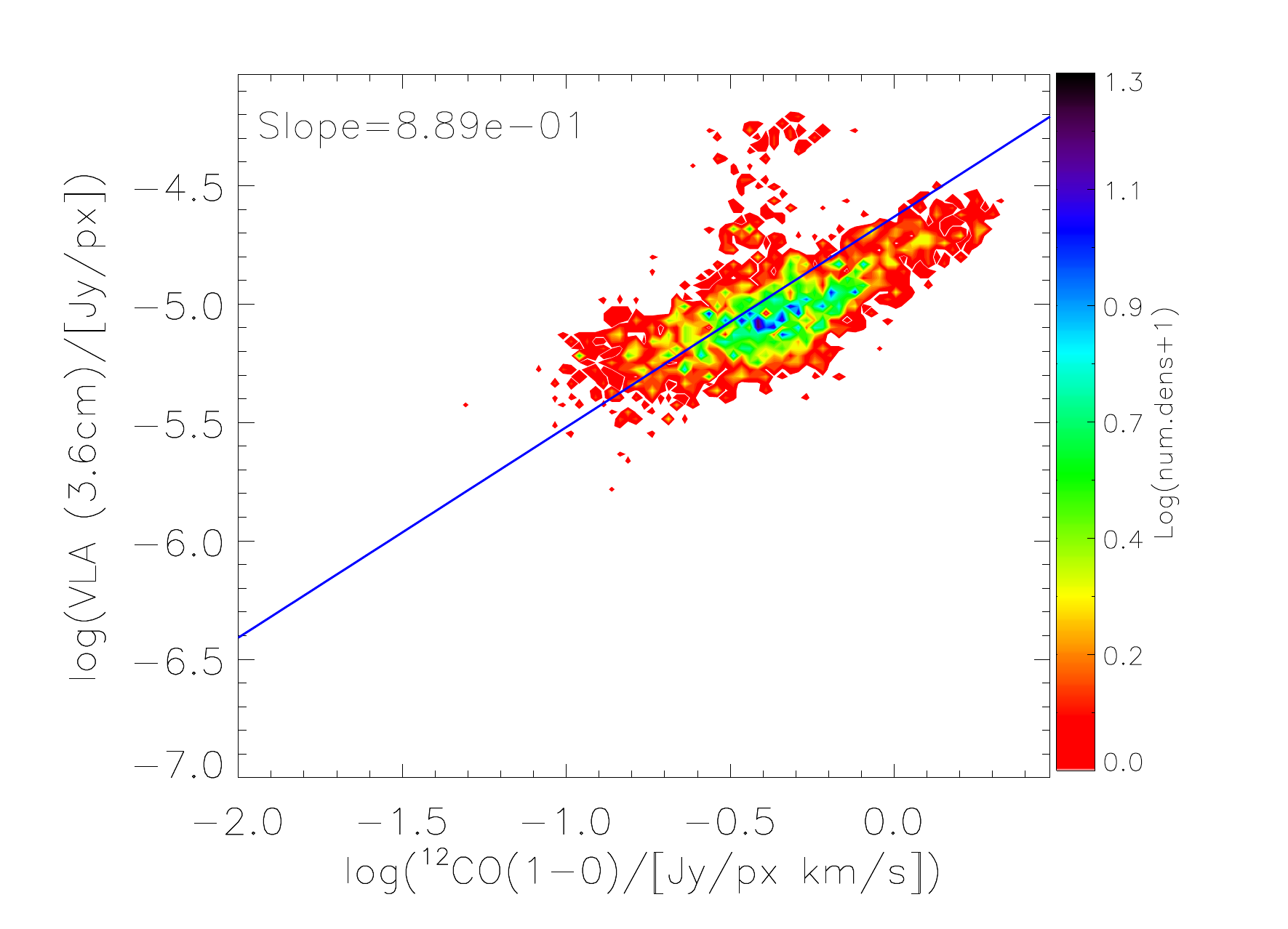}\includegraphics[angle=0]{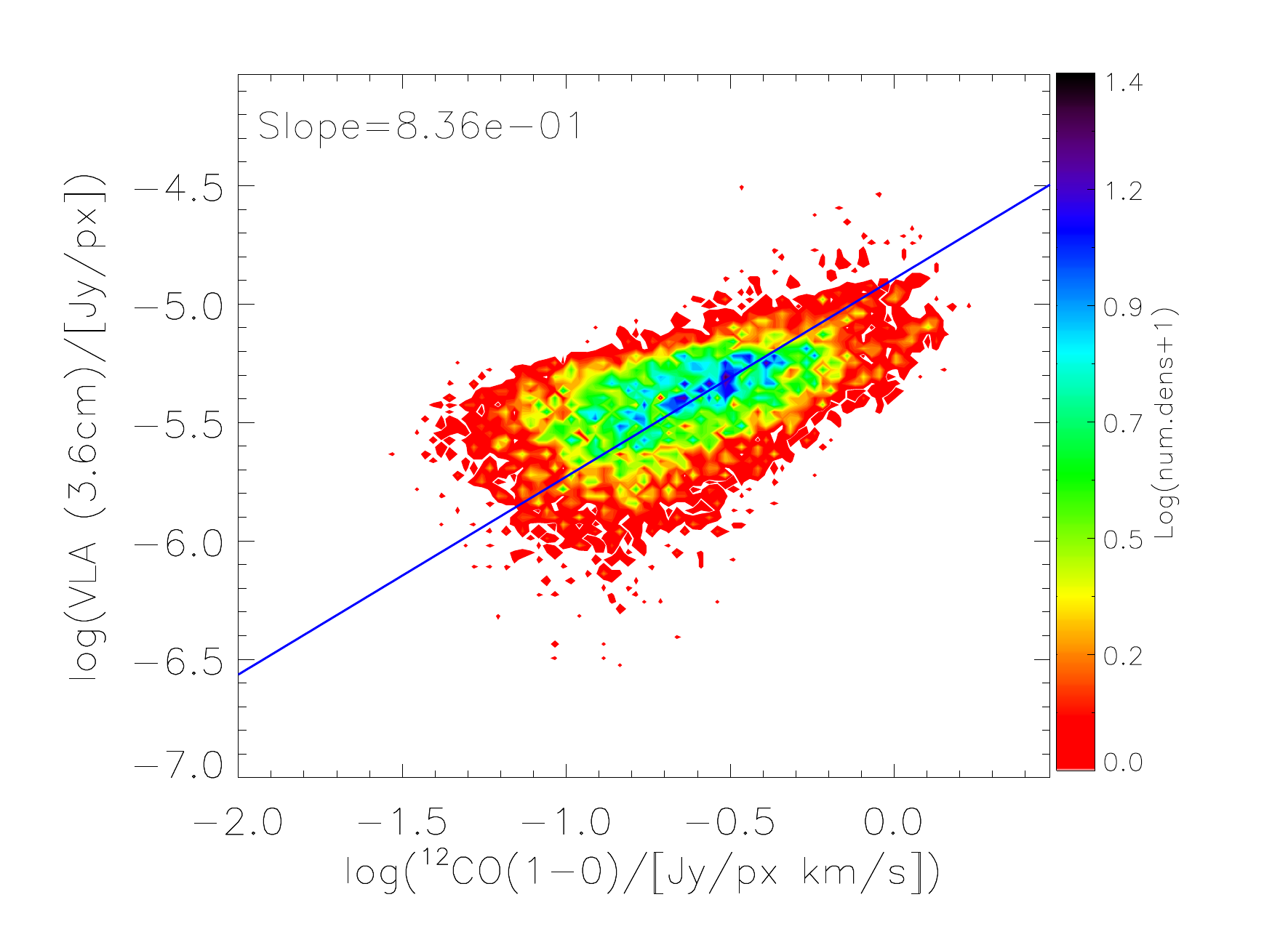}\includegraphics[angle=0]{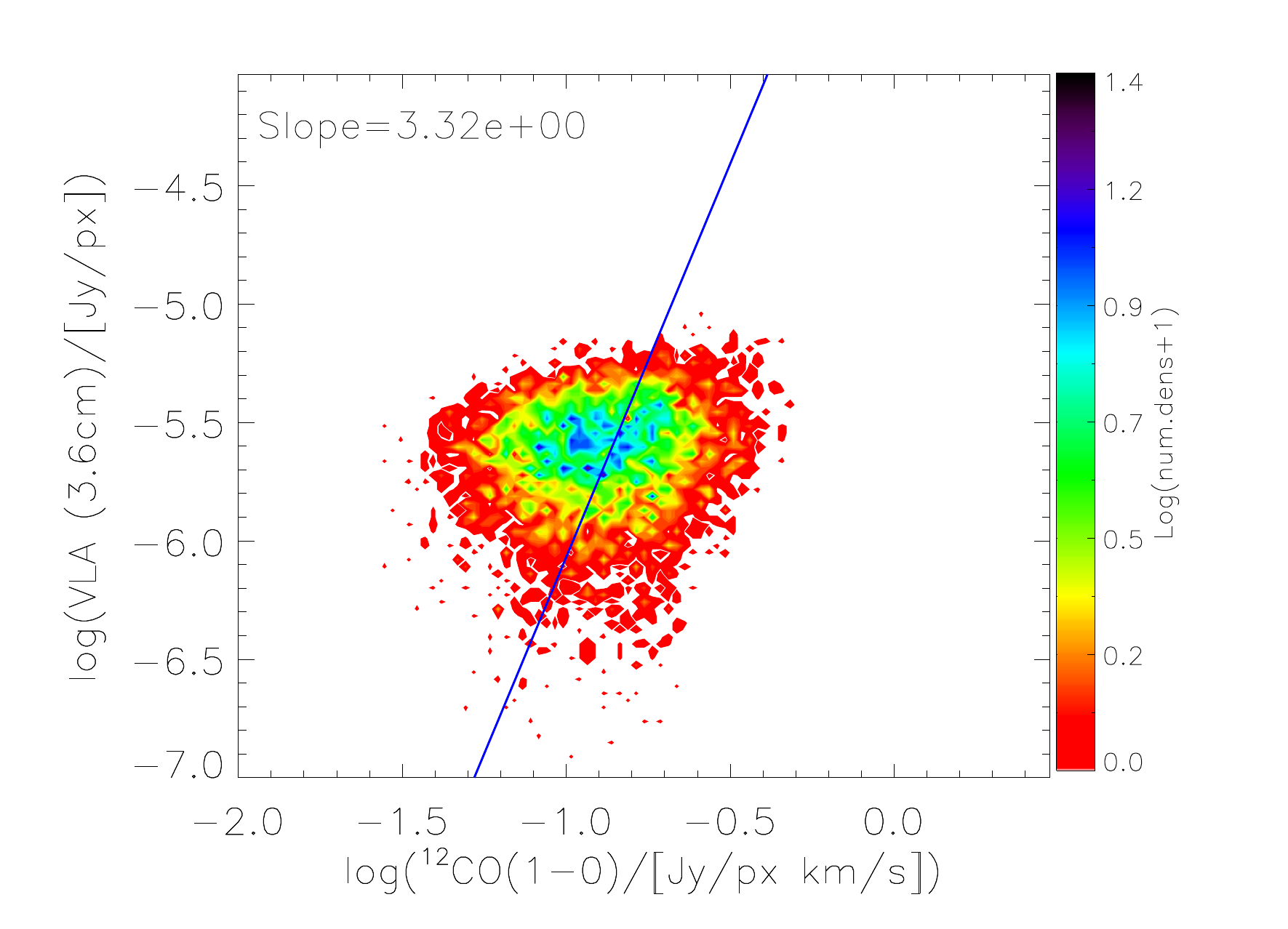}}
\resizebox{0.9\hsize}{!}{\includegraphics[angle=0]{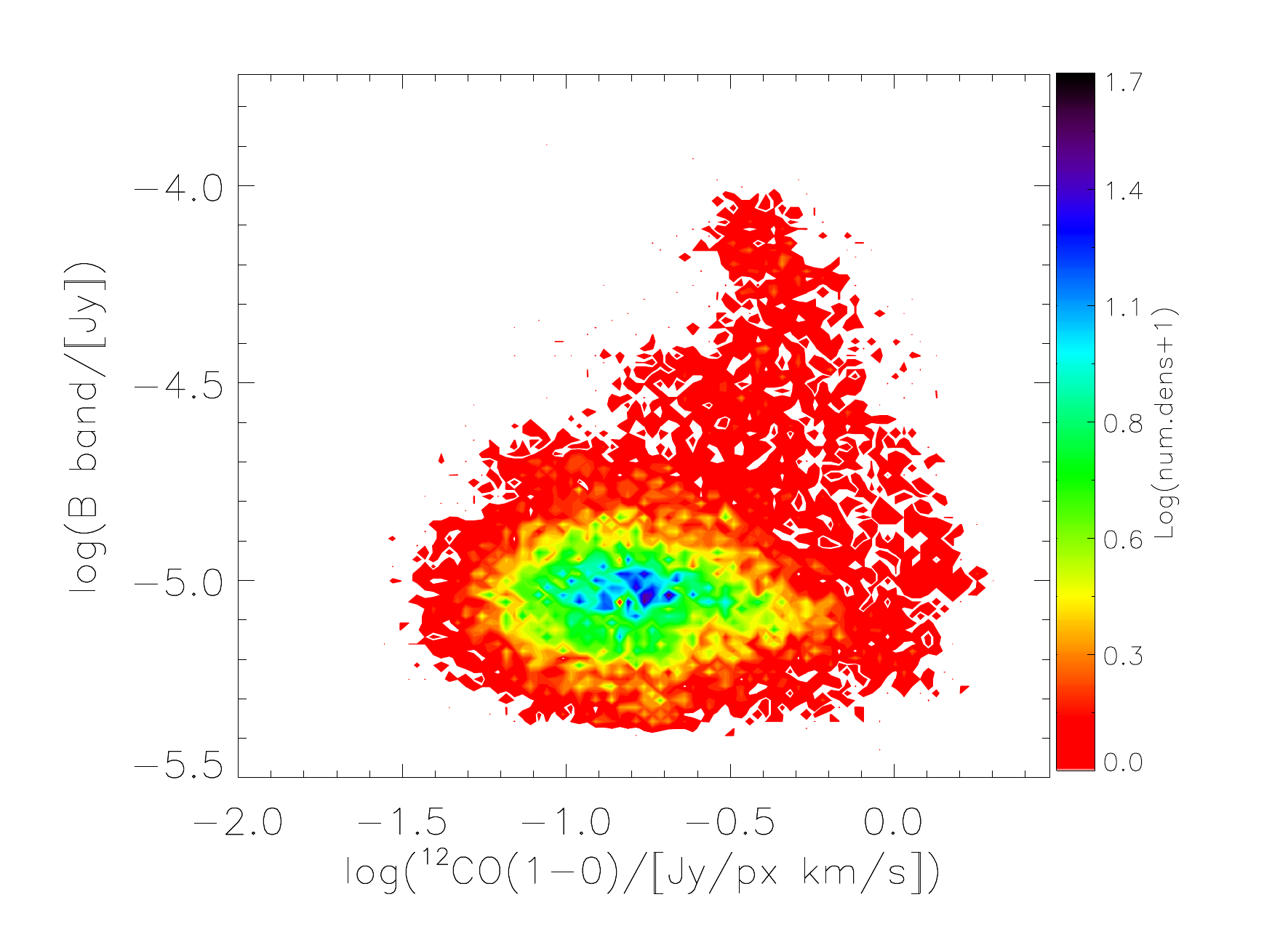}\includegraphics[angle=0]{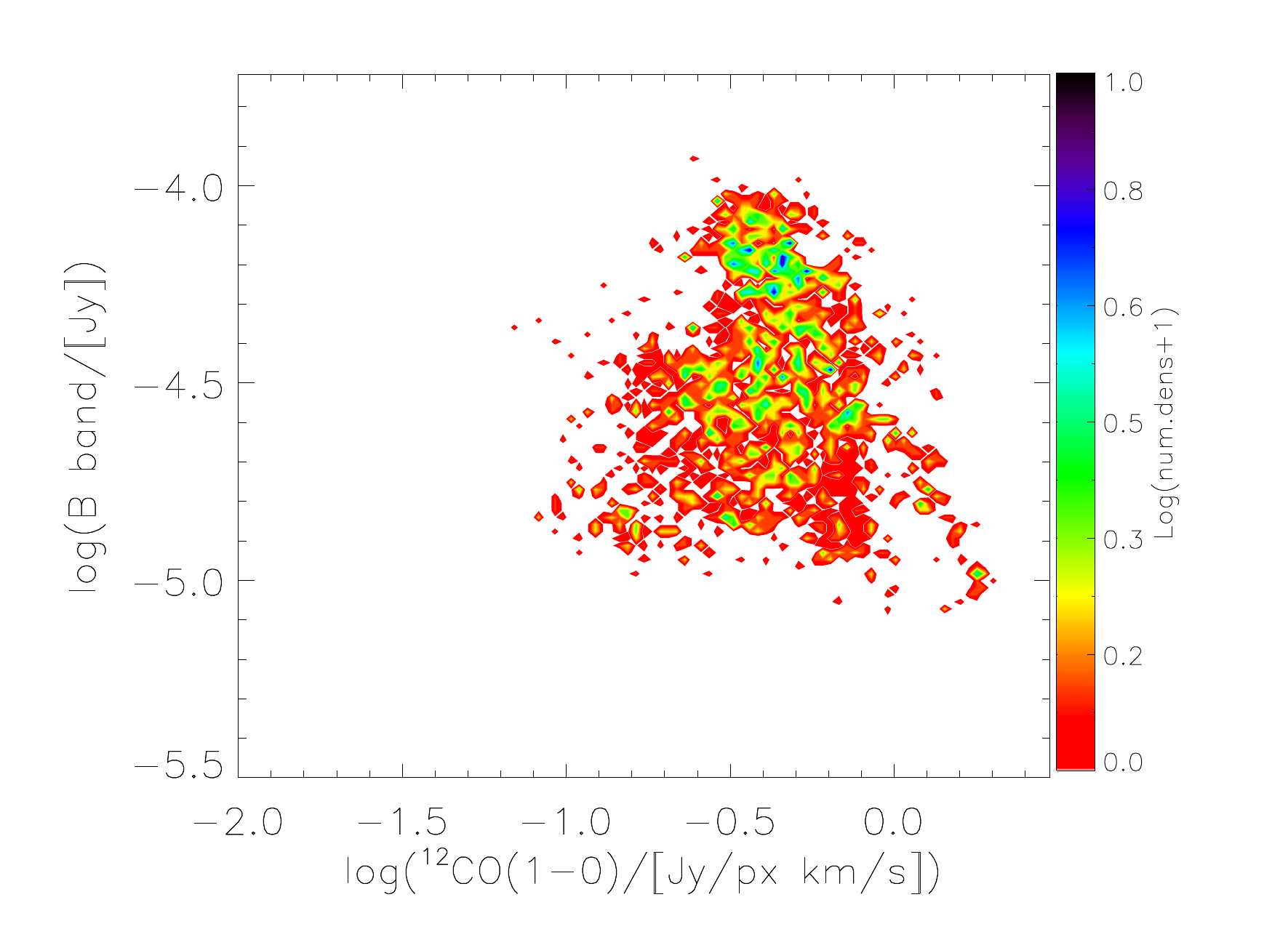}\includegraphics[angle=0]{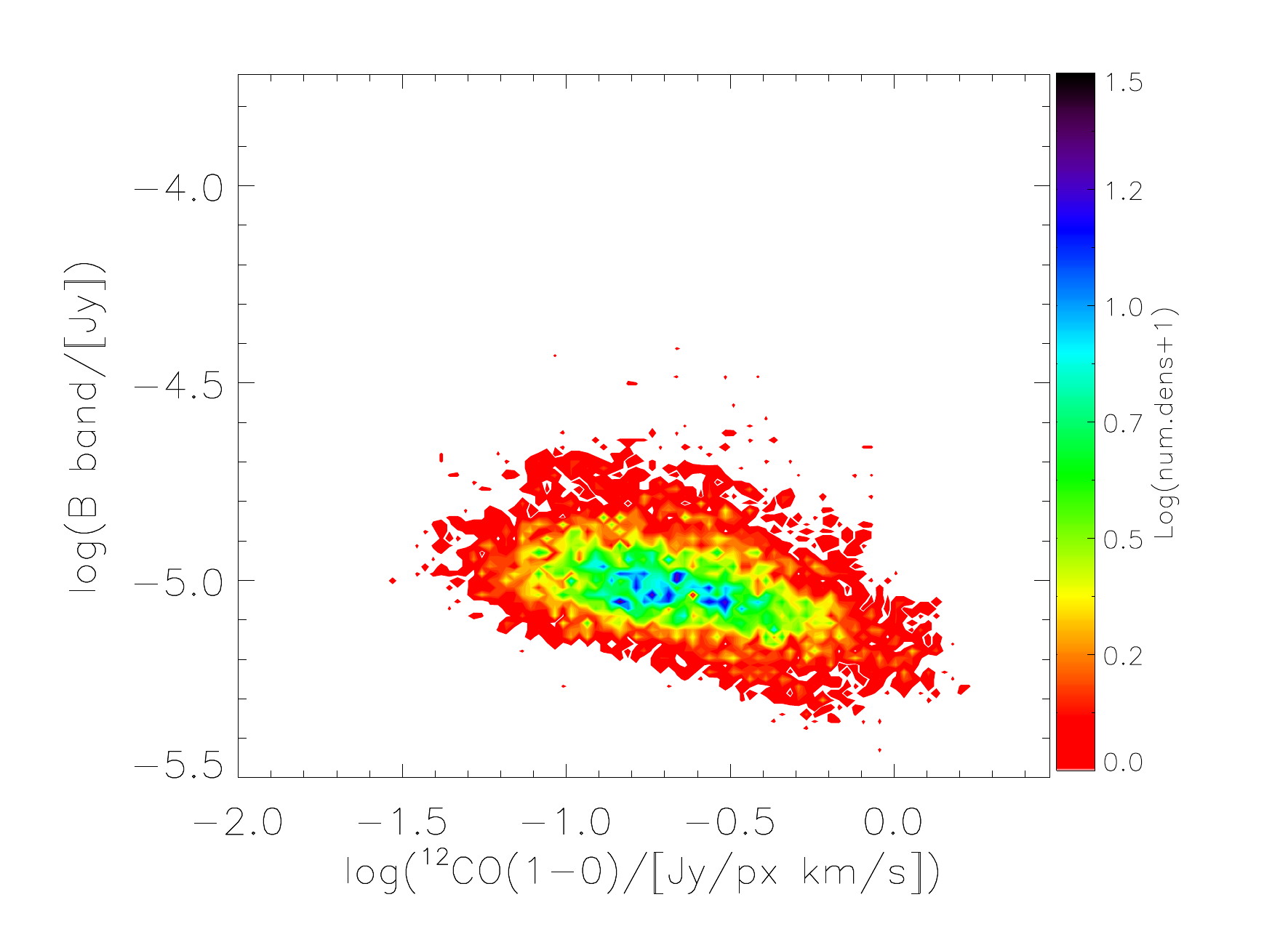}\includegraphics[angle=0]{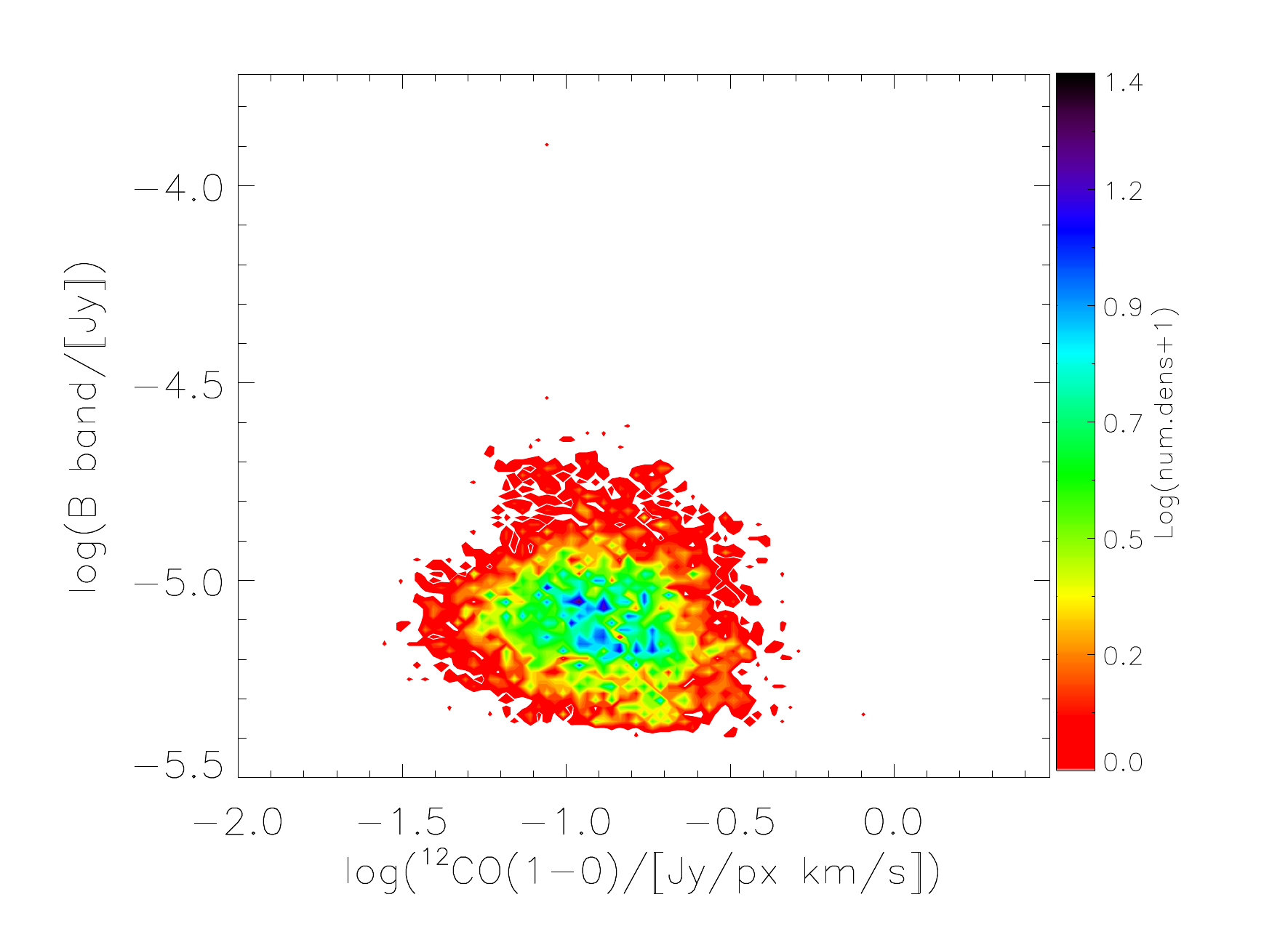}}\\
\resizebox{0.9\hsize}{!}{\includegraphics[angle=0]{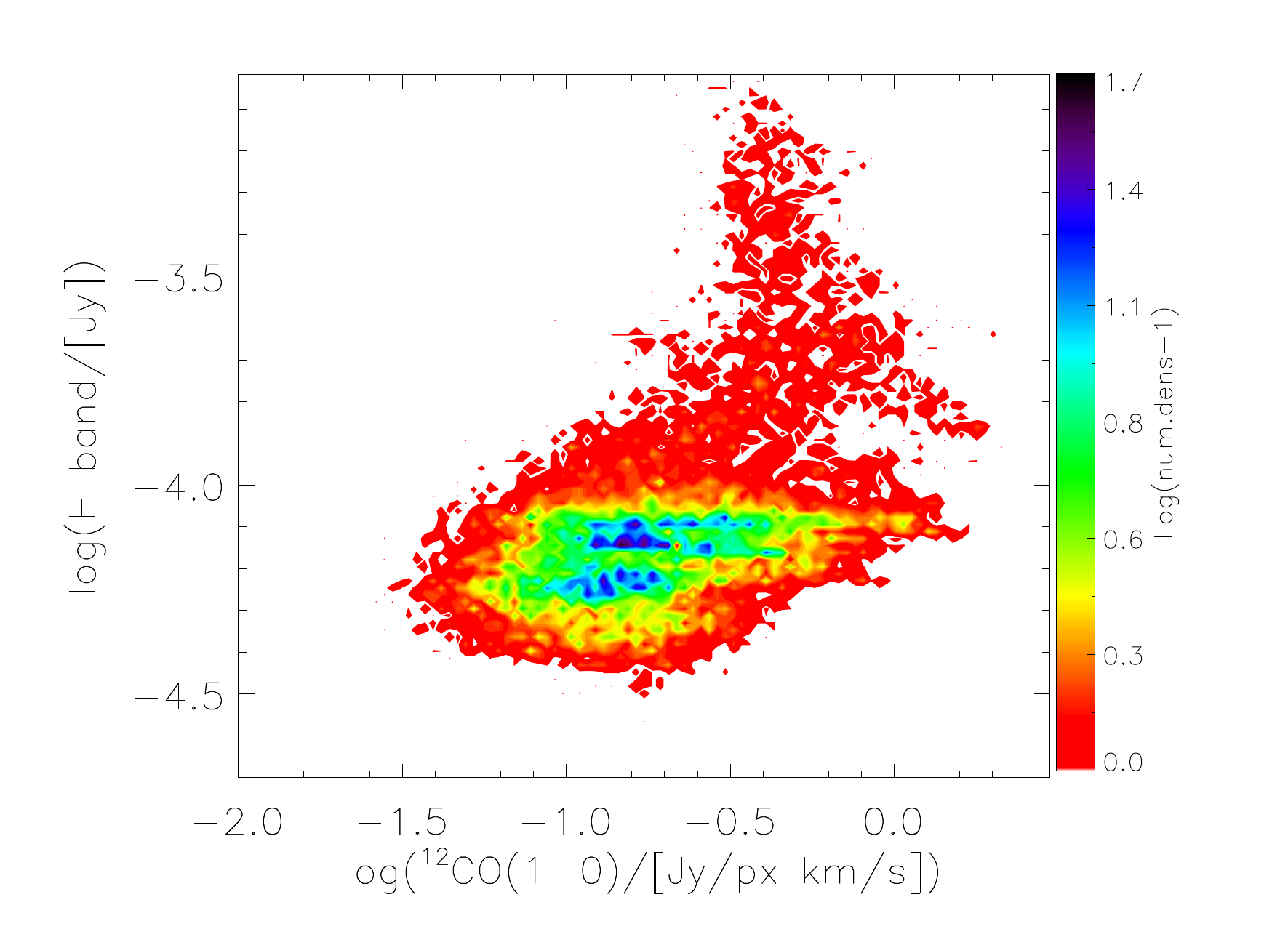}\includegraphics[angle=0]{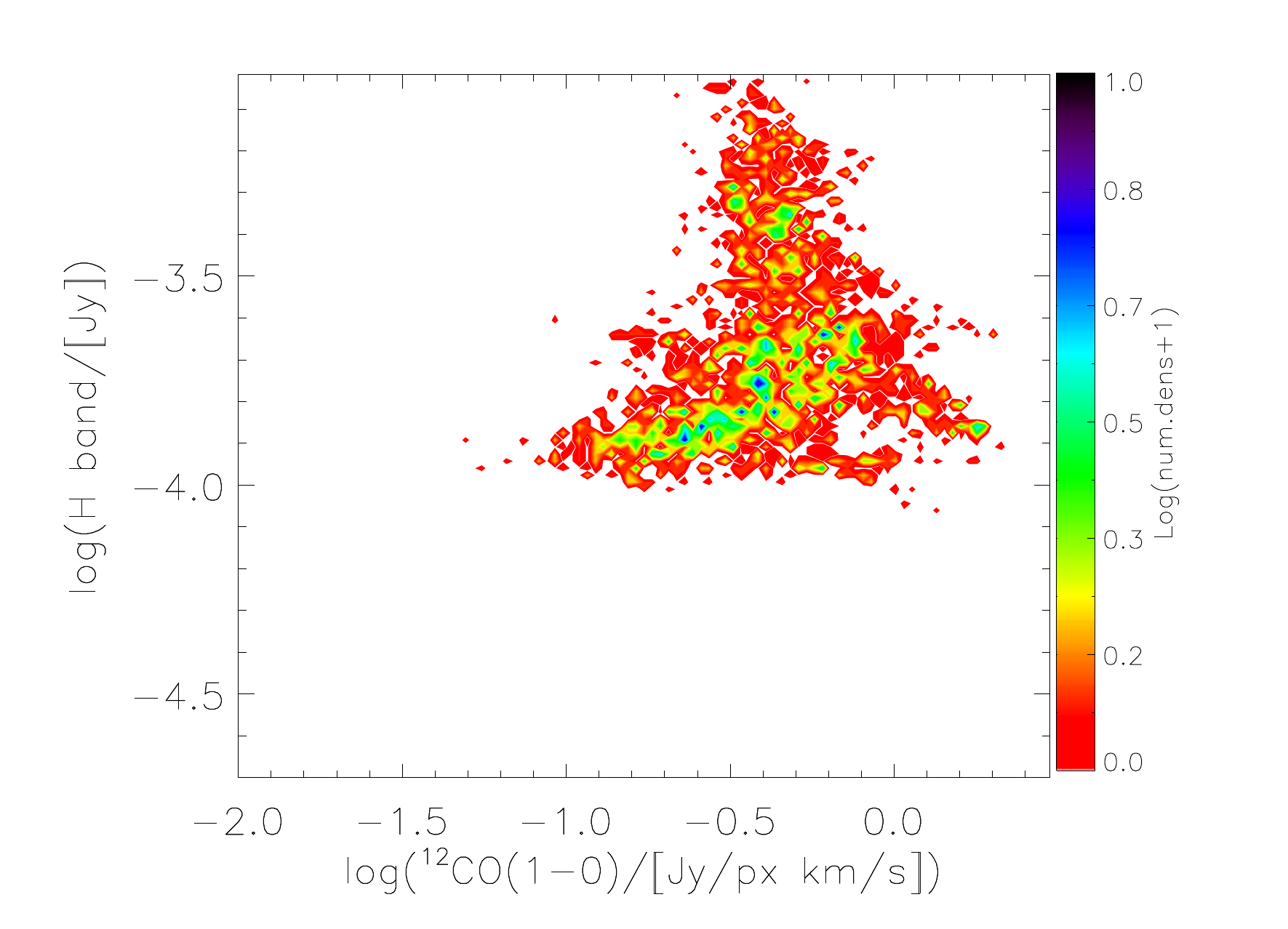}\includegraphics[angle=0]{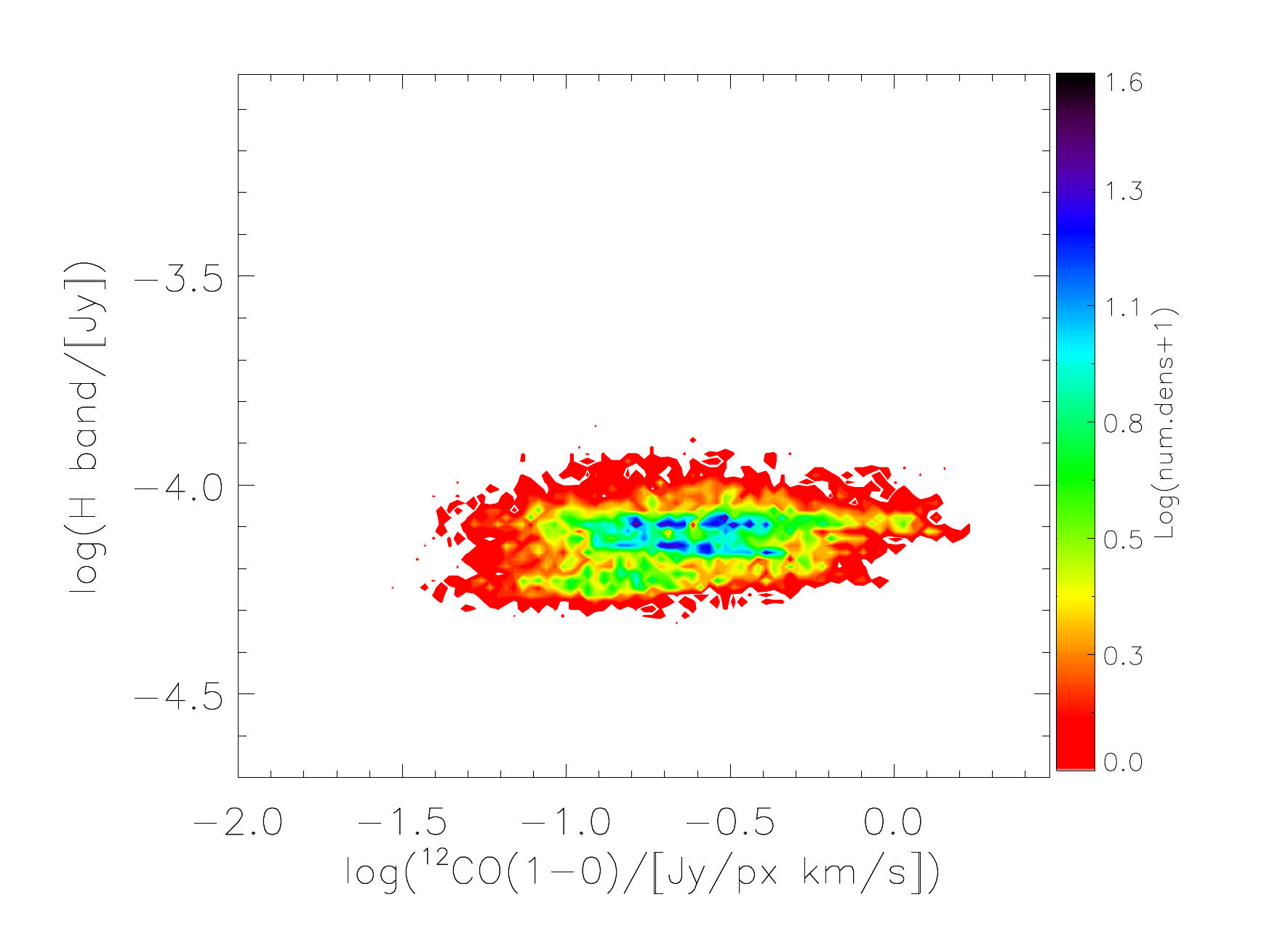}\includegraphics[angle=0]{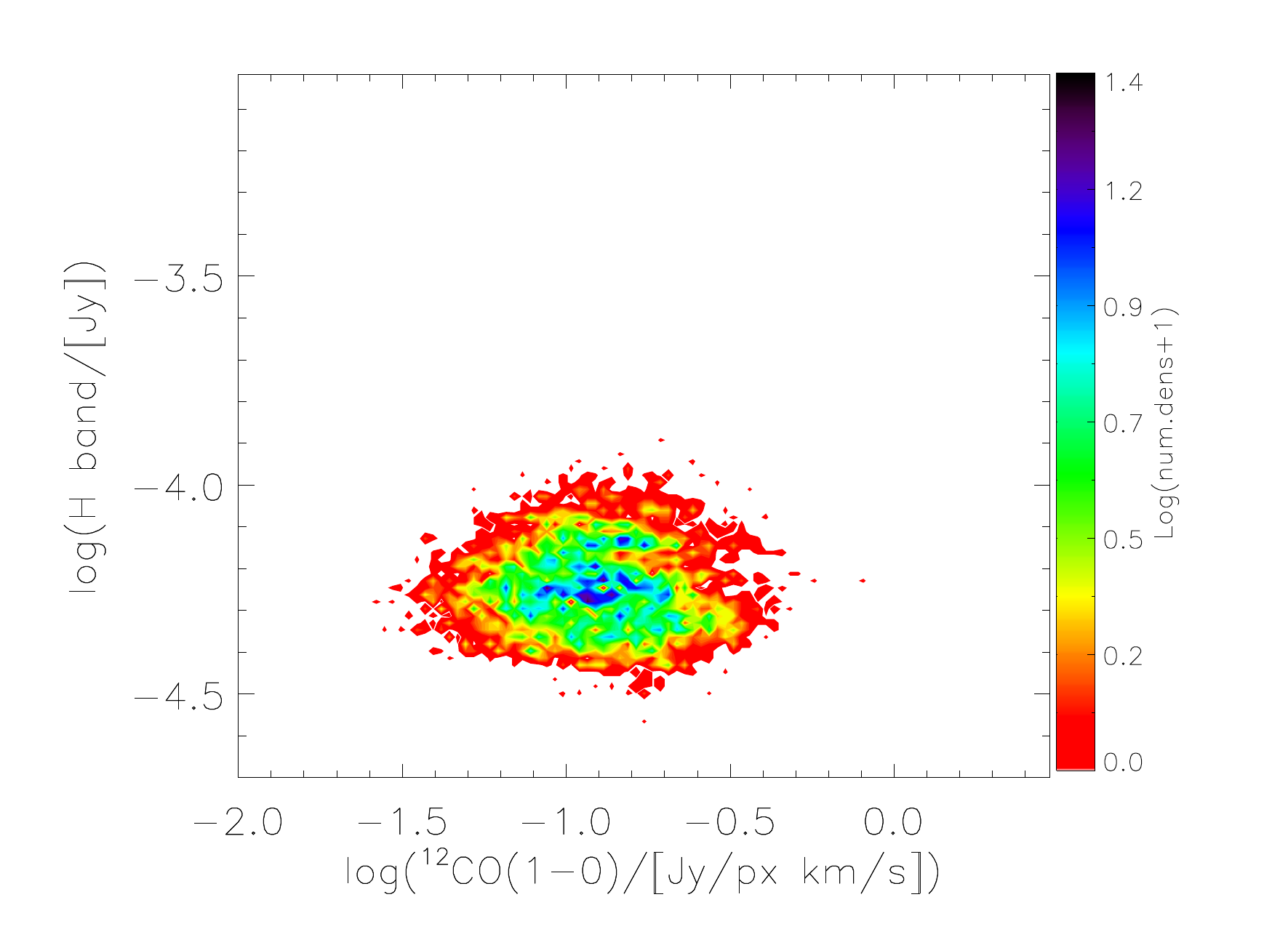}}
\resizebox{0.9\hsize}{!}{\includegraphics[angle=0]{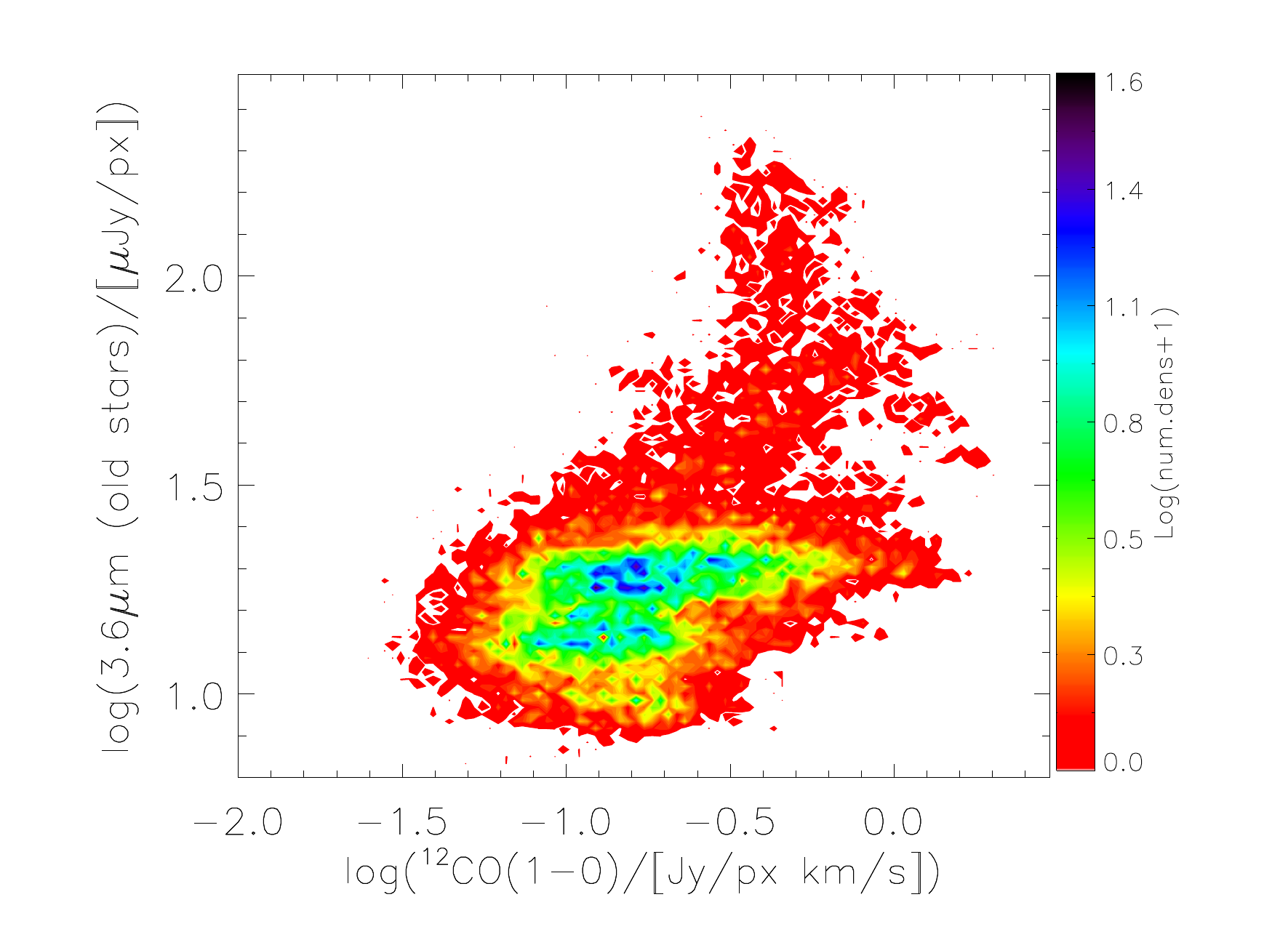}\includegraphics[angle=0]{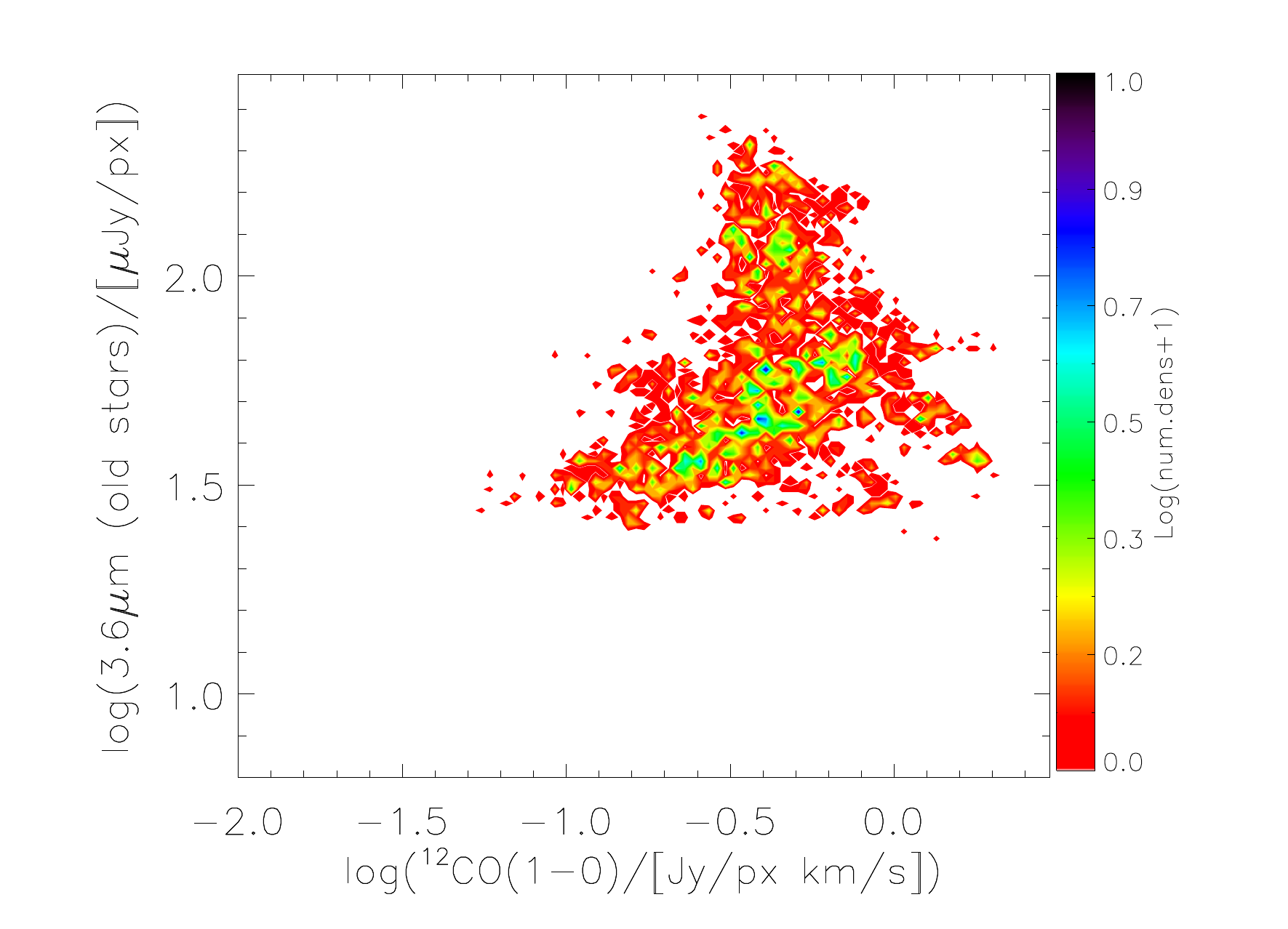}\includegraphics[angle=0]{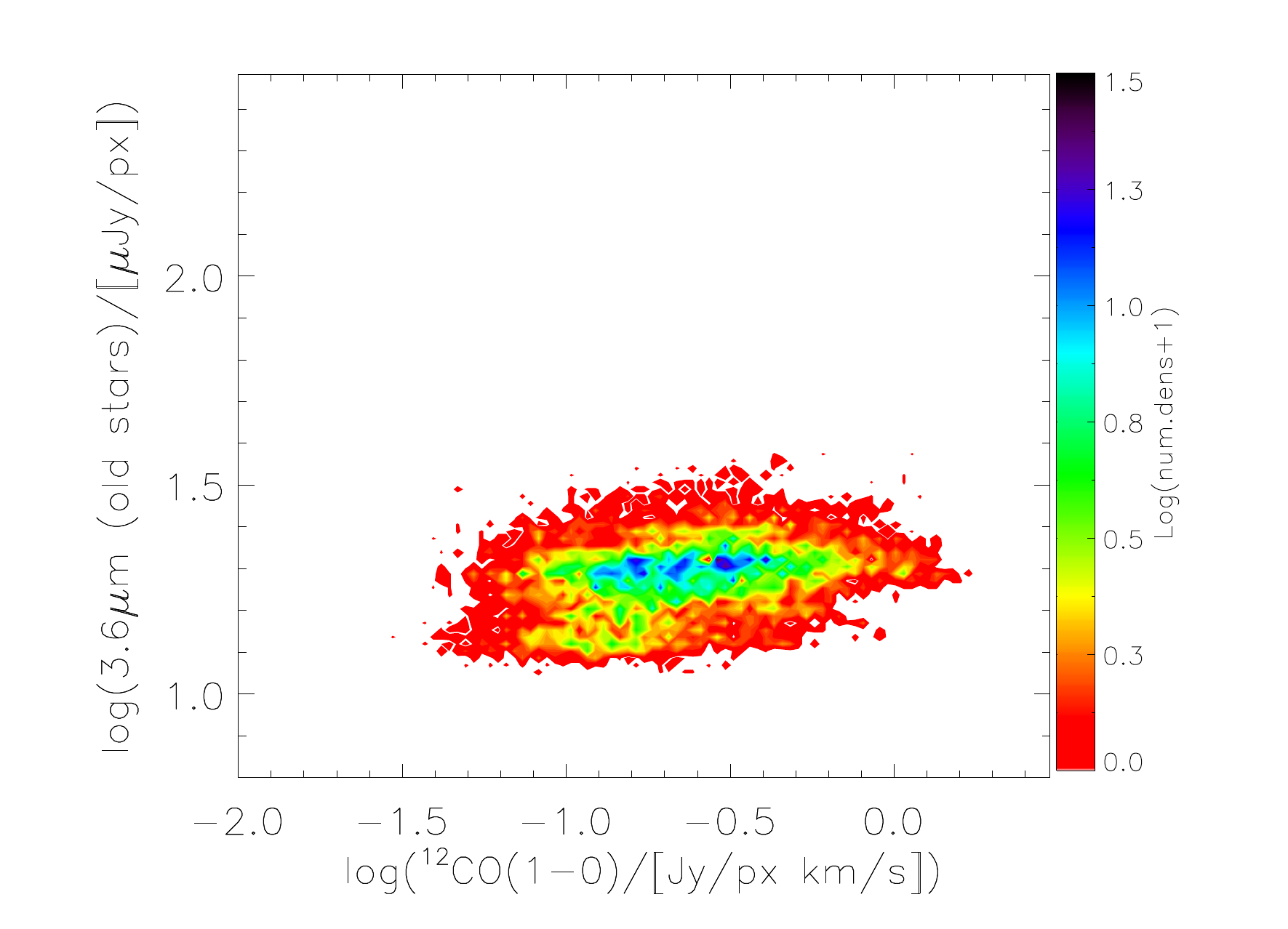}\includegraphics[angle=0]{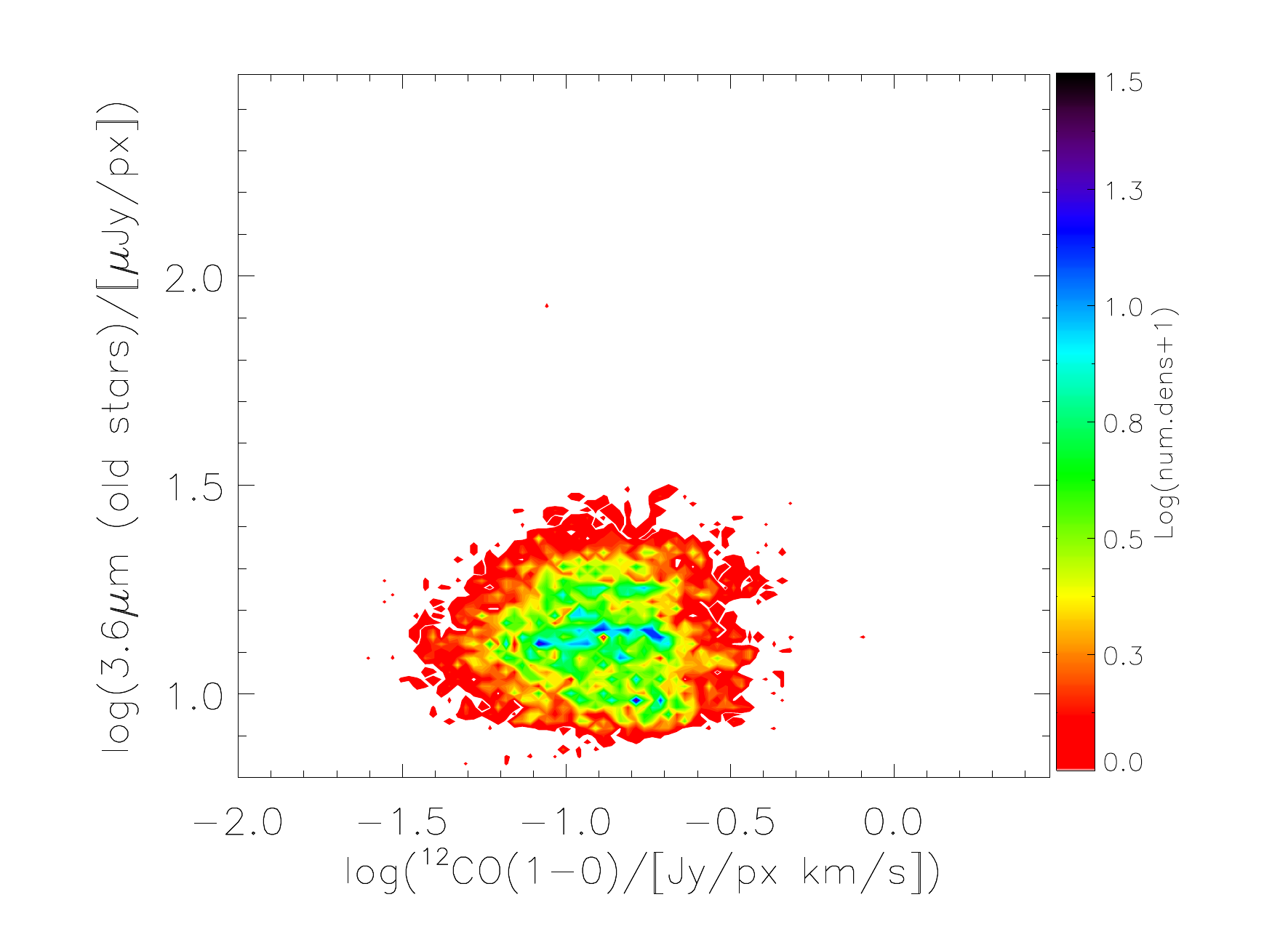}}
\end{center}
\caption{Pixel-by-pixel comparison at 3.0'' resolution of the \coone ~emission versus tracers of stellar light
({\it from top to bottom}):
HST H$\alpha$ emission, 3.6\,cm radio continuum, HST B band, HST H band, and the
stellar emission at 3.6$\mu$m. 
The distribution is shown ({\it from left to right}) 
for the full PAWS FoV, and separately for the central 40'',
the spiral arms and the inter-arm region (as defined in Fig. \ref{fig:co_2d}). The density is given on a logarithmic color scale.
\label{fig:sf_pix}}
\end{figure}

From the second moment we can obtain a measure of the
intrinsic thickness of the arms. For arms with azimuthal intensity
profiles that are roughly gaussian, the convolved profile has a
measured width of $\sigma$=$\sqrt{\sigma_{CO}^2+\sigma_{\lambda}^2}$,
where $\sigma_{CO}$ and $\sigma_{\lambda}$ are the intrinsic widths of
the arms traced by CO emission or at wavelength $\lambda$.
We verified that the width at 95\% maximum of the
CO-brightness auto correlation profile (CO crossed with itself)
corresponds well to the width estimated by eye from the morphology
of CO brightness. We therefore measure the convolved arm width as
$w$$\sim$0.64 $\sigma$, where $\sigma$ is the 2nd moment of the
profile. We adopt this width as the uncertainty on our measurement of
$\phi$ (see Fig. \ref{fig:ism_cross} and \ref{fig:sf_cross}).

%Results are presented in Figures A-C.Typical cc-profiles are
%presented in Figure A.  In Figure B we plot radial profiles of the
%angular offset $\phi$ between the CO emission and the emission at
%variety of wavelengths. The strength of the correlation between CO
%and $\lambda$ is shown in Figure C.
%\bibliography{xxx}

\clearpage

%%%%%%%%%%%%%%%%%%%%%%%%%%%%%%%%
%%% Fig. 1 - CO  - 2D view

\clearpage

%%%%%%%%%%%%%%%%%%%%%%%%%%%%%%%%
%%% Fig. 2 - CO vs. ISM tracers - 2D view

\clearpage

%%%%%%%%%%%%%%%%%%%%%%%%%%%%%%%%
%%% Fig 3 - CO vs. PAH tracers - 2D view

\clearpage

%%%%%%%%%%%%%%%%%%%%%%%%%%%%%%%%
%%% Fig. 4 - CO vs. ISM tracers - polar view

\clearpage

%%%%%%%%%%%%%%%%%%%%%%%%%%%%%%%%
%%% Fig. 5 - CO vs. ISM tracers - polar cross-correlation

\clearpage

%%%%%%%%%%%%%%%%%%%%%%%%%%%%%%%%
%%% Fig. 6 - CO vs. ISM  tracers - pixel-by-pixel view in log

\clearpage

%%%%%%%%%%%%%%%%%%%%%%%%%%%%%%%%
%%% Fig. 7 - CO vs. PAH tracers - pixel-by-pixel view in log

\clearpage

%%%%%%%%%%%%%%%%%%%%%%%%%%%%%%%%
%%% Fig. 8 - CO vs. SFR tracers - 2D view

\clearpage

%%%%%%%%%%%%%%%%%%%%%%%%%%%%%%%%
%%% Fig. 9 - CO vs. gravitational potential - 2D view

\clearpage

%%%%%%%%%%%%%%%%%%%%%%%%%%%%%%%%
%%% Fig. 10 - CO vs. SFR tracers - polar view

%%%%%%%%%%%%%%%%%%%%%%%%%%%%%%%%
%%% Fig. 11 - CO vs. ISM tracers - polar cross-correlation

\clearpage

%%%%%%%%%%%%%%%%%%%%%%%%%%%%%%%%
%%%% Fig. 12  - CO vs. SFR tracers - pixel-by-pixel in log

\clearpage

%%%%%%%%%%%%%%%%%%%%%%%%%%%%%%%%
%%%% Fig. 13  - CO & ISM tracers - cross-correlation profiles

\begin{figure}
\begin{center}
\includegraphics[width=80mm]{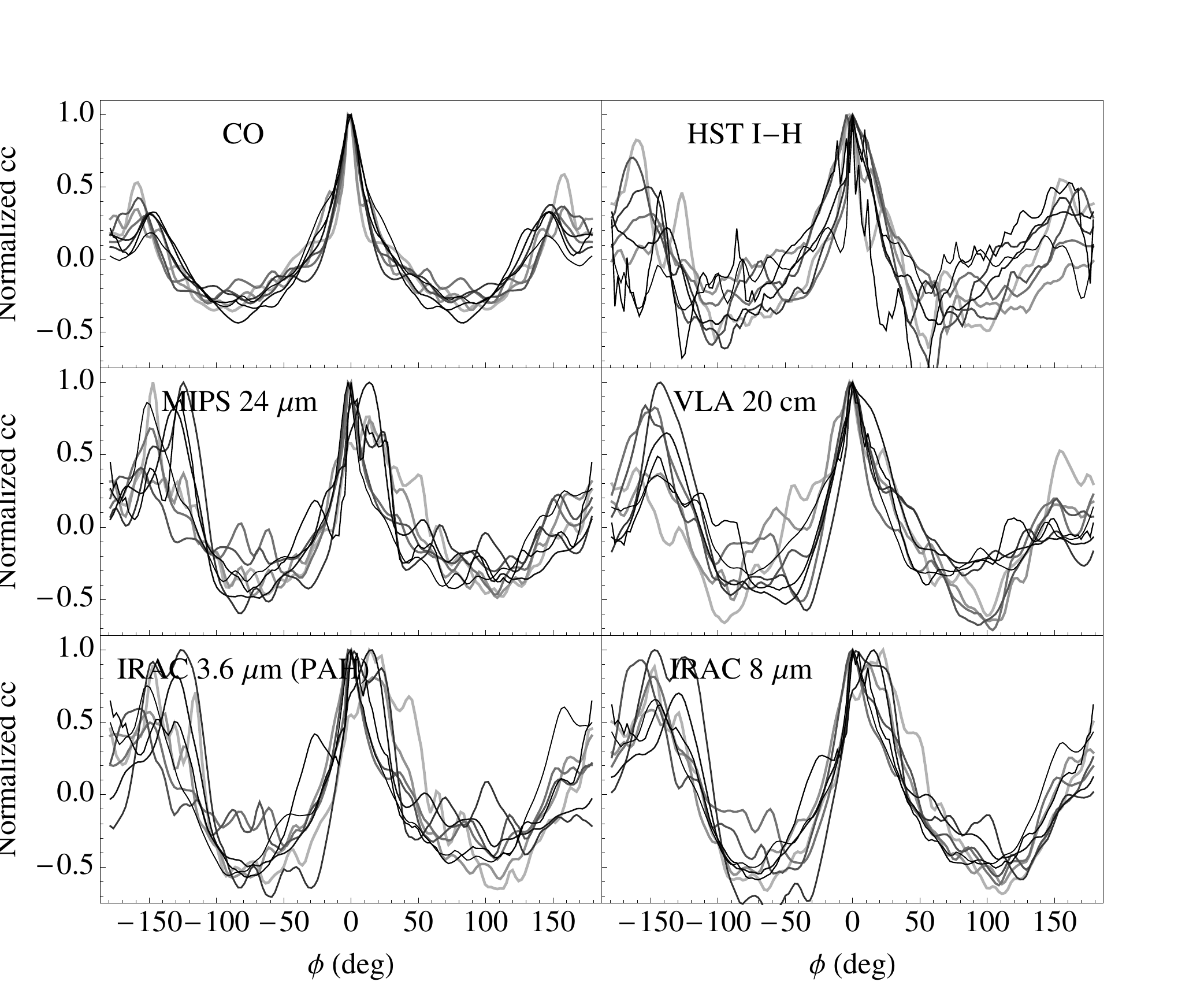}
\end{center}
\caption{Azimuthal profiles of the polar cross-correlation between \coone ~intensity and other ISM tracers drawn in the radial range of 
$\rm 30\as\,<\,r\,<80\as$ in steps of 4.5$\as$:
\coone ~intensity ({\it top left}), HST $I-H$ color ({\it top right}), 
MIPS 24\,$\mu$m image ({\it middle left}), VLA 20\,cm continuum ({\it middle right}),
non-stellar continuum at IRAC 3.6\,$\mu$m and 8\,$\mu$m ({\it bottom}). 
The radial position of each profile is indicated by gray-scale and line thickness, from small (black and thin line) to large 
(light gray and thick line) galacto-centric radius. Each profile is normalized to the maximum difference between the 
measured cross-correlation (cc) profile and the average ($\rm <cc>$) signal at that radius. 
\label{fig:ism_cc}}
\end{figure}

\clearpage

%%%%%%%%%%%%%%%%%%%%%%%%%%%%%%%%
%%%% Fig. 14  - CO & stellar tracers - cross-correlation profiles

\begin{figure}
\begin{center}
\includegraphics[width=180mm]{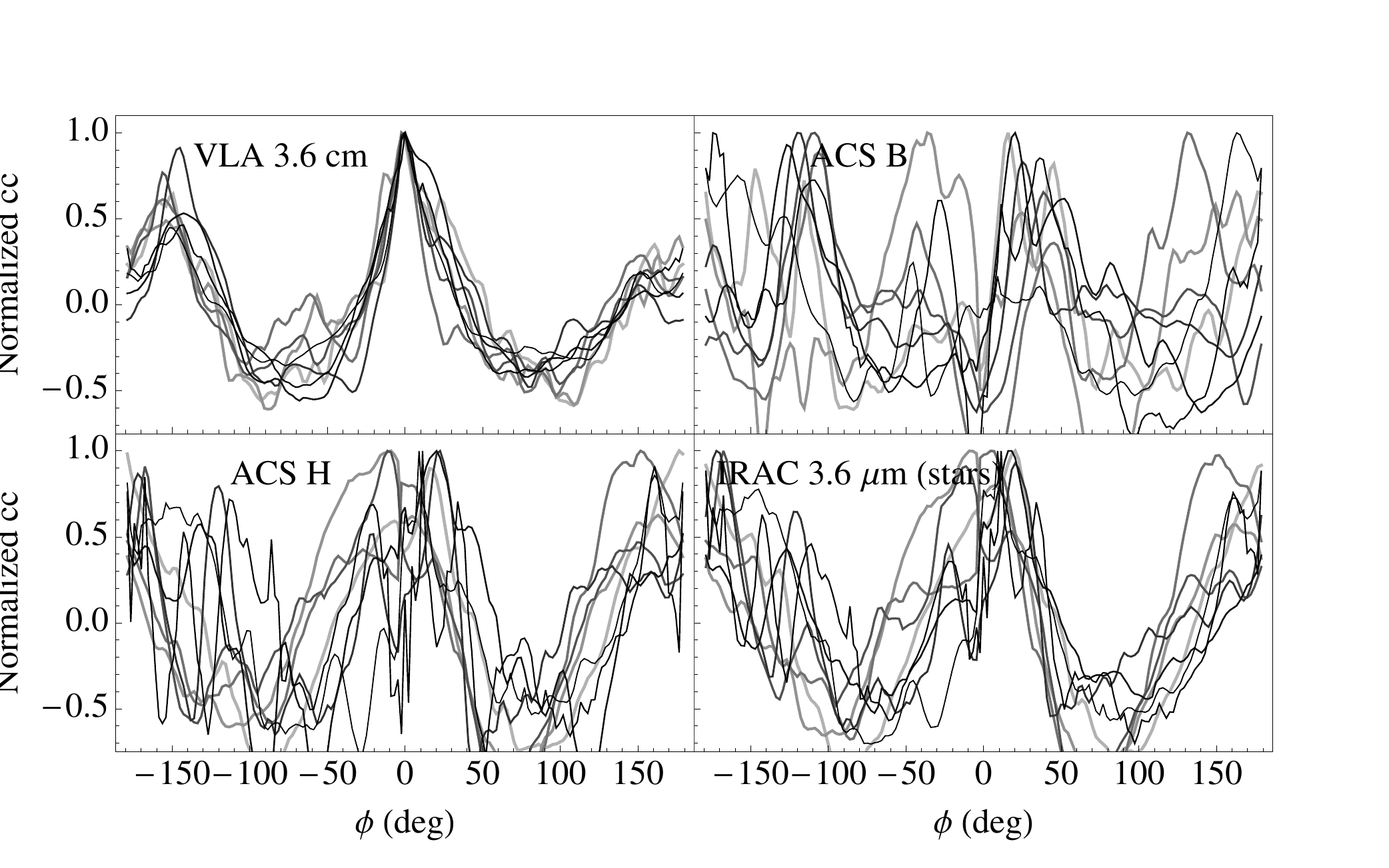}
\end{center}
\caption{Azimuthal profiles of the polar cross-correlation between \coone ~intensity and SFR and stellar population 
tracers drawn in the radial range of $\rm 30\as\,<\,r\,<80\as$ in steps of 4.5$\as$:
VLA 3.6\,cm continuum ({\it top left}), HST B band ({\it top right}), HST H band ({\it bottom left}),
 and the contaminant-corrected IRAC 3.6\,$\mu$m emission ({\it bottom right}). 
The radial position of each profile is indicated by gray-scale and line thickness, from small (black and thin line) to large 
(light gray and thick line) galactocentric radius. Each profile is normalized to the maximum difference between the 
measured cross-correlation (cc) profile and the average ($\rm <cc>$) signal at that radius. \label{fig:sf_cc}}
\end{figure}

\clearpage

%%%%%%%%%%%%%%%%%%%%%%%%%%%%%%%%
%%%%%%%% Tab. 1 - basic M51 parameters

\begin{deluxetable}{lcl}
\tabletypesize{\normalsize}
\tablecaption{Basic Parameters for M\,51 (NGC\,5194)}
\tablehead{
\colhead{Parameter} &
\colhead{Value} &
\colhead{Comments} 
}
\startdata
Hubble type & SA(s)bc pec & NED \\
Center & 13:29:52.7087  +47:11:42.789 & (J2000), (1)\\
$\rm v_{sys} (LSR)$ & 471.7$\pm$0.3\,km/s& radio convention, (2)\\
Distance & 7.6\,Mpc &  1$\as$=36.8\,pc, (3)\\
Inclination & 22$\pm$5 & (4) \\
Position angle & 172$\pm$3 & (4)\\
Stellar mass $\rm M_{\star}$ &  $\rm 3.6\times10^{10}\,\msun$ & (5) \\
Atomic gas mass $\rm M_{HI}$ & $\rm 2.8\times10^9\,\msun$ & (5)\\
Molecular gas mass $\rm M_{mol}$ & $\rm 6.2\times10^9\,\msun$ & (2)\\
Cold gas mass $\rm M_{gas}$ & $\rm 9.0\times10^9\,\msun$ & \\
$f_{gas} = M_{gas}/(M_{gas}+M_{\star})$ & 0.2 &\\
\enddata
% Stellar mass $\rm M_{\star}$ & $\rm 3.6\times10^{10}\.\msun$ & (5) \\
%% Text for table notes should follow after the \enddata but before
%% the \end{deluxetable}. Make sure there is at least one \tablenotemark
%% in the table for each \tablenotetext.
\tablecomments{Basic parameters adopted for M\,51 (NGC\,5194) and used in all PAWS related publications.
References:
(1) \cite{hagiwara07},
(2) \cite{shetty07},
(3) \cite{ciardullo02},
(4) Colombo et al. (in prep.),
(5) \cite{leroy08}
\label{tab:m51}}
\end{deluxetable}

\clearpage

%%%%%%%%%%%%%%%%%%%%%%%%%%%%%%%%
%%%%%%%%  Tab. 2 - multi-lambda data 

\begin{deluxetable}{llrrll}
\tabletypesize{\footnotesize}
\tablecaption{Multi-$\lambda$ data of M\,51 used}
\tablehead{
\colhead{$\lambda$/filter/line} &
\colhead{Instrument} &
\colhead{Resolution} &
\colhead{Pixel scale} &
\colhead{Source} &
\colhead{Comments} 
\\
& &
\colhead{$\as \times \as$} &
\colhead{$\as/px$} & &
}
\startdata
CO(1-0) & PdBI+30m & 1.16$\times$0.97 & 0.3 & PAWS & $\rm PA \sim 73\dg$, (1 \\
J band & 2MASS & 2.5 & 1 & 2MASS & used as astrometric reference \\
B band & HST ACS & 0.1 & 0.05 & HST Heritage & offset of 0.1$\as$ \& -0.4$\as$, (2)\\
V band & HST ACS & 0.1 & 0.05 & HST Heritage & offset of 0.1$\as$ \& -0.4$\as$, (2)\\
I band & HST ACS & 0.1 & 0.05 & HST Heritage & offset of 0.1$\as$ \& -0.4$\as$, (2)\\
H$\alpha$ & HST ACS & 0.1 & 0.05 & HST Heritage & offset of 0.1$\as$ \& -0.4$\as$, (2)\\
H band & HST NICMOS & 0.2 & 0.05 & M. Regan & offset 0.1$\as$ \& -0.4$\as$\\
i band & SDSS & 1.4 & 0.4 & DRS7 & montage of 6 tiles (see \S \ref{subsec:astrom})\\
H$\alpha$ & KPNO & $\sim$1.6 & 0.305 & SINGS & SINGS, (3)\\
3.6$\mu$m & SST IRAC & 1.90\tablenotemark{a} & 0.75 & S$^4$G &  (4)\\
3.6$\mu$m & SST IRAC & 1.90\tablenotemark{a} & 0.75 & ICA &  (5)\\
4.5$\mu$m & SST IRAC & 1.81\tablenotemark{a} & 0.75 & S$^4$G &  (4)\\
5.8$\mu$m & SST IRAC & 2.11\tablenotemark{a} & 0.75 & SINGS &  DR5, (3)\\
8.0$\mu$m & SST IRAC & 2.82\tablenotemark{a} & 0.75 & SINGS &  DR5, (3)\\
24$\mu$m  & SST MIPS & 6.43\tablenotemark{a}  & 1.5 & SINGS & DR5, offset of 0.25$\as$ \& 0.5$\as$, (3)\\
24$\mu$m  & SST MIPS &  $\sim$2.4 & 0.4 & HiRes  & offset of 0.25$\as$ \& 0.5$\as$, (6) \\
70$\mu$m & SST MIPS & 18.74\tablenotemark{a}  & 4.5 & SINGS & DR5, Decl. offset of 0.375$\as$, (3) \\
160$\mu$m  & SST MIPS & 38.78\tablenotemark{a}  & 9.0 & SINGS & DR5, Decl. offset of 0.375$\as$, (3) \\
70$\mu$m & HSO PACS & 5.67 & 1.0 & HSO archive & offset of -2.5$\as$ \& -2.0$\as$, (7)\\
160$\mu$m & HSO PACS & 11.18\tablenotemark{a} &1.0 & HSO archive & offset of -2.5$\as$ \& -2.0$\as$, (7)\\
CII & HSO PACS & $\sim$11.5 & 1.762 & HSO archive & offset of 2.64$\as$ and -1.76$\as$, (8)\\
FUV & GALEX & 4.48\tablenotemark{a}  & 1.5 & archive & offset of 0.225$\as$ \& 0.525$\as$, (9) \\
NUV & GALEX & 5.05\tablenotemark{a}  & 1.5 & archive & offset of 0.225$\as$ \& 0.525$\as$, (9) \\
3.6cm & VLA+Eff & 2.4 & 0.4 & & (6)\\
6cm & VLA+Eff & 2.0 & 0.4 & & (6) \\
20cm & VLA+Eff & 1.4 & 0.4 & & (6)\\
HI & VLA & 5.78$\times$5.56 & 1.5 & THINGS & $\rm PA -68.0\dg$, (10)\\
\enddata
\tablenotetext{a}{\cite{aniano11}}
\tablecomments{Overview of the data used here and astrometric modification applied. See text for 
details on the astrometric corrections (\S \ref{subsec:astrom}) and on the source and processing of the data in Section \S \ref{sec:data}. References:
(1) \cite{pety13};
(2) \cite{mutchler05};
(3) \cite{kennicutt03};
(4) \cite{sheth10};
(5) \cite{meidt12};
(6) \cite{dumas11};
(7) \cite{mentuch12};
(8) courtesy of K. Croxall;
(9) courtesy of F. Bigiel;
(10) \cite{walter08}.
\label{tab:data}}
\end{deluxetable}

\end{document}